# *Ab Initio* Calculations of XUV Ground and Excited States for First-Row Transition Metal Oxides

*Isabel M. Klein, Alex Krotz†, Jonathan M. Michelsen, Scott K. Cushing\**

Division of Chemistry and Chemical Engineering, California Institute of Technology, Pasadena, CA 91125, USA.

Transient X-ray spectroscopies have become ubiquitous in studying photoexcited dynamics in solar energy materials due to their sensitivity to carrier occupations and local chemical or structural dynamics. The interpretation of solid-state photoexcited dynamics, however, is complicated by the core-hole perturbation and the resulting many-body dynamics. Here, an *ab initio*, Bethe-Salpeter equation (BSE) approach is developed that can incorporate photoexcited state effects for solid-state materials. The extreme ultraviolet (XUV) absorption spectra for the ground, photoexcited, and thermally expanded states of first row transition metal oxides – $TiO_2$, α-$Cr_2O_3$, β-$MnO_2$, α-$Fe_2O_3$, $Co_3O_4$, NiO, CuO, and ZnO – are calculated to demonstrate the accuracy of this approach. The theory is used to decompose the core-valence excitons into the separate components of the X-ray transition Hamiltonian for each of the transition metal oxides investigated. The decomposition provides a physical intuition about the origins of XUV spectral features as well as how the spectra will change following photoexcitation. The method is easily generalized to other K, L, M, and N edges to provide a general approach for analyzing transient X-ray absorption or reflection data.

## 1. Introduction

Transient extreme ultraviolet (XUV) spectroscopy has been used to measure electron and hole populations, charge-transfer dynamics in multilayer junctions, electron-phonon coupling, and polaron state formation in a variety of materials.[1–9] The relatively low energy (10 – 150 eV) of the XUV transitions makes the core-level transition sensitive to delocalized valence states, which while providing new insights, also adds additional complexity to interpreting measurements. The measurements of transition metal $M_{2,3}$ edges are particularly popular because they provide information about the oxidation state, hybridization, coordination geometry, and spin state of the transition metal.[4,10–15] Theoretical approaches to predicting ground and excited-state XUV transitions, however, are still in development, especially for solids, where complex band structures, core-hole screening, and other many-body effects can obscure the underlying carrier and lattice dynamics.[11,16–20]

Previous theoretical work has used a variety of approaches to model and understand measured XUV spectra. These methods have included density functional theory (DFT), wave function theory, and response and polarization propagator theories.[21] Some of these methods, including static exchange, transition potential DFT (TP-DFT), and the core-valence separation (CVS) approximation, were specifically designed for the simulation of X-ray spectroscopies. However, appropriate integration of both core-excited and valence-excited states has proven difficult.[21–26] Other approaches, including time-dependent DFT (TDDFT) methods, multiconfigurational wave function theory, and coupled cluster methods, are more generalizable and can be used to predict X-ray transitions.[19–21,27–31] Real time TDDFT (RT-TDDFT) has been used to successfully evaluate photoexcited perturbations to X-ray edges for a variety of materials, including α-Fe$_2$O$_3$ and α-Cr$_2$O$_3$.[32–35] The strength of RT-TDDFT methods is that they can determine the spectral response from a single time propagation.[21] That said, RT-TDDFT methods are computationally expensive, which makes computing picosecond and longer timescales photodynamics or back-extracting information from measured dynamics difficult.

Transition metal $M_{2,3}$ edges challenge these methods for several reasons. The $M_{2,3}$ edges of transition metal oxides have strong angular momentum and exchange effects. A commonly used and highly successful approach for modeling the $M_{2,3}$ edge in XUV spectra has been a semi-empirical atomic multiplet theoretical approach, Charge Transfer Multiplet program for X-ray Absorption Spectroscopy (CTM4XAS).[4,12,18,36–39] This method has been successfully applied to transition metal oxide spectra, but the method is not *ab initio* and does not accurately capture the many-body effects that are important in solids. Incorporating excited state effects and dynamics into CTM4XAS calculations also requires assumptions about changes in oxidation state and crystal field parameters to align simulations with experiment.[4,12,18,28,36–38]

In this paper, we explore the validity of a Bethe-Salpeter equation (BSE) approach for predicting the XUV spectrum of various transition metal oxides in both the ground and excited states.[40] Based on the Obtaining Core Excitations from *Ab initio* electronic structure and the NIST BSE solver (OCEAN) framework, which has previously been applied to various K and L edges, we include new capabilities to the code that allow the XUV spectrum to be related to the underlying band structure, as well as an adiabatic approximation to include photoexcited dynamics.[41–47] With the given modifications, the X-ray transition Hamiltonian is deconvoluted into its constituent parts, shedding light on the fundamental origins of the X-ray absorption spectra. The range of calculations performed here can serve as a guide for future X-ray measurements of transition metals, as well as other complex, angular momentum coupled peaks. Moreover, these methods provide a general and facile approach for modeling time-resolved X-ray spectra for synchrotron, table-top XUV, or free electron laser measurements.[41–47]

## 2. Methods

The theoretical approach is based on modifications to the OCEAN code, previously reported in reference 40.[40,48,49] The OCEAN code has also been verified for ground state K, L, M, and N edge calculations, as well as for M and N edges in our own work.[3,43,44,46–48,50,50–53] Here, the OCEAN code is modified to project components of the X-ray transition Hamiltonian onto the band structure, as well as to include photoexcited carrier distributions, band gap renormalization, and lattice effects in the calculations. Given the relatively low computational expense of the underlying OCEAN code, our modified package can be used to back-extract these quantities from measured spectra by predicting multiple possible configurations such as time-dependent electron and hole distributions.

More technically, the initial core-level spectra for the ground and excited states of each transition metal oxide are calculated in three steps. The core-level transition matrix elements are first determined by projecting an all-electron atomic calculation of the core levels onto a density functional theory (DFT+U) calculation of the valence wavefunctions using a projector augmented wave (PAW)-style optimal projector functions (OPFs) with a scissor correction of the band gap.[48,49,54] Errors in the use of DFT for transition metal oxides are mitigated by the inclusion of the Hubbard U term to correct the self-interaction error of the GGA functionals and the scissor correction that accounts for overestimations of band gaps.[55,56] Following DFT+U calculations, screening of the core electrons in the presence of the core-hole is determined self-consistently.[49,54] Atomic multiplet effects are included through core-hole spin-orbit splitting and exchange interactions.[48] Finally, the transition matrix elements between the core wavefunction and the valence wavefunctions are calculated using the specified photon operator. The ground state absorption spectrum is obtained by iteratively solving the BSE using a Haydock recursive algorithm.[49,54] Alternatively, the real-space wavefunction of the core-valence excitons can be calculated using a Generalized Minimal Residual (GMRES) method, which allows the core-valence exciton and constituent components to be projected onto the material's band structure.

Quantitatively, the XUV spectrum relies on a BSE Hamiltonian for the calculation of the transition rate and XUV cross section. This transition rate can be calculated with equation 1,

$$\Gamma_O(\omega, q) = 2\pi \sum_F |\langle I|\hat{O}(\omega, q)|F\rangle|^2 \delta(\omega + E_I - E_F)\delta(q + k_I - k_F) \quad (1)$$

where $\Gamma_O$ is the overall transition rate and $\hat{O}$ is the many-body electron-photon interaction. Summing over all possible excited states, the transition probability is calculated with equation 2,

$$\Gamma_O(\omega, q) = -Im[\langle I|\hat{O}(\omega, q) G_2(\omega, q) \hat{O}^\dagger(\omega, q)|I\rangle] \quad (2)$$

where $G_2$ is the two-particle Green's function. The k-space dependent overall transition probability calculated from equation 2 can be projected onto the band structure, while the energy-dependent transition probability can be plotted on the calculated absorption spectrum. Equation 2 is used to calculate the XUV absorption spectrum for all input wavefunctions. The effective BSE Hamiltonian can be defined,

$$G_2 = [\omega - H_{BSE}]^{-1} \quad (3)$$

$$H_{BSE} = H_e - H_h + V_X - V_D \quad (4)$$

where $H_e$ is the electron Hamiltonian, $H_h$ is the core-hole Hamiltonian, and $V_X$ and $V_D$ are the exchange and direct interaction terms, respectively. These terms are separable due to the linearity of the BSE Hamiltonian, and the magnitudes of each can be projected onto the band structure. The hole Hamiltonian is defined by

$$H_h = E_h - i\Gamma_j + \chi_j \quad (5)$$

where $E_h$ is the hole binding energy, $\chi_j$ is the spin-orbit splitting and $\Gamma_j$ is the broadening due to the different lifetimes of angular momentum split edges. $H_h$ contains information about the angular momentum splitting and coupling because both $\chi_j$ and $\Gamma_j$ depend on the total angular momentum state $j$. The direct interaction is the screened Coulomb scattering between the excited electron and core hole,

$$V_D = \hat{a}_c^\dagger(r,\sigma)\hat{a}_h(r',\sigma')W(r,r',\omega)\hat{a}_c(r,\sigma)\hat{a}_h^\dagger(r',\sigma') \qquad (6)$$

where the screening of this interaction ($W$) is determined by the dielectric response of the system,

$$W(r,r',\omega) = \int d^3r'' \frac{\epsilon^{-1}(r,r'',\omega)}{|r''-r'|} \qquad (7)$$

The direct term, $V_D$, provides insight into the magnitude of the screening contribution, while the exchange interaction,

$$V_X = \hat{a}_c^\dagger(r,\sigma)\hat{a}_h(r',\sigma')\frac{1}{|r-r'|}\hat{a}_c(r',\sigma)\hat{a}_h^\dagger(r,\sigma') \qquad (8)$$

involves core-hole operators at both $r$ and $r'$ and sheds light onto the exchange between wavefunction configurations.

The core-valence exciton wavefunction, $|\psi\rangle$, can be given as an expansion in the basis of Kohn-Sham states, $|n,k\rangle$, where $n$ is the valence band index and $k$ is the wavevector. The projection, $D$, of $|\psi\rangle$ onto the band structure for a given term of the BSE Hamiltonian is given by

$$D = |\langle\psi|n,k\rangle\langle n,k|H_{BSE}|\psi\rangle|^2 \qquad (9)$$

where the projection operator is $|n,k\rangle\langle n,k|$. The projection is then interpolated in reciprocal-space along the k-path and plotted on the band structure.

The projection maps the calculated XUV spectrum onto the ground state band structure. This projection can be calculated for each of the separable terms of the BSE Hamiltonian, allowing for the spin-orbit angular momentum coupling, core-hole screening, exchange effects, and higher-order BSE terms described in equations 3 – 8 to be separated. Excited state effects are included under the adiabatic approximation that the core-hole lifetime is shorter than any of the modeled excited state electron and phonon scattering processes. Since the excited state predictions occur at a minimum of tens of femtoseconds, this assumption is valid, however it would not hold true for few-cycle optical pump experiments or attosecond effects.[48] For the state filling calculations that model photoexcitation on short timescales, the valence and conduction band occupations are modified to reflect a change in carrier occupation at the band edges (Figures S6, S19, S32, S44, S57, S70, S82, and S95). To model thermal isotropic lattice expansion that occurs from acoustic phonons, the unit cells were isotopically expanded between 0.3 and 1.2 % (Table S3), corresponding to a temperature increase from 300 K to 650 K, and the OCEAN calculation was run with the expanded lattice.[57] This temperature range was chosen to simulate the expansion experienced by these materials following thermalization of photoexcited carriers. As reported elsewhere, polaron states can also be included in the underlying DFT calculation to predict their effect.[40] In all cases, the differential transient XUV spectra are calculated by subtracting the calculated ground state spectrum from the relevant excited state spectrum. Full details of ground state and excited state calculations are provided in the SI.

### 3. Results and Discussion

The rest of the paper is divided into the following sub-sections for clarity. Initial calculations are performed to demonstrate the robustness of the OCEAN technique in modeling the $M_{2,3}$ edge XUV absorption spectra of transition metal oxides in Section 3.1, *Ground State Validation.* Next, in Section 3.2 on *Ground State Hamiltonian Discussion*, we decompose the ground state X-ray transition Hamiltonians into their constituent parts and explain how the angular momentum coupling, core-hole screening, and exchange effects underpin the XUV spectra for the transition metal oxides investigated here. The methods are extended in Section 3.3, *Excited State Validation*, to include photoexcited state effect. Finally, a discussion of how both the ground state and excited state Hamiltonian contributions influence the transient spectral features is undertaken in Section 3.4 on *Excited State Trends and Hamiltonian Discussion*. In each discussion, the relationships between physical characteristics of transition metal oxides, Hamiltonian contributions, and XUV spectral features are explained.

## 3.1 Ground State Validation

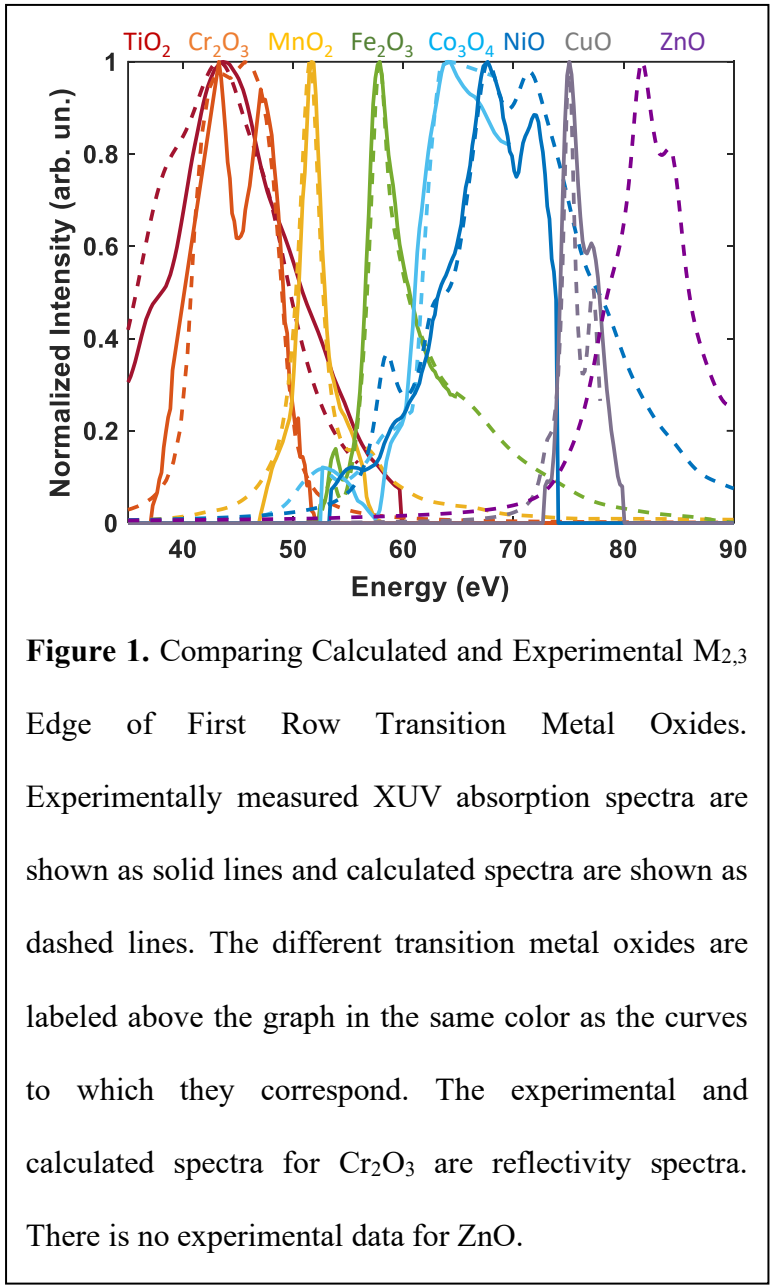

**Figure 1.** Comparing Calculated and Experimental $M_{2,3}$ Edge of First Row Transition Metal Oxides. Experimentally measured XUV absorption spectra are shown as solid lines and calculated spectra are shown as dashed lines. The different transition metal oxides are labeled above the graph in the same color as the curves to which they correspond. The experimental and calculated spectra for $Cr_2O_3$ are reflectivity spectra. There is no experimental data for ZnO.

Comparisons between the measured and calculated ground state $M_{2,3}$ edge absorption spectra are shown in Figure 1 and highlight the accuracy of the BSE approach. Figure 1 shows the calculated ground state $M_{2,3}$ edge absorption spectra (dashed lines) of rutile $TiO_2$, α-$Cr_2O_3$, β-$MnO_2$, α-$Fe_2O_3$, $Co_3O_4$, NiO, CuO, and ZnO compared to experimentally measured spectra (solid lines).[1,4,10,18,58–60] $VO_2$ is excluded as the near room temperature phase change dynamics under thermal and optical excitation are outside the scope of this study.[61] Only an experimental reflectivity spectrum was available for α-$Cr_2O_3$.[4] No experimental data was available for comparison to ZnO. The OCEAN calculated spectra are broadened with an energy-dependent Gaussian with a high energy Fano correction for direct comparison with previous reports.[3,47] The broadening method adequately accounts for the different lifetimes of the angular momentum and atomic multiplet split peaks, as well as the Fano-type line shape of the $M_{2,3}$ edge that arises from the many-body renormalization due to the half-filled d orbitals in transition metal oxides.[62] The broadening scheme is fully discussed in the SI, along with the unbroadened calculated spectra (Figures S3, S15, S28, S41, S53, S66, S79, and S91).

A first approximation of an XUV absorption spectrum is the dipole-allowed transitions to the unoccupied 3d density of states. However, this approximation does not consider the core-hole perturbation of the final transition state. An XUV absorption spectrum (eq. 2) comes from a combination of spin-orbit angular momentum coupling (eq. 5), core-hole screening (eq. 6), and exchange effects on the core-hole excited state (eq. 8) that all act to perturb the final transition DOS.[16,20] The angular momentum contribution refers to the spin-orbit coupling of the core-hole and the valence state, exchange effects arise from the exchange interaction between the core and valence wavefunctions in the final core-hole excited state, and the screening term describes the ability of the valence electrons to screen the resultant core-valence exciton.[63] Combined, these components are what influence the formation of the core-valence excitons in the BSE calculation and thus the calculated spectra. Understanding these sub-terms is therefore critical to understanding the origins of the XUV spectra.

### 3.2 Ground State Hamiltonian Discussion

The OCEAN code was modified to project the magnitude of the linearly separable BSE Hamiltonian components onto the band structure (eq. 9). The full projections are shown in Figures S8, S21, S34, S46, S59, S72, S84, and S97, however it is more instructive to look at the relative magnitude of each component, summed across k-space. The fractional contribution of each Hamiltonian component is plotted in Figure 2. Moving across the 3d row, from $TiO_2$ to $Co_3O_4$, the relative contribution of the angular momentum coupling to the overall BSE Hamiltonian increases while the importance of the screening and exchange components decrease. Then, from $Co_3O_4$ to ZnO, these trends invert and the importance of the screening and exchange components increase while the contribution of the angular momentum coupling term decreases.

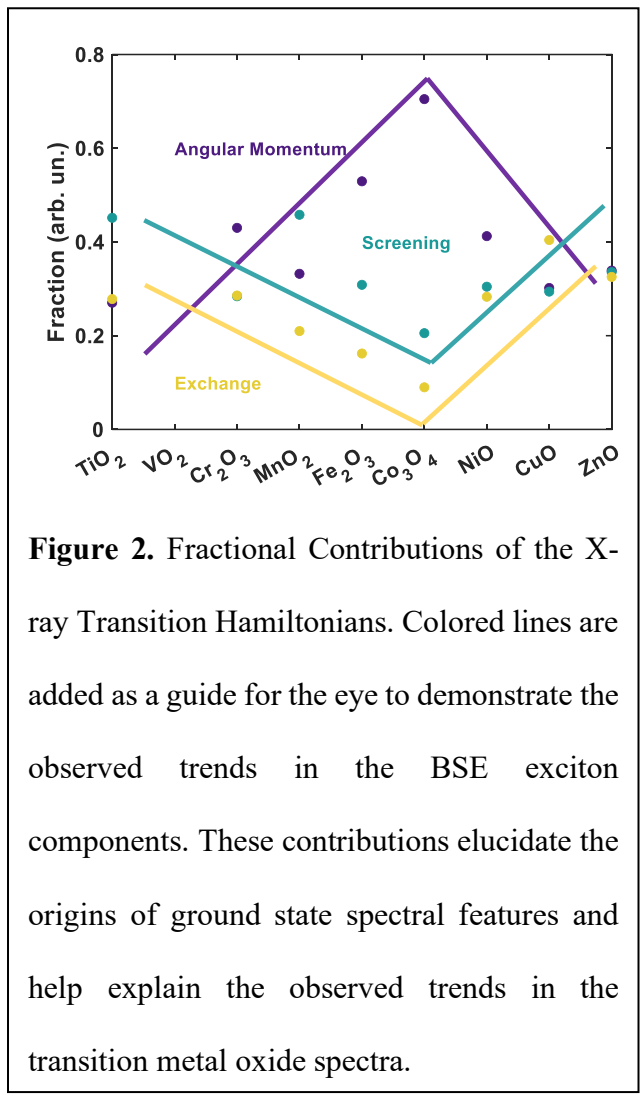

**Figure 2.** Fractional Contributions of the X-ray Transition Hamiltonians. Colored lines are added as a guide for the eye to demonstrate the observed trends in the BSE exciton components. These contributions elucidate the origins of ground state spectral features and help explain the observed trends in the transition metal oxide spectra.

The qualitative trends in Figure 2 follow the filling of the d-orbitals, flipping at the half-filled point. The trend in the angular momentum can be understood in terms of unpaired valence electrons. From $TiO_2$ to $Co_3O_4$, the number of unpaired electrons in the 3d band increases as the formal d count increases from $d^0$ to $d^6/d^7$ (Figure S102b). The increase in the number of unpaired electrons increases the angular momentum coupling between the core-hole and valence state. From NiO to ZnO, the formal d count continues to increase from $d^8$ to $d^{10}$, leaving fewer unpaired electrons in the 3d band, leading to decreased angular momentum coupling.

Angular momentum coupling is usually the dominant Hamiltonian contribution to the XUV spectra as seen in Figure 2.

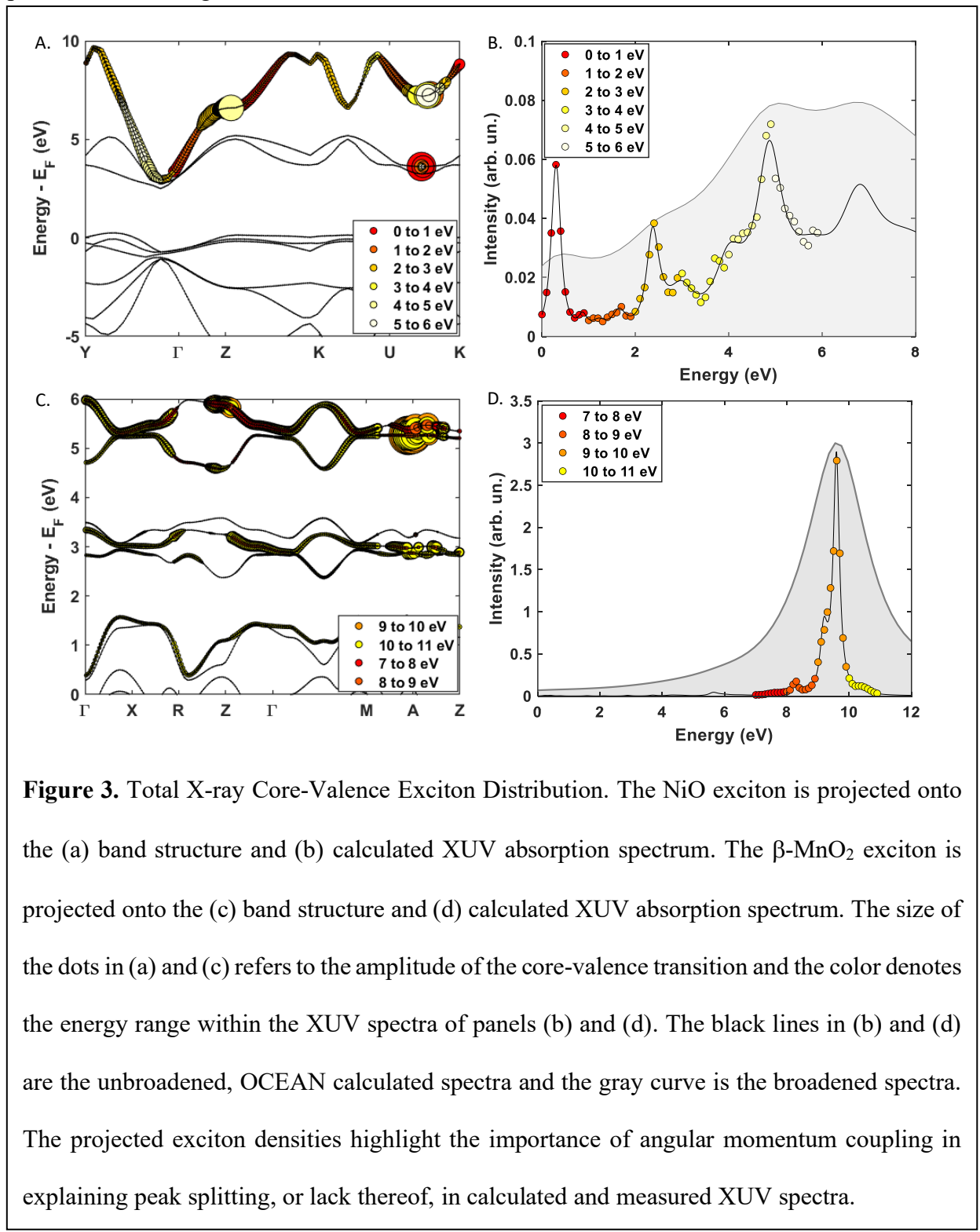

**Figure 3.** Total X-ray Core-Valence Exciton Distribution. The NiO exciton is projected onto the (a) band structure and (b) calculated XUV absorption spectrum. The β-$MnO_2$ exciton is projected onto the (c) band structure and (d) calculated XUV absorption spectrum. The size of the dots in (a) and (c) refers to the amplitude of the core-valence transition and the color denotes the energy range within the XUV spectra of panels (b) and (d). The black lines in (b) and (d) are the unbroadened, OCEAN calculated spectra and the gray curve is the broadened spectra. The projected exciton densities highlight the importance of angular momentum coupling in explaining peak splitting, or lack thereof, in calculated and measured XUV spectra.

To demonstrate the impact of angular momentum coupling, Figure 3 shows the energy-decomposed core-valence exciton distribution for NiO as projected onto the NiO band structure and calculated spectrum. Comparing the band structure in Figure 3a to the XUV absorption in Figure 3b shows that angular momentum coupling redistributes core-level transitions across a large energy range, even though the transitions are primarily from a single conduction band. While the exciton density that corresponds to a lower energy peak in the spectrum may correspond to a lower energy band, this linearity is inconsistent. The exciton density is nonlinearly distributed throughout the band structure, as can be seen by comparing the color maps. Underlying peaks (black trace, Figure 3b) are often hidden by core-hole and experimental broadening, which lead to very broad spectra (grey curve, Figure 3b) that may otherwise be misinterpreted as simply the dipole-allowed density of states. The findings for NiO can be contrasted with those for CuO (Figures S79 – 81), where the peaks in the XUV spectrum arise from transitions to different bands in the conduction band and do more closely mirror the dipole-allowed density of states. Figure 3 highlights the difficulty of interpreting XUV spectra without theoretical support.[14]

The trends in screening can also be understood in terms of the d-electron filling, although from a different perspective than the angular momentum coupling. As the d-states are filled, transition metal oxides in the middle of the periodic table have more localized 3d electrons and lower carrier mobilities, as compared to the unfilled or completely filled $TiO_2$ and ZnO that have more delocalized valence electrons and higher carrier mobilities.[64–66] Delocalized carriers better screen the core-valence exciton, reducing angular momentum splitting within the XUV spectra. In addition, from $\alpha$-$Fe_2O_3$ to $Co_3O_4$ to NiO, the M-O bond covalency increases, leading to increased hybridization, and ultimately increased screening.[12] This phenomenon is seen in the spectra of β-$MnO_2$ (Figure 3d). Based on its d-count, β-$MnO_2$ should have strong angular momentum splitting, like α-Fe$_2$O$_3$, but strong screening is observed through the decomposition of the X-ray transition Hamiltonian (Figure S34b). The effects of screening are apparent when the energy-decomposed core-valence exciton distribution for β-MnO$_2$ is projected onto the band structure (Figure 3c) and absorption spectrum (Figure 3d), and clearly result in a single peak, even before experimental broadening is considered.

The exchange interaction follows the trend in screening, not angular momentum, as it depends on the number of possible configurations created by exchange of the core-hole and valence-electron.[67] As the d-count increases from TiO$_2$ to α-Fe$_2$O$_3$, the number of possible configurations also increases. From α-Fe$_2$O$_3$ to ZnO, as the d-count continues to increase, additional valence electrons block potential configurations, decreasing the exchange effects. In contrast to screening, exchange effects act to redistribute the core-valence exciton transitions across the band structure. Even though a material like TiO$_2$ or ZnO may have relatively minimal angular momentum coupling, there is still a redistribution of the underlying d-orbital conduction bands in the XUV spectra due to exchange effects (see Figures S3 and S91).

### 3.3 Excited State Validation

The ground state Hamiltonian contributions form an intuition about how photoexcitation perturbs the XUV spectra. The accuracy of the excited state approximation is first verified in Figure 4 by comparing the initial photoexcited differential spectra measured experimentally (yellow) with simulated state filling (purple). For these materials, the photoexcited state is often modelled as a ligand-to-metal charge transfer (LMCT) state, where the electron is localized at the metal center. More accurate and generalizable, we take the approach of changing the occupation of the band structure to replicate photoexcited carrier distributions.[9,14,40] The experimental and theoretical transient XUV spectra immediately after photoexcitation (Figure 4) agree with good accuracy for $TiO_2$, $\alpha\text{-}Fe_2O_3$, $Co_3O_4$, and NiO using this approach. The experimental (grey) picosecond timescale dynamics are also shown. The acoustic phonons created during optical phonon decay and electron-phonon scattering on the 10's to 100's of picoseconds timescale lead to a thermally expanded lattice, the calculated differential for which is shown in the green line. In these experiments, as compared to semiconductors like Si, Ge, and ZnTe, polarons are dominant on the 1-10 picosecond timescale, so the thermally expanded spectra do not match the long timescales experimental data, but these spectra are still given for reference.[68] The predictions for all the studied materials are given in Figure 5, along with the calculated ground state spectra for reference.

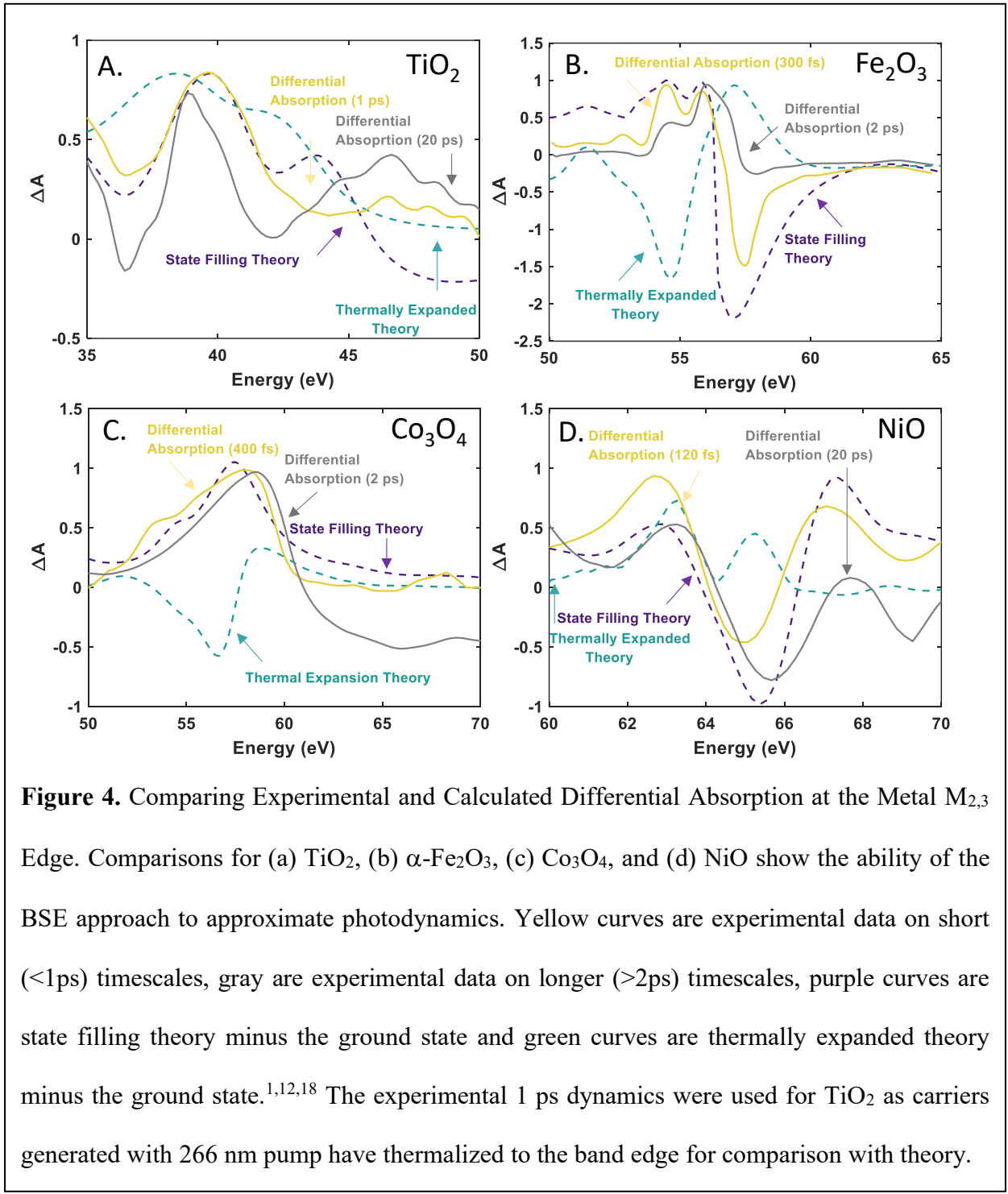

**Figure 4.** Comparing Experimental and Calculated Differential Absorption at the Metal $M_{2,3}$ Edge. Comparisons for (a) $TiO_2$, (b) $\alpha$-$Fe_2O_3$, (c) $Co_3O_4$, and (d) NiO show the ability of the BSE approach to approximate photodynamics. Yellow curves are experimental data on short (<1ps) timescales, gray are experimental data on longer (>2ps) timescales, purple curves are state filling theory minus the ground state and green curves are thermally expanded theory minus the ground state.[1,12,18] The experimental 1 ps dynamics were used for $TiO_2$ as carriers generated with 266 nm pump have thermalized to the band edge for comparison with theory.

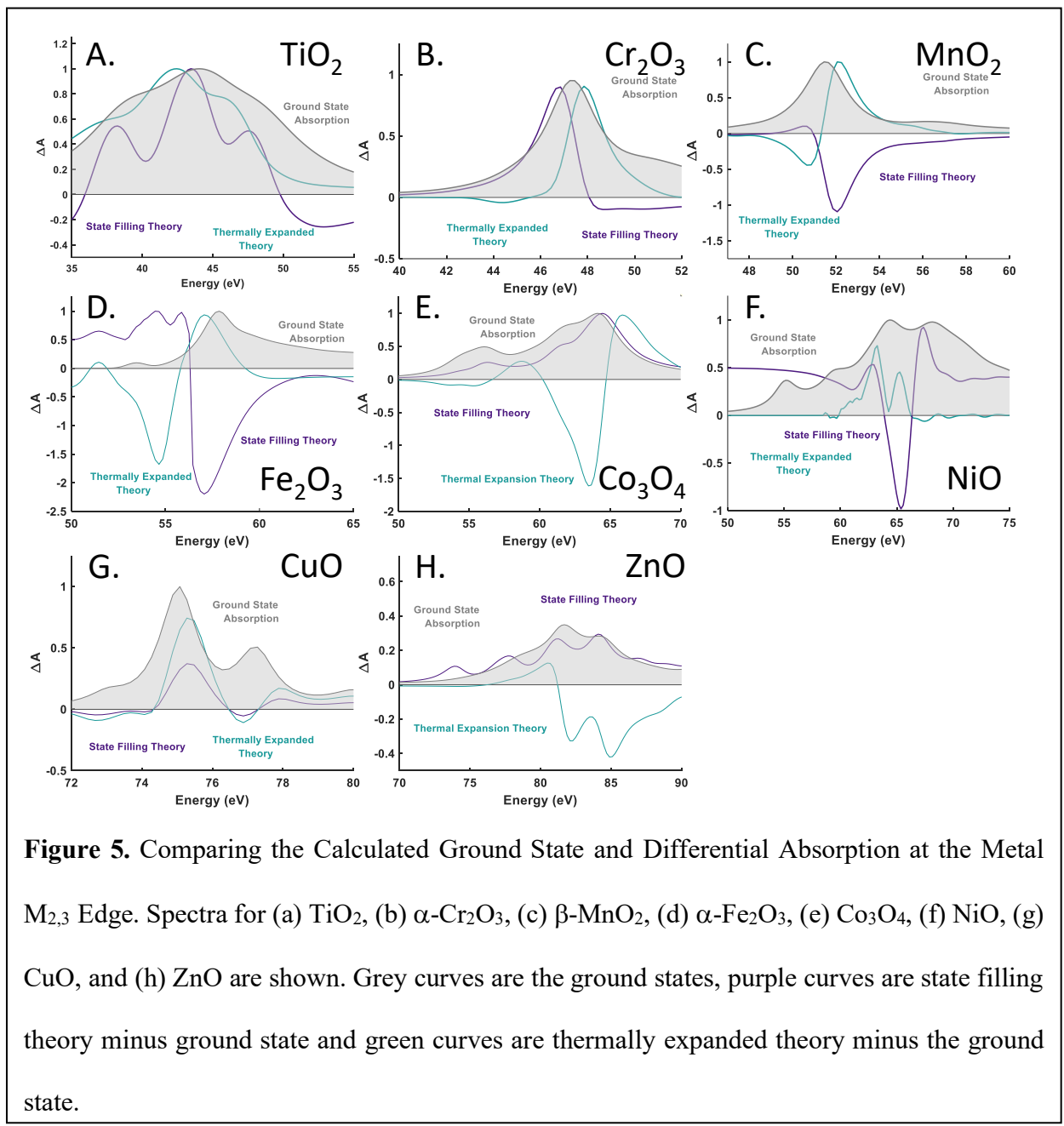

**Figure 5.** Comparing the Calculated Ground State and Differential Absorption at the Metal $M_{2,3}$ Edge. Spectra for (a) $TiO_2$, (b) α-$Cr_2O_3$, (c) β-$MnO_2$, (d) α-$Fe_2O_3$, (e) $Co_3O_4$, (f) NiO, (g) CuO, and (h) ZnO are shown. Grey curves are the ground states, purple curves are state filling theory minus ground state and green curves are thermally expanded theory minus the ground state.

**3.4 Excited State Trends and Hamiltonian Discussion**

In general, the changes in the angular momentum and exchange effects are the largest contributors to the excited state spectra (Figure 6a) for all compounds. The LMCT transition common to transition metal oxides is the main reason for this observed trend; as the photoexcited electron transfers from the predominantly O 2p valence bands to the predominantly metal 3d conduction bands, the number of unpaired electrons changes, perturbing the peak splitting. From $d^0$ to $d^{10}$, adding an extra electron to the 3d band has a decreasing impact on the core-hole screening (Figure 6a). Meanwhile, adding the extra electron has an increasing effect on the angular momentum coupling and exchange interactions with increasing d-orbital occupation. Unsurprisingly, the exchange interaction becomes a dominant term in the excited state because of the change in the possible configurations. Combined, the interplay of increased angular momentum coupling and exchange interactions versus decreasing screening leads to the observed photoexcited peak shifts and splitting in Figure 5 (purple traces) and the overall negative to positive center of mass peak shift trend upon photoexcitation shown in Figure 6c.

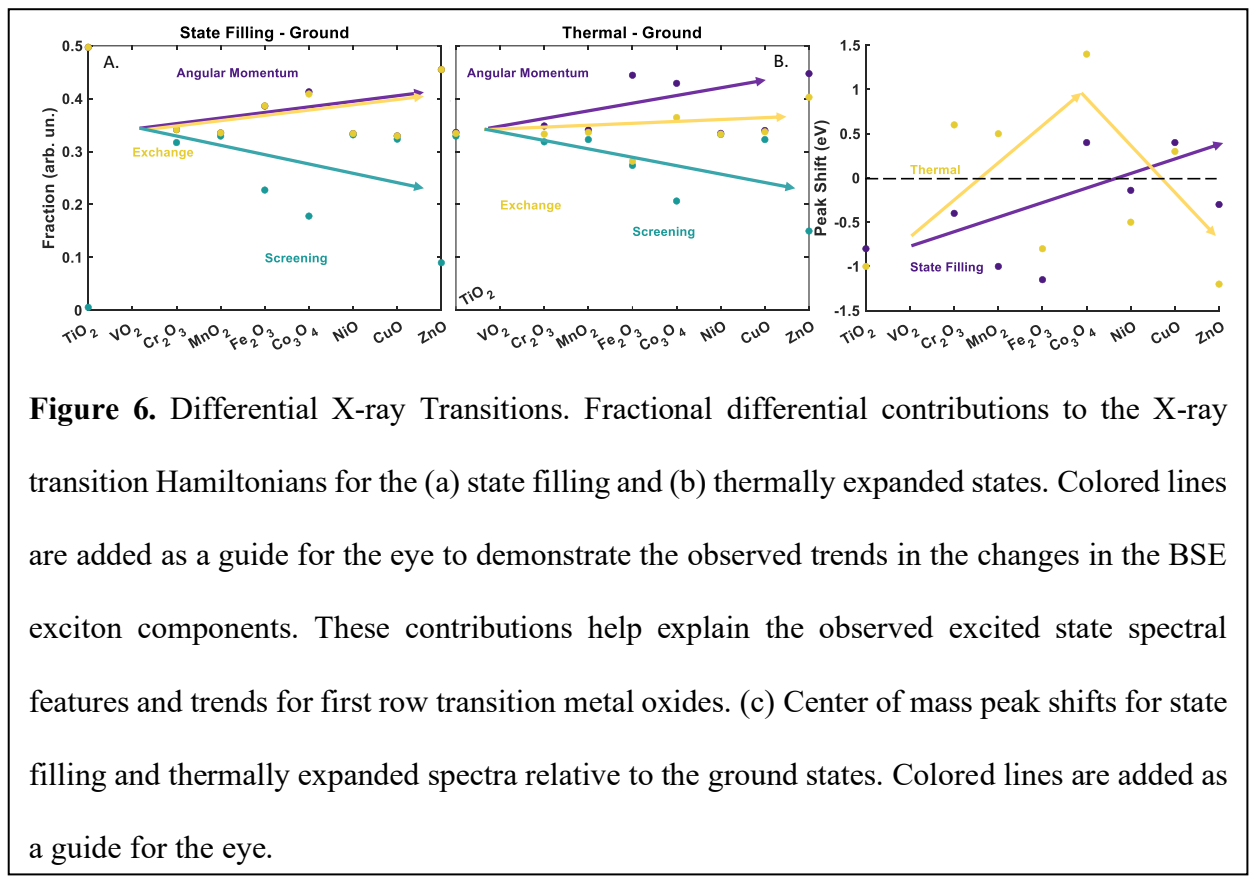

**Figure 6.** Differential X-ray Transitions. Fractional differential contributions to the X-ray transition Hamiltonians for the (a) state filling and (b) thermally expanded states. Colored lines are added as a guide for the eye to demonstrate the observed trends in the changes in the BSE exciton components. These contributions help explain the observed excited state spectral features and trends for first row transition metal oxides. (c) Center of mass peak shifts for state filling and thermally expanded spectra relative to the ground states. Colored lines are added as a guide for the eye.

Some special comments should be made about the observation of state filling effects. In most materials studied to date with delocalized valence states – Si, Ge, ZnTe – an increase or decrease in absorption is measured because X-ray transitions are blocked or allowed due to changes in the photoexcited carrier distribution in the band structure.[3,5,6,69] In all these predicted compounds, however, the change in angular momentum and exchange effects dominate the photoexcitation, distorting the XUV spectra, and acting to block the creation of distinct state filling trends. Instead, the energy and carrier density must be judged by the overall spectral shift and the increase or decrease in peak splitting and amplitude, depending on the material. ZnO and $TiO_2$ are of particular interest because, while they have highly mobile carriers, they experience the largest change in angular momentum coupling from photoexcitation because there are no unpaired electrons in the ground states of these materials. While exchange effects and angular momentum coupling played a small role in the ground state spectra (Figure 2), these components dominate the changes in the photoexcited spectrum (Figure 6a). Resultantly, a negative shift of the overall spectra is the only indicator of the excitation energy and carrier density, and the extraction of electron and hole energies versus time would have to be extracted by calculating the XUV spectrum at each time point.

The change in the XUV spectrum after thermal expansion follows a different trend than the photoexcitation. Again, however, the observed trend is not surprising as the d-band filling remains unchanged, so the trends should mimic the ground state Hamiltonian (Figure 2) as they do (Figure 6b). Instead, the dominant effect in thermal expansion is that the decreased metal-ligand wavefunction overlap will lead to more localized d-states and a smaller crystal field. Accordingly, angular momentum coupling will dominate as screening and exchange interactions decrease (Figure 6b). Again, thermal expansion in these materials leads to a counterintuitive trend as compared to the delocalized semiconductors measured in the past (Si, Ge, ZnTe).[3,5,6,69] In these materials, heating leads to a decreased bandgap and the spectra have been measured to redshift as a result. However, as the thermally expanded transients in Figure 5 and the center of mass peak shifts in Figure 6c show, the dominant change in angular momentum coupling leads to complex spectral features that do not always follow the redshift seen in traditional semiconductors. This finding highlights the importance of not making assumptions when analyzing measured spectra as the measured spectral signatures could be falsely assigned.

## 4. Conclusion

There are several key takeaways from this study. First, a Bethe-Salpeter equation approach is proven a facile method for calculating the challenging M edges of the transition metal oxides. This study supports the universality of our excited state approximation, as proven in previous studies on simpler semiconductors like Si, Ge, and ZnTe. The underlying DFT calculation allows easy inclusion of phase changes, polarons, and other structural dynamics into the excited state calculation while the changes in state filling are addressed in the BSE stage. Further work will involve incorporating a full GW self-energy calculation to better understand the screening and band gap renormalization effects. Moreover, decomposing the X-ray transition Hamiltonian into its fundamental components and projecting it onto the band structure gives a physical intuition for the otherwise complex XUV spectra. The methods used here should be generalizable to higher energy X-ray edges and therefore prove powerful for understanding tabletop, synchrotron, and X-ray free electron laser experiments.

## ASSOCIATED CONTENT

**Supporting Information**.

The following files are available free of charge.

Computational details and outputs, full Hamiltonian decompositions of core-valence excitons for all states, excited state spectra (file type, i.e., PDF)

## AUTHOR INFORMATION

**Corresponding Author**

*Scott Kevin Cushing. Email: scushing@caltech.edu

**Present Addresses**

†Department of Chemistry, Northwestern University, 2145 Sheridan Road, Evanston, Illinois 60208, USA.

**Author Contributions**

All authors have given approval to the final version of the manuscript.

**Funding Sources**

This material is based upon work supported by the Air Force Office of Scientific Research under award number FA9550-21-1-0022. I.M.K. was supported by an NSF Graduate Research Fellowship (DGE-1745301).

**Notes**

Any opinions, findings, and conclusions or recommendations expressed in this material are those of the author(s) and do not necessarily reflect the views of the United States Air Force. The authors declare no competing financial interest.

REFERENCES

(1) Cushing, S. K.; Porter, I. J.; Roulet, B. R. de; Lee, A.; Marsh, B. M.; Szoke, S.; Vaida, M. E.; Leone, S. R. Layer-Resolved Ultrafast Extreme Ultraviolet Measurement of Hole Transport in a Ni-TiO2-Si Photoanode. *Science Advances* **2020**, *6* (14), eaay6650. https://doi.org/10.1126/sciadv.aay6650.

(2) Biswas, S.; Husek, J.; Londo, S.; Fugate, E. A.; Baker, L. R. Identifying the Acceptor State in NiO Hole Collection Layers: Direct Observation of Exciton Dissociation and Interfacial Hole Transfer across a $Fe_2O_3$/NiO Heterojunction. *Phys. Chem. Chem. Phys.* **2018**, *20* (38), 24545–24552. https://doi.org/10.1039/C8CP04502J.

(3) Cushing, S. K.; Zürch, M.; Kraus, P. M.; Carneiro, L. M.; Lee, A.; Chang, H.-T.; Kaplan, C. J.; Leone, S. R. Hot Phonon and Carrier Relaxation in Si(100) Determined by Transient Extreme Ultraviolet Spectroscopy. *Structural Dynamics* **2018**, *5* (5), 054302. https://doi.org/10.1063/1.5038015.

(4) Biswas, S.; Husek, J.; Baker, L. R. Elucidating Ultrafast Electron Dynamics at Surfaces Using Extreme Ultraviolet (XUV) Reflection–Absorption Spectroscopy. *Chem. Commun.* **2018**, *54* (34), 4216–4230. https://doi.org/10.1039/C8CC01745J.

(5) Zürch, M.; Chang, H.-T.; Borja, L. J.; Kraus, P. M.; Cushing, S. K.; Gandman, A.; Kaplan, C. J.; Oh, M. H.; Prell, J. S.; Prendergast, D.; Pemmaraju, C. D.; Neumark, D. M.; Leone, S. R. Direct and Simultaneous Observation of Ultrafast Electron and Hole Dynamics in Germanium. *Nature Communications* **2017**, *8* (1), 1–11. https://doi.org/10.1038/ncomms15734.

(6) Zürch, M.; Chang, H.-T.; Kraus, P. M.; Cushing, S. K.; Borja, L. J.; Gandman, A.; Kaplan, C. J.; Oh, M. H.; Prell, J. S.; Prendergast, D.; Pemmaraju, C. D.; Neumark, D. M.; Leone, S. R. Ultrafast Carrier Thermalization and Trapping in Silicon-Germanium Alloy Probed by Extreme Ultraviolet Transient Absorption Spectroscopy. *Structural Dynamics* **2017**, *4* (4), 044029. https://doi.org/10.1063/1.4985056.

(7) Lin, M.-F.; Verkamp, M. A.; Leveillee, J.; Ryland, E. S.; Benke, K.; Zhang, K.; Weninger, C.; Shen, X.; Li, R.; Fritz, D.; Bergmann, U.; Wang, X.; Schleife, A.; Vura-Weis, J. Carrier-Specific Femtosecond XUV Transient Absorption of $PbI_2$ Reveals Ultrafast Nonradiative Recombination. *J. Phys. Chem. C* **2017**, *121* (50), 27886–27893. https://doi.org/10.1021/acs.jpcc.7b11147.

(8) Kaplan, C. J.; Kraus, P. M.; Ross, A. D.; Zürch, M.; Cushing, S. K.; Jager, M. F.; Chang, H.-T.; Gullikson, E. M.; Neumark, D. M.; Leone, S. R. Femtosecond Tracking of Carrier Relaxation in Germanium with Extreme Ultraviolet Transient Reflectivity. *Phys. Rev. B* **2018**, *97* (20), 205202. https://doi.org/10.1103/PhysRevB.97.205202.

(9) Cirri, A.; Husek, J.; Biswas, S.; Baker, L. R. Achieving Surface Sensitivity in Ultrafast XUV Spectroscopy: M2,3-Edge Reflection–Absorption of Transition Metal Oxides. *J. Phys. Chem. C* **2017**, *121* (29), 15861–15869. https://doi.org/10.1021/acs.jpcc.7b05127.

(10) Wang, H.; Young, A. T.; Guo, J.; Cramer, S. P.; Friedrich, S.; Braun, A.; Gu, W. Soft X-Ray Absorption Spectroscopy and Resonant Inelastic X-Ray Scattering Spectroscopy below 100 EV: Probing First-Row Transition-Metal M-Edges in Chemical Complexes. *J Synchrotron Radiat* **2013**, *20* (Pt 4), 614–619. https://doi.org/10.1107/S0909049513003142.

(11) Ryland, E. S.; Zhang, K.; Vura-Weis, J. Sub-100 Fs Intersystem Crossing to a Metal-Centered Triplet in Ni(II)OEP Observed with M-Edge XANES. *J. Phys. Chem. A* **2019**, *123* (25), 5214–5222. https://doi.org/10.1021/acs.jpca.9b03376.

(12) Biswas, S.; Husek, J.; Londo, S.; Baker, L. R. Highly Localized Charge Transfer Excitons in Metal Oxide Semiconductors. *Nano Lett.* **2018**, *18* (2), 1228–1233. https://doi.org/10.1021/acs.nanolett.7b04818.

(13) Zhang, K.; Lin, M.-F.; Ryland, E. S.; Verkamp, M. A.; Benke, K.; de Groot, F. M. F.; Girolami, G. S.; Vura-Weis, J. Shrinking the Synchrotron: Tabletop Extreme Ultraviolet Absorption of Transition-Metal Complexes. *J. Phys. Chem. Lett.* **2016**, *7* (17), 3383–3387. https://doi.org/10.1021/acs.jpclett.6b01393.

(14) Vura-Weis, J.; Jiang, C.-M.; Liu, C.; Gao, H.; Lucas, J. M.; de Groot, F. M. F.; Yang, P.; Alivisatos, A. P.; Leone, S. R. Femtosecond $M_{2,3}$-Edge Spectroscopy of Transition-Metal Oxides: Photoinduced Oxidation State Change in α-$Fe_2O_3$. *J. Phys. Chem. Lett.* **2013**, *4* (21), 3667–3671. https://doi.org/10.1021/jz401997d.

(15) Vaida, M. E.; Leone, S. R. Femtosecond Extreme Ultraviolet Photoemission Spectroscopy: Observation of Ultrafast Charge Transfer at the n-TiO2/p-Si(100) Interface with Controlled TiO2 Oxygen Vacancies. *J. Phys. Chem. C* **2016**, *120* (5), 2769–2776. https://doi.org/10.1021/acs.jpcc.5b11161.

(16) Liu, H.; Klein, I. M.; Michelsen, J. M.; Cushing, S. K. Element-Specific Electronic and Structural Dynamics Using Transient XUV and Soft X-Ray Spectroscopy. *Chem* **2021**, *7* (10), 2569–2584. https://doi.org/10.1016/j.chempr.2021.09.005.

(17) Zhang, K.; Ash, R.; Girolami, G. S.; Vura-Weis, J. Tracking the Metal-Centered Triplet in Photoinduced Spin Crossover of Fe(Phen)32+ with Tabletop Femtosecond M-Edge X-Ray Absorption Near-Edge Structure Spectroscopy. *J. Am. Chem. Soc.* **2019**, *141* (43), 17180–17188. https://doi.org/10.1021/jacs.9b07332.

(18) Carneiro, L. M.; Cushing, S. K.; Liu, C.; Su, Y.; Yang, P.; Alivisatos, A. P.; Leone, S. R. Excitation-Wavelength-Dependent Small Polaron Trapping of Photoexcited Carriers in α-Fe2O3. *Nature Mater* **2017**, *16* (8), 819–825. https://doi.org/10.1038/nmat4936.

(19) de Groot, F. High-Resolution X-Ray Emission and X-Ray Absorption Spectroscopy. *Chem. Rev.* **2001**, *101* (6), 1779–1808. https://doi.org/10.1021/cr9900681.

(20) Groot, F. de. Multiplet Effects in X-Ray Spectroscopy. *Coordination Chemistry Reviews* **2005**, *249* (1–2), 31–63. https://doi.org/10.1016/j.ccr.2004.03.018.

(21) Norman, P.; Dreuw, A. Simulating X-Ray Spectroscopies and Calculating Core-Excited States of Molecules. *Chem. Rev.* **2018**, *118* (15), 7208–7248. https://doi.org/10.1021/acs.chemrev.8b00156.

(22) Ekimova, M.; Quevedo, W.; Szyc, Ł.; Iannuzzi, M.; Wernet, P.; Odelius, M.; Nibbering, E. T. J. Aqueous Solvation of Ammonia and Ammonium: Probing Hydrogen Bond Motifs with FT-IR and Soft X-Ray Spectroscopy. *J. Am. Chem. Soc.* **2017**, *139* (36), 12773–12783. https://doi.org/10.1021/jacs.7b07207.

(23) Cederbaum, L. S.; Domcke, W.; Schirmer, J. Many-Body Theory of Core Holes. *Phys. Rev. A* **1980**, *22* (1), 206–222. https://doi.org/10.1103/PhysRevA.22.206.

(24) Barth, A.; Schirmer, J. Theoretical Core-Level Excitation Spectra of N$_2$ and CO by a New Polarisation Propagator Method. *J. Phys. B: Atom. Mol. Phys.* **1985**, *18* (5), 867–885. https://doi.org/10.1088/0022-3700/18/5/008.

(25) Triguero, L.; Pettersson, L. G. M.; Ågren, H. Calculations of Near-Edge x-Ray-Absorption Spectra of Gas-Phase and Chemisorbed Molecules by Means of Density-Functional and Transition-Potential Theory. *Phys. Rev. B* **1998**, *58* (12), 8097–8110. https://doi.org/10.1103/PhysRevB.58.8097.

(26) Stöhr, J. NEXAFS Spectroscopy. In *NEXAFS Spectroscopy*; Stöhr, J., Ed.; Springer Series in Surface Sciences; Springer: Berlin, Heidelberg, 1992; pp 1–7. https://doi.org/10.1007/978-3-662-02853-7_1.

(27) Reinholdt, P.; Vidal, M. L.; Kongsted, J.; Iannuzzi, M.; Coriani, S.; Odelius, M. Nitrogen K-Edge X-Ray Absorption Spectra of Ammonium and Ammonia in Water Solution: Assessing the Performance of Polarizable Embedding Coupled Cluster Methods. *J. Phys. Chem. Lett.* **2021**, *12* (36), 8865–8871. https://doi.org/10.1021/acs.jpclett.1c02031.

(28) Montorsi, F.; Segatta, F.; Nenov, A.; Mukamel, S.; Garavelli, M. Soft X-Ray Spectroscopy Simulations with Multiconfigurational Wave Function Theory: Spectrum Completeness, Sub-EV Accuracy, and Quantitative Reproduction of Line Shapes. *J. Chem. Theory Comput.* **2022**, *18* (2), 1003–1016. https://doi.org/10.1021/acs.jctc.1c00566.

(29) Gao, S.-P.; Pickard, C. J.; Payne, M. C.; Zhu, J.; Yuan, J. Theory of Core-Hole Effects in $1s$ Core-Level Spectroscopy of the First-Row Elements. *Phys. Rev. B* **2008**, *77* (11), 115122. https://doi.org/10.1103/PhysRevB.77.115122.

(30) Cederbaum, L. S.; Domcke, W. A Many-body Approach to the Vibrational Structure in Molecular Electronic Spectra. I. Theory. *J. Chem. Phys.* **1976**, *64* (2), 603–611. https://doi.org/10.1063/1.432250.

(31) Runge, E.; Gross, E. K. U. Density-Functional Theory for Time-Dependent Systems. *Phys. Rev. Lett.* **1984**, *52* (12), 997–1000. https://doi.org/10.1103/PhysRevLett.52.997.

(32) Kolesov, G.; Kolesov, B. A.; Kaxiras, E. Polaron-Induced Phonon Localization and Stiffening in Rutile ${\mathrm{TiO}}_{2}$. *Phys. Rev. B* **2017**, *96* (19), 195165. https://doi.org/10.1103/PhysRevB.96.195165.

(33) Lian, C.; Ali, Z. A.; Kwon, H.; Wong, B. M. Indirect but Efficient: Laser-Excited Electrons Can Drive Ultrafast Polarization Switching in Ferroelectric Materials. *J. Phys. Chem. Lett.* **2019**, *10* (12), 3402–3407. https://doi.org/10.1021/acs.jpclett.9b01046.

(34) Wang, Y.; Lopata, K.; Chambers, S. A.; Govind, N.; Sushko, P. V. Optical Absorption and Band Gap Reduction in (Fe1–XCrx)2O3 Solid Solutions: A First-Principles Study. *J. Phys. Chem. C* **2013**, *117* (48), 25504–25512. https://doi.org/10.1021/jp407496w.

(35) Konecny, L.; Vicha, J.; Komorovsky, S.; Ruud, K.; Repisky, M. Accurate X-Ray Absorption Spectra near L- and M-Edges from Relativistic Four-Component Damped Response Time-Dependent Density Functional Theory. *Inorg. Chem.* **2022**, *61* (2), 830–846. https://doi.org/10.1021/acs.inorgchem.1c02412.

(36) Husek, J.; Cirri, A.; Biswas, S.; Baker, L. R. Surface Electron Dynamics in Hematite (α-Fe2O3): Correlation between Ultrafast Surface Electron Trapping and Small Polaron Formation †Electronic Supplementary Information (ESI) Available. See DOI:

10.1039/C7sc02826a. *Chem Sci* **2017**, *8* (12), 8170–8178. https://doi.org/10.1039/c7sc02826a.

(37) Biswas, S.; Wallentine, S.; Bandaranayake, S.; Baker, L. R. Controlling Polaron Formation at Hematite Surfaces by Molecular Functionalization Probed by XUV Reflection-Absorption Spectroscopy. *J. Chem. Phys.* **2019**, *151* (10), 104701. https://doi.org/10.1063/1.5115163.

(38) Porter, I. J.; Cushing, S. K.; Carneiro, L. M.; Lee, A.; Ondry, J. C.; Dahl, J. C.; Chang, H.-T.; Alivisatos, A. P.; Leone, S. R. Photoexcited Small Polaron Formation in Goethite (α-FeOOH) Nanorods Probed by Transient Extreme Ultraviolet Spectroscopy. *J. Phys. Chem. Lett.* **2018**, *9* (14), 4120–4124. https://doi.org/10.1021/acs.jpclett.8b01525.

(39) Stavitski, E.; de Groot, F. M. F. The CTM4XAS Program for EELS and XAS Spectral Shape Analysis of Transition Metal L Edges. *Micron* **2010**, *41* (7), 687–694. https://doi.org/10.1016/j.micron.2010.06.005.

(40) Klein, I. M.; Liu, H.; Nimlos, D.; Krotz, A.; Cushing, S. K. *Ab Initio* Prediction of Excited-State and Polaron Effects in Transient XUV Measurements of α-$Fe_2O_3$. *J. Am. Chem. Soc.* **2022**, jacs.2c03994. https://doi.org/10.1021/jacs.2c03994.

(41) Soldatov, M. A.; Martini, A.; Bugaev, A. L.; Pankin, I.; Medvedev, P. V.; Guda, A. A.; Aboraia, A. M.; Podkovyrina, Y. S.; Budnyk, A. P.; Soldatov, A. A.; Lamberti, C. The Insights from X-Ray Absorption Spectroscopy into the Local Atomic Structure and Chemical Bonding of Metal–Organic Frameworks. *Polyhedron* **2018**, *155*, 232–253. https://doi.org/10.1016/j.poly.2018.08.004.

(42) Guda, S. A.; Guda, A. A.; Soldatov, M. A.; Lomachenko, K. A.; Bugaev, A. L.; Lamberti, C.; Gawelda, W.; Bressler, C.; Smolentsev, G.; Soldatov, A. V.; Joly, Y. Optimized Finite Difference Method for the Full-Potential XANES Simulations: Application to Molecular Adsorption Geometries in MOFs and Metal–Ligand Intersystem Crossing Transients. *J. Chem. Theory Comput.* **2015**, *11* (9), 4512–4521. https://doi.org/10.1021/acs.jctc.5b00327.

(43) Vinson, J.; Jach, T.; Müller, M.; Unterumsberger, R.; Beckhoff, B. Quasiparticle Lifetime Broadening in Resonant X-Ray Scattering of ${\mathrm{NH}}_{4}{\mathrm{NO}}_{3}$. *Phys. Rev. B* **2016**, *94* (3), 035163. https://doi.org/10.1103/PhysRevB.94.035163.

(44) Li, L.; Zhang, R.; Vinson, J.; Shirley, E. L.; Greeley, J. P.; Guest, J. R.; Chan, M. K. Y. Imaging Catalytic Activation of CO2 on Cu2O (110): A First-Principles Study. *Chem. Mater.* **2018**, *30* (6), 1912–1923. https://doi.org/10.1021/acs.chemmater.7b04803.

(45) Woicik, J. C.; Weiland, C.; Rumaiz, A. K.; Brumbach, M.; Quackenbush, N. F.; Ablett, J. M.; Shirley, E. L. Revealing Excitonic Processes and Chemical Bonding in $\mathrm{Mo}{\mathrm{S}}_{2}$ by X-Ray Spectroscopy. *Phys. Rev. B* **2018**, *98* (11), 115149. https://doi.org/10.1103/PhysRevB.98.115149.

(46) Geondzhian, A.; Gilmore, K. Demonstration of Resonant Inelastic X-Ray Scattering as a Probe of Exciton-Phonon Coupling. *Phys. Rev. B* **2018**, *98* (21), 214305. https://doi.org/10.1103/PhysRevB.98.214305.

(47) Attar, A. R.; Chang, H.-T.; Britz, A.; Zhang, X.; Lin, M.-F.; Krishnamoorthy, A.; Linker, T.; Fritz, D.; Neumark, D. M.; Kalia, R. K.; Nakano, A.; Ajayan, P.; Vashishta, P.; Bergmann, U.; Leone, S. R. Simultaneous Observation of Carrier-Specific Redistribution and Coherent Lattice Dynamics in 2H-MoTe2 with Femtosecond Core-Level Spectroscopy. *ACS Nano* **2020**, *14* (11), 15829–15840. https://doi.org/10.1021/acsnano.0c06988.

(48) Vinson, J.; Rehr, J. J.; Kas, J. J.; Shirley, E. L. Bethe-Salpeter Equation Calculations of Core Excitation Spectra. *Phys. Rev. B* **2011**, *83* (11), 115106. https://doi.org/10.1103/PhysRevB.83.115106.

(49) Gilmore, K.; Vinson, J.; Shirley, E. L.; Prendergast, D.; Pemmaraju, C. D.; Kas, J. J.; Vila, F. D.; Rehr, J. J. Efficient Implementation of Core-Excitation Bethe–Salpeter Equation Calculations. *Computer Physics Communications* **2015**, *197*, 109–117. https://doi.org/10.1016/j.cpc.2015.08.014.

(50) Kvashnina, K. O.; Butorin, S. M. High-Energy Resolution X-Ray Spectroscopy at Actinide M4,5 and Ligand K Edges: What We Know, What We Want to Know, and What We Can Know. *Chem. Commun.* **2022**, *58* (3), 327–342. https://doi.org/10.1039/D1CC04851A.

(51) Nattino, F.; Marzari, N. Operando XANES from First-Principles and Its Application to Iridium Oxide. *Phys. Chem. Chem. Phys.* **2020**, *22* (19), 10807–10818. https://doi.org/10.1039/C9CP06726D.

(52) Charles, N.; Yu, Y.; Giordano, L.; Jung, R.; Maglia, F.; Shao-Horn, Y. Toward Establishing Electronic and Phononic Signatures of Reversible Lattice Oxygen Oxidation in Lithium Transition Metal Oxides For Li-Ion Batteries. *Chem. Mater.* **2020**, *32* (13), 5502–5514. https://doi.org/10.1021/acs.chemmater.0c00245.

(53) Wansleben, M.; Vinson, J.; Holfelder, I.; Kayser, Y.; Beckhoff, B. Valence-to-Core X-Ray Emission Spectroscopy of Ti, TiO, and $TiO_2$ by Means of a Double Full-Cylinder Crystal von Hamos Spectrometer. *X-Ray Spectrometry* **2019**, *48* (2), 102–106. https://doi.org/10.1002/xrs.3000.

(54) Vinson, J. Bethe-Salpeter Equation Approach for Calculations of X-Ray Spectra. Thesis, University of Washington, Seattle, WA, 2013.

(55) Piccinin, S. The Band Structure and Optical Absorption of Hematite (α-$Fe_2O_3$): A First-Principles GW-BSE Study. *Phys. Chem. Chem. Phys.* **2019**, *21* (6), 2957–2967. https://doi.org/10.1039/C8CP07132B.

(56) Anisimov, V. I.; Zaanen, J.; Andersen, O. K. Band Theory and Mott Insulators: Hubbard *U* Instead of Stoner *I*. *Phys. Rev. B* **1991**, *44* (3), 943–954. https://doi.org/10.1103/PhysRevB.44.943.

(57) Smyth, J. R.; Jacobsen, S. D.; Hazen, R. M. Comparative Crystal Chemistry of Dense Oxide Minerals. *Reviews in Mineralogy and Geochemistry* **2000**, *41* (1), 157–186. https://doi.org/10.2138/rmg.2000.41.6.

(58) Agarwal, B. K.; Givens, M. P. Soft X-Ray Absorption by Manganese and Manganese Oxide. *Journal of Physics and Chemistry of Solids* **1958**, *6* (2–3), 178–179. https://doi.org/10.1016/0022-3697(58)90092-1.

(59) Jiang, C.-M.; Baker, L. R.; Lucas, J. M.; Vura-Weis, J.; Alivisatos, A. P.; Leone, S. R. Characterization of Photo-Induced Charge Transfer and Hot Carrier Relaxation Pathways in Spinel Cobalt Oxide ($Co_3O_4$). *J. Phys. Chem. C* **2014**, *118* (39), 22774–22784. https://doi.org/10.1021/jp5071133.

(60) Chiuzbăian, S. G.; Ghiringhelli, G.; Dallera, C.; Grioni, M.; Amann, P.; Wang, X.; Braicovich, L.; Patthey, L. Localized Electronic Excitations in NiO Studied with Resonant Inelastic X-Ray Scattering at the Ni M Threshold: Evidence of Spin Flip. *Phys. Rev. Lett.* **2005**, *95* (19), 197402. https://doi.org/10.1103/PhysRevLett.95.197402.

(61) Basu, R.; Srihari, V.; Sardar, M.; Srivastava, S. K.; Bera, S.; Dhara, S. Probing Phase Transition in $VO_2$ with the Novel Observation of Low-Frequency Collective Spin Excitation. *Sci Rep* **2020**, *10* (1), 1977. https://doi.org/10.1038/s41598-020-58813-x.

(62) Luo, H. G.; Xiang, T.; Wang, X. Q.; Su, Z. B.; Yu, L. Fano Resonance for Anderson Impurity Systems. *Phys. Rev. Lett.* **2004**, *92* (25), 256602. https://doi.org/10.1103/PhysRevLett.92.256602.

(63) van der Heide, P. A. W. Multiplet Splitting Patterns Exhibited by the First Row Transition Metal Oxides in X-Ray Photoelectron Spectroscopy. *Journal of Electron Spectroscopy and Related Phenomena* **2008**, *164* (1), 8–18. https://doi.org/10.1016/j.elspec.2008.04.001.

(64) Lany, S. Semiconducting Transition Metal Oxides. *J. Phys.: Condens. Matter* **2015**, *27* (28), 283203. https://doi.org/10.1088/0953-8984/27/28/283203.

(65) Tokura, Y. Correlated-Electron Physics in Transition-Metal Oxides. *Physics Today* **2003**, *56* (7), 50–55. https://doi.org/10.1063/1.1603080.

(66) Hozoi, L. Localized States in Transition Metal Oxides. Thesis fully internal (DIV), University of Groningen, Groningen, 2003.

(67) Gagliardi, L.; Truhlar, D. G.; Li Manni, G.; Carlson, R. K.; Hoyer, C. E.; Bao, J. L. Multiconfiguration Pair-Density Functional Theory: A New Way To Treat Strongly Correlated Systems. *Acc. Chem. Res.* **2017**, *50* (1), 66–73. https://doi.org/10.1021/acs.accounts.6b00471.

(68) Reticcioli, M.; Diebold, U.; Kresse, G.; Franchini, C. Small Polarons in Transition Metal Oxides. In *Handbook of Materials Modeling: Applications: Current and Emerging Materials*; Andreoni, W., Yip, S., Eds.; Springer International Publishing: Cham, 2019; pp 1–39. https://doi.org/10.1007/978-3-319-50257-1_52-1.

(69) Cushing, S. K.; Lee, A.; Porter, I. J.; Carneiro, L. M.; Chang, H.-T.; Zürch, M.; Leone, S. R. Differentiating Photoexcited Carrier and Phonon Dynamics in the Δ, L, and Γ Valleys of Si(100) with Transient Extreme Ultraviolet Spectroscopy. *J. Phys. Chem. C* **2019**, *123* (6), 3343–3352. https://doi.org/10.1021/acs.jpcc.8b10887.

*Ab Initio* **Calculations of XUV Ground and Excited States for First-Row Transition Metal Oxides**

Isabel M. Klein[1], Alex Krotz[1], Jonathan M. Michelsen[1], Scott K. Cushing[1,*]

[1]Division of Chemistry and Chemical Engineering, California Institute of Technology, Pasadena, CA 91125, USA.

[*]Corresponding author. Email: scushing@caltech.edu

**Contents:**

1.  **Explanation of SI Layout**

Given the extensive nature of the data generated in preparation of this comparative study, a summary of each section is provided here.

**2. Computational Methods**

   a. Computational Parameters

   The general information for performing both QuantumEspresso and OCEAN calculations is given here. Specific details for individual transition metal oxides are also included.

   b. Fundamental Equations Underlying the OCEAN Code

   A detailed description of the equations underlying the OCEAN code, including the modifications made in this work to allow for the projection and decomposition of the core-valence excitons.

   c. Energy Dependent Broadening Code

   A detailed description of the Matlab code used to broaden the OCEAN calculated absorption spectra. Specific values used in the broadening code for each transition metal oxide are also provided.

**3 – 10. Transition Metal Oxides**

For each transition metal oxide investigated herein, calculations were performed for a ground state, a state-filling excited state, and a thermally expanded excited state. Because the calculations for each of these states for each transition metal oxide follow the same general outline, summaries and descriptions of the calculations are provided here, while the outputs and results for specific transition metal oxides are provided in sections S3 – S10.

a. Structural Data for Calculations

   i. Ground State and State Blocking

   Calculations for both the ground state and the state-filling excited state of the transition metal oxide were performed on the unit cell provided in this section.

   ii. Thermally Expanded Lattice

   Calculations for the thermally expanded excited state of the transition metal oxide were performed on the unit cell provided here. The specific change that models the thermal expansion is in the scaling of the primitive vectors.

b. Ground State Calculations

   i. Band Structure and DOS

   Initial DFT calculations on the ground state of each transition metal oxide were performed to verify computational parameters used in this study. The OCEAN computational program is not yet able to use hybrid pseudopotentials, and as such it was important to compare the band structure and DOS shown here to previous computational work.[1] Partial DOS with orbital breakdowns are shown to fully understand the DOS and the excitations calculated in the OCEAN calculations. As commented earlier, a scissor shift was used for to correct the band gap because hybrid pseudopotentials could not be used. Since only the conduction band states are needed for the final calculations, the calculation can still be accurate even without a hybrid functional. Accordingly, adjusting the U parameter in the DFT+U calculations had little effect on the final XUV absorption spectrum.

### ii. Ground State Spectrum

The ground state XUV absorption spectra were calculated and broadened by convoluting the calculated XUV absorption spectrum with an energy-dependent Gaussian, as described in S2c.[2,3]

### iii. Ground State GMRES Energy Decomposition

The ground state XUV core-valence exciton densities were calculated for specific energy ranges and overlaid on the XUV absorption spectra and band structure for each transition metal oxide. To calculate the exciton density for these specific energy ranges, a GMRES calculation was used, restricting the energy to a 1 eV range. The range was changed to allow the entire region around the Fermi energy, and around the $M_{2,3}$ edge under investigation, to be calculated.

## c. Excited State Calculations

### i. State filling band diagrams

The electron and hole occupations resulting from excitation to the conduction band minimum were used in the OCEAN calculation to generate model state filling XUV absorption spectra.

### ii. Full Spectrum

Excited state spectra were calculated as described in detail in the text, using the same computational parameters as the ground state calculation. Excited state broadening was performed consistent with previously described ground state broadening in S2c.

## d. Hamiltonian Decompositions

### i. Total Exciton Comparisons

Comparisons between the total core-valence exciton density for the three states (ground, state-filling, and thermally expanded) for each transition metal oxide modeled in this study are shown in this section. Additionally, comparisons of the differential between the total core-valence exciton density for the excited states of the transition metal oxide and the ground state modeled in this study are shown here.

ii. Hamiltonian Decomposition of Exciton Components for ground, state filling, and thermally expanded models

> Hamiltonian decomposition of the core-valence exciton for the ground state, state filling, and thermally expanded excited states of each transition metal oxide are presented here. These decompositions are projected onto the band structure and are used to qualitatively compare trends between states, as well as quantify the differences between the excited states and the ground state Hamiltonian contributions. All contributions for one state are normalized relative to each other to facilitate direct comparison. The angular momentum, screening, and exchange effects are discussed in the main text. The high-order bubble and ladder many-body interactions relate to the formation and renormalization of the core-valence exciton.
>
> Differential Hamiltonian decompositions for the state filling and thermally expanded excited states relative to the ground state are used to quantitatively determine the importance of different components in the transient changes observed experimentally. All contributions are normalized relative to each other to facilitate direct, quantitative comparison.

## 10. Comparisons

a. Thermal Expansion Calculations

> The percent expansion from of the crystal lattice for each transition metal oxide when increasing the temperature from 300K to 650K is given here.

b. Physical Properties for Hamiltonian Contribution Explanations

> The trends in the physical properties of the transition metal oxides that are used in explaining the trends in the Hamiltonian contributions are given here.

# 1. Computational Methods

a. Computational Input Parameters

Geometry optimization and DFT calculations were performed with the Quantum ESPRESSO package using Perdew-Burke-Ernzerhof (PBE) pseudopotentials under the generalized gradient approximation (GGA). The associated ground state wavefunctions were constructed using a plane wave basis set with components up to a kinetic energy cutoff of 250 Ry. Reciprocal space was sampled using an 4x4x4 Gamma-centered mesh with a 0.02 eV Gaussian smearing of orbital occupancies. DFT simulations were performed on unit cells of each of the transition metal oxides investigated in this study. Self-consistent calculations were performed to a convergence of $10^{-6}$ eV/atom and forces on ions under $10^{-3}$ eV/Å. DFT-BSE calculations were conducted using the same parameters. Additionally, an 8x8x8 k-point grid for the screening mesh, 100 bands, and a cutoff radius of 4.5 Bohr were used in the DFT-BSE calculations of the XUV absorption spectra.[1,4]

**Table 1.** Computational Parameters of Transition Metal Oxides

| Transition Metal Oxide | Dielectric Constant | Hubbard U | Scissor Correction (eV) | Metal Oxidation |
|---|---|---|---|---|
| $TiO_2$ | 6 | | | +4 |
| $Cr_2O_3$ | 11 | 3 | -2 | +3 |
| $MnO_2$ | 10 | 4 | | +4 |
| $Fe_2O_3$ | 25 | 4 | -3 | +3 |
| $Co_3O_4$ | 12.8 | | +2 | +2/+3 |
| NiO | 11.9 | | | +2 |
| CuO | 6.46 | | | +2 |
| ZnO | 10.4 | 13 | | +2 |

b. Fundamental Equations Underlying the OCEAN (DFT-BSE) Code

*Part I.* The calculated core-valence transition rate can be calculated following equation (1) below,

$$\Gamma_O(\omega, q) = 2\pi \sum_F |\langle I|\hat{O}(\omega, q)|F\rangle|^2 \delta(\omega + E_I - E_F)\delta(q + k_I - k_F) \quad (1)$$

where $\Gamma_O$ is the overall transition rate and O is the many-body electron-photon interaction. Summing over all possible excited states, $F$, the transition probability, or XUV cross section, is calculated. Given the cumbersome nature of this summation and the completeness of the sum over all final states, equations 2 and 3 can be used.

$$\Gamma_O(\omega, q) = 2\pi \sum_F \langle I|\hat{O}(\omega, q)|F\rangle\langle F|O^\dagger(\omega, q)|I\rangle \delta(\omega + E_I - E_F)\delta(q + k_I - k_F) \quad (2)$$

$$\Gamma_O(\omega, q) = -Im[\langle I|\hat{O}(\omega, q)G_2(\omega, q)\hat{O}^\dagger(\omega, q)|I\rangle] \quad (3)$$

$G_2$ is the two-particle Green's function, which is related to BSE Hamiltonian,

$$G_2 = [\omega - H_{BSE}]^{-1} \quad (4)$$

$$H_{BSE} = H_e - H_h + V_X - V_D \tag{5}$$

where $H_e$ is the electron Hamiltonian, $H_h$ is the hole Hamiltonian, and $V_X$ and $V_D$ are the exchange and direct interaction terms, respectively. These components of the BSE Hamiltonian can be addressed separately and provide insight into the different effects discussed in the manuscript.

The electron Hamiltonian is built up of Kohn-Sham orbitals and includes solid-state effects. $H_e$ also includes self-energy effects by adding a self-energy correction to the single-particle energies. Adopting a quasi-particle picture of the system, and treating the many-body states as perturbations of the non-interacting system, the many-body Hamiltonian can be divided into a single-particle and an interaction term,

$$H_e = H_0 + \Sigma = H_0 + i\hbar GW \tag{6}$$

where $\Sigma$ is the electron self-energy operator and contains information about all interactions between the electron and the rest of the electrons in the system, $H_0$ is the non-interacting Hamiltonian, $G$ is Green's function and $W$ is the screened coulomb interaction. Self-energy calculations are often carried out using the Kohn-Sham orbitals as a basis for the non-interacting Green's function.

The hole Hamiltonian,

$$H_h = E_h - i\Gamma_j + \chi_j \tag{7}$$

where $E_h$ is the hole binding energy, $\chi_j$ is the spin-orbit splitting and $\Gamma_j$ is the life-time broadening, contains information about the angular momentum splitting and coupling, as both $\chi_j$ and $\Gamma_j$ depend on the total angular momentum state $j$.

The direct interaction term provides insight into the screened Coulomb scattering between the excited electron and core-hole,

$$V_D = \hat{a}_c^\dagger(r,\sigma)\hat{a}_h(r',\sigma')W(r,r',\omega)\hat{a}_c(r,\sigma)\hat{a}_h^\dagger(r',\sigma') \tag{8}$$

The screening of this interaction, $W$, is determined by the dielectric response of the system, $\epsilon^{-1}$, where

$$W(r,r',\omega) = \int d^3r'' \frac{\epsilon^{-1}(r,r'',\omega)}{r''-r'\vee} \tag{9}$$

The exchange interaction,

$$V_X = \hat{a}_c^\dagger(r,\sigma)\hat{a}_h(r',\sigma')\frac{1}{r-r'\vee\hat{a}_c(r',\sigma)\hat{a}_h^\dagger(r,\sigma')} \tag{10}$$

involves core-hole operators at both r and r' and is therefore limited to a finite region due to the local nature of the core-hole. The decomposition of these components shows how the BSE Hamiltonian can provide insight into the different effects that influence the core-valence exciton and XUV absorption spectrum.

*Part II.* There are two ways to understand how the core-valence exciton density and the associated components are projected onto the band structure. First, we can look to equations 1 and 2, where it is clear that the transition rate and probability, and therefore the core-valence exciton density, is dependent on the k-points of the initial and final states under investigation. The transition probability between initial and final states is calculated for each k-point and then this k-space dependent core-exciton density is projected onto the band structure. The core-valence exciton density can be further decomposed into the components of the BSE Hamiltonian. While it is not as straightforward as adding all the respective terms together, the BSE Hamiltonian can be decomposed to evaluate the magnitudes of the influence of angular momentum spin-orbit coupling, core-hole screening, and exchange effects on the XUV absorption. The k-point dependence of the transition probability and the magnitudes of these effects allows us to plot the core-valence exciton densities directly on the band structure.

Another way to understand the projection of the core-valence exciton density and associated components onto the band structure is as follows. The core-valence exciton wavefunction, $|\psi\rangle$, is given as an expansion in the basis of Kohn-Sham states, $|n, k\rangle$, given by the DFT calculations, where *n* is the valence band index and *k* is the wavevector. The projection, *D*, of $|\psi\rangle$ onto the band structure for a given term of the BSE Hamiltonian, *H,* is given by

$$D = |\langle\psi|n, k\rangle\langle n, k|H_{BSE}|\psi\rangle|^2 \tag{11}$$

where the projection operator is $|n, k\rangle\langle n, k|$. The projection values are then interpolated in reciprocal-space along the paths of the band structure diagram and plotted on the band structure.

*Part III.* Once the core-valence exciton and individual contribution densities are calculated, the relative magnitudes of these contributions can be compared. In comparing the projections of the exciton onto the band structures of each transition metal oxide, the colored bubbles for the ground state calculation are normalized and thus the size of the different bubbles can be compared to determine the relative strengths of different contributions within a single state of the system. To compare trends of these components between different states, the differential of the core-valence exciton density between the ground state and the three excited states under investigation is calculated. The magnitude of the differential allows us to compare trends between different states and determine how much different contributions change between the ground state and excited states, thus providing information regarding the origins of measured excited state changes.

c. Energy Dependent Broadening

```
lor = @(E,w,E0) (1/2*pi).*w./((E-E0).^2+(w/2).^2);
lor = @(E,w,E0) lor(E,w,E0)/trapz(E,lor(E,w,E0));
lin_broad_ocean = zeros(size(energy));

for i = 1:size(energy,2);
   lin_broad_ocean=lor(energy,w,energy(i)).*ocean_intp(i);
   lin_broad_ocean= lin_broad_ocean + lin_broad_ocean;
```

ocean_broad_lin = lin_broad_ocean + ocean_intp;
End

**Table 2.** Lorentzian Peak Widths used for Transition Metal Oxide Broadening

| Transition Metal Oxide | Lorentzian Width (w) |
|---|---|
| $TiO_2$ | 4 |
| $Cr_2O_3$ | 2 |
| $MnO_2$ | 2 |
| $Fe_2O_3$ | 0.5 |
| $Co_3O_4$ | 2.5 |
| NiO | 2 |
| CuO | 1 |
| ZnO | 2 |

While the broadening scheme above was used for all the transition metal oxides, an additional GW stretch was used to correct the energy scaling for $TiO_2$ and NiO.

## 2. TiO$_2$

a. Structural Data for Calculations

i. Ground State and State Blocking

**Unit Cell Parameters, (bohr)**

{9.0989  9.0989  5.9151}

**Primitive Vectors**

{1.0 0.0 0.0
 0.0 1.0 0.0
 0.0 0.0 1.0}

**Reduced coordinates, ( x, y, z )**

Ti  0.0000  0.0000  0.0000

Ti  0.5000  0.5000  0.5000

O   0.304809167  0.304809167  0.0000

O  -0.304809167 -0.304809167  0.0000

O   0.804801130  0.195198870  0.5000

O   0.195198870  0.804801130  0.5000

ii. Thermally Expanded Lattice

**Unit Cell Parameters, (bohr)**

{9.1899  9.1899  5.9742}

**Primitive Vectors**

{1.0 0.0 0.0
 0.0 1.0 0.0
 0.0 0.0 1.0}

**Reduced coordinates, ( x, y, z )**

Ti  0.0000  0.0000  0.0000

Ti  0.5000  0.5000  0.5000

O   0.304809167  0.304809167  0.0000

O  -0.304809167 -0.304809167  0.0000

O   0.804801130  0.195198870  0.5000

O   0.195198870  0.804801130  0.5000

b. Ground State Calculations

i. Band Structure and DOS

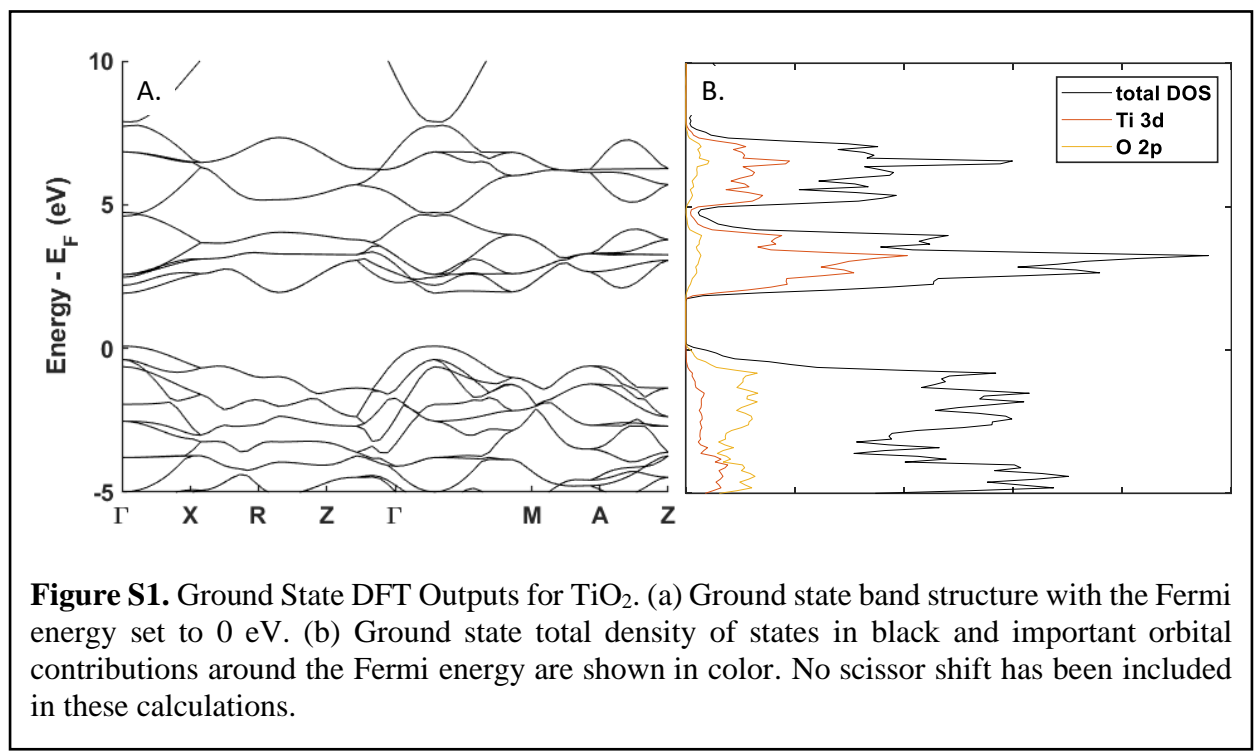

**Figure S1.** Ground State DFT Outputs for $TiO_2$. (a) Ground state band structure with the Fermi energy set to 0 eV. (b) Ground state total density of states in black and important orbital contributions around the Fermi energy are shown in color. No scissor shift has been included in these calculations.

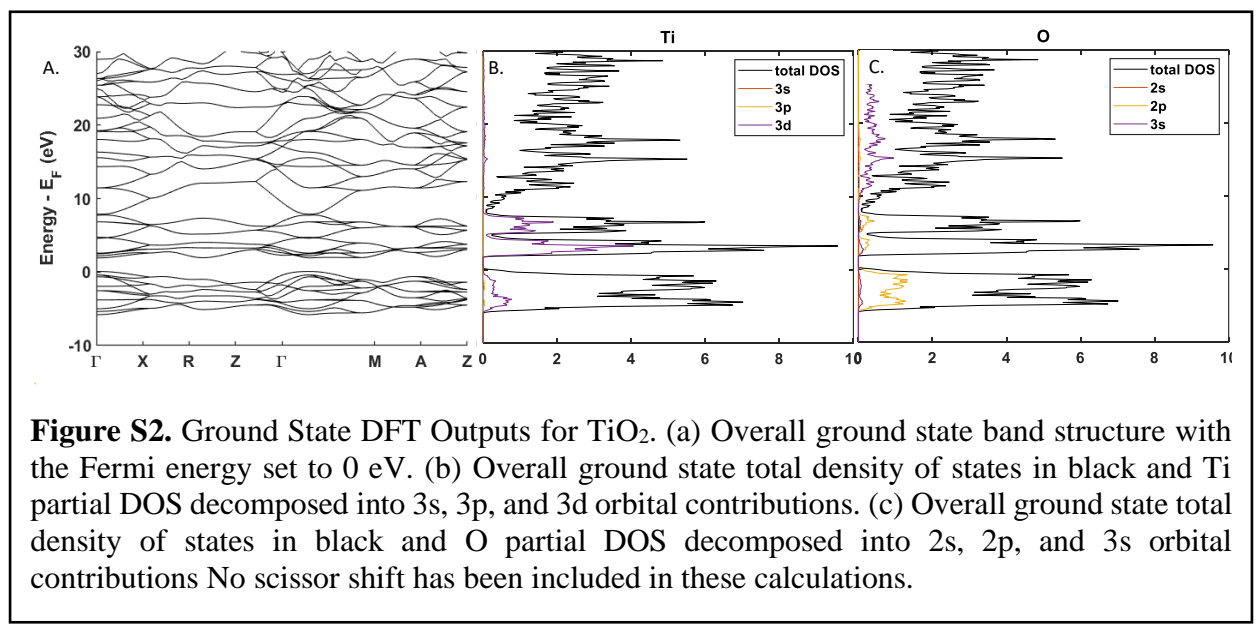

**Figure S2.** Ground State DFT Outputs for $TiO_2$. (a) Overall ground state band structure with the Fermi energy set to 0 eV. (b) Overall ground state total density of states in black and Ti partial DOS decomposed into 3s, 3p, and 3d orbital contributions. (c) Overall ground state total density of states in black and O partial DOS decomposed into 2s, 2p, and 3s orbital contributions No scissor shift has been included in these calculations.

## ii. Ground State Spectrum[5]

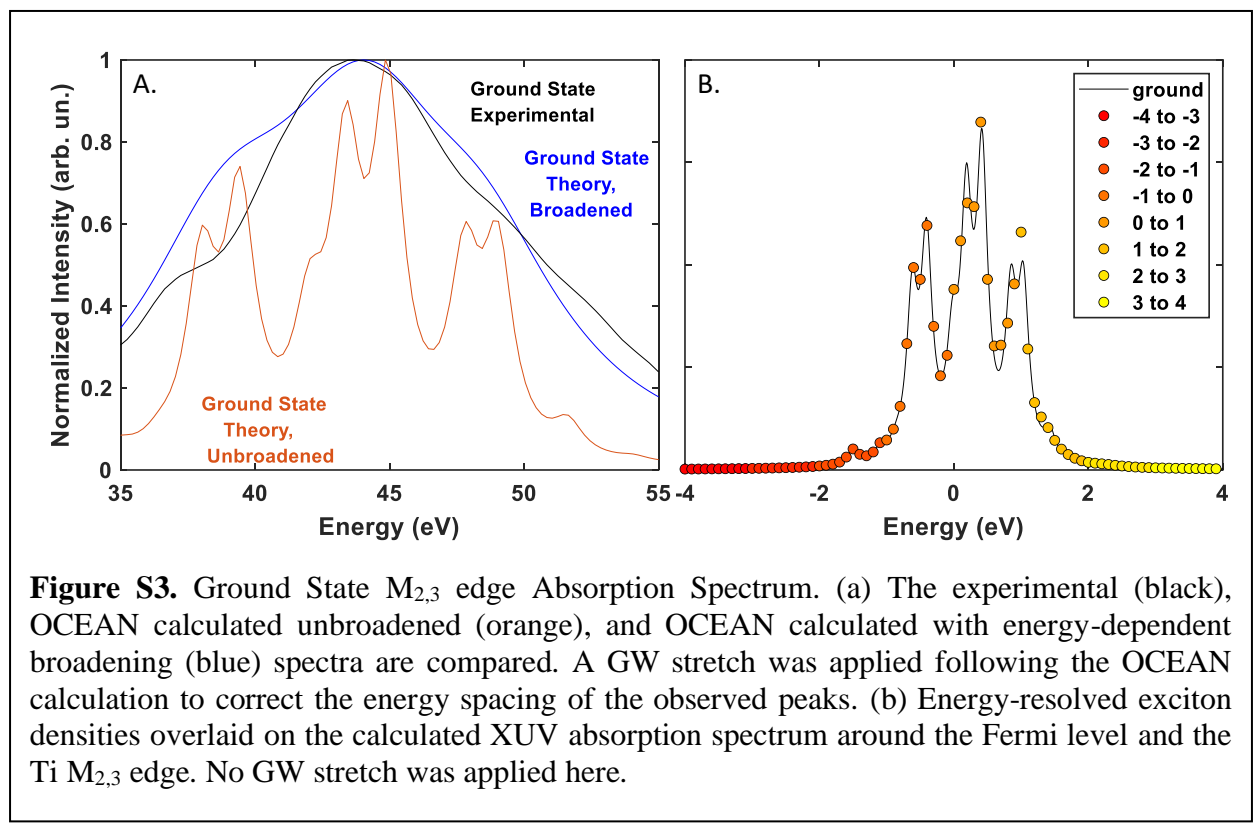

**Figure S3.** Ground State $M_{2,3}$ edge Absorption Spectrum. (a) The experimental (black), OCEAN calculated unbroadened (orange), and OCEAN calculated with energy-dependent broadening (blue) spectra are compared. A GW stretch was applied following the OCEAN calculation to correct the energy spacing of the observed peaks. (b) Energy-resolved exciton densities overlaid on the calculated XUV absorption spectrum around the Fermi level and the Ti $M_{2,3}$ edge. No GW stretch was applied here.

iii. Ground State GMRES Energy Decomposition

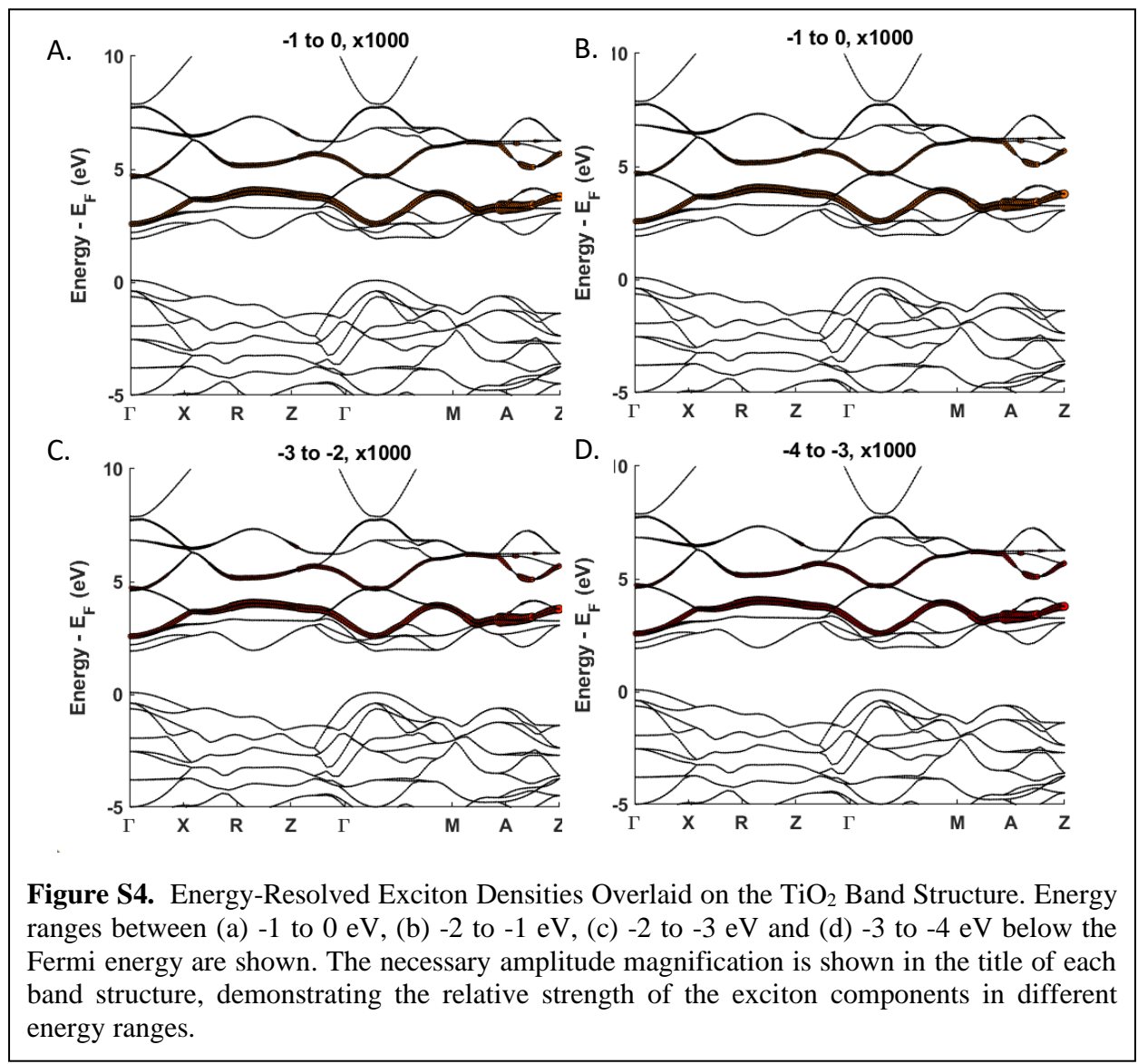

**Figure S4.** Energy-Resolved Exciton Densities Overlaid on the TiO$_2$ Band Structure. Energy ranges between (a) -1 to 0 eV, (b) -2 to -1 eV, (c) -2 to -3 eV and (d) -3 to -4 eV below the Fermi energy are shown. The necessary amplitude magnification is shown in the title of each band structure, demonstrating the relative strength of the exciton components in different energy ranges.

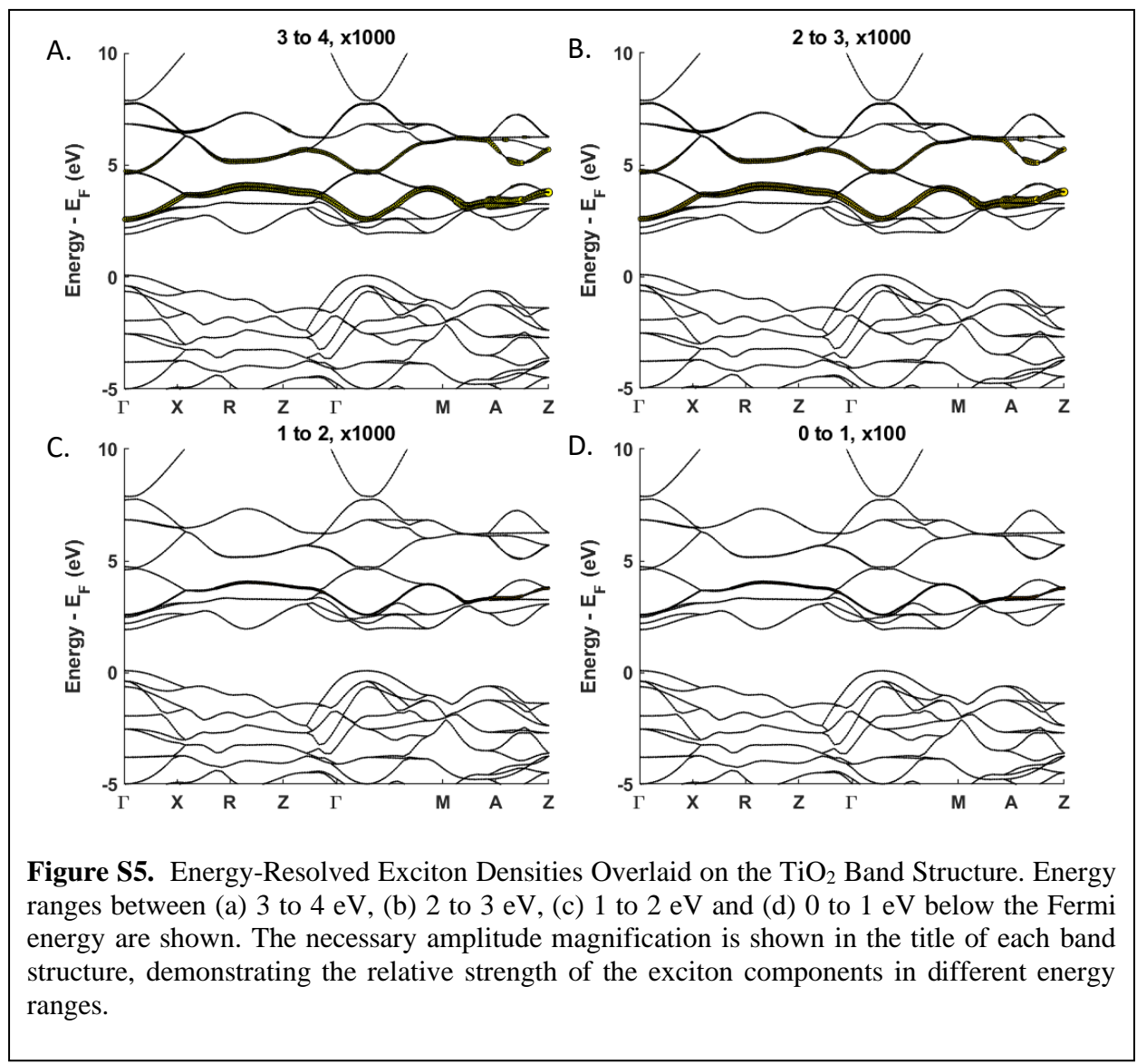

**Figure S5.** Energy-Resolved Exciton Densities Overlaid on the TiO$_2$ Band Structure. Energy ranges between (a) 3 to 4 eV, (b) 2 to 3 eV, (c) 1 to 2 eV and (d) 0 to 1 eV below the Fermi energy are shown. The necessary amplitude magnification is shown in the title of each band structure, demonstrating the relative strength of the exciton components in different energy ranges.

c. Excited State Calculations

    i. State filling band diagrams and Full Spectra

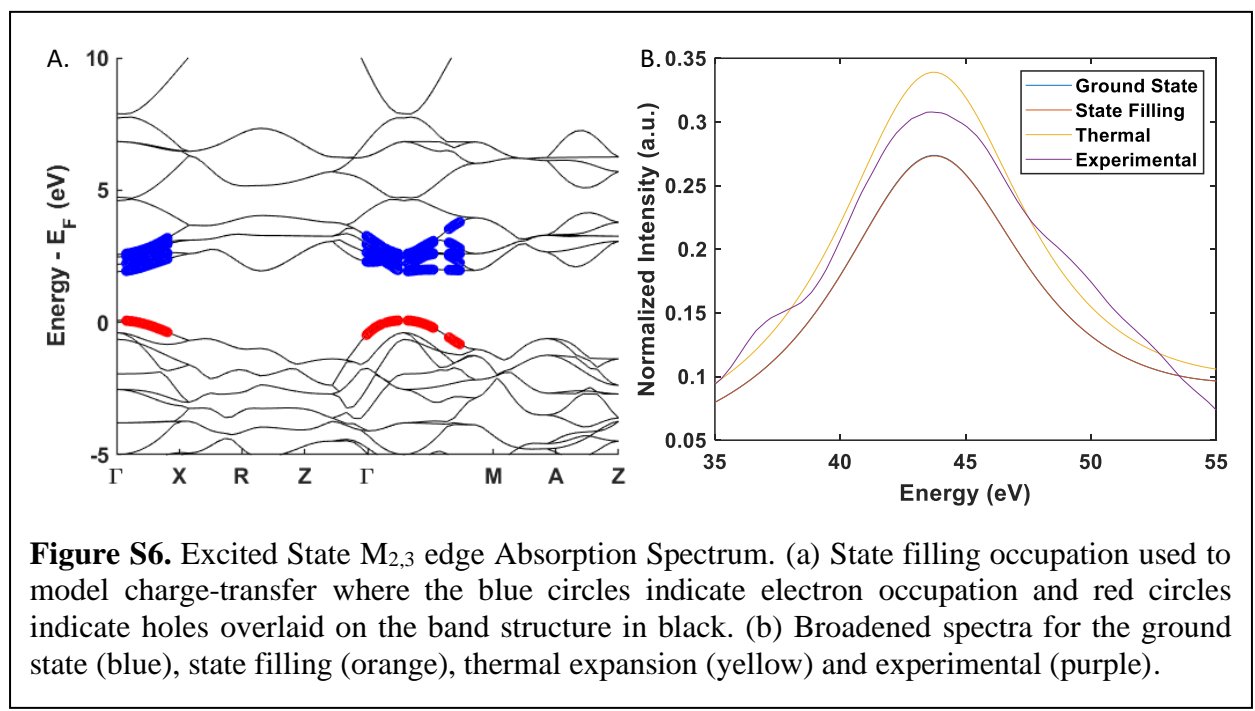

**Figure S6.** Excited State $M_{2,3}$ edge Absorption Spectrum. (a) State filling occupation used to model charge-transfer where the blue circles indicate electron occupation and red circles indicate holes overlaid on the band structure in black. (b) Broadened spectra for the ground state (blue), state filling (orange), thermal expansion (yellow) and experimental (purple).

d. Hamiltonian Decompositions

   i. Total Exciton Comparisons

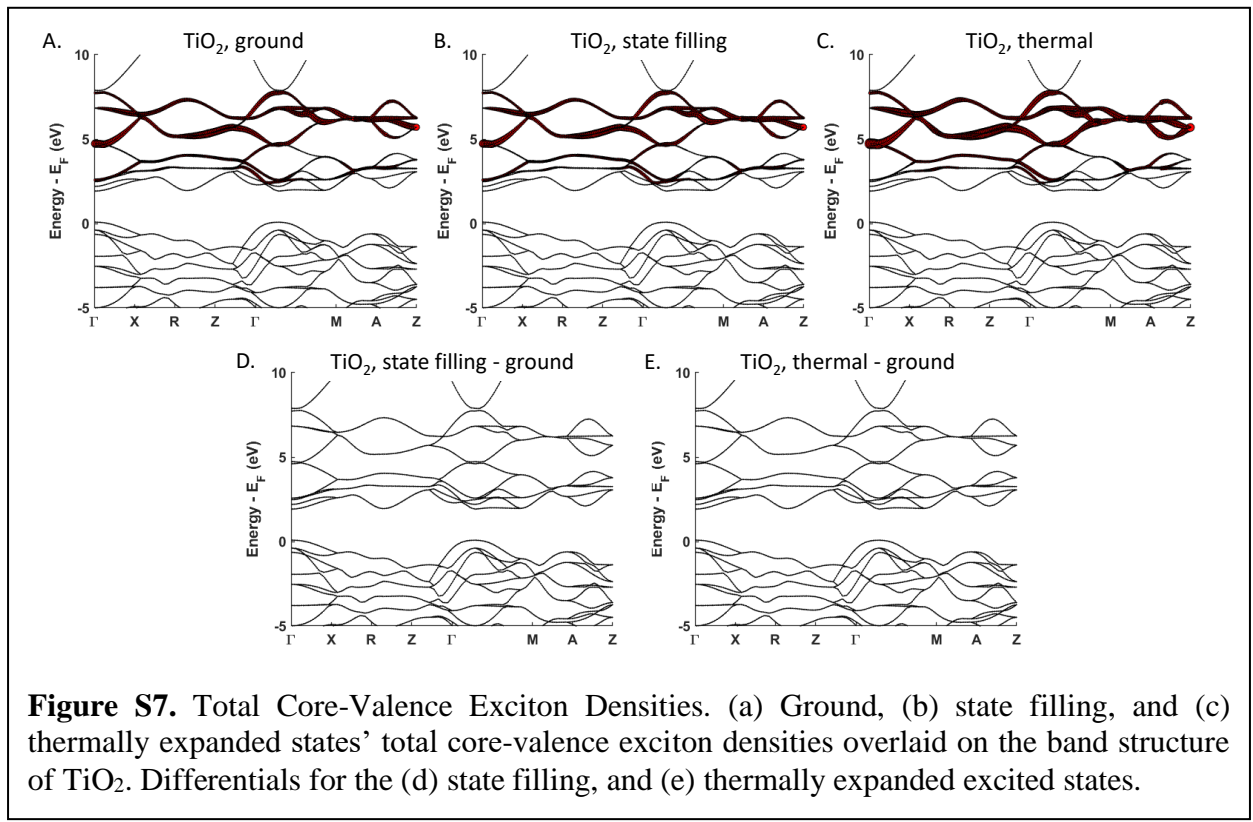

**Figure S7.** Total Core-Valence Exciton Densities. (a) Ground, (b) state filling, and (c) thermally expanded states' total core-valence exciton densities overlaid on the band structure of $TiO_2$. Differentials for the (d) state filling, and (e) thermally expanded excited states.

ii. Hamiltonian Decomposition of Exciton Components for ground, state filling, and thermally expanded models.

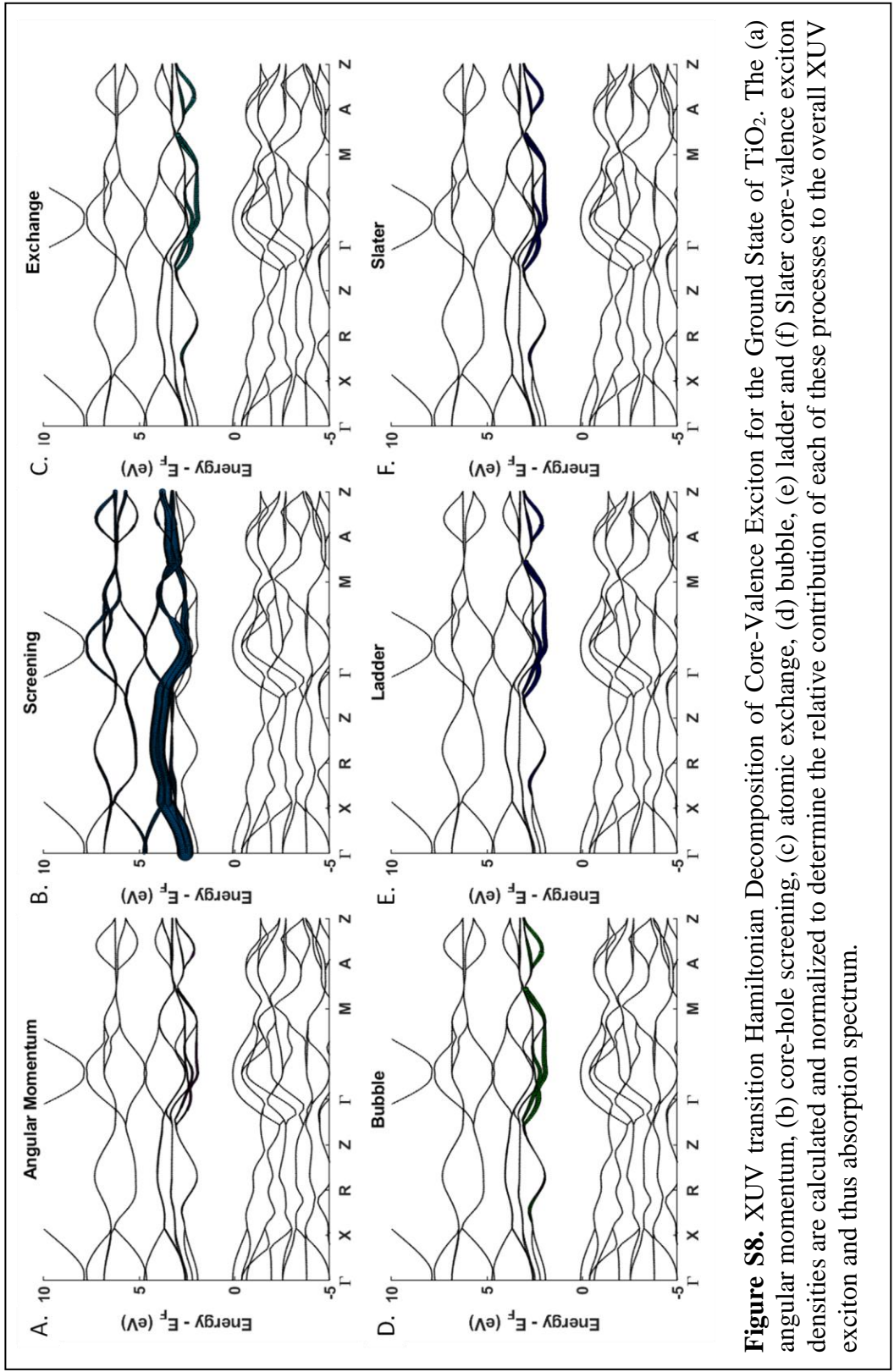

**Figure S8.** XUV transition Hamiltonian Decomposition of Core-Valence Exciton for the Ground State of TiO$_2$. The (a) angular momentum, (b) core-hole screening, (c) atomic exchange, (d) bubble, (e) ladder and (f) Slater core-valence exciton densities are calculated and normalized to determine the relative contribution of each of these processes to the overall XUV exciton and thus absorption spectrum.

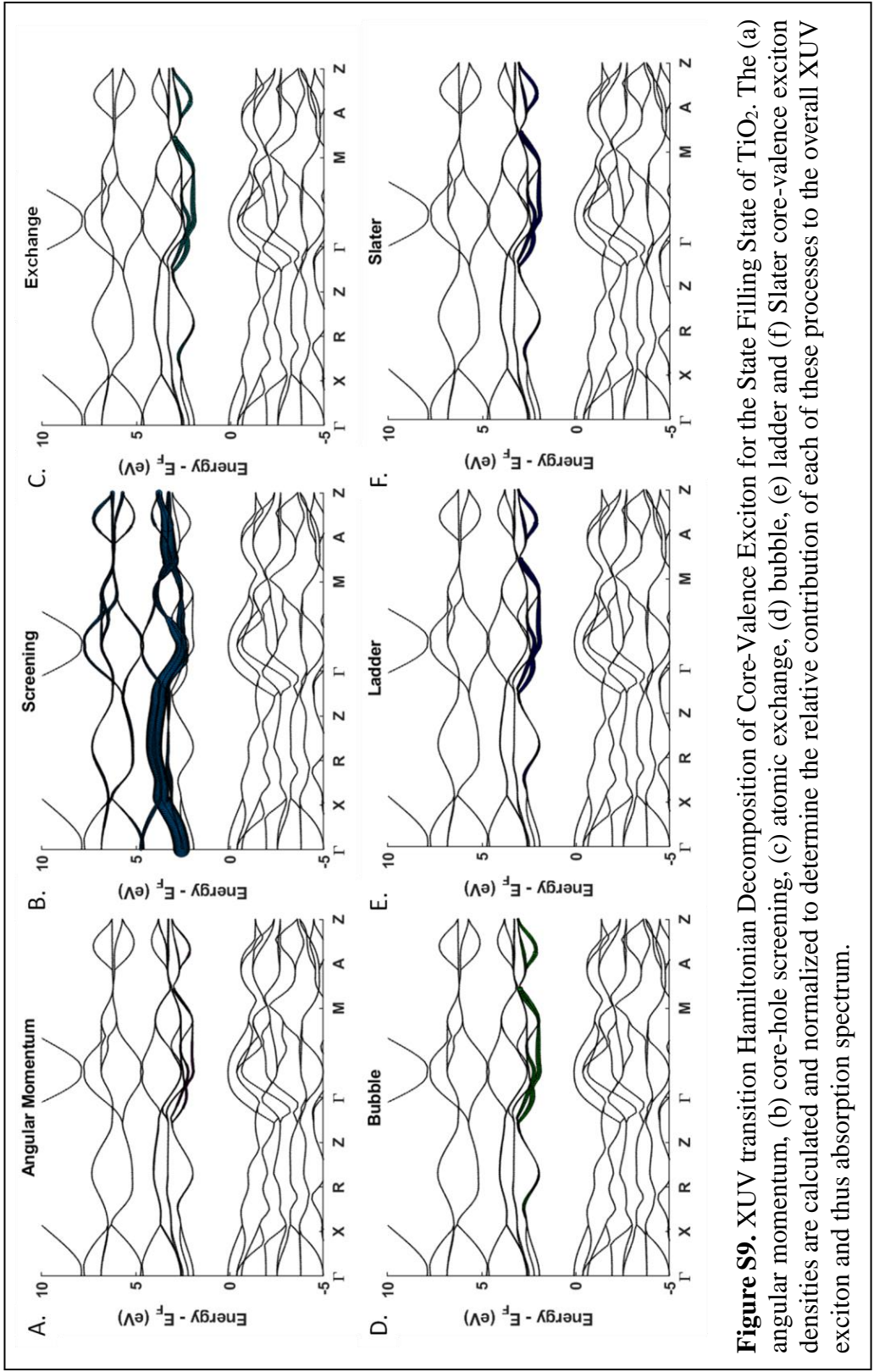

**Figure S9.** XUV transition Hamiltonian Decomposition of Core-Valence Exciton for the State Filling State of TiO$_2$. The (a) angular momentum, (b) core-hole screening, (c) atomic exchange, (d) bubble, (e) ladder and (f) Slater core-valence exciton densities are calculated and normalized to determine the relative contribution of each of these processes to the overall XUV exciton and thus absorption spectrum.

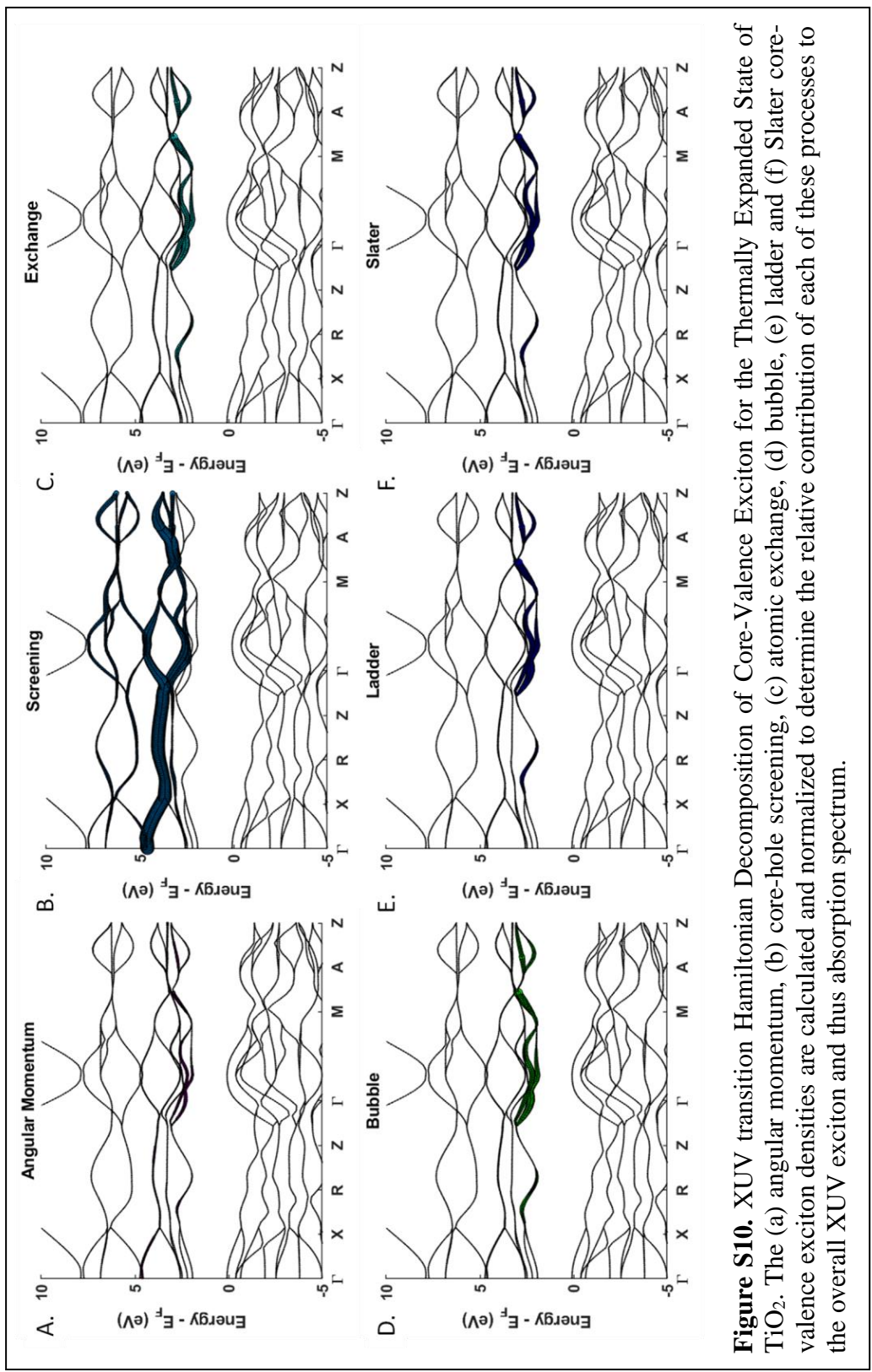

**Figure S10.** XUV transition Hamiltonian Decomposition of Core-Valence Exciton for the Thermally Expanded State of $TiO_2$. The (a) angular momentum, (b) core-hole screening, (c) atomic exchange, (d) bubble, (e) ladder and (f) Slater core-valence exciton densities are calculated and normalized to determine the relative contribution of each of these processes to the overall XUV exciton and thus absorption spectrum.

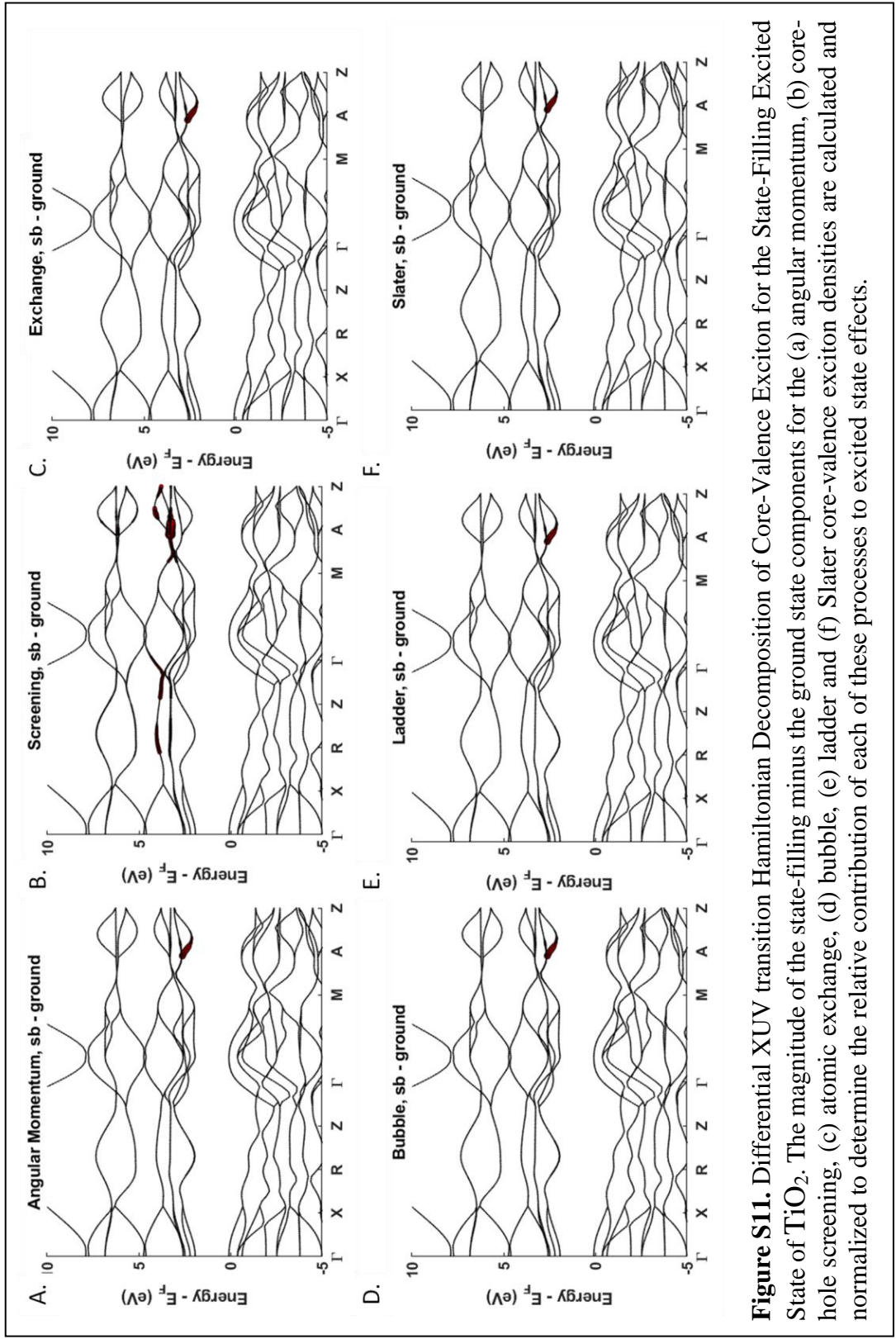

**Figure S11.** Differential XUV transition Hamiltonian Decomposition of Core-Valence Exciton for the State-Filling Excited State of $TiO_2$. The magnitude of the state-filling minus the ground state components for the (a) angular momentum, (b) core-hole screening, (c) atomic exchange, (d) bubble, (e) ladder and (f) Slater core-valence exciton densities are calculated and normalized to determine the relative contribution of each of these processes to excited state effects.

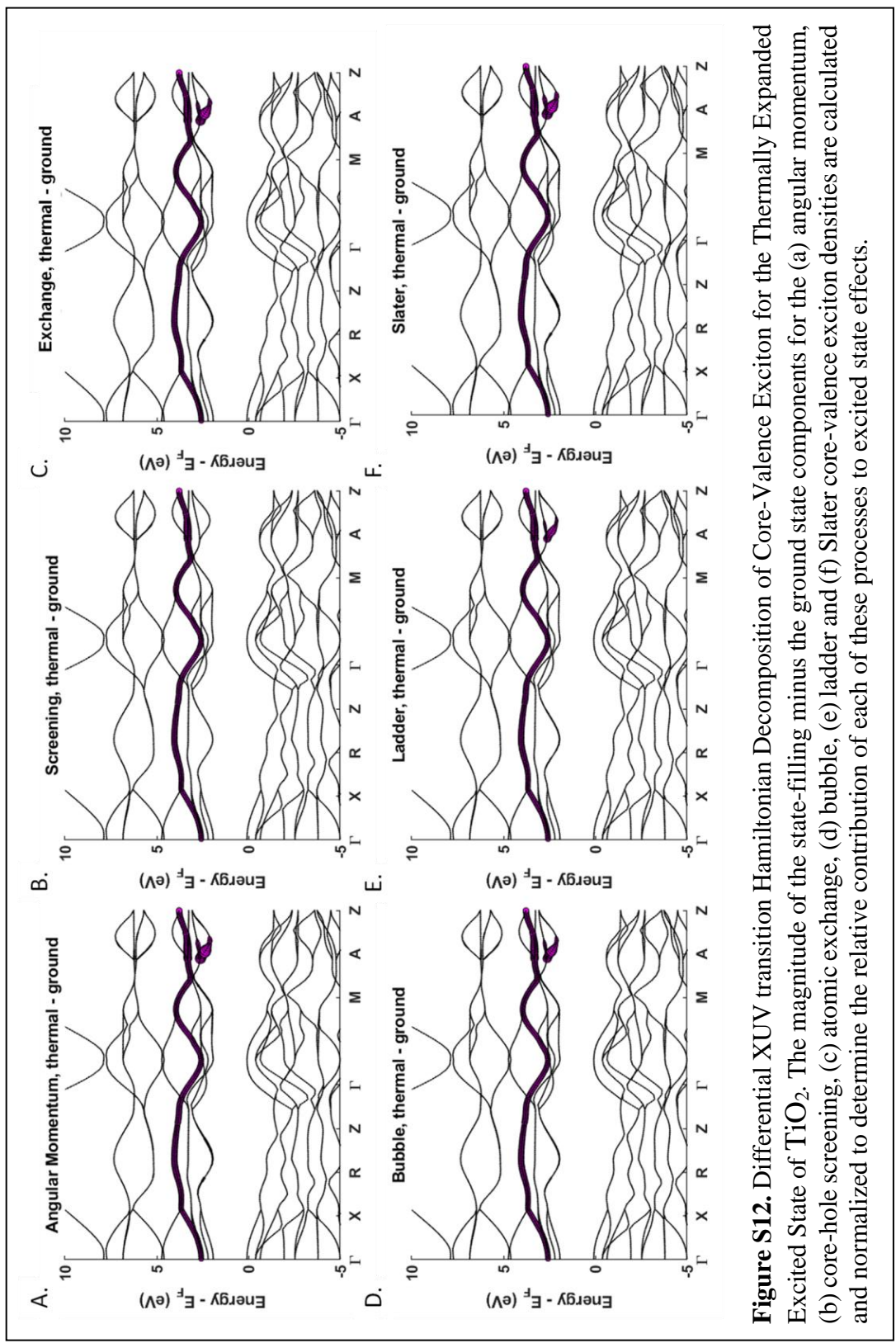

**Figure S12.** Differential XUV transition Hamiltonian Decomposition of Core-Valence Exciton for the Thermally Expanded Excited State of $TiO_2$. The magnitude of the state-filling minus the ground state components for the (a) angular momentum, (b) core-hole screening, (c) atomic exchange, (d) bubble, (e) ladder and (f) Slater core-valence exciton densities are calculated and normalized to determine the relative contribution of each of these processes to excited state effects.

## 2. $Cr_2O_3$

As the only available experimental spectrum of $Cr_2O_3$ was a reflectivity spectrum, the following Matlab code was used to calculate the reflectivity spectrum of $Cr_2O_3$ from the OCEAN calculated absorption spectrum.

```matlab
Cr2O3_exp = importdata('Cr2O3.csv');
files = dir('absspct_Cr_ground.0001_3p_01');
num_cf = size(files,1);
eps1 = 0;
eps2 = 0;
for i = 1:size(files,1)
   [E,eps1i,eps2i] = import_ocean_epsilon(files(i).name);
   eps1 = eps1 + eps1i;
   eps2 = eps2 + eps2i;
end
eps1 = eps1./num_cf;
eps2 = eps2./num_cf;
figure(1);
plot(E,eps1);
hold on;
%plot(energy,eps2);
title('absorption');
theta = 8*pi/180;
eps_sb = eps2 + eps1.*1i;
n_sb = sqrt(eps_sb);
R_sb = abs( (sqrt(1-(1./n_sb.*sin(theta)).^2)-n_sb.*cos(theta))./(sqrt(1-(1./n_sb.*sin(theta)).^2)+n_sb.*cos(theta)) ).^2;
figure(2);
plot(E,R_sb);
hold on;
title('reflection');
```

a. Structural Data for Calculations

i. Ground State and State Blocking

**Unit Cell Parameters (bohr)**

{9.48  9.48  26.72}

**Primitive Vectors**

{0.5  -0.2887  0.333

0    0.5773  0.333

-0.5  -0.2887  0.333}

**Reduced coordinates, ( x, y, z )**

Cr1  0.344311154  0.344311154  0.344311154

Cr1  0.844318706  0.844318706  0.844318706

Cr2  0.155681294  0.155681294  0.155681294

Cr2  0.655688846  0.655688846  0.655688846

O    0.936543866  0.563452102  0.249999406

O    0.563452102  0.249999406  0.936543866

O    0.249999406  0.936543866  0.563452102

O    0.063456134  0.436547898  0.750000594

O    0.436547898  0.750000594  0.063456134

O    0.750000594  0.063456134  0.436547898

ii. Thermally Expanded Lattice

**Unit Cell Parameters (bohr)**

{9.54  9.52  26.89}

**Primitive Vectors**

{0.5  -0.2887  0.333

0    0.5773  0.333

-0.5  -0.2887  0.333}

**Reduced coordinates, ( x, y, z )**

Cr1  0.344311154  0.344311154  0.344311154

Cr1  0.844318706  0.844318706  0.844318706

Cr2  0.155681294  0.155681294  0.155681294

Cr2  0.655688846  0.655688846  0.655688846

O    0.936543866  0.563452102  0.249999406

O    0.563452102  0.249999406  0.936543866

O    0.249999406  0.936543866  0.563452102

O    0.063456134  0.436547898  0.750000594

O    0.436547898  0.750000594  0.063456134

O    0.750000594  0.063456134  0.436547898

## i. Band Structure and DOS

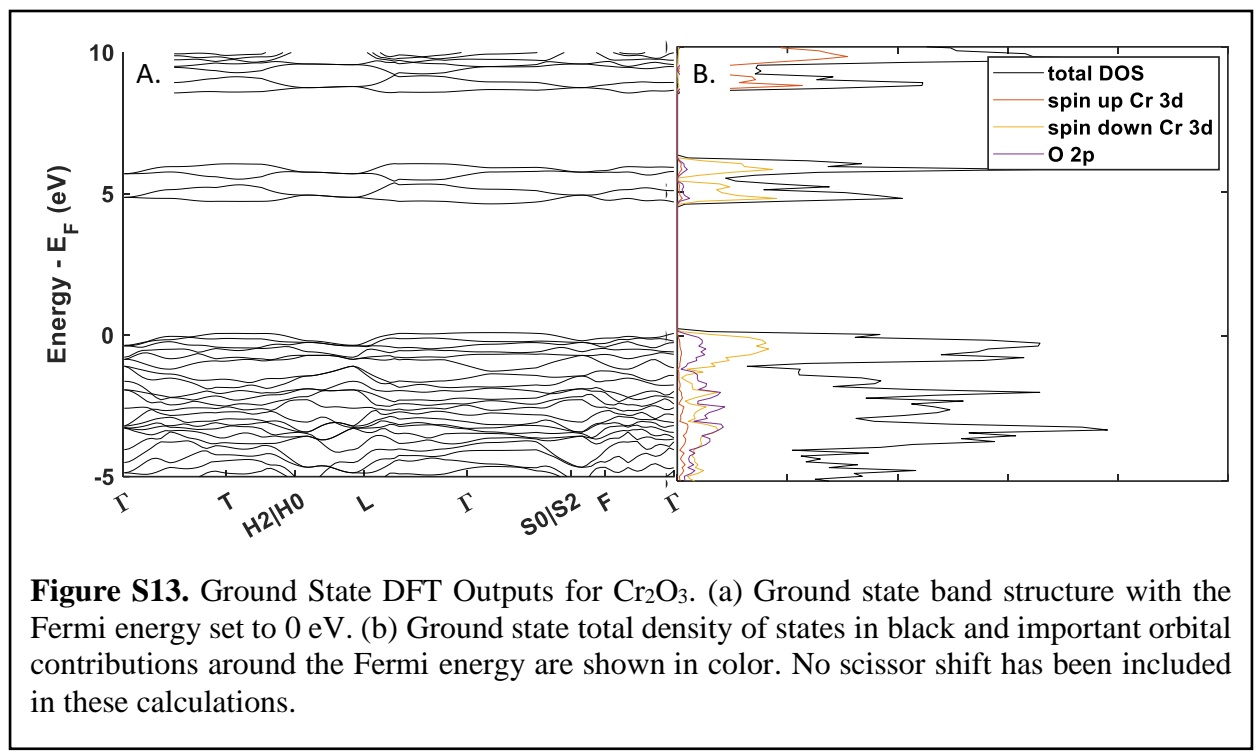

**Figure S13.** Ground State DFT Outputs for $Cr_2O_3$. (a) Ground state band structure with the Fermi energy set to 0 eV. (b) Ground state total density of states in black and important orbital contributions around the Fermi energy are shown in color. No scissor shift has been included in these calculations.

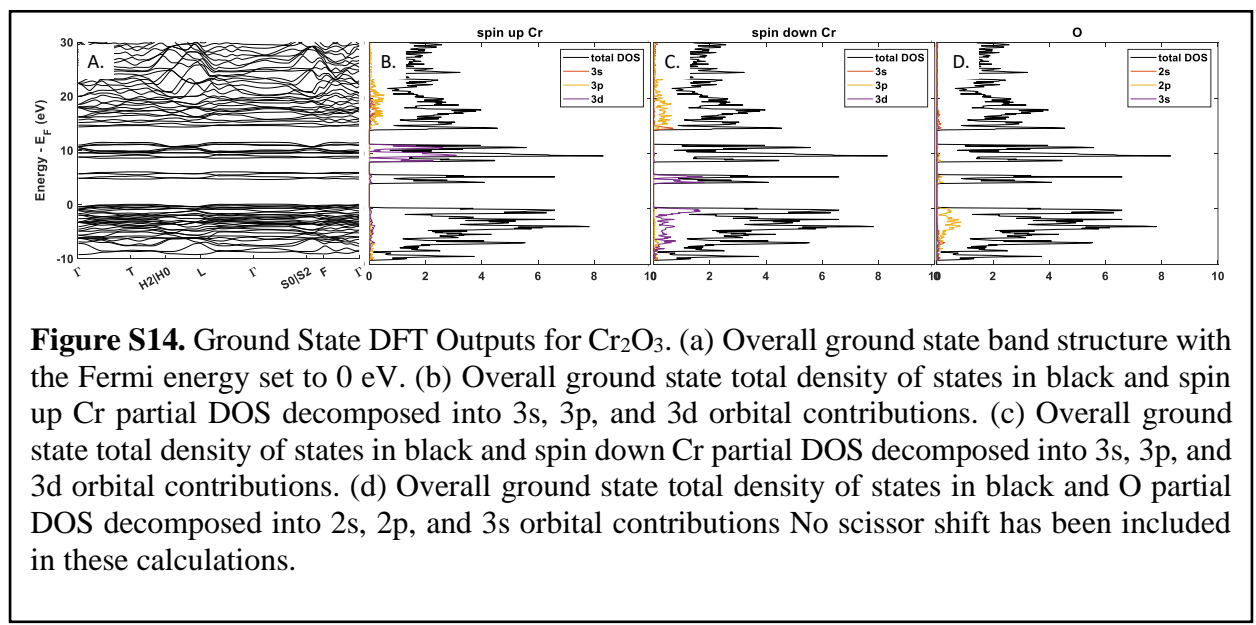

**Figure S14.** Ground State DFT Outputs for $Cr_2O_3$. (a) Overall ground state band structure with the Fermi energy set to 0 eV. (b) Overall ground state total density of states in black and spin up Cr partial DOS decomposed into 3s, 3p, and 3d orbital contributions. (c) Overall ground state total density of states in black and spin down Cr partial DOS decomposed into 3s, 3p, and 3d orbital contributions. (d) Overall ground state total density of states in black and O partial DOS decomposed into 2s, 2p, and 3s orbital contributions No scissor shift has been included in these calculations.

## ii. Ground State Spectrum[6]

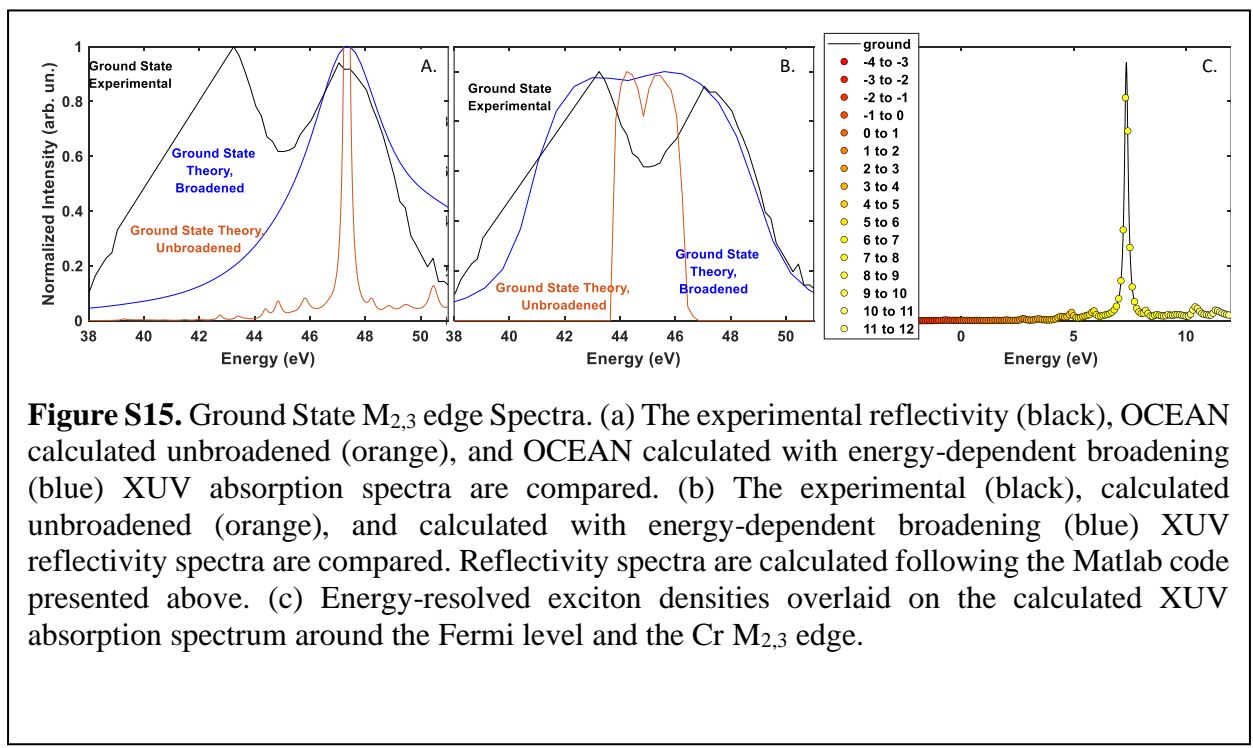

**Figure S15.** Ground State $M_{2,3}$ edge Spectra. (a) The experimental reflectivity (black), OCEAN calculated unbroadened (orange), and OCEAN calculated with energy-dependent broadening (blue) XUV absorption spectra are compared. (b) The experimental (black), calculated unbroadened (orange), and calculated with energy-dependent broadening (blue) XUV reflectivity spectra are compared. Reflectivity spectra are calculated following the Matlab code presented above. (c) Energy-resolved exciton densities overlaid on the calculated XUV absorption spectrum around the Fermi level and the Cr $M_{2,3}$ edge.

iii. Ground State GMRES Energy Decomposition

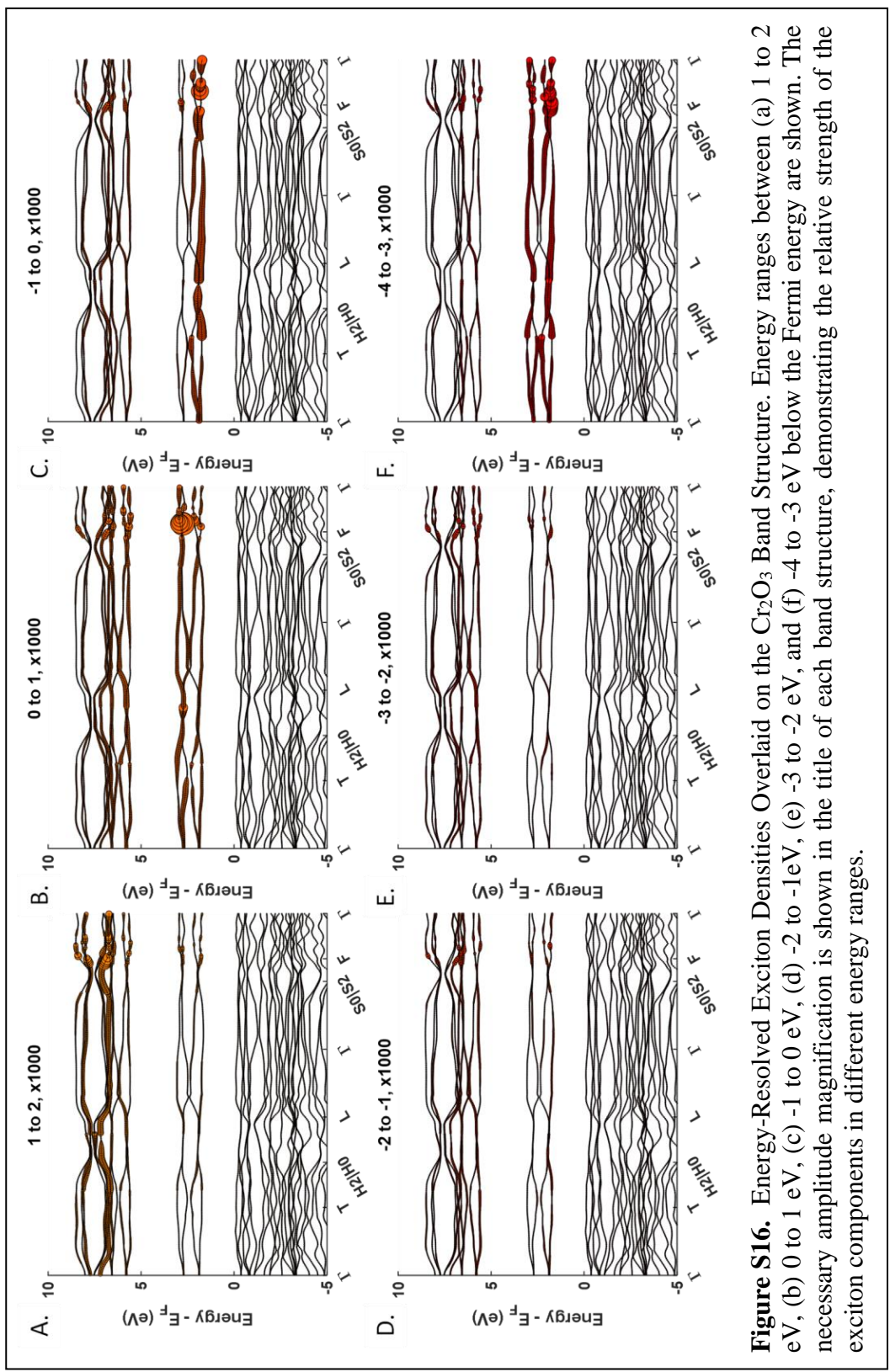

**Figure S16.** Energy-Resolved Exciton Densities Overlaid on the Cr$_2$O$_3$ Band Structure. Energy ranges between (a) 1 to 2 eV, (b) 0 to 1 eV, (c) -1 to 0 eV, (d) -2 to -1eV, (e) -3 to -2 eV, and (f) -4 to -3 eV below the Fermi energy are shown. The necessary amplitude magnification is shown in the title of each band structure, demonstrating the relative strength of the exciton components in different energy ranges.

**Figure S17.** Energy-Resolved Exciton Densities Overlaid on the $Cr_2O_3$ Band Structure. Energy ranges between (a) 11 to 12 eV, (b) 10 to 11 eV, (c) 9 to 10 eV, (d) 8 to 9 eV, (e) 7 to 8 eV, and (f) 6 to 7 eV below the Fermi energy are shown. The necessary amplitude magnification is shown in the title of each band structure, demonstrating the relative strength of the exciton components in different energy ranges.

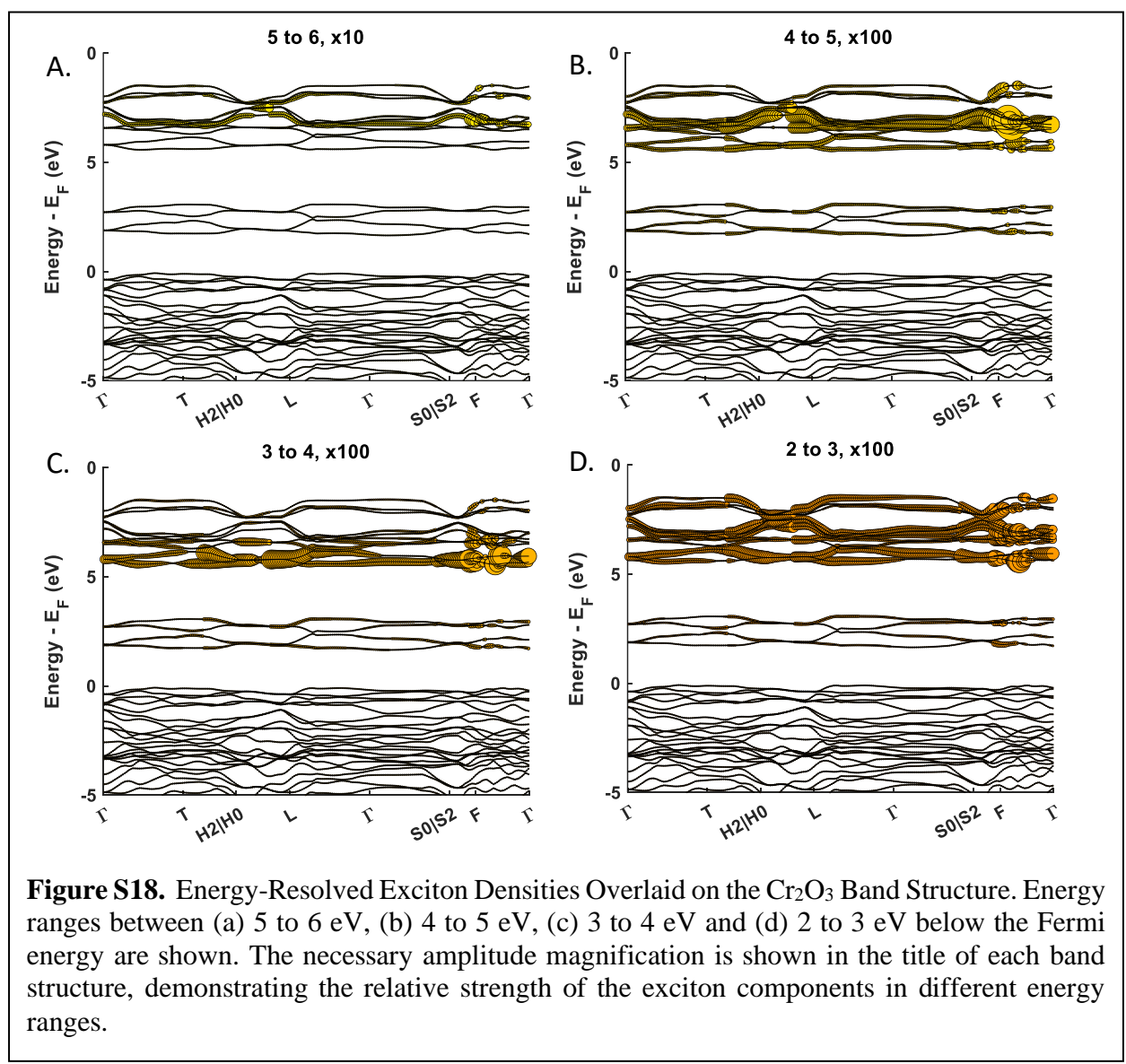

**Figure S18.** Energy-Resolved Exciton Densities Overlaid on the $Cr_2O_3$ Band Structure. Energy ranges between (a) 5 to 6 eV, (b) 4 to 5 eV, (c) 3 to 4 eV and (d) 2 to 3 eV below the Fermi energy are shown. The necessary amplitude magnification is shown in the title of each band structure, demonstrating the relative strength of the exciton components in different energy ranges.

c. Excited State Calculations

    i. State filling band diagrams and Full Spectra

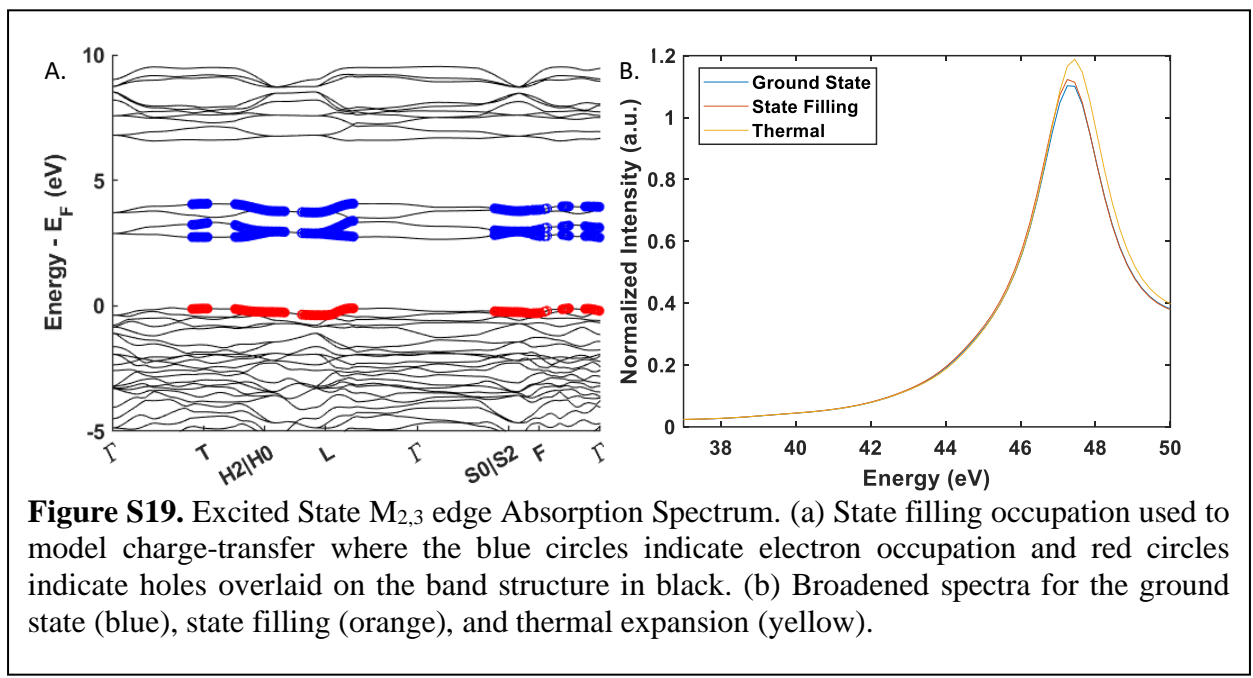

**Figure S19.** Excited State $M_{2,3}$ edge Absorption Spectrum. (a) State filling occupation used to model charge-transfer where the blue circles indicate electron occupation and red circles indicate holes overlaid on the band structure in black. (b) Broadened spectra for the ground state (blue), state filling (orange), and thermal expansion (yellow).

### d. Hamiltonian Decompositions

#### i. Total Exciton Comparisons

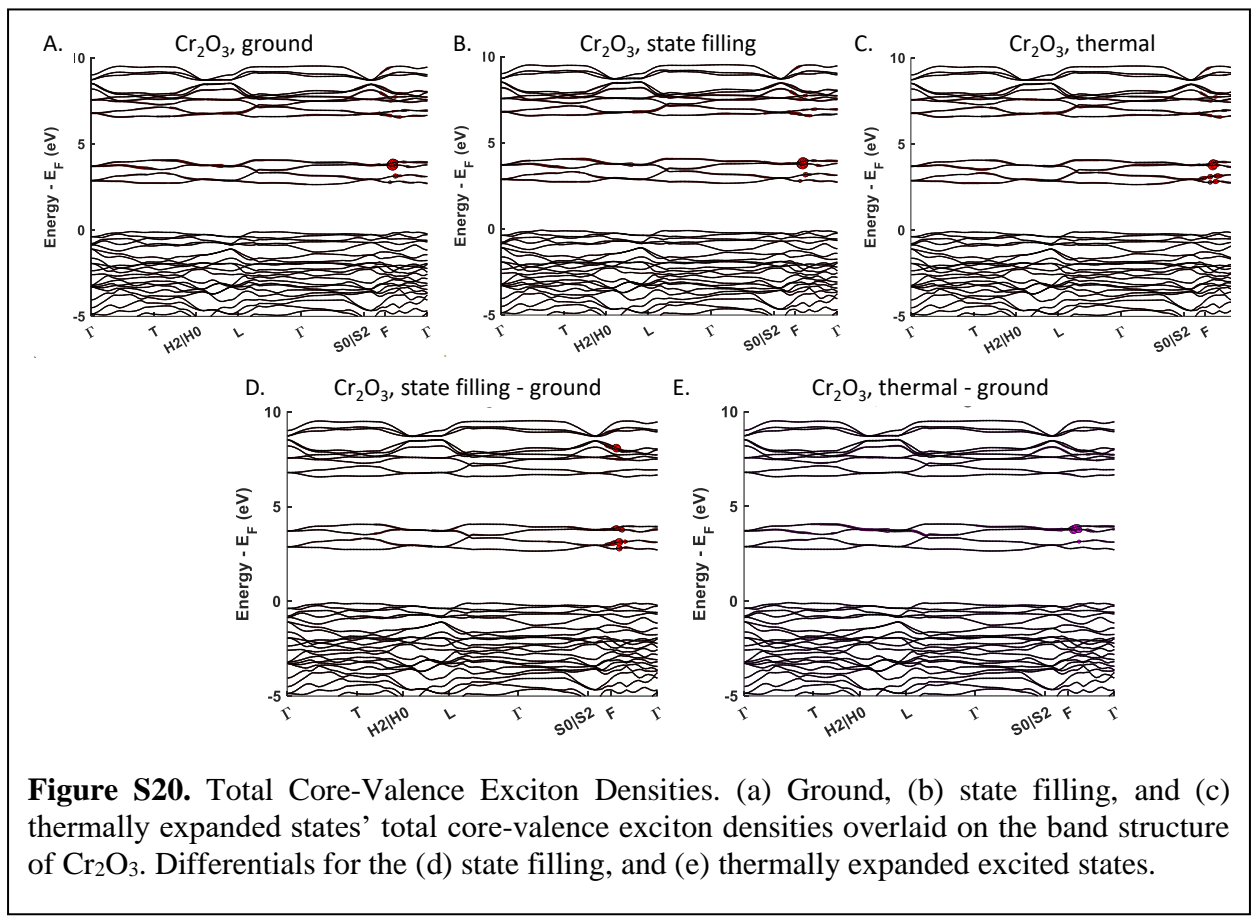

**Figure S20.** Total Core-Valence Exciton Densities. (a) Ground, (b) state filling, and (c) thermally expanded states' total core-valence exciton densities overlaid on the band structure of $Cr_2O_3$. Differentials for the (d) state filling, and (e) thermally expanded excited states.

ii. Hamiltonian Decomposition of Exciton Components for ground, state filling, and thermally expanded models.

**Figure S21.** XUV transition Hamiltonian Decomposition of Core-Valence Exciton for the Ground State of $Cr_2O_3$. The (a) angular momentum, (b) core-hole screening, (c) atomic exchange, (d) bubble, (e) ladder and (f) Slater core-valence exciton densities are calculated and normalized to determine the relative contribution of each of these processes to the overall XUV exciton and thus absorption spectrum.

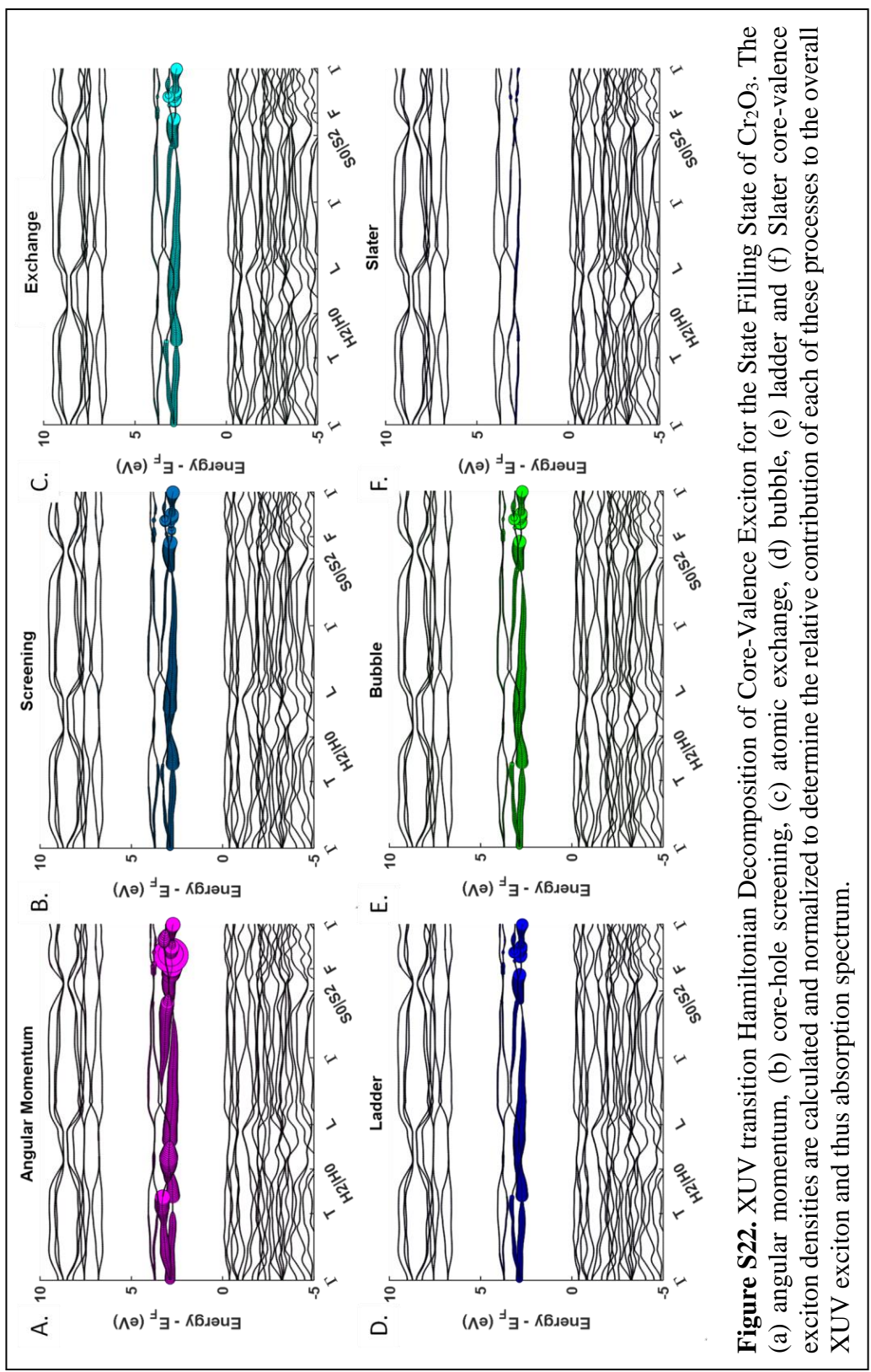

**Figure S22.** XUV transition Hamiltonian Decomposition of Core-Valence Exciton for the State Filling State of $Cr_2O_3$. The (a) angular momentum, (b) core-hole screening, (c) atomic exchange, (d) bubble, (e) ladder and (f) Slater core-valence exciton densities are calculated and normalized to determine the relative contribution of each of these processes to the overall XUV exciton and thus absorption spectrum.

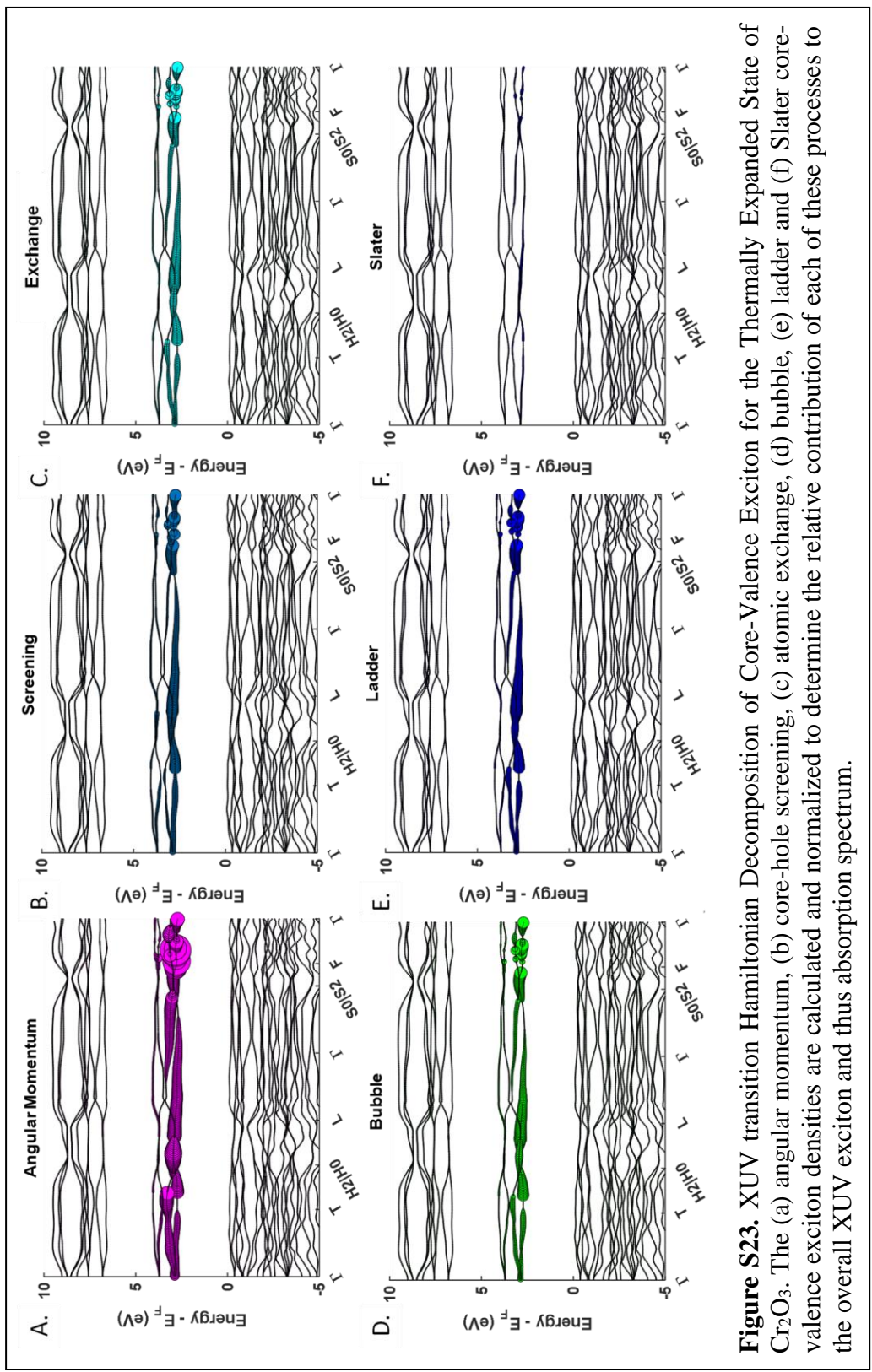

**Figure S23.** XUV transition Hamiltonian Decomposition of Core-Valence Exciton for the Thermally Expanded State of $Cr_2O_3$. The (a) angular momentum, (b) core-hole screening, (c) atomic exchange, (d) bubble, (e) ladder and (f) Slater core-valence exciton densities are calculated and normalized to determine the relative contribution of each of these processes to the overall XUV exciton and thus absorption spectrum.

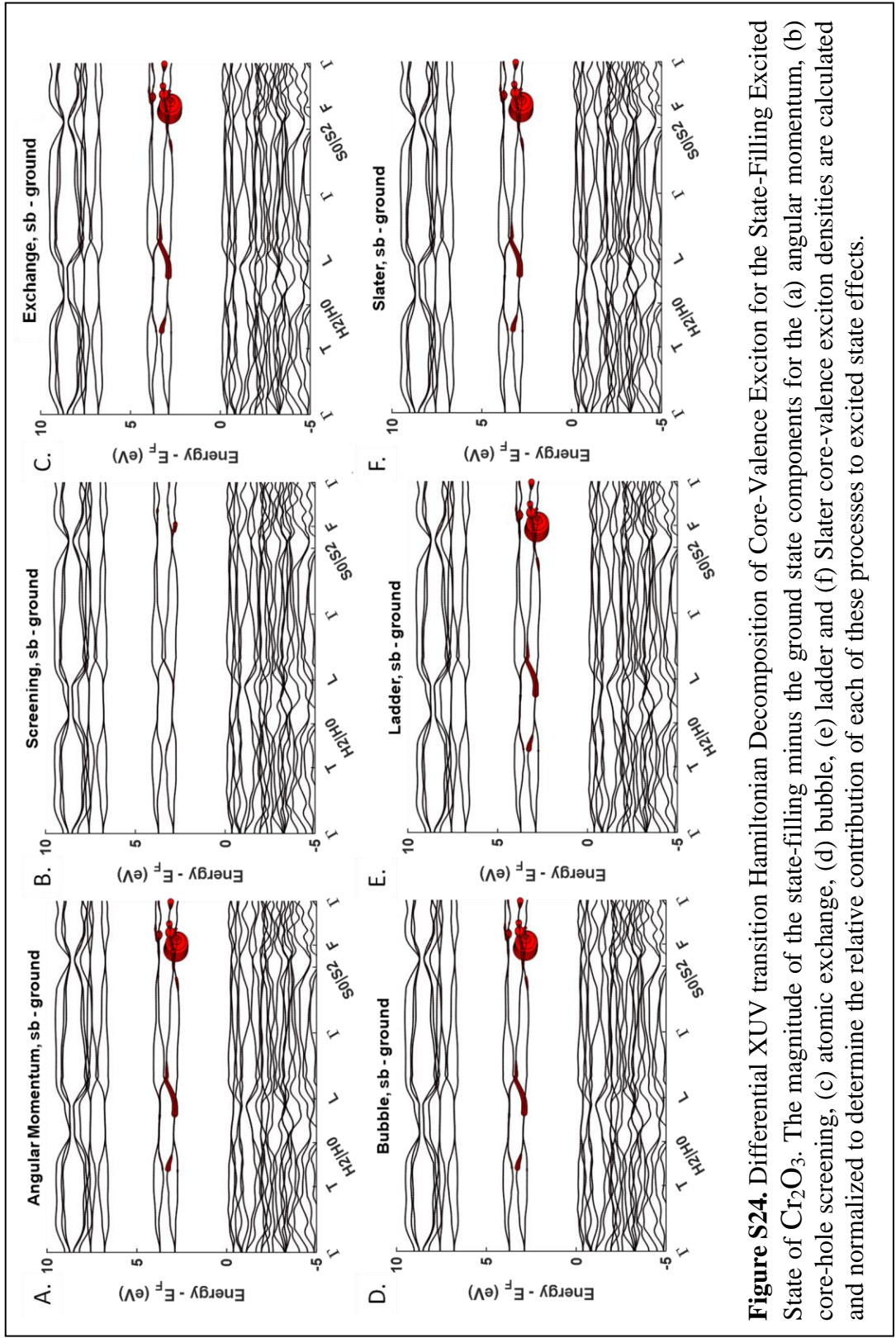

**Figure S24.** Differential XUV transition Hamiltonian Decomposition of Core-Valence Exciton for the State-Filling Excited State of $Cr_2O_3$. The magnitude of the state-filling minus the ground state components for the (a) angular momentum, (b) core-hole screening, (c) atomic exchange, (d) bubble, (e) ladder and (f) Slater core-valence exciton densities are calculated and normalized to determine the relative contribution of each of these processes to excited state effects.

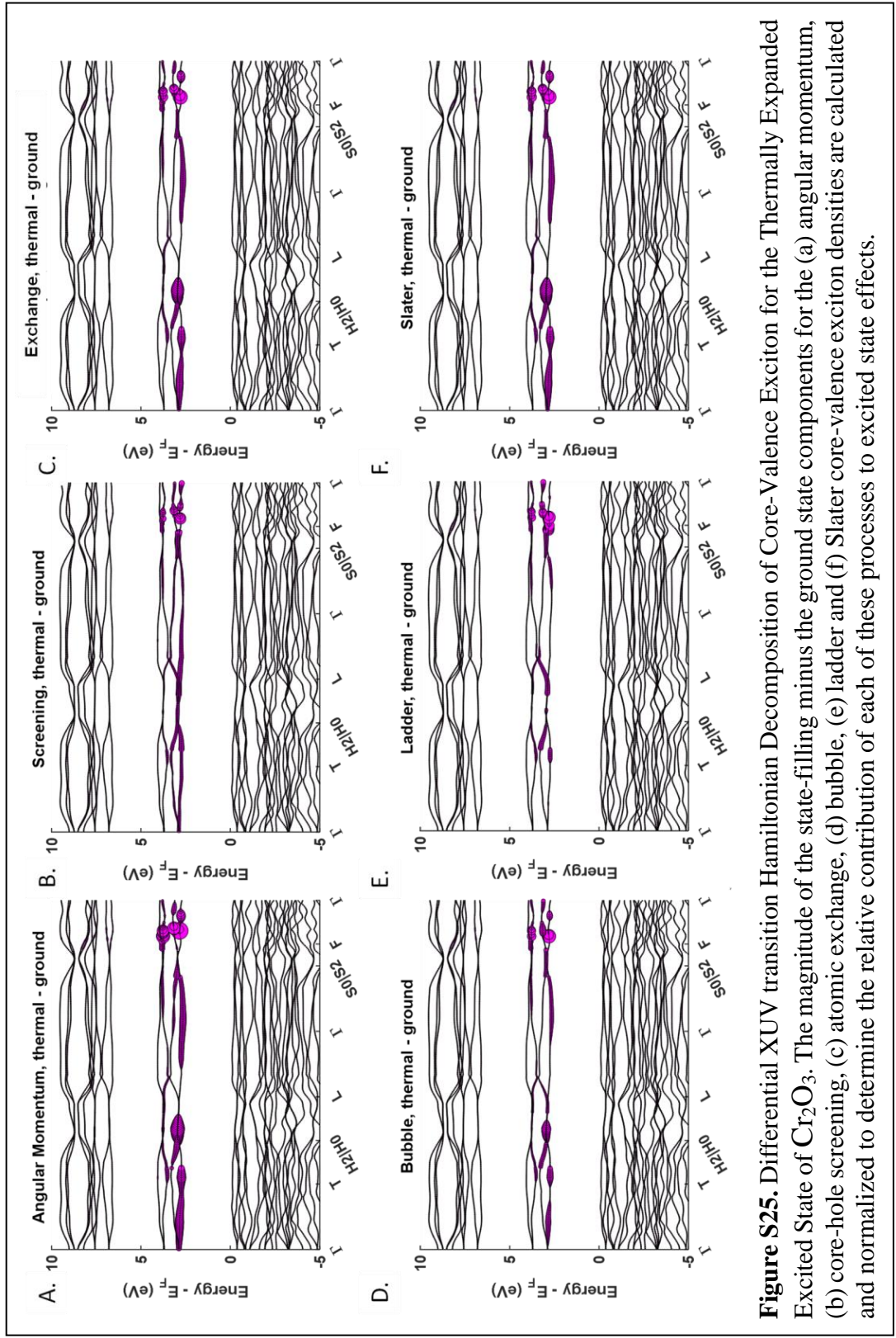

**Figure S25.** Differential XUV transition Hamiltonian Decomposition of Core-Valence Exciton for the Thermally Expanded Excited State of $Cr_2O_3$. The magnitude of the state-filling minus the ground state components for the (a) angular momentum, (b) core-hole screening, (c) atomic exchange, (d) bubble, (e) ladder and (f) Slater core-valence exciton densities are calculated and normalized to determine the relative contribution of each of these processes to excited state effects.

## 4. MnO$_2$

a. Structural Data for Calculations

i. Ground State and State Blocking

**Unit Cell Parameters (bohr)**

{8.4518  8.4518  5.4896}

**Primitive Vectors**

{1   0   0

0   1   0

0   0   1}

**Reduced coordinates, ( x, y, z )**

Mn1  0.000000000   0.000000000   0.000000000

Mn2  0.500000000   0.500000000   0.500000000

O    0.293959177   0.293959177   -0.000000000

O    0.203610439   0.796389561   0.500000000

O    0.706040823   0.706040823   -0.000000000

O    0.796389561   0.203610439   0.500000000

ii. Thermally Expanded Lattice

**Unit Cell Parameters (bohr)**

{8.4729 8.4729 5.5033}

**Primitive Vectors**

{1   0   0

0   1   0

0   0   1}

**Reduced coordinates, ( x, y, z )**

Mn1  0.000000000   0.000000000   0.000000000

Mn2  0.500000000   0.500000000   0.500000000

O    0.293959177   0.293959177   -0.000000000

O    0.203610439   0.796389561   0.500000000

O    0.706040823   0.706040823   -0.000000000

O    0.796389561   0.203610439   0.500000000

b. Ground State Calculations

i. Band Structure and DOS

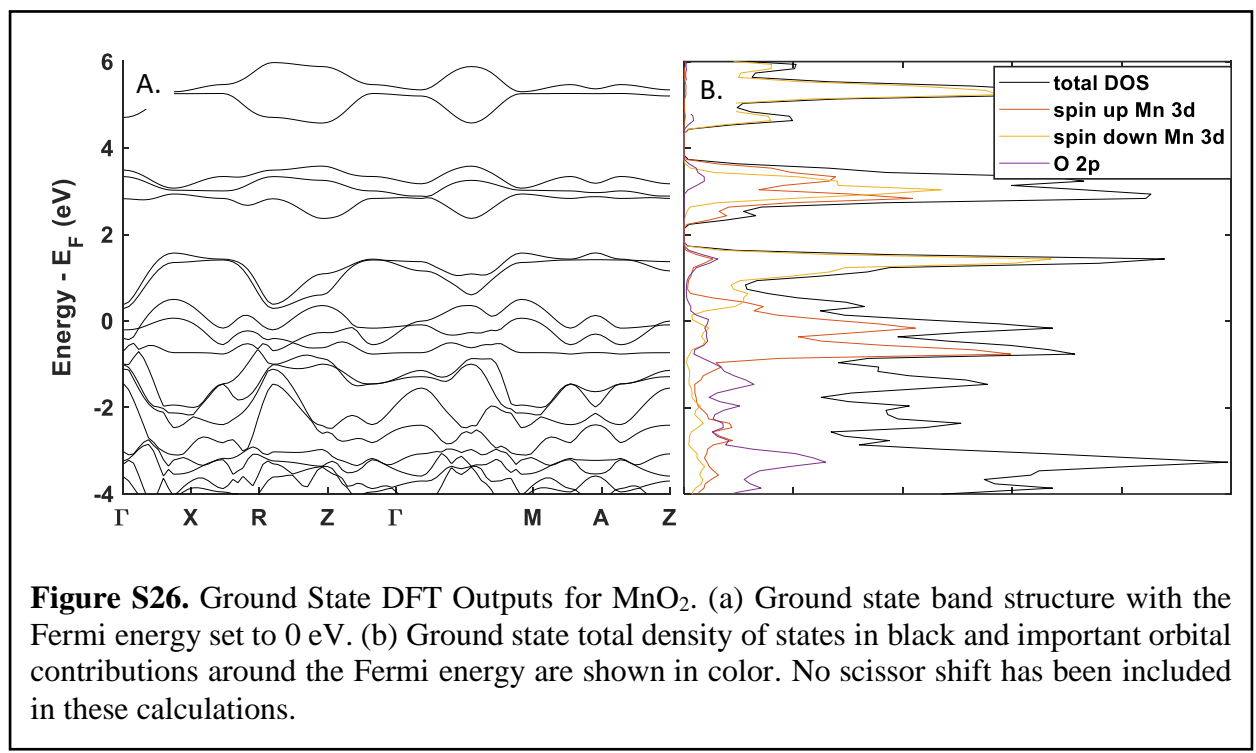

**Figure S26.** Ground State DFT Outputs for $MnO_2$. (a) Ground state band structure with the Fermi energy set to 0 eV. (b) Ground state total density of states in black and important orbital contributions around the Fermi energy are shown in color. No scissor shift has been included in these calculations.

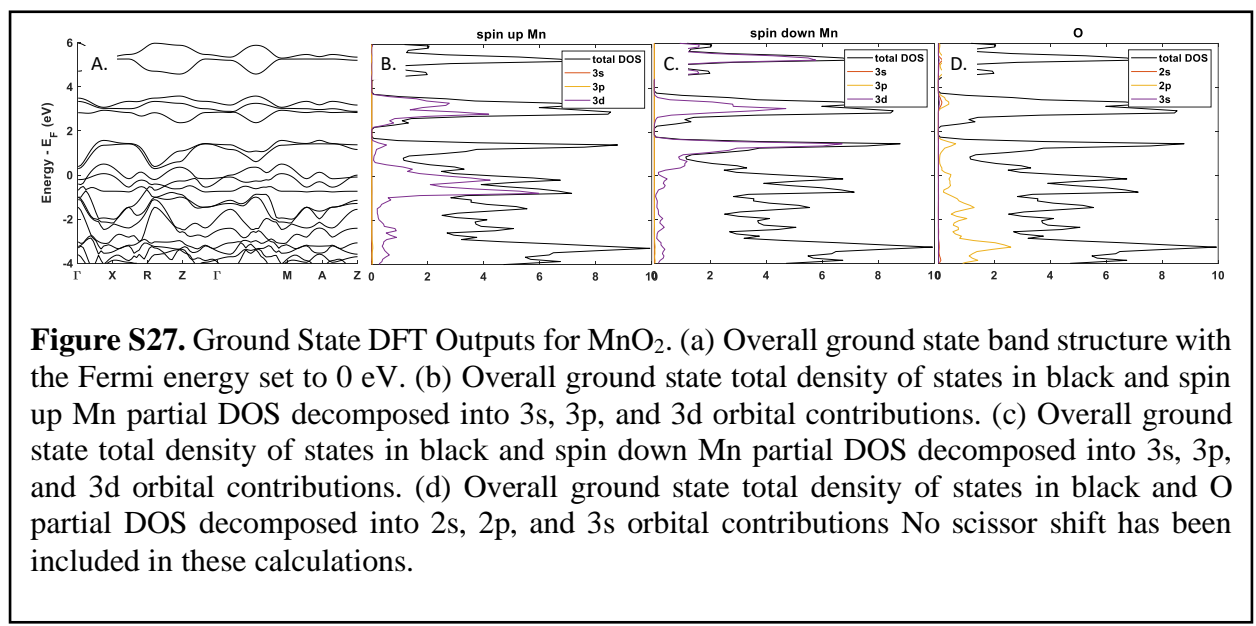

**Figure S27.** Ground State DFT Outputs for $MnO_2$. (a) Overall ground state band structure with the Fermi energy set to 0 eV. (b) Overall ground state total density of states in black and spin up Mn partial DOS decomposed into 3s, 3p, and 3d orbital contributions. (c) Overall ground state total density of states in black and spin down Mn partial DOS decomposed into 3s, 3p, and 3d orbital contributions. (d) Overall ground state total density of states in black and O partial DOS decomposed into 2s, 2p, and 3s orbital contributions No scissor shift has been included in these calculations.

ii. Ground State Spectrum[7]

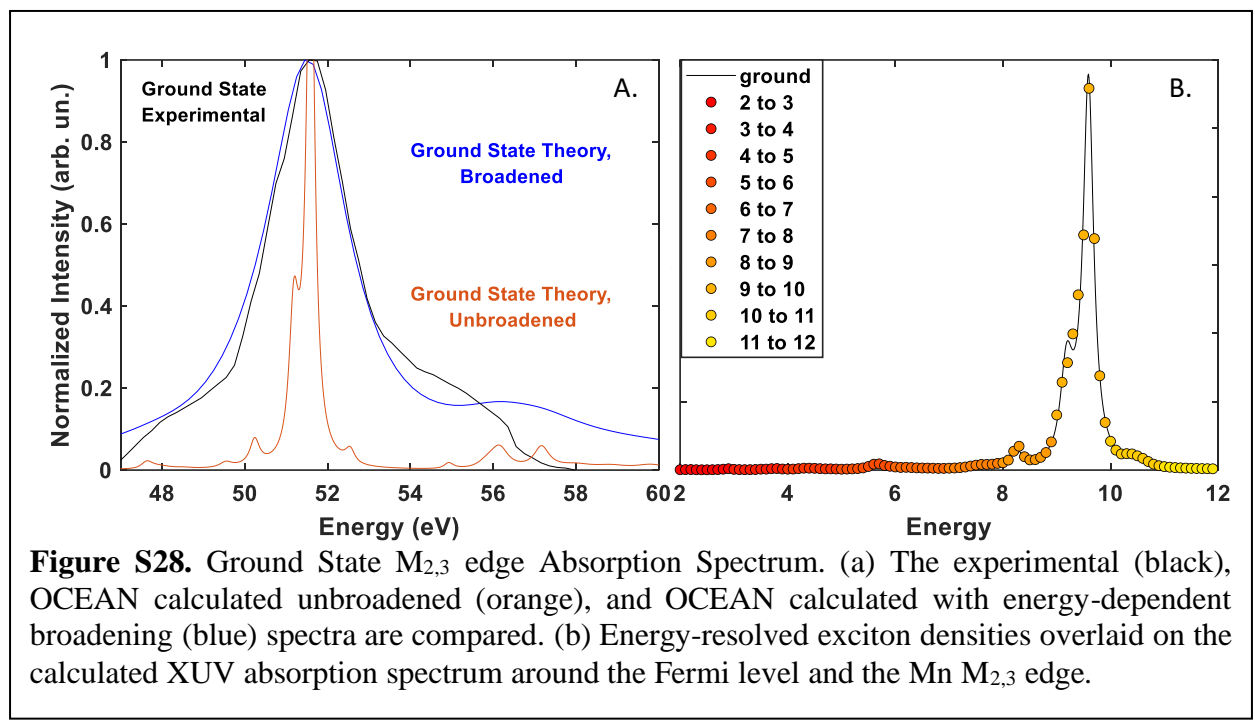

**Figure S28.** Ground State $M_{2,3}$ edge Absorption Spectrum. (a) The experimental (black), OCEAN calculated unbroadened (orange), and OCEAN calculated with energy-dependent broadening (blue) spectra are compared. (b) Energy-resolved exciton densities overlaid on the calculated XUV absorption spectrum around the Fermi level and the Mn $M_{2,3}$ edge.

iii. Ground State GMRES Energy Decomposition

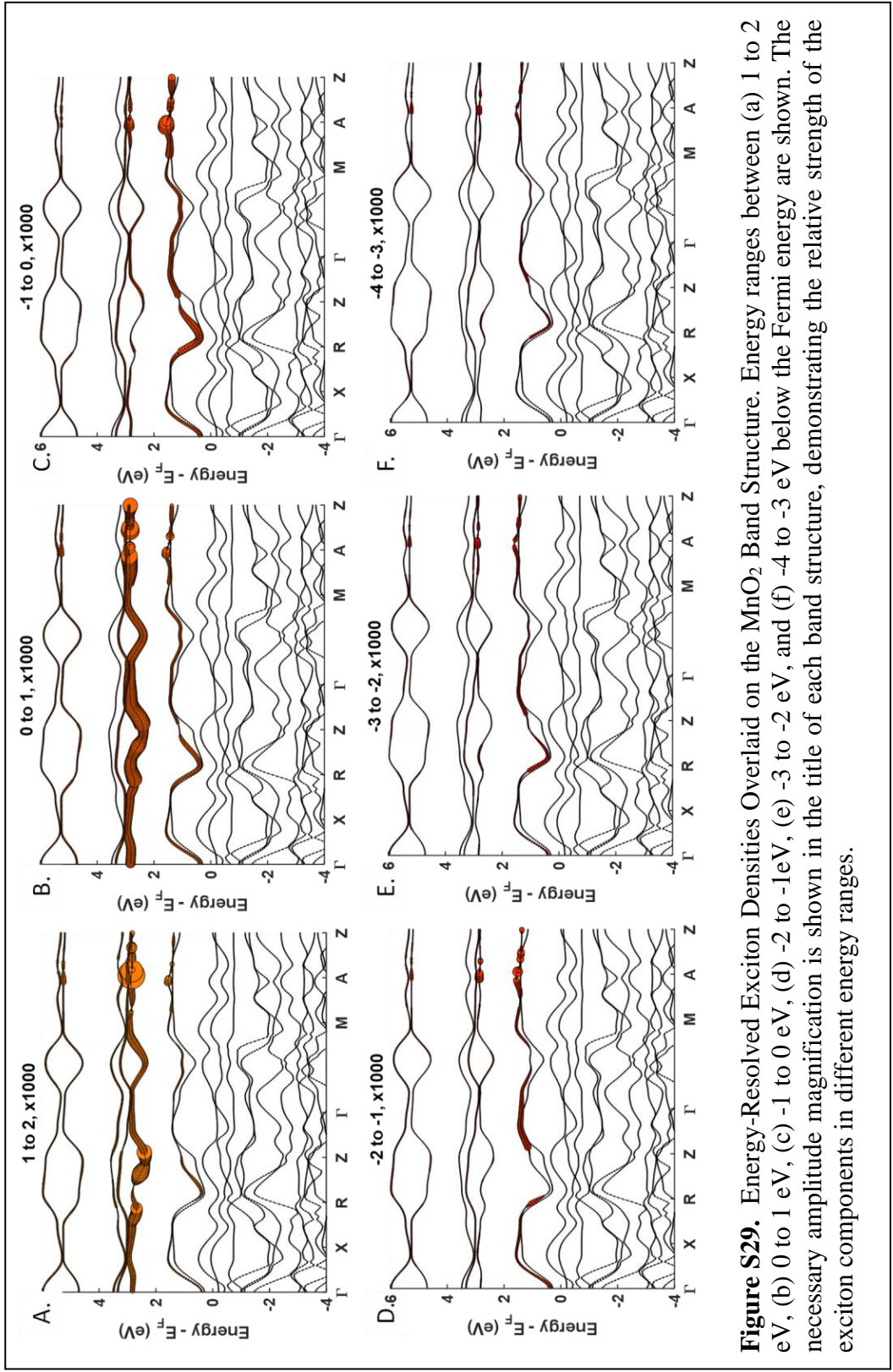

**Figure S29.** Energy-Resolved Exciton Densities Overlaid on the MnO$_2$ Band Structure. Energy ranges between (a) 1 to 2 eV, (b) 0 to 1 eV, (c) -1 to 0 eV, (d) -2 to -1eV, (e) -3 to -2 eV, and (f) -4 to -3 eV below the Fermi energy are shown. The necessary amplitude magnification is shown in the title of each band structure, demonstrating the relative strength of the exciton components in different energy ranges.

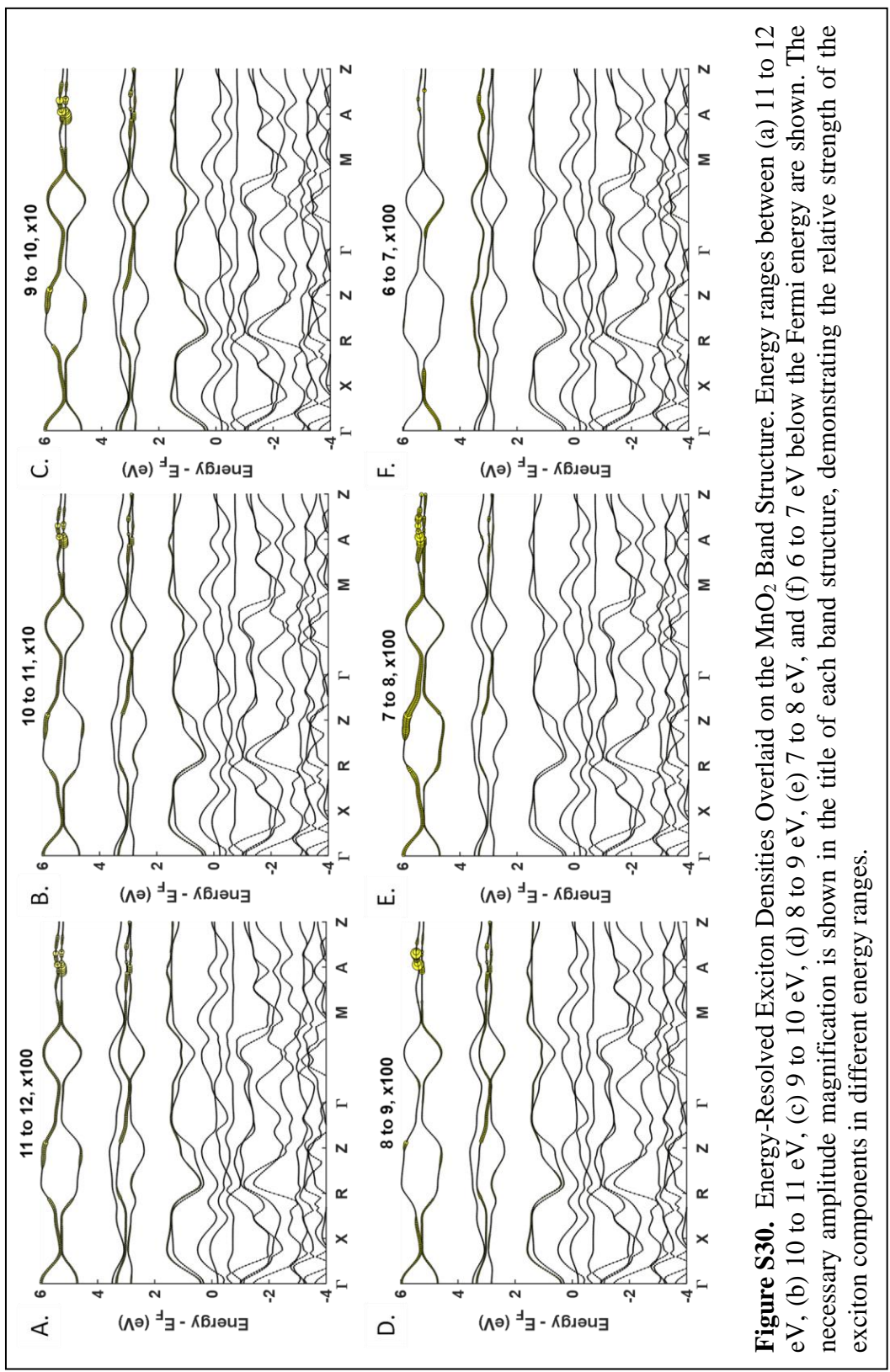

**Figure S30.** Energy-Resolved Exciton Densities Overlaid on the MnO$_2$ Band Structure. Energy ranges between (a) 11 to 12 eV, (b) 10 to 11 eV, (c) 9 to 10 eV, (d) 8 to 9 eV, (e) 7 to 8 eV, and (f) 6 to 7 eV below the Fermi energy are shown. The necessary amplitude magnification is shown in the title of each band structure, demonstrating the relative strength of the exciton components in different energy ranges.

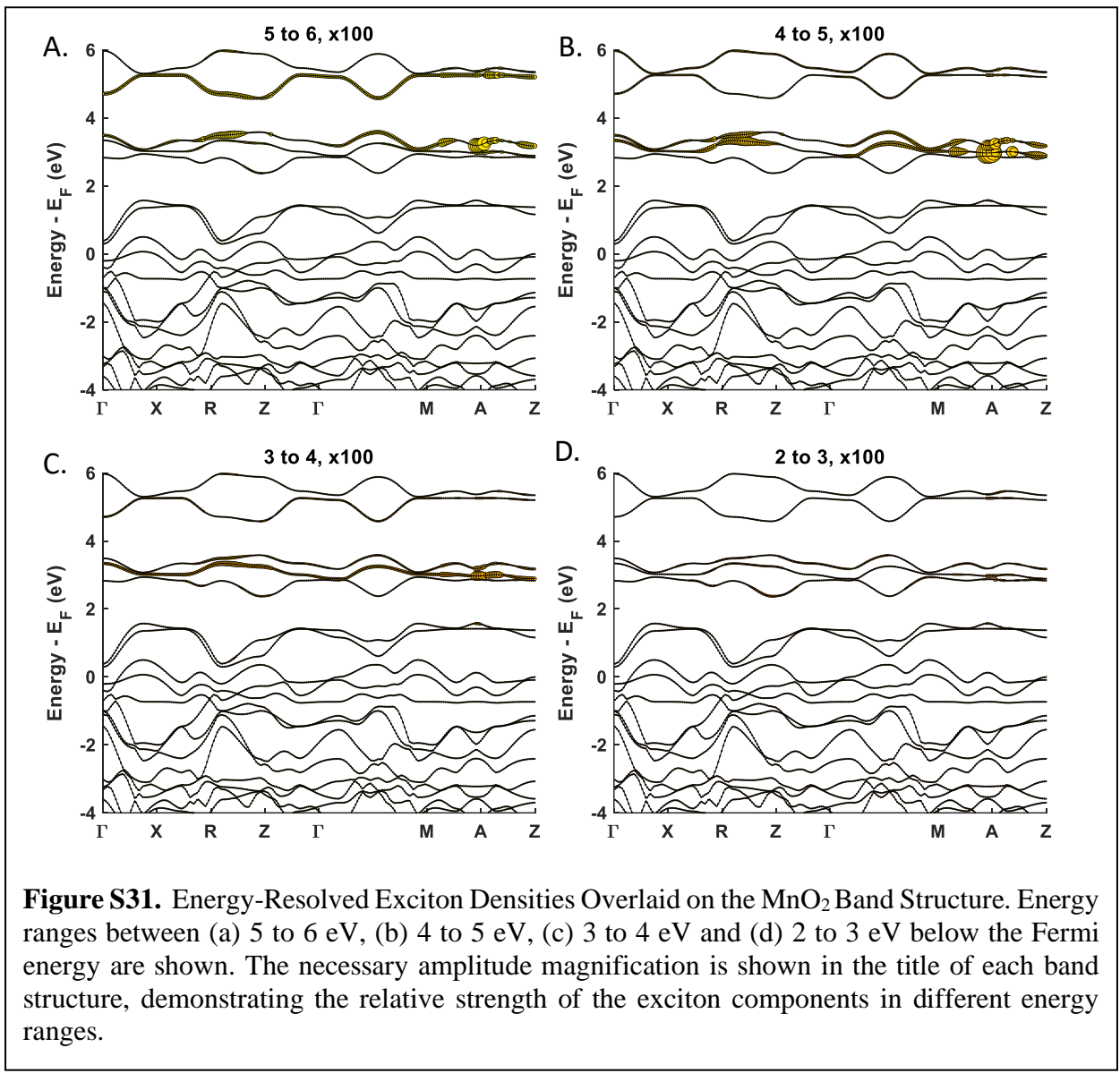

**Figure S31.** Energy-Resolved Exciton Densities Overlaid on the MnO$_2$ Band Structure. Energy ranges between (a) 5 to 6 eV, (b) 4 to 5 eV, (c) 3 to 4 eV and (d) 2 to 3 eV below the Fermi energy are shown. The necessary amplitude magnification is shown in the title of each band structure, demonstrating the relative strength of the exciton components in different energy ranges.

c. Excited State Calculations

   i. State filling band diagrams and Full Spectra

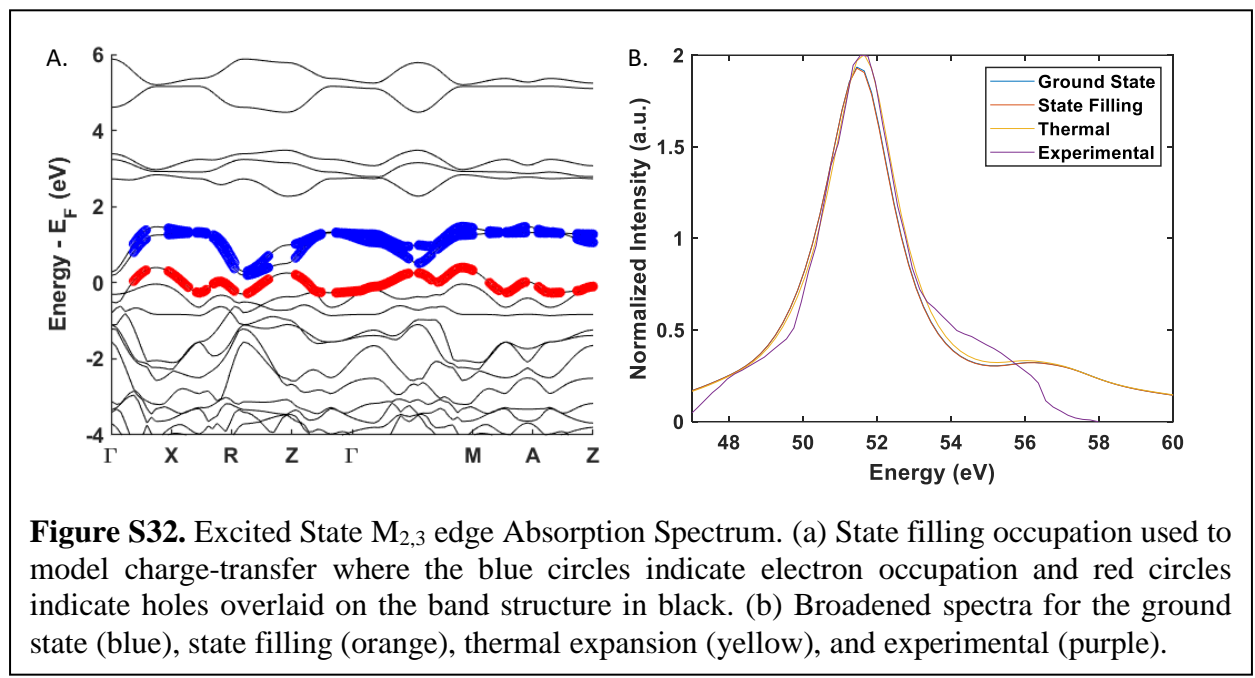

**Figure S32.** Excited State M$_{2,3}$ edge Absorption Spectrum. (a) State filling occupation used to model charge-transfer where the blue circles indicate electron occupation and red circles indicate holes overlaid on the band structure in black. (b) Broadened spectra for the ground state (blue), state filling (orange), thermal expansion (yellow), and experimental (purple).

d. Hamiltonian Decompositions

    i. Total Exciton Comparisons

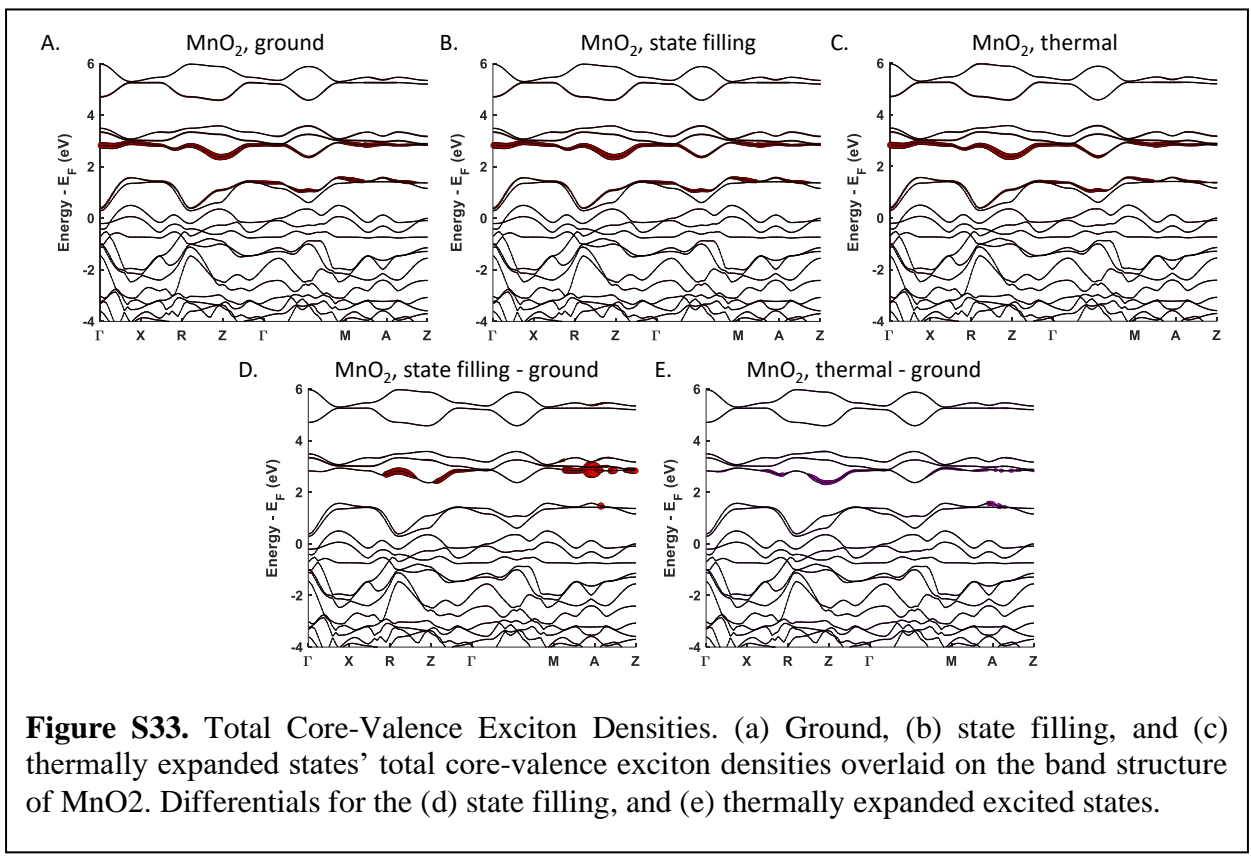

**Figure S33.** Total Core-Valence Exciton Densities. (a) Ground, (b) state filling, and (c) thermally expanded states' total core-valence exciton densities overlaid on the band structure of MnO2. Differentials for the (d) state filling, and (e) thermally expanded excited states.

ii. Hamiltonian Decomposition of Exciton Components for ground, state filling, and thermally expanded models.

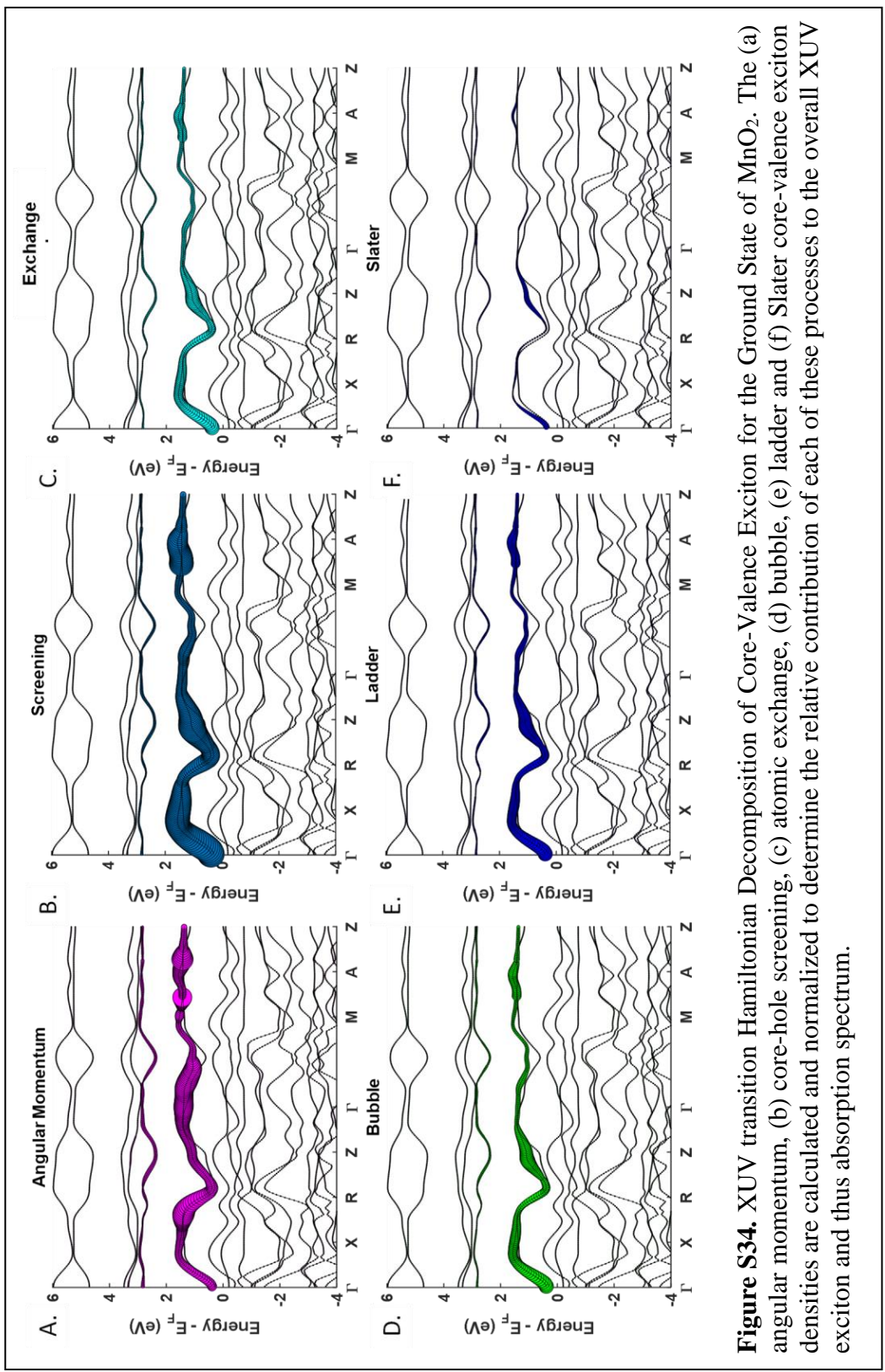

**Figure S34.** XUV transition Hamiltonian Decomposition of Core-Valence Exciton for the Ground State of MnO$_2$. The (a) angular momentum, (b) core-hole screening, (c) atomic exchange, (d) bubble, (e) ladder and (f) Slater core-valence exciton densities are calculated and normalized to determine the relative contribution of each of these processes to the overall XUV exciton and thus absorption spectrum.

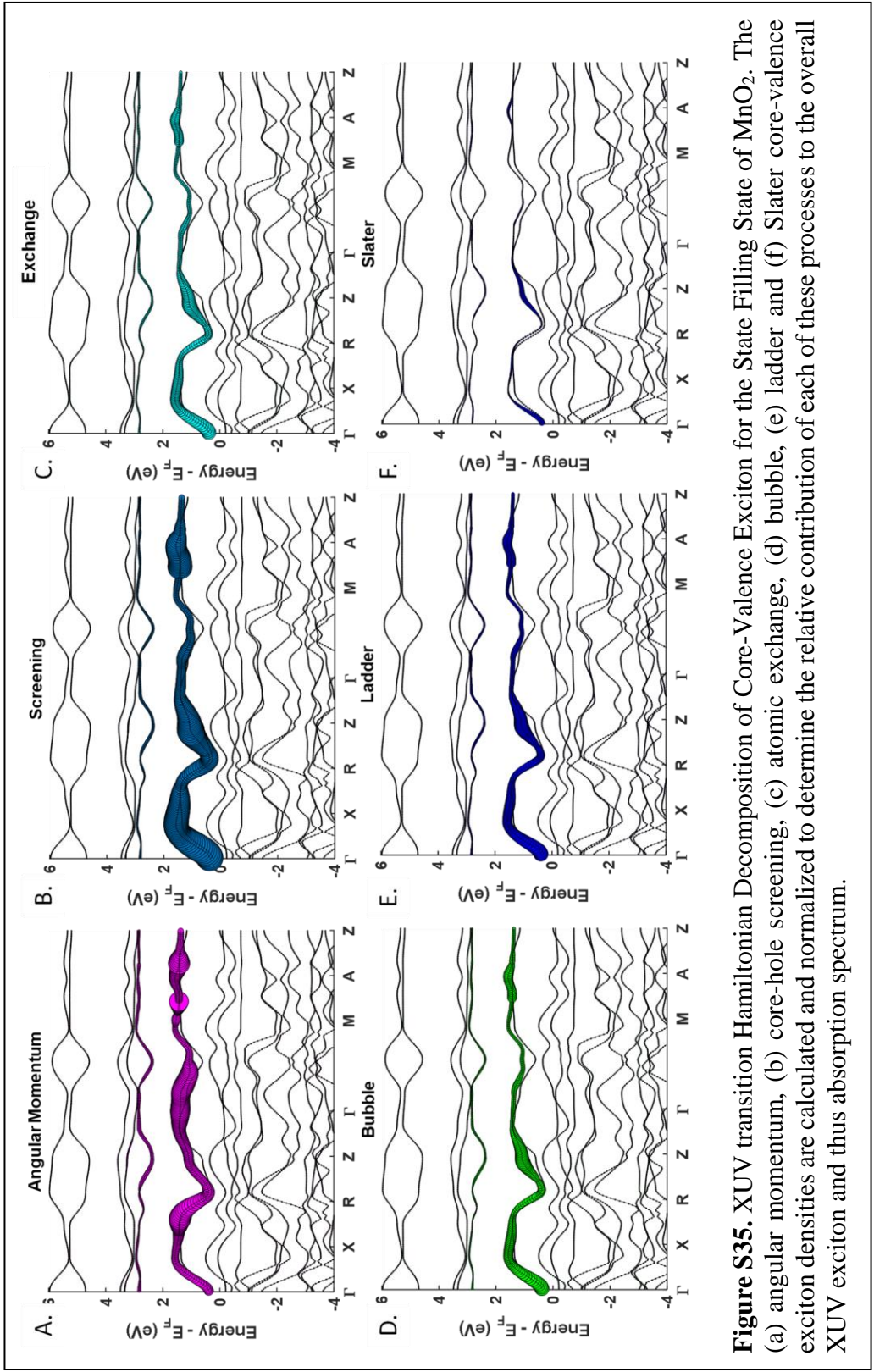

**Figure S35.** XUV transition Hamiltonian Decomposition of Core-Valence Exciton for the State Filling State of $MnO_2$. The (a) angular momentum, (b) core-hole screening, (c) atomic exchange, (d) bubble, (e) ladder and (f) Slater core-valence exciton densities are calculated and normalized to determine the relative contribution of each of these processes to the overall XUV exciton and thus absorption spectrum.

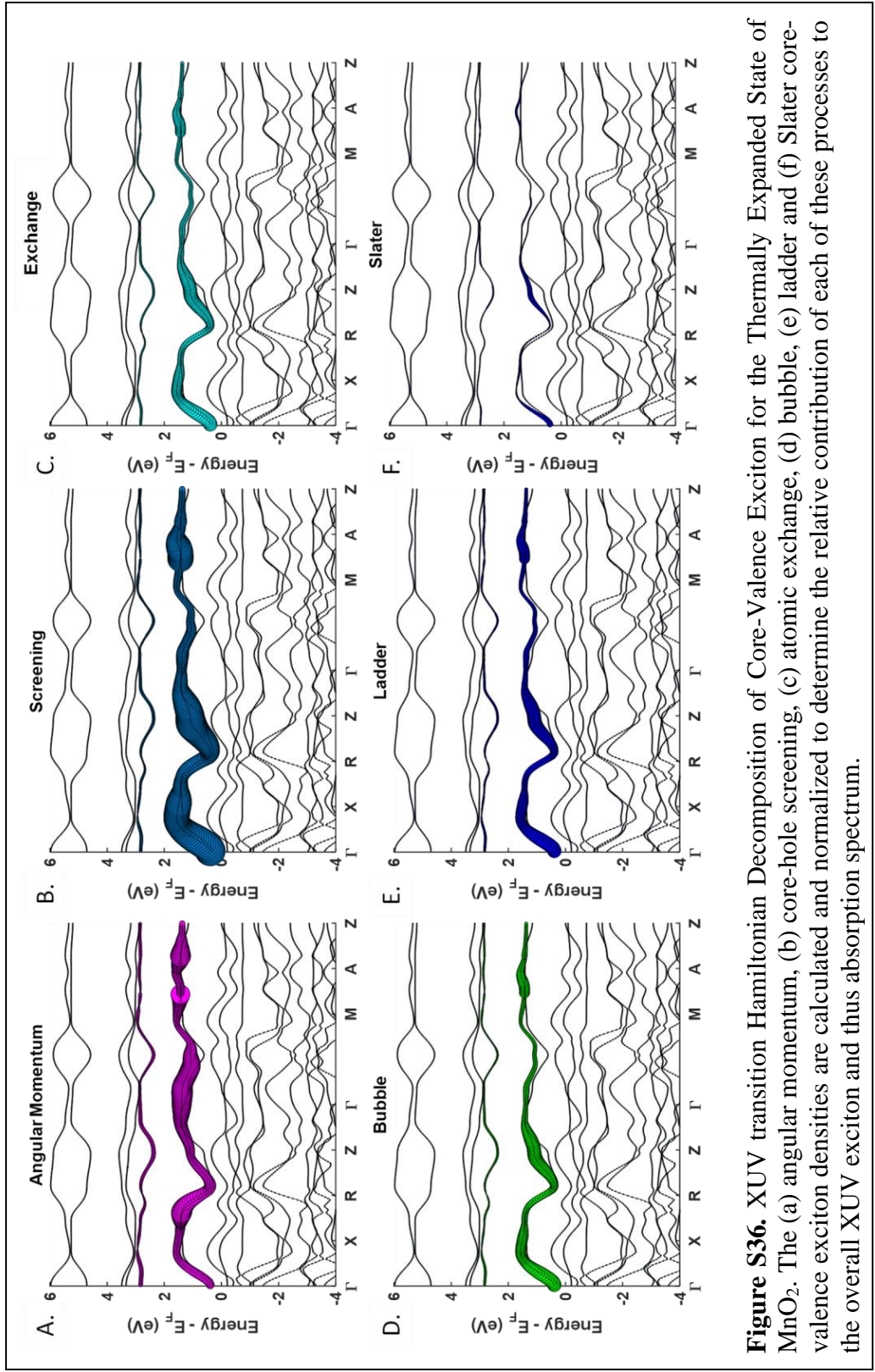

**Figure S36.** XUV transition Hamiltonian Decomposition of Core-Valence Exciton for the Thermally Expanded State of $MnO_2$. The (a) angular momentum, (b) core-hole screening, (c) atomic exchange, (d) bubble, (e) ladder and (f) Slater core-valence exciton densities are calculated and normalized to determine the relative contribution of each of these processes to the overall XUV exciton and thus absorption spectrum.

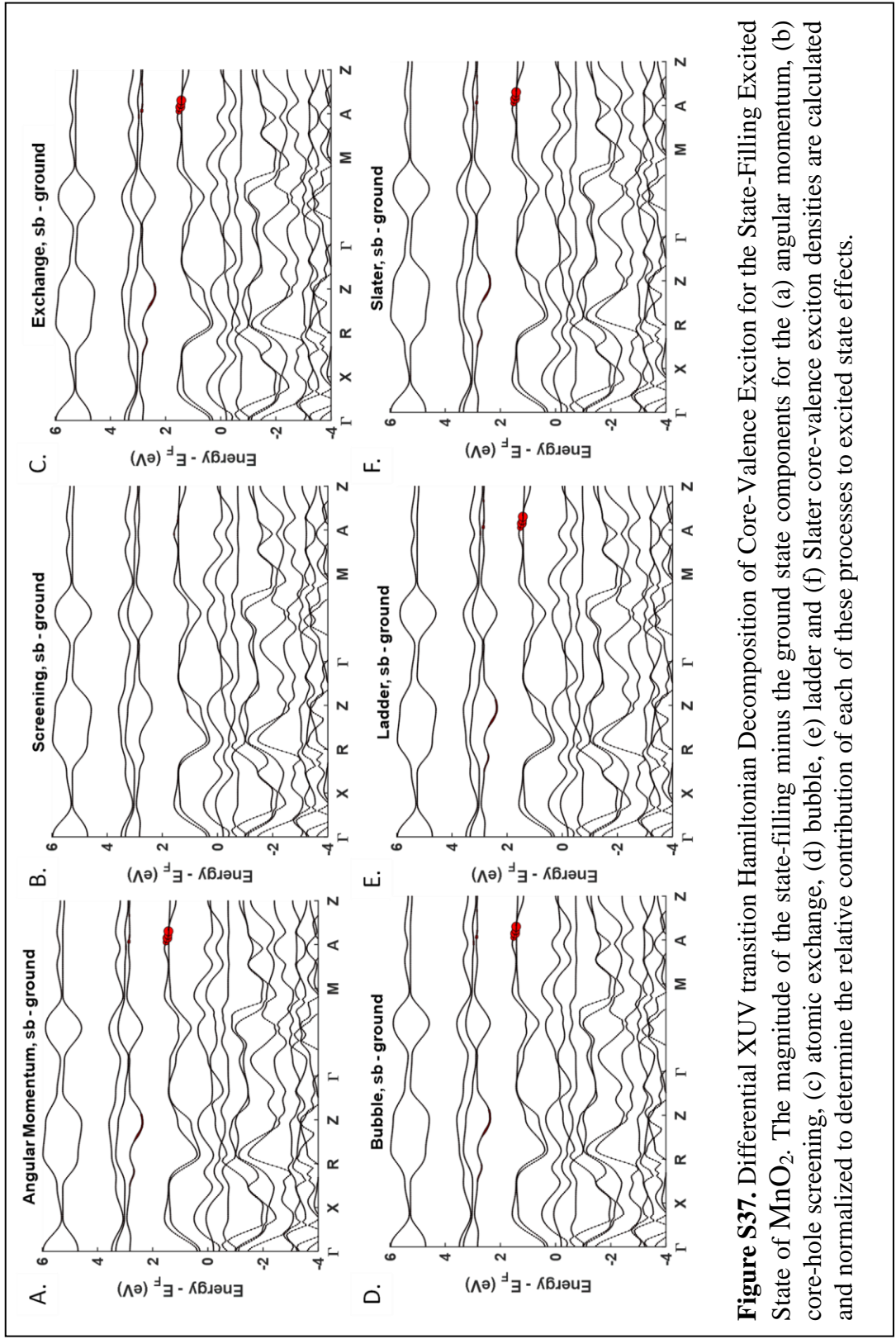

**Figure S37.** Differential XUV transition Hamiltonian Decomposition of Core-Valence Exciton for the State-Filling Excited State of $MnO_2$. The magnitude of the state-filling minus the ground state components for the (a) angular momentum, (b) core-hole screening, (c) atomic exchange, (d) bubble, (e) ladder and (f) Slater core-valence exciton densities are calculated and normalized to determine the relative contribution of each of these processes to excited state effects.

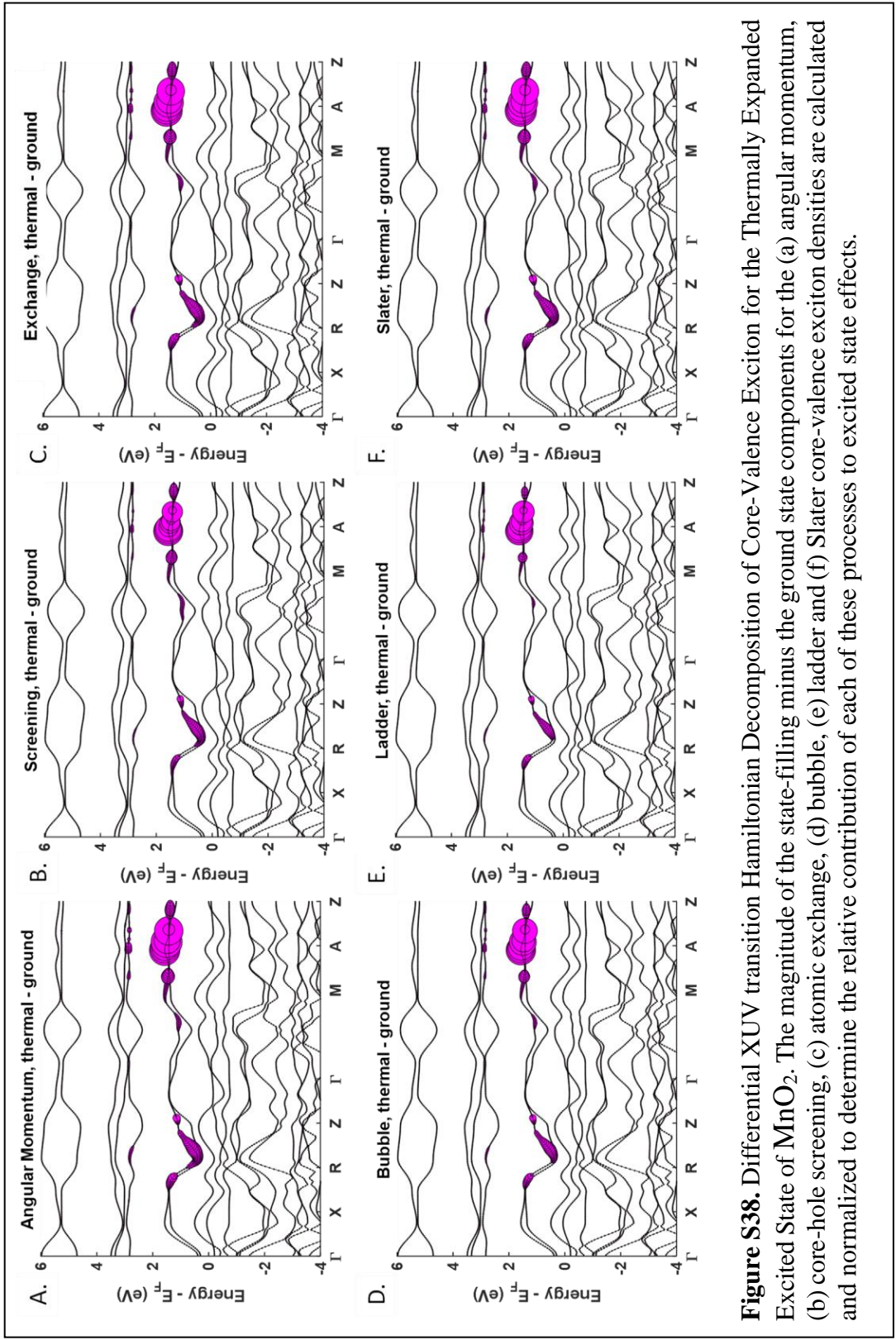

**Figure S38.** Differential XUV transition Hamiltonian Decomposition of Core-Valence Exciton for the Thermally Expanded Excited State of $MnO_2$. The magnitude of the state-filling minus the ground state components for the (a) angular momentum, (b) core-hole screening, (c) atomic exchange, (d) bubble, (e) ladder and (f) Slater core-valence exciton densities are calculated and normalized to determine the relative contribution of each of these processes to excited state effects.

## 5. α-Fe$_2$O$_3$

a. Structural Data for Calculations

i. Ground State and State Blocking

**Unit Cell Parameters**

{10.2423  10.2423  10.2423}

**Primitive Vectors**

{1.0000000000  0.0000000000  0.0000000000

0.5695664700  0.8219452757  0.0000000000

0.5695664700  0.2982686482  0.7659176521}

**Reduced coordinates, ( x, y, z ),**

Fe1  0.1433915  0.1433904  0.1433906

Fe1  0.8566085  0.8566096  0.8566094

Fe2  0.3566087  0.3566104  0.3566106

Fe2  0.6433913  0.6433896  0.6433894

O    0.7500005  0.4472258  0.0527735

O    0.9472262  0.2499995  0.5527741

O    0.4472260  0.0527734  0.7500002

O    0.2499995  0.5527742  0.9472265

O    0.0527738  0.7500005  0.4472259

O    0.5527740  0.9472266  0.2499998

ii. Thermally Expanded Lattice

**Unit Cell Parameters**

{10.327  10.327  10.327}

**Primitive Vectors**

{1.0000000000  0.0000000000  0.0000000000

0.5695664700  0.8219452757  0.0000000000

0.5695664700  0.2982686482  0.7659176521}

**Reduced coordinates, ( x, y, z )**

Fe1  0.1433915  0.1433904  0.1433906

Fe1  0.8566085  0.8566096  0.8566094

Fe2  0.3566087  0.3566104  0.3566106

Fe2  0.6433913  0.6433896  0.6433894

O    0.7500005  0.4472258  0.0527735

O    0.9472262  0.2499995  0.5527741

O    0.4472260  0.0527734  0.7500002

O    0.2499995  0.5527742  0.9472265

O    0.0527738  0.7500005  0.4472259

O    0.5527740  0.9472266  0.2499998

b. Ground State Calculations

i. Band Structure and DOS

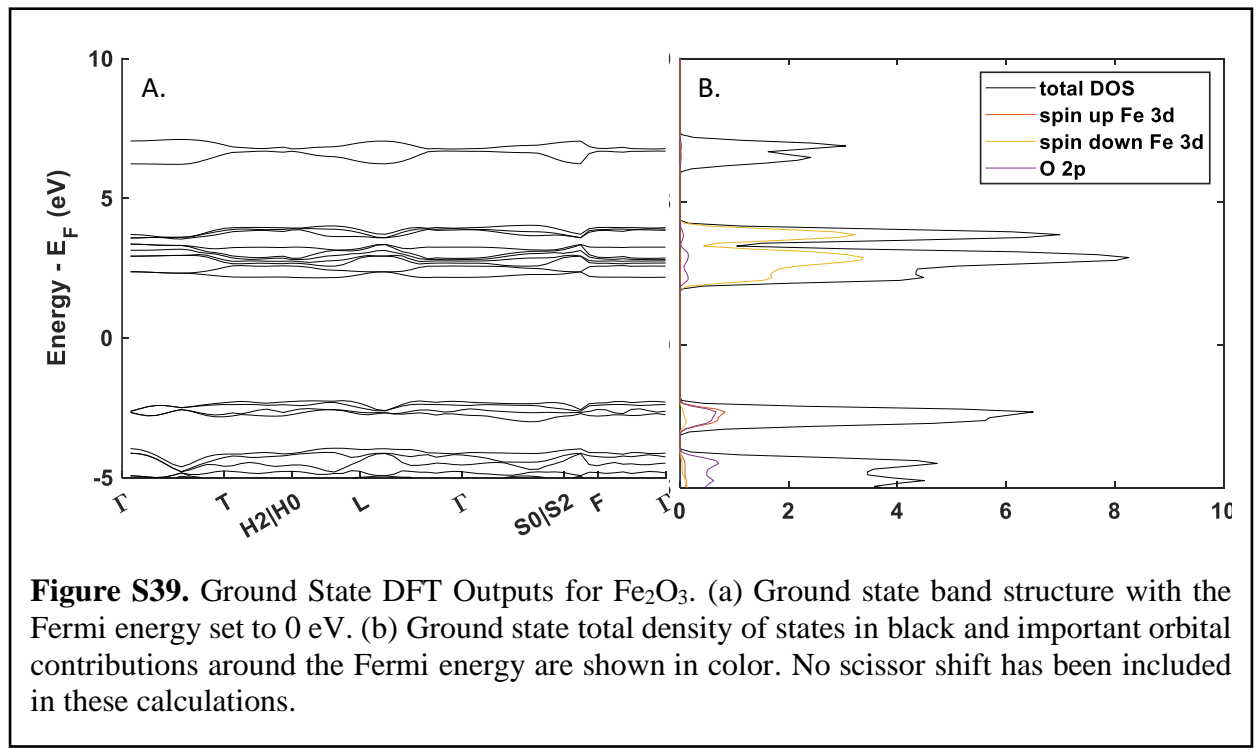

**Figure S39.** Ground State DFT Outputs for $Fe_2O_3$. (a) Ground state band structure with the Fermi energy set to 0 eV. (b) Ground state total density of states in black and important orbital contributions around the Fermi energy are shown in color. No scissor shift has been included in these calculations.

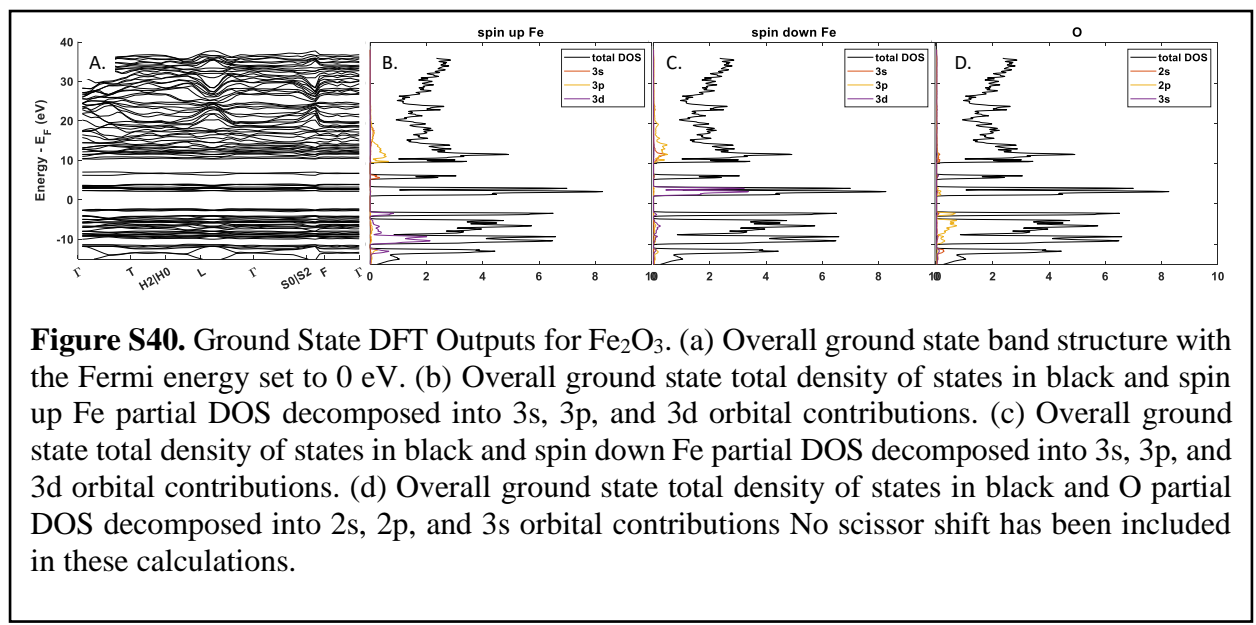

**Figure S40.** Ground State DFT Outputs for $Fe_2O_3$. (a) Overall ground state band structure with the Fermi energy set to 0 eV. (b) Overall ground state total density of states in black and spin up Fe partial DOS decomposed into 3s, 3p, and 3d orbital contributions. (c) Overall ground state total density of states in black and spin down Fe partial DOS decomposed into 3s, 3p, and 3d orbital contributions. (d) Overall ground state total density of states in black and O partial DOS decomposed into 2s, 2p, and 3s orbital contributions No scissor shift has been included in these calculations.

ii. Ground State Spectrum[8]

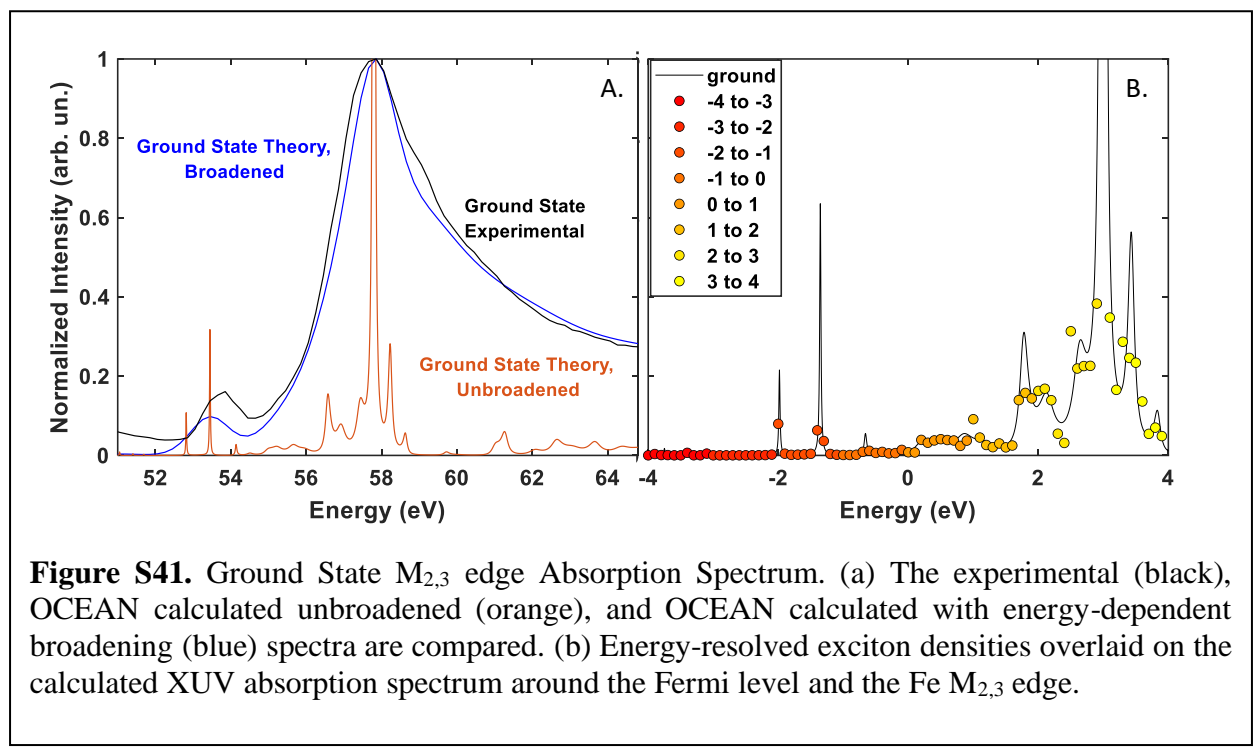

**Figure S41.** Ground State $M_{2,3}$ edge Absorption Spectrum. (a) The experimental (black), OCEAN calculated unbroadened (orange), and OCEAN calculated with energy-dependent broadening (blue) spectra are compared. (b) Energy-resolved exciton densities overlaid on the calculated XUV absorption spectrum around the Fermi level and the Fe $M_{2,3}$ edge.

iii. Ground State GMRES Energy Decomposition

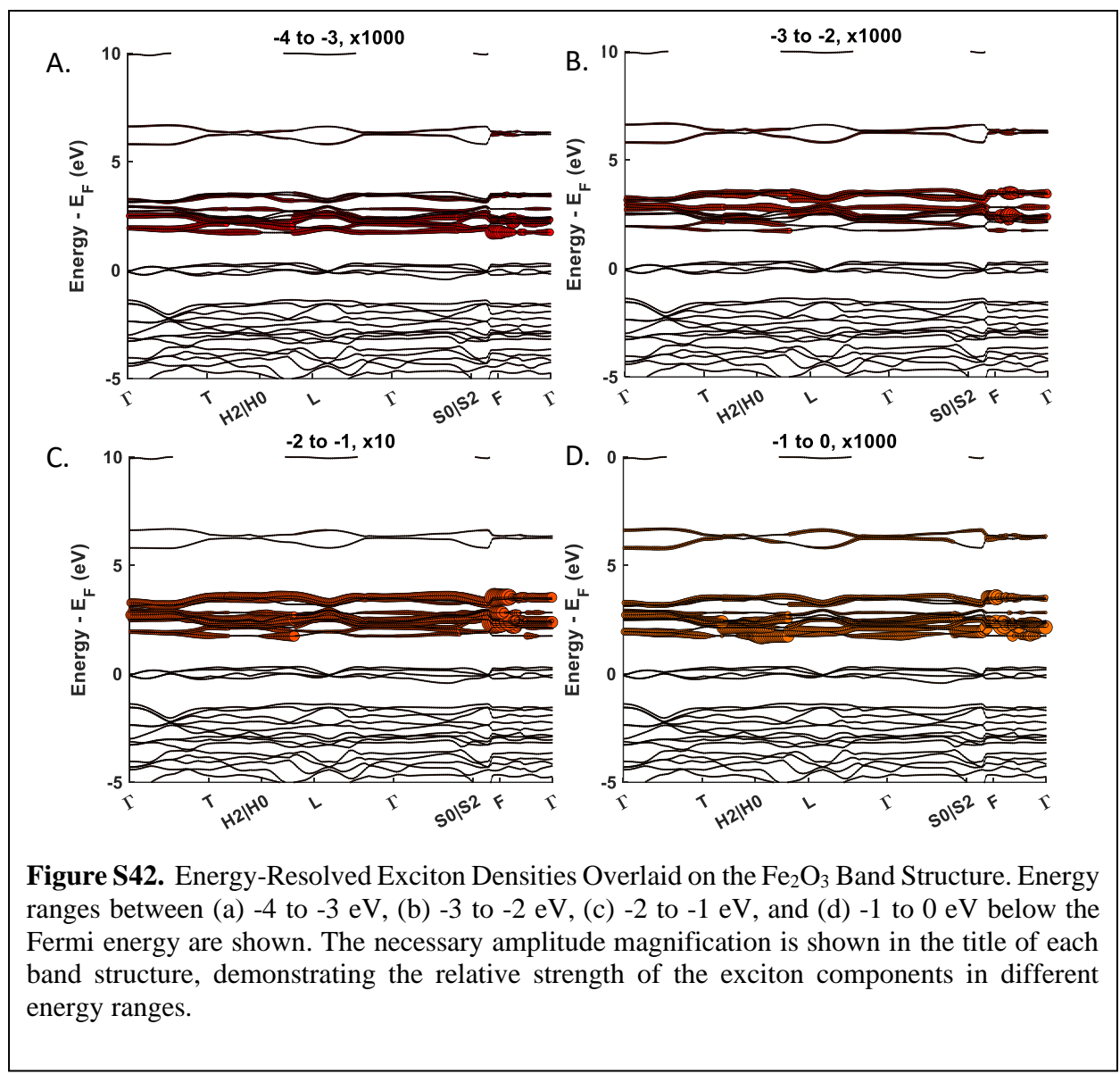

**Figure S42.** Energy-Resolved Exciton Densities Overlaid on the Fe$_2$O$_3$ Band Structure. Energy ranges between (a) -4 to -3 eV, (b) -3 to -2 eV, (c) -2 to -1 eV, and (d) -1 to 0 eV below the Fermi energy are shown. The necessary amplitude magnification is shown in the title of each band structure, demonstrating the relative strength of the exciton components in different energy ranges.

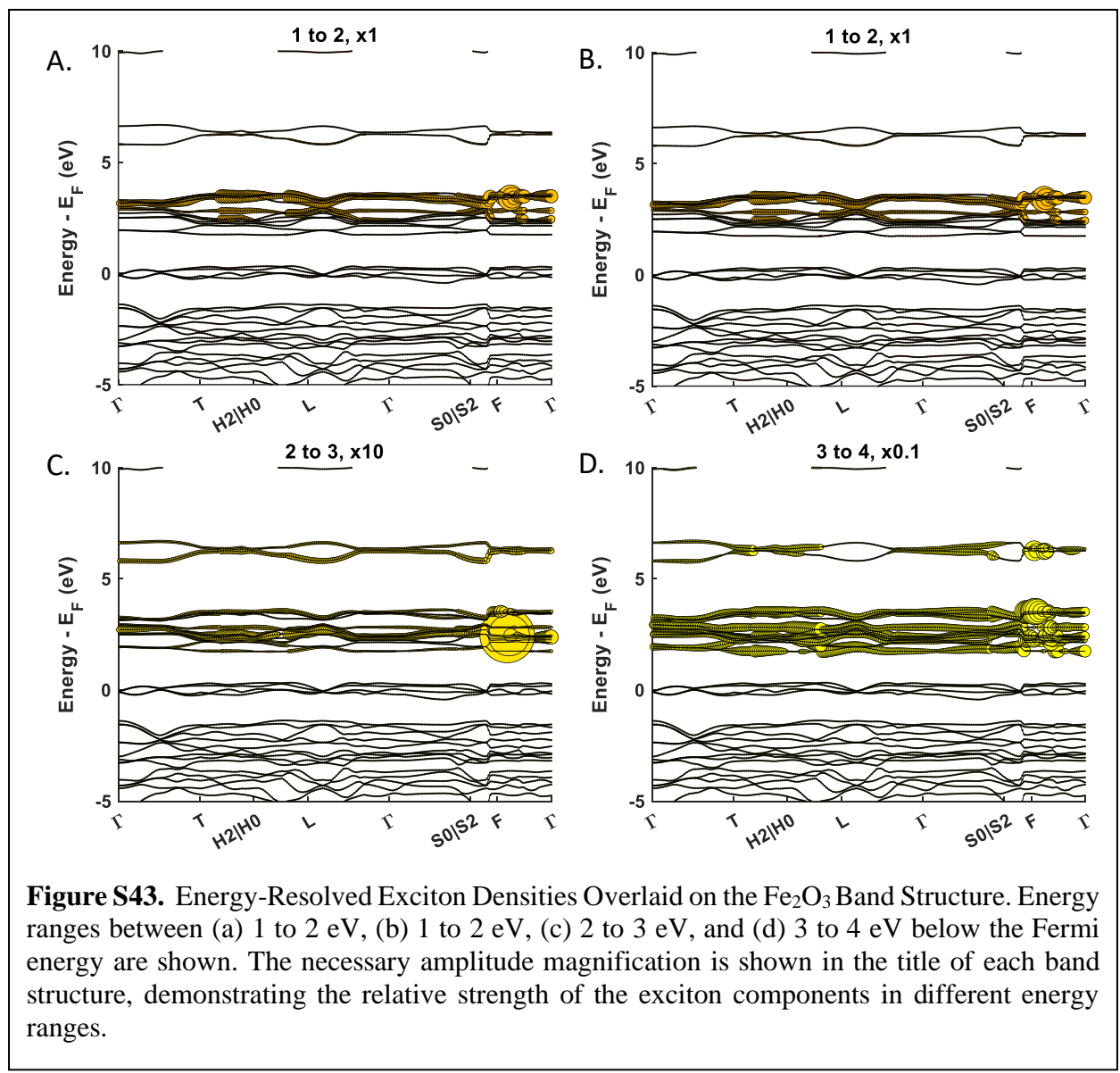

**Figure S43.** Energy-Resolved Exciton Densities Overlaid on the $Fe_2O_3$ Band Structure. Energy ranges between (a) 1 to 2 eV, (b) 1 to 2 eV, (c) 2 to 3 eV, and (d) 3 to 4 eV below the Fermi energy are shown. The necessary amplitude magnification is shown in the title of each band structure, demonstrating the relative strength of the exciton components in different energy ranges.

c. Excited State Calculations

   i. State filling band diagrams and Full Spectra

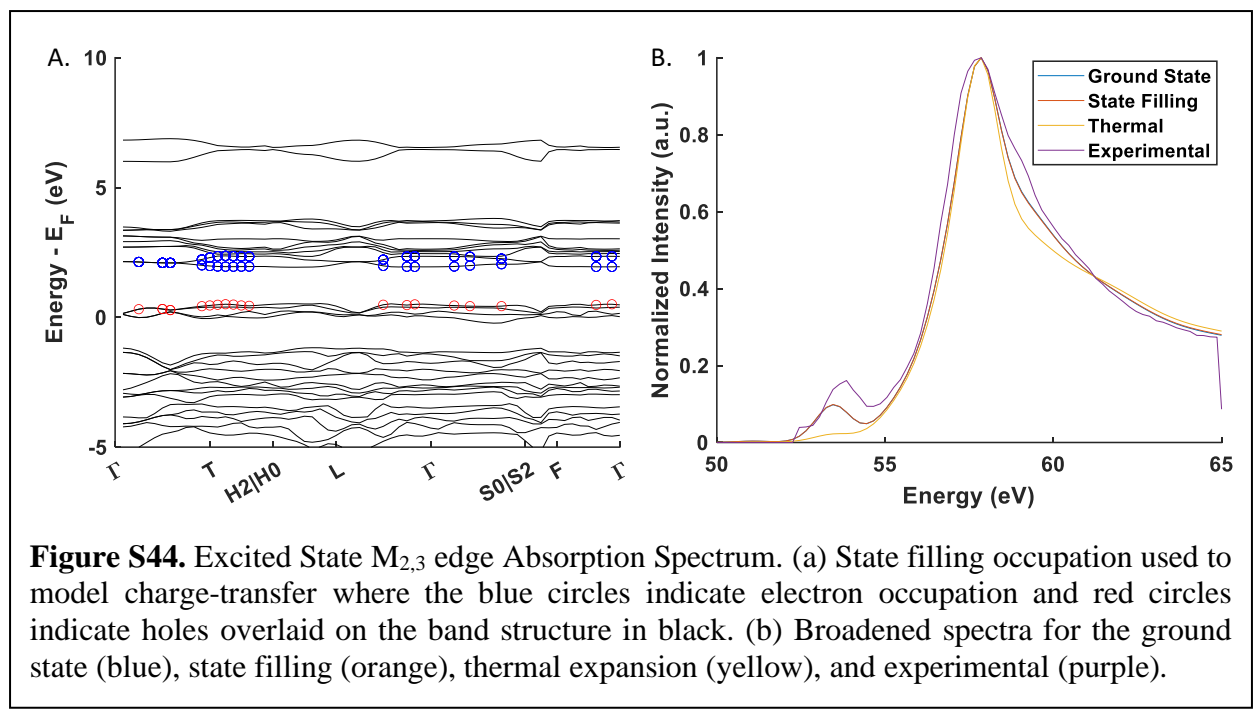

**Figure S44.** Excited State $M_{2,3}$ edge Absorption Spectrum. (a) State filling occupation used to model charge-transfer where the blue circles indicate electron occupation and red circles indicate holes overlaid on the band structure in black. (b) Broadened spectra for the ground state (blue), state filling (orange), thermal expansion (yellow), and experimental (purple).

### d. Hamiltonian Decompositions

#### i. Total Exciton Comparisons

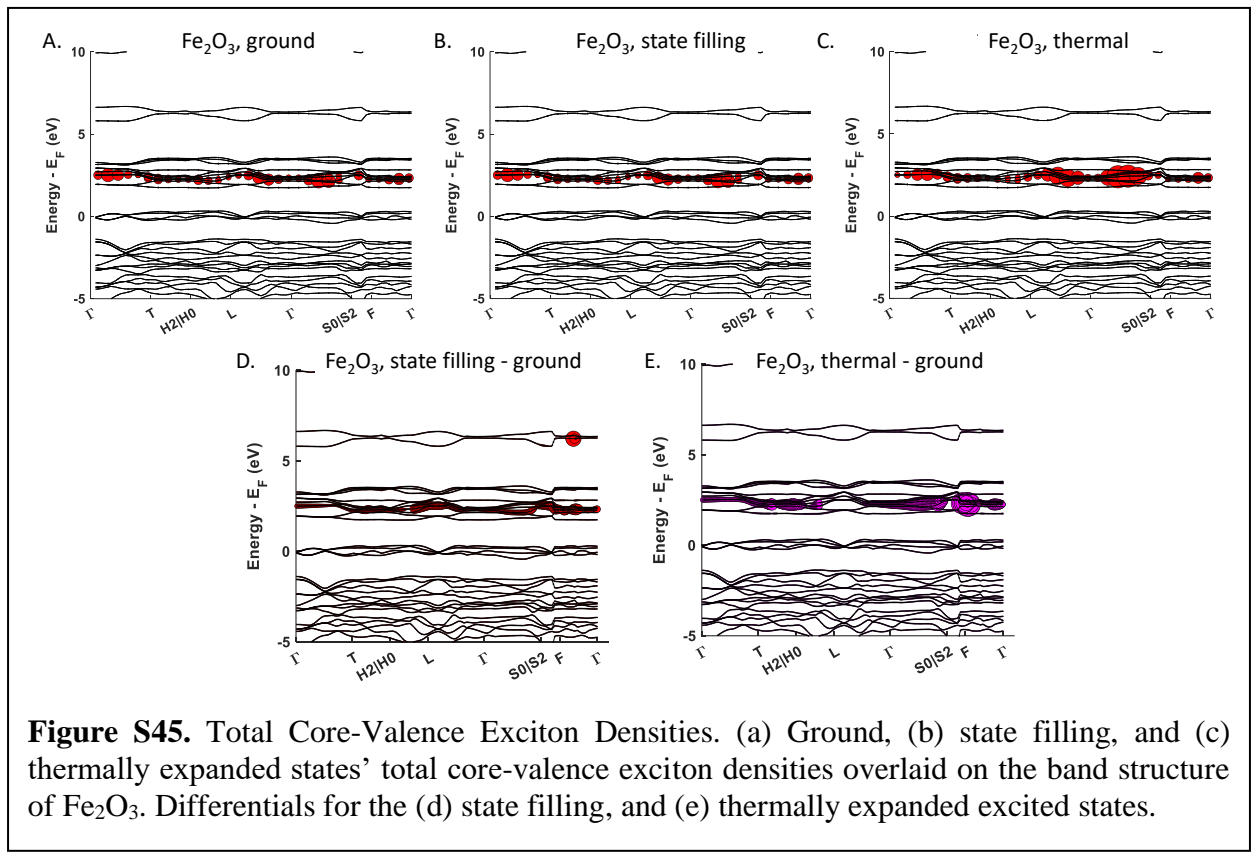

**Figure S45.** Total Core-Valence Exciton Densities. (a) Ground, (b) state filling, and (c) thermally expanded states' total core-valence exciton densities overlaid on the band structure of $Fe_2O_3$. Differentials for the (d) state filling, and (e) thermally expanded excited states.

ii. Hamiltonian Decomposition of Exciton Components for ground, state filling, and thermally expanded models.

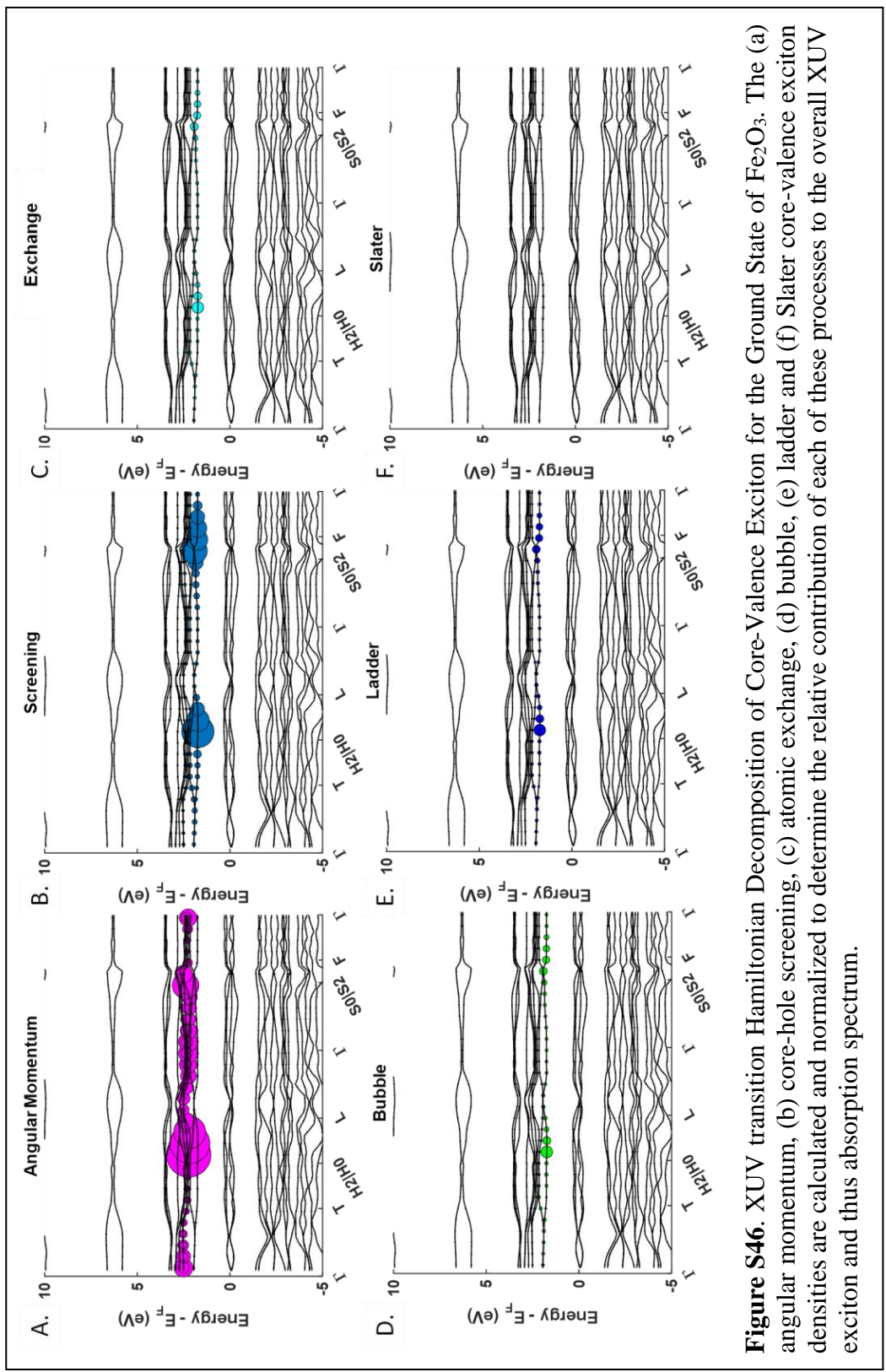

**Figure S46.** XUV transition Hamiltonian Decomposition of Core-Valence Exciton for the Ground State of Fe$_2$O$_3$. The (a) angular momentum, (b) core-hole screening, (c) atomic exchange, (d) bubble, (e) ladder and (f) Slater core-valence exciton densities are calculated and normalized to determine the relative contribution of each of these processes to the overall XUV exciton and thus absorption spectrum.

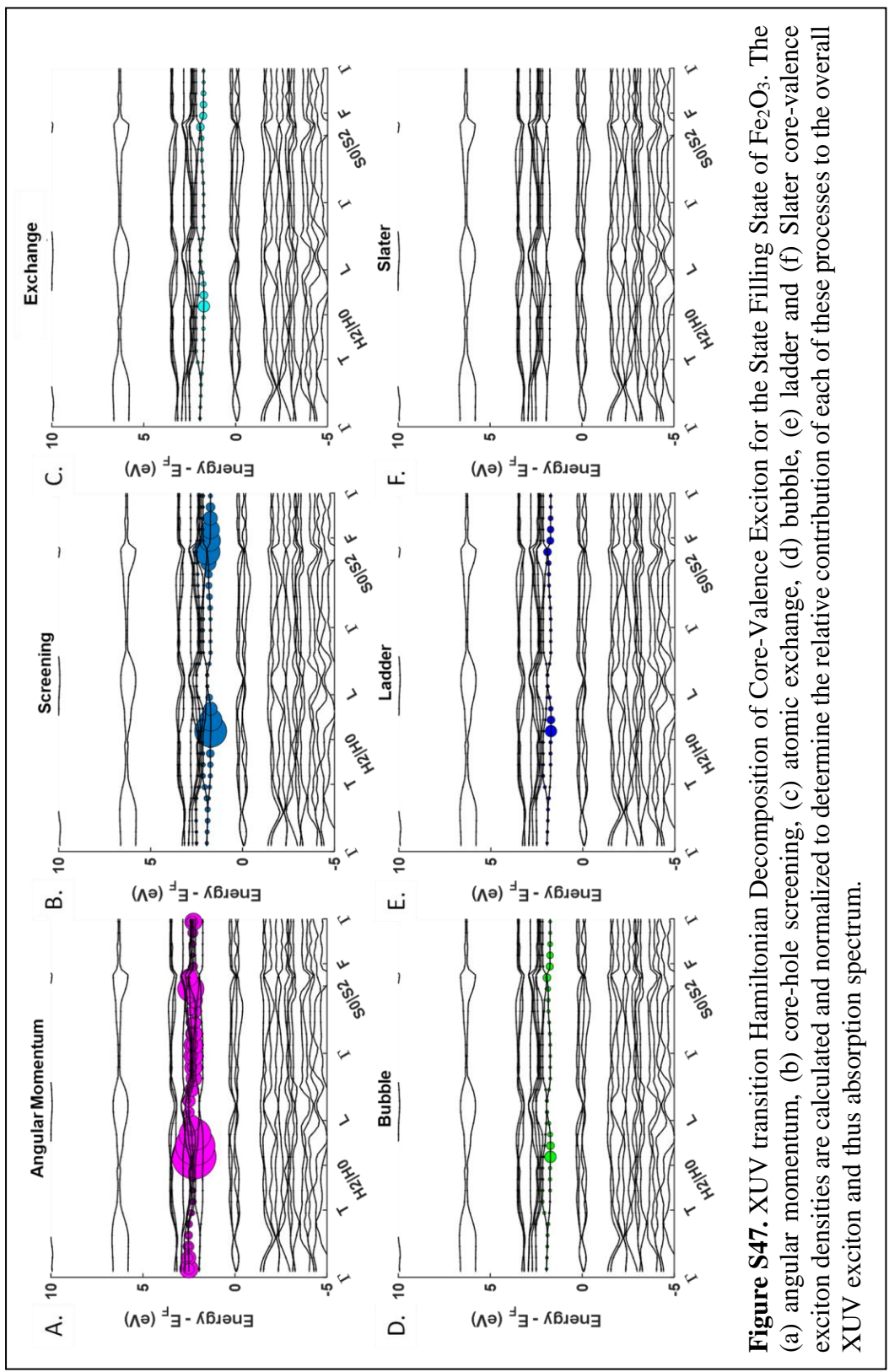

**Figure S47.** XUV transition Hamiltonian Decomposition of Core-Valence Exciton for the State Filling State of $Fe_2O_3$. The (a) angular momentum, (b) core-hole screening, (c) atomic exchange, (d) bubble, (e) ladder and (f) Slater core-valence exciton densities are calculated and normalized to determine the relative contribution of each of these processes to the overall XUV exciton and thus absorption spectrum.

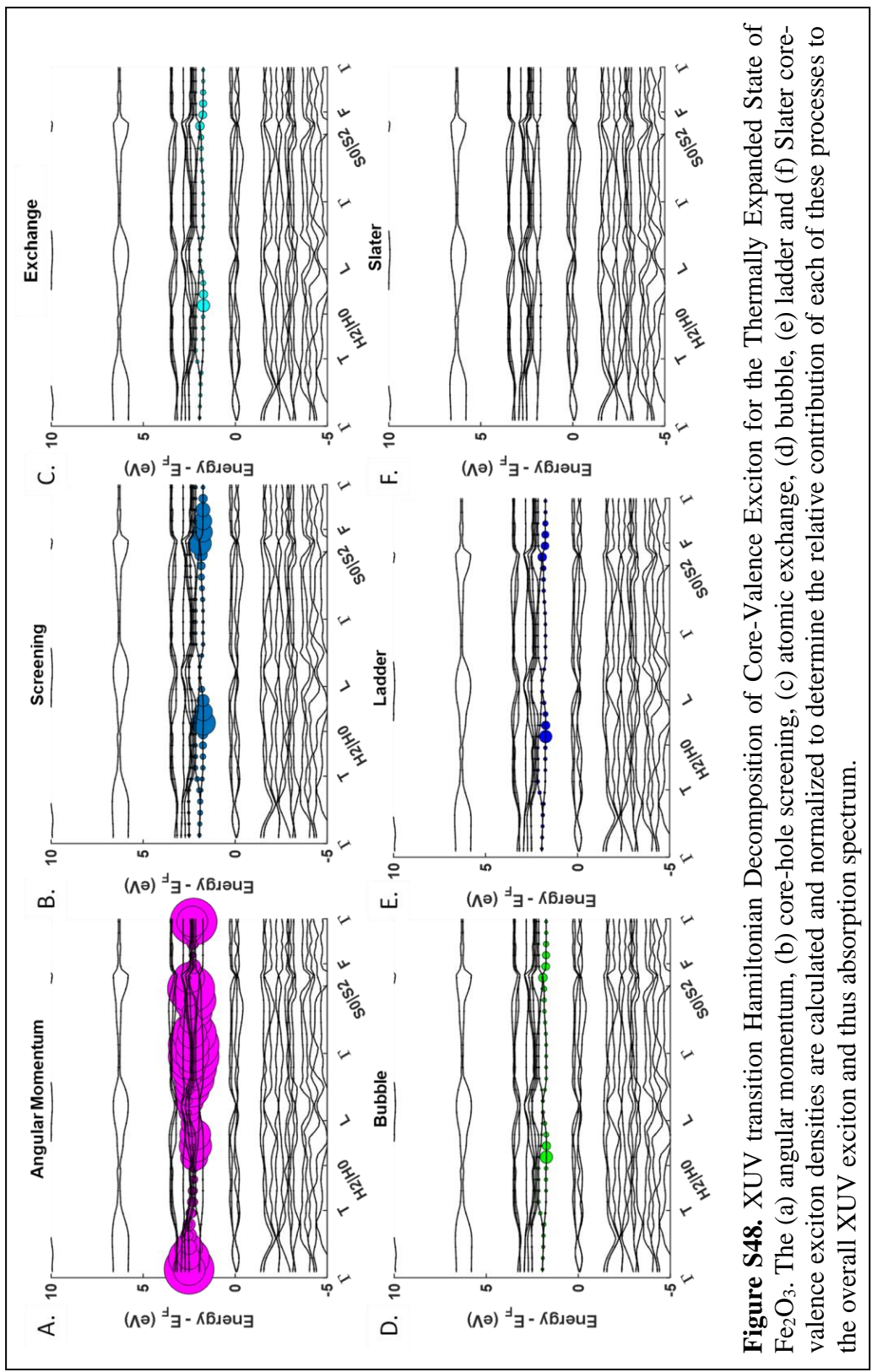

**Figure S48.** XUV transition Hamiltonian Decomposition of Core-Valence Exciton for the Thermally Expanded State of $Fe_2O_3$. The (a) angular momentum, (b) core-hole screening, (c) atomic exchange, (d) bubble, (e) ladder and (f) Slater core-valence exciton densities are calculated and normalized to determine the relative contribution of each of these processes to the overall XUV exciton and thus absorption spectrum.

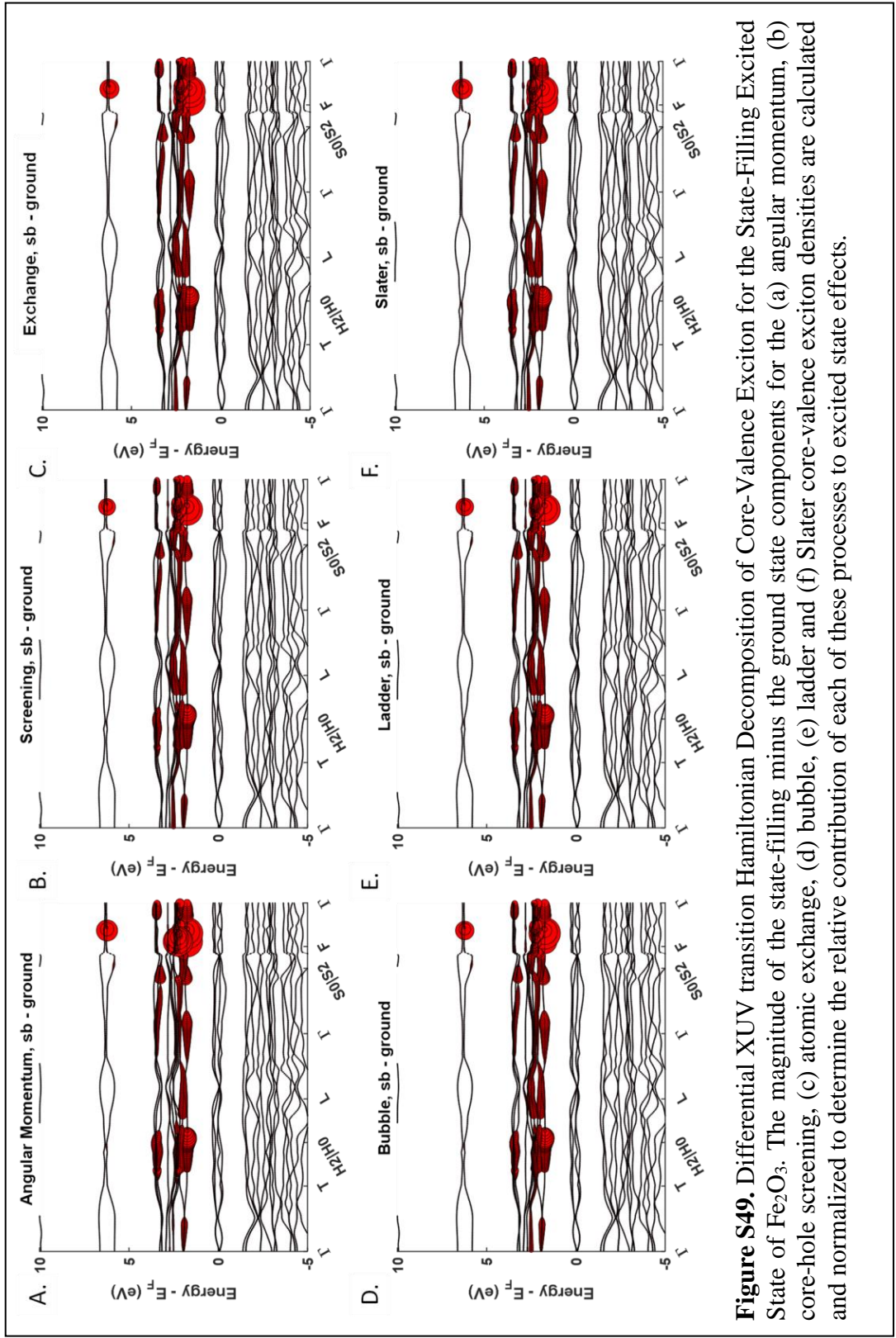

**Figure S49.** Differential XUV transition Hamiltonian Decomposition of Core-Valence Exciton for the State-Filling Excited State of $Fe_2O_3$. The magnitude of the state-filling minus the ground state components for the (a) angular momentum, (b) core-hole screening, (c) atomic exchange, (d) bubble, (e) ladder and (f) Slater core-valence exciton densities are calculated and normalized to determine the relative contribution of each of these processes to excited state effects.

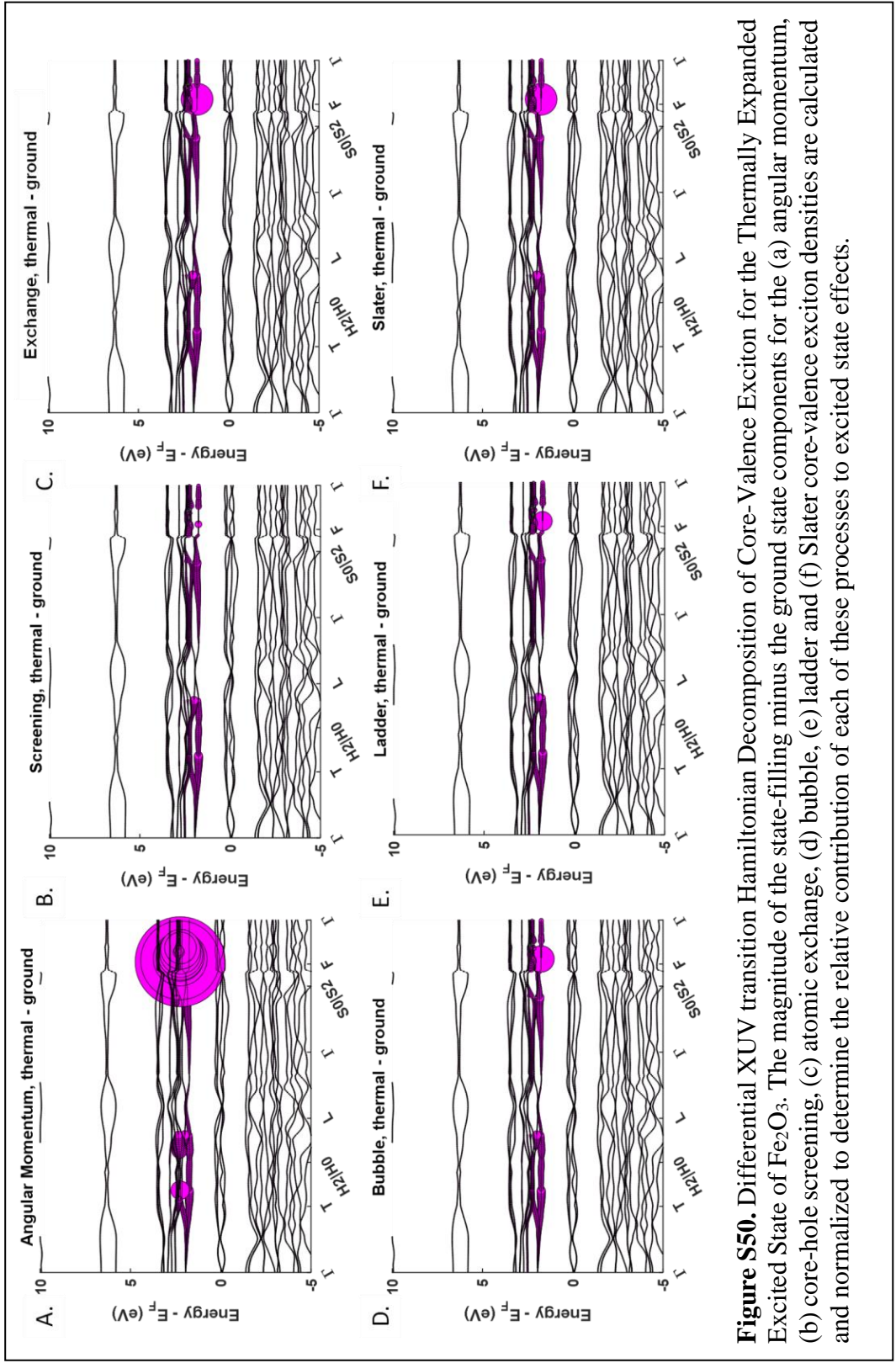

**Figure S50.** Differential XUV transition Hamiltonian Decomposition of Core-Valence Exciton for the Thermally Expanded Excited State of $Fe_2O_3$. The magnitude of the state-filling minus the ground state components for the (a) angular momentum, (b) core-hole screening, (c) atomic exchange, (d) bubble, (e) ladder and (f) Slater core-valence exciton densities are calculated and normalized to determine the relative contribution of each of these processes to excited state effects.

## 6. $Co_3O_4$

a. Structural Data for Calculations

i. Ground State and State Blocking

**Unit Cell Parameters (bohr)**

{11.221  11.221  11.221 }

**Primitive Vectors**

{0.0  0.5  0.5

 0.5  0.0  0.5

 0.5  0.5  0.0}

**Reduced coordinates, ( x, y, z )**

Co1 0.125  0.125  0.125

Co1 0.875  0.875  0.875

Co1 0.500  0.500  0.500

Co2 0.500  0.500  0.000

Co2 0.500  0.000  0.500

Co2 0.000  0.500  0.500

O 0.2642  0.2642  0.2642

O 0.2642  0.2642  -0.2926

O 0.2642  -0.2926  0.2642

O -0.2926  0.2642  0.2642

O -0.2642  -0.2642  1.2926

O -0.2642  -0.2642  -0.2642

O -0.2642  1.2926  -0.2642

O 1.2926  -0.2642  -0.2642

ii. Thermally Expanded Lattice

**Unit Cell Parameters (bohr)**

{11.258  11.258  11.258 }

**Primitive Vectors**

{0.0  0.5  0.5

 0.5  0.0  0.5

 0.5  0.5  0.0}

**Reduced coordinates, ( x, y, z )**

Co1 0.125  0.125  0.125

Co1 0.875  0.875  0.875

Co1 0.500  0.500  0.500

Co2 0.500  0.500  0.000

Co2 0.500  0.000  0.500

Co2 0.000  0.500  0.500

O 0.2642  0.2642  0.2642

O 0.2642  0.2642  -0.2926

O 0.2642  -0.2926  0.2642

O -0.2926  0.2642  0.2642

O -0.2642  -0.2642  1.2926

O -0.2642  -0.2642  -0.2642

O -0.2642  1.2926  -0.2642

O 1.2926  -0.2642  -0.2642

b. Ground State Calculations

i. Band Structure and DOS

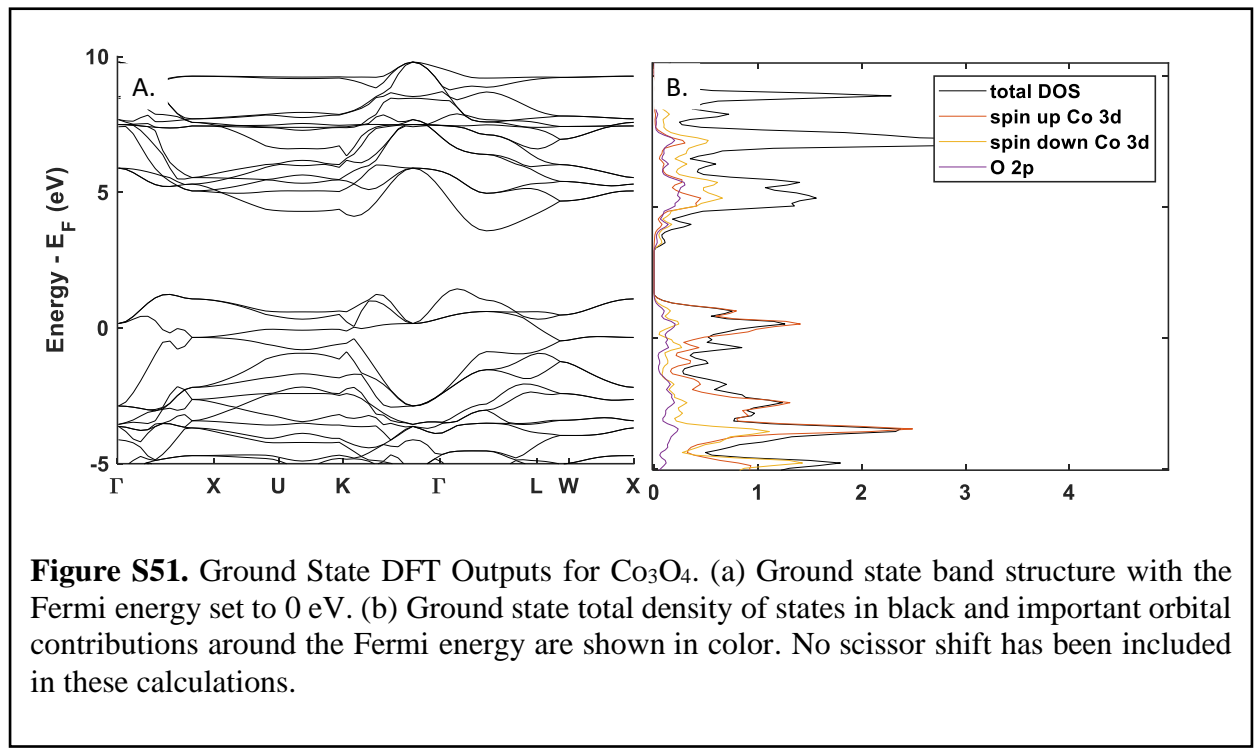

**Figure S51.** Ground State DFT Outputs for $Co_3O_4$. (a) Ground state band structure with the Fermi energy set to 0 eV. (b) Ground state total density of states in black and important orbital contributions around the Fermi energy are shown in color. No scissor shift has been included in these calculations.

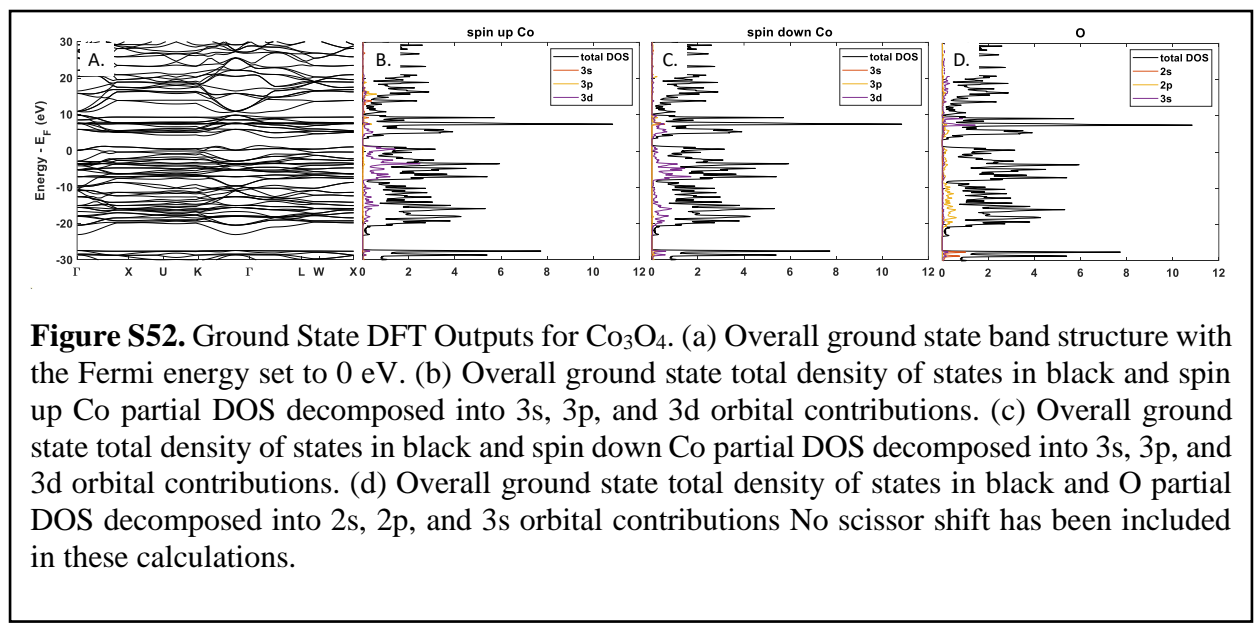

**Figure S52.** Ground State DFT Outputs for $Co_3O_4$. (a) Overall ground state band structure with the Fermi energy set to 0 eV. (b) Overall ground state total density of states in black and spin up Co partial DOS decomposed into 3s, 3p, and 3d orbital contributions. (c) Overall ground state total density of states in black and spin down Co partial DOS decomposed into 3s, 3p, and 3d orbital contributions. (d) Overall ground state total density of states in black and O partial DOS decomposed into 2s, 2p, and 3s orbital contributions No scissor shift has been included in these calculations.

## ii. Ground State Spectrum[9]

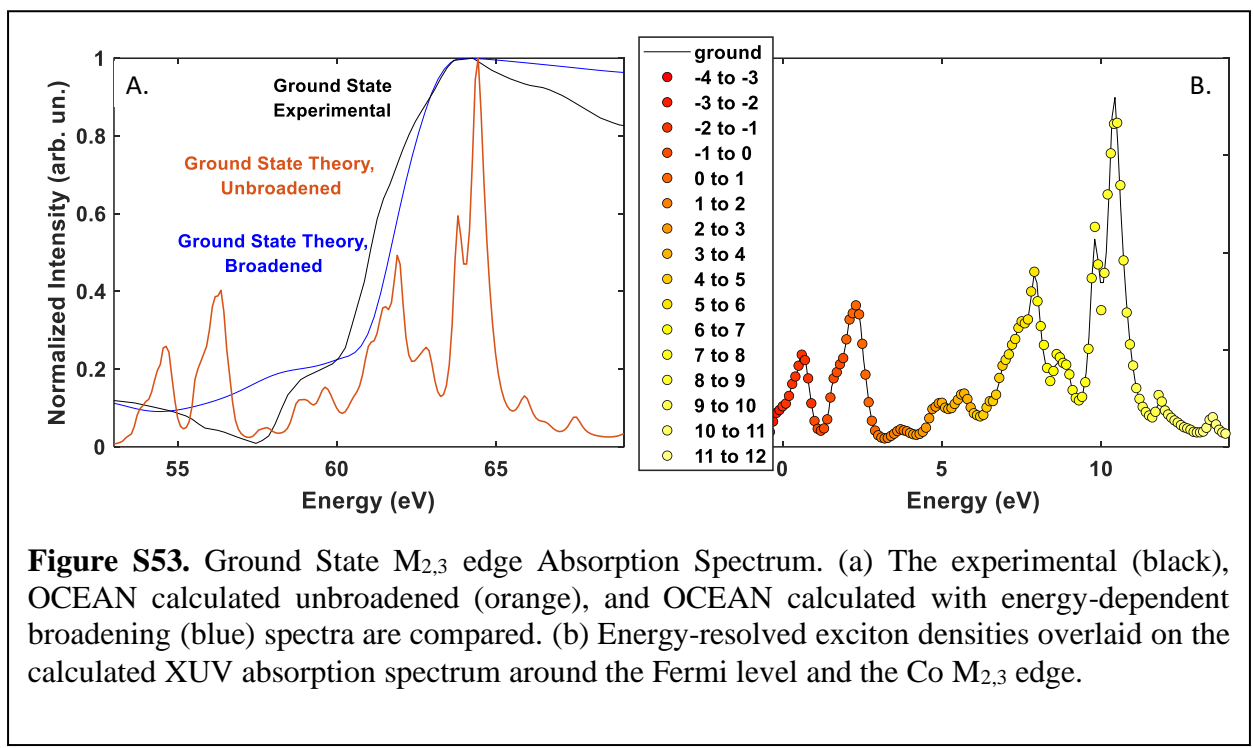

**Figure S53.** Ground State $M_{2,3}$ edge Absorption Spectrum. (a) The experimental (black), OCEAN calculated unbroadened (orange), and OCEAN calculated with energy-dependent broadening (blue) spectra are compared. (b) Energy-resolved exciton densities overlaid on the calculated XUV absorption spectrum around the Fermi level and the Co $M_{2,3}$ edge.

iii. Ground State GMRES Energy Decomposition

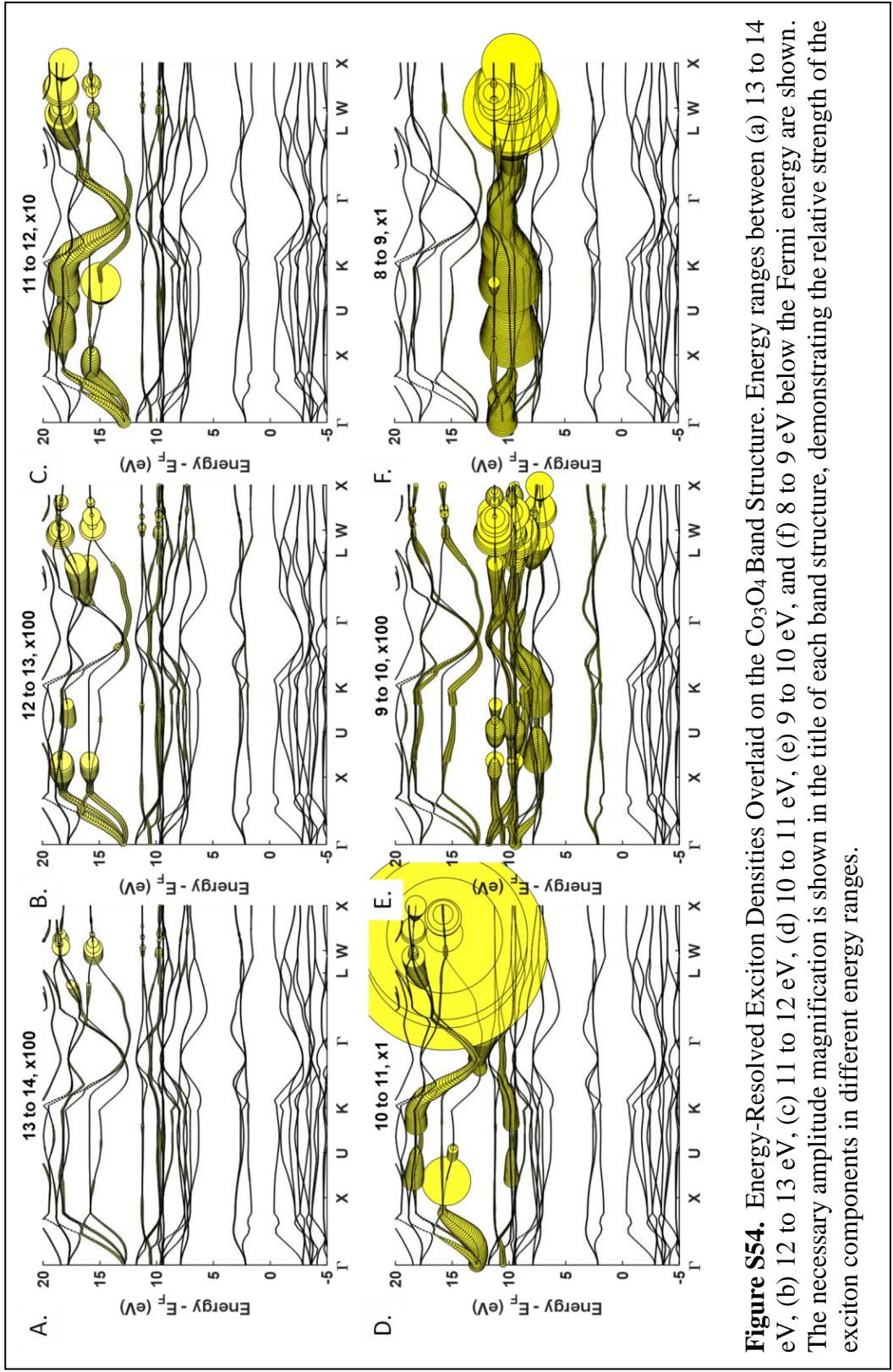

**Figure S54.** Energy-Resolved Exciton Densities Overlaid on the $Co_3O_4$ Band Structure. Energy ranges between (a) 13 to 14 eV, (b) 12 to 13 eV, (c) 11 to 12 eV, (d) 10 to 11 eV, (e) 9 to 10 eV, and (f) 8 to 9 eV below the Fermi energy are shown. The necessary amplitude magnification is shown in the title of each band structure, demonstrating the relative strength of the exciton components in different energy ranges.

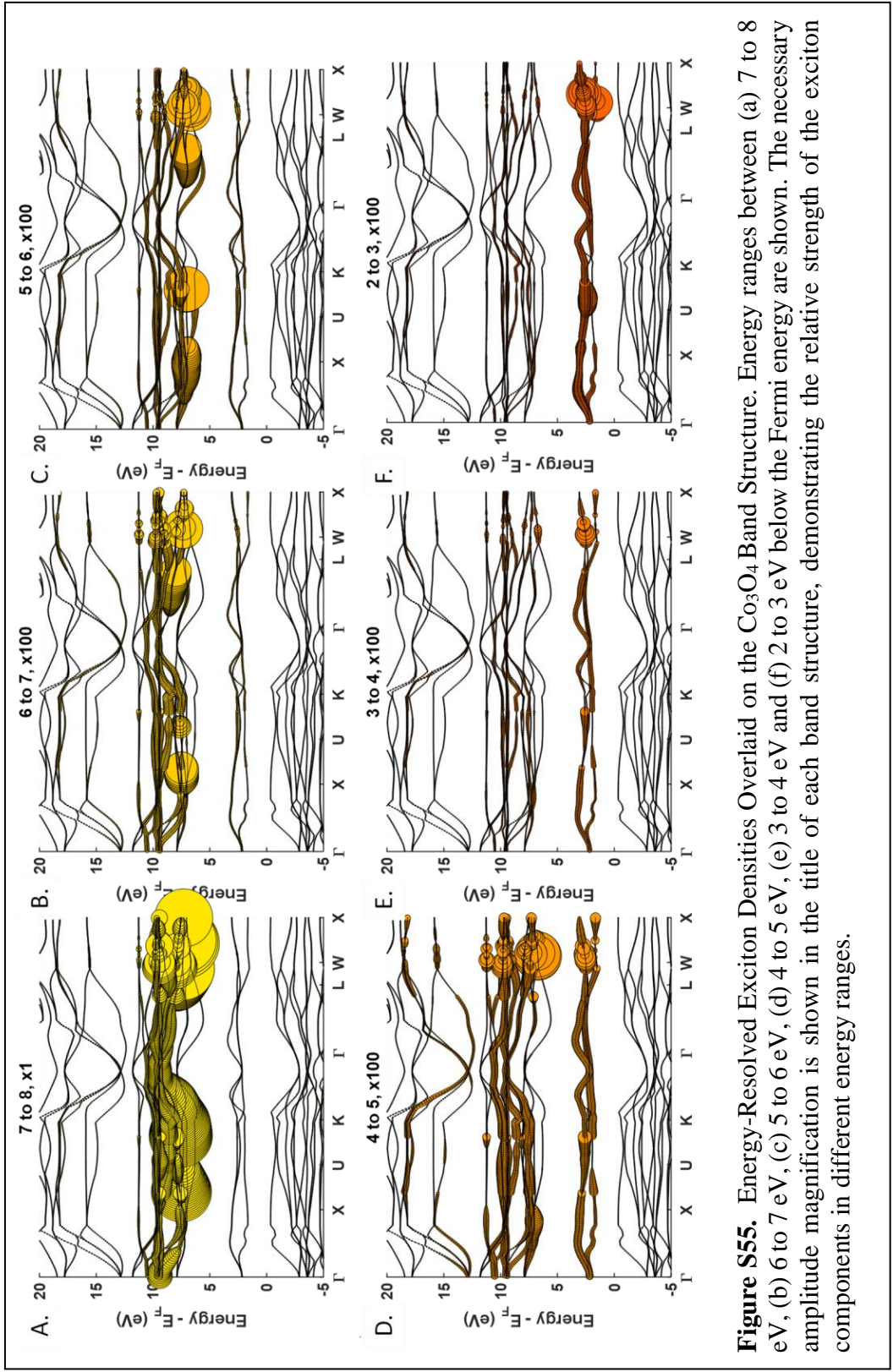

**Figure S55.** Energy-Resolved Exciton Densities Overlaid on the $Co_3O_4$ Band Structure. Energy ranges between (a) 7 to 8 eV, (b) 6 to 7 eV, (c) 5 to 6 eV, (d) 4 to 5 eV, (e) 3 to 4 eV and (f) 2 to 3 eV below the Fermi energy are shown. The necessary amplitude magnification is shown in the title of each band structure, demonstrating the relative strength of the exciton components in different energy ranges.

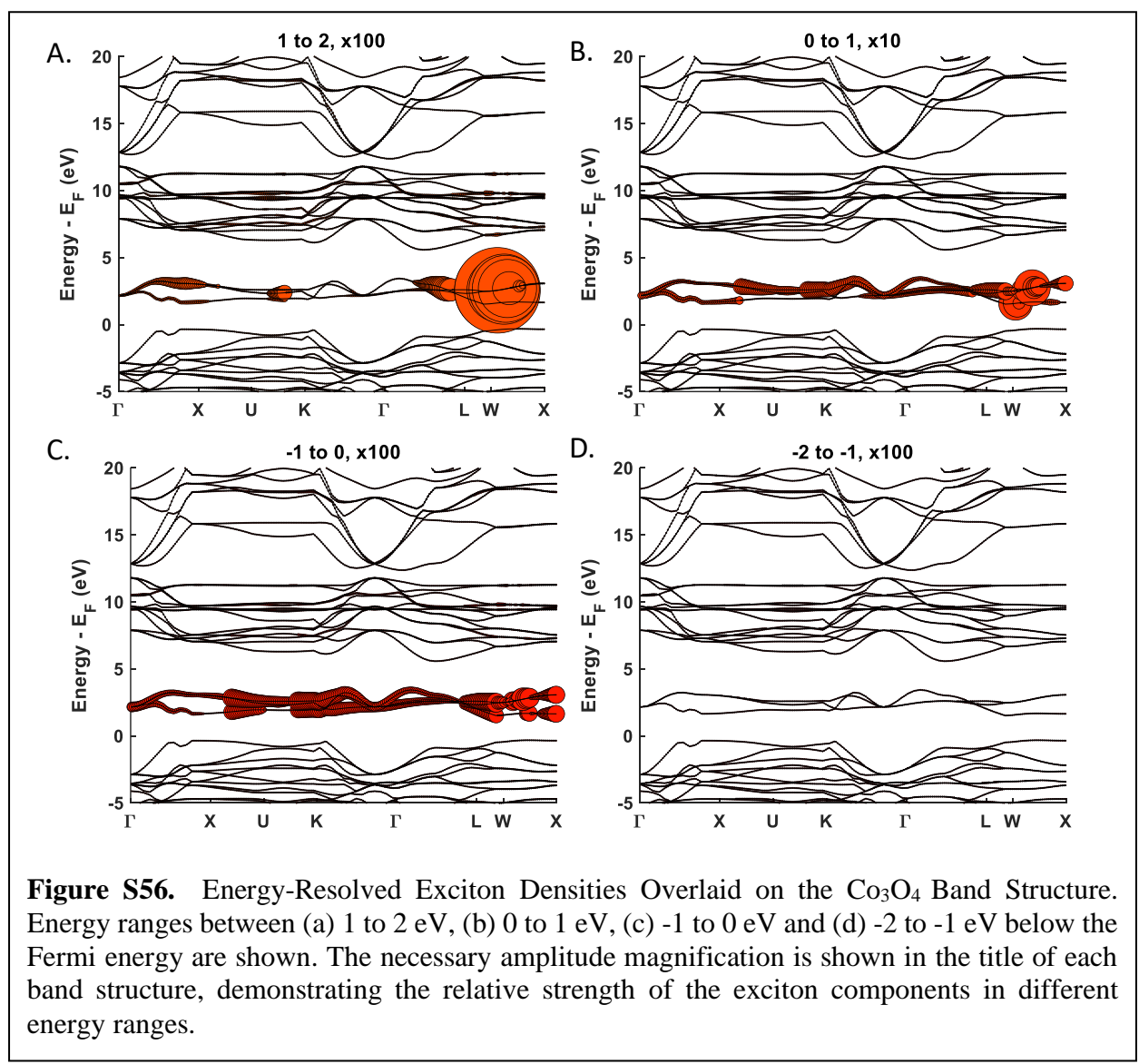

**Figure S56.** Energy-Resolved Exciton Densities Overlaid on the Co$_3$O$_4$ Band Structure. Energy ranges between (a) 1 to 2 eV, (b) 0 to 1 eV, (c) -1 to 0 eV and (d) -2 to -1 eV below the Fermi energy are shown. The necessary amplitude magnification is shown in the title of each band structure, demonstrating the relative strength of the exciton components in different energy ranges.

c. Excited State Calculations

    i. State filling band diagrams and Full Spectra

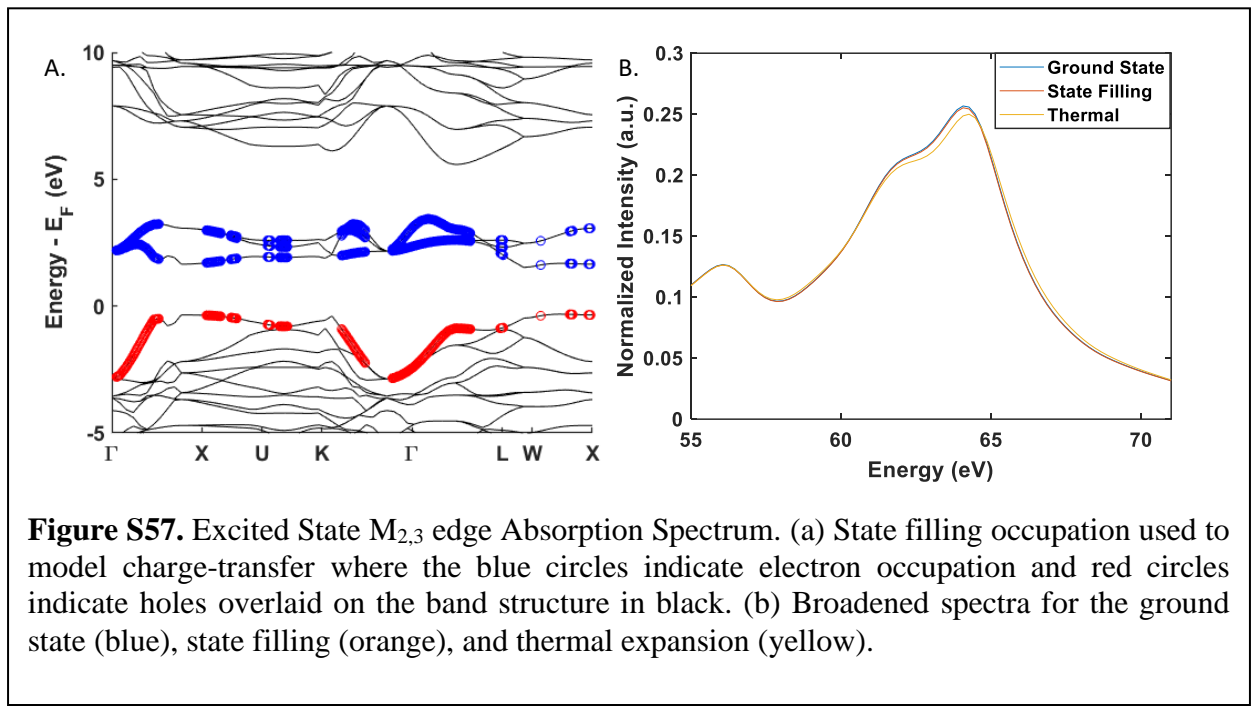

**Figure S57.** Excited State $M_{2,3}$ edge Absorption Spectrum. (a) State filling occupation used to model charge-transfer where the blue circles indicate electron occupation and red circles indicate holes overlaid on the band structure in black. (b) Broadened spectra for the ground state (blue), state filling (orange), and thermal expansion (yellow).

### d. Hamiltonian Decompositions

#### i. Total Exciton Comparisons

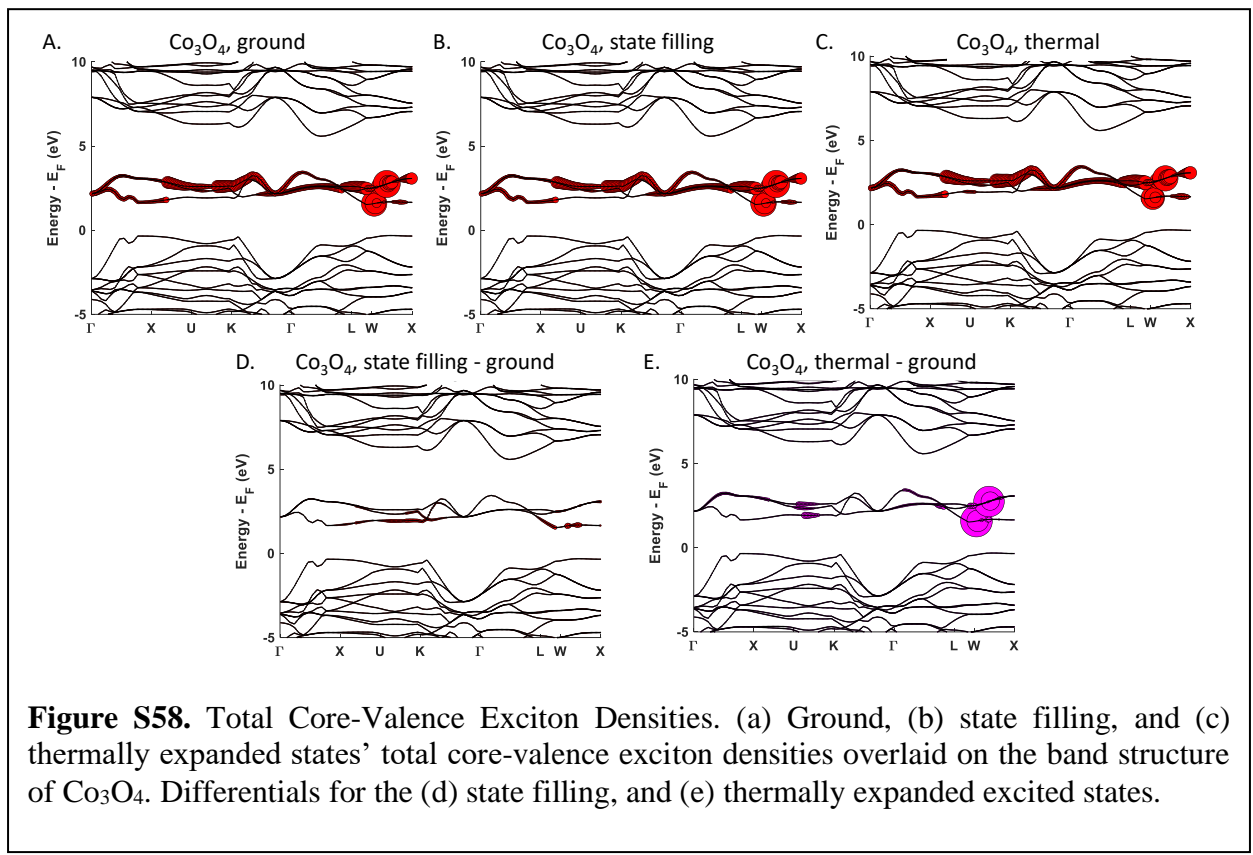

**Figure S58.** Total Core-Valence Exciton Densities. (a) Ground, (b) state filling, and (c) thermally expanded states' total core-valence exciton densities overlaid on the band structure of $Co_3O_4$. Differentials for the (d) state filling, and (e) thermally expanded excited states.

ii. Hamiltonian Decomposition of Exciton Components for ground, state filling, and thermally expanded models.

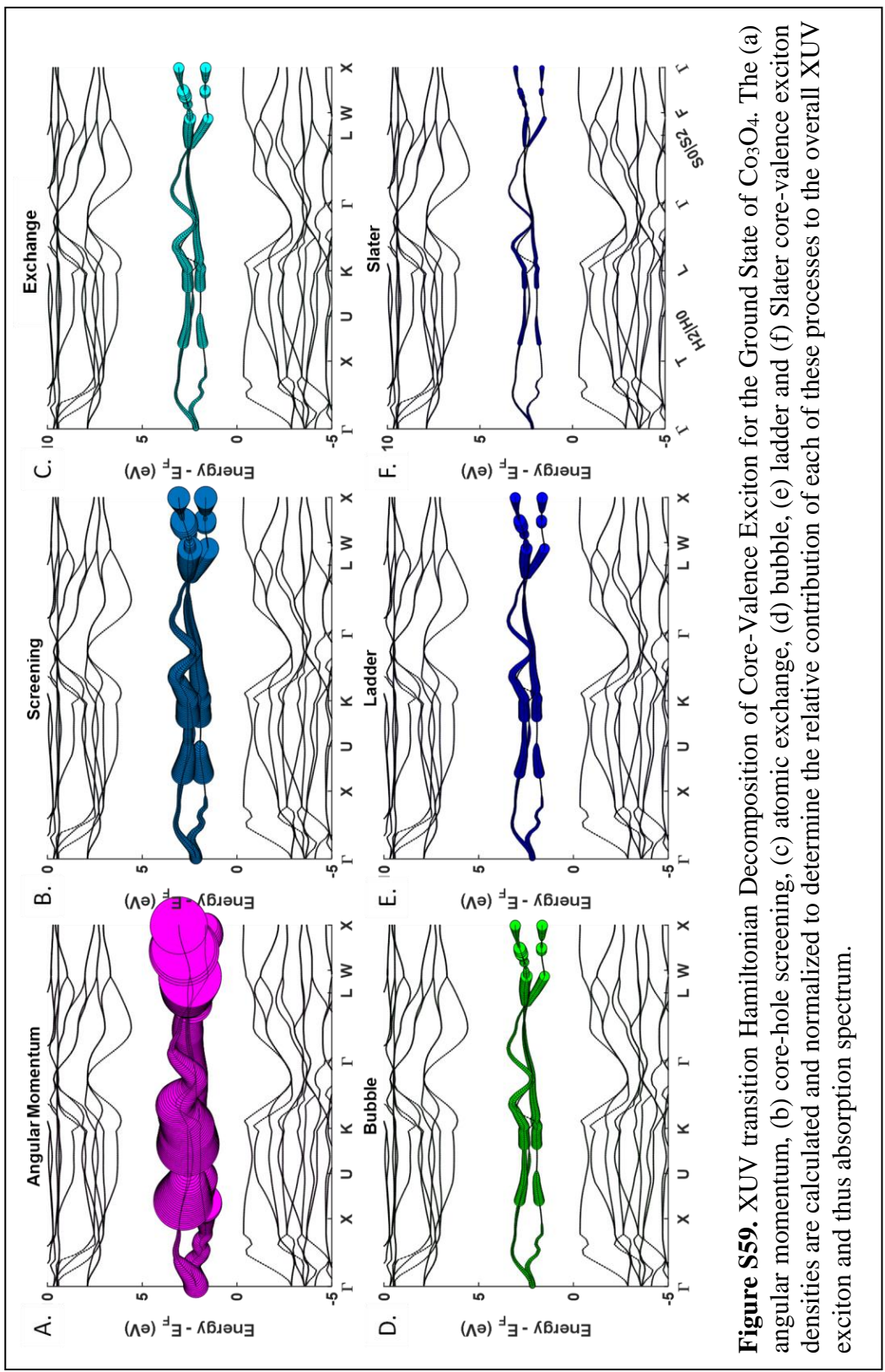

**Figure S59.** XUV transition Hamiltonian Decomposition of Core-Valence Exciton for the Ground State of $Co_3O_4$. The (a) angular momentum, (b) core-hole screening, (c) atomic exchange, (d) bubble, (e) ladder and (f) Slater core-valence exciton densities are calculated and normalized to determine the relative contribution of each of these processes to the overall XUV exciton and thus absorption spectrum.

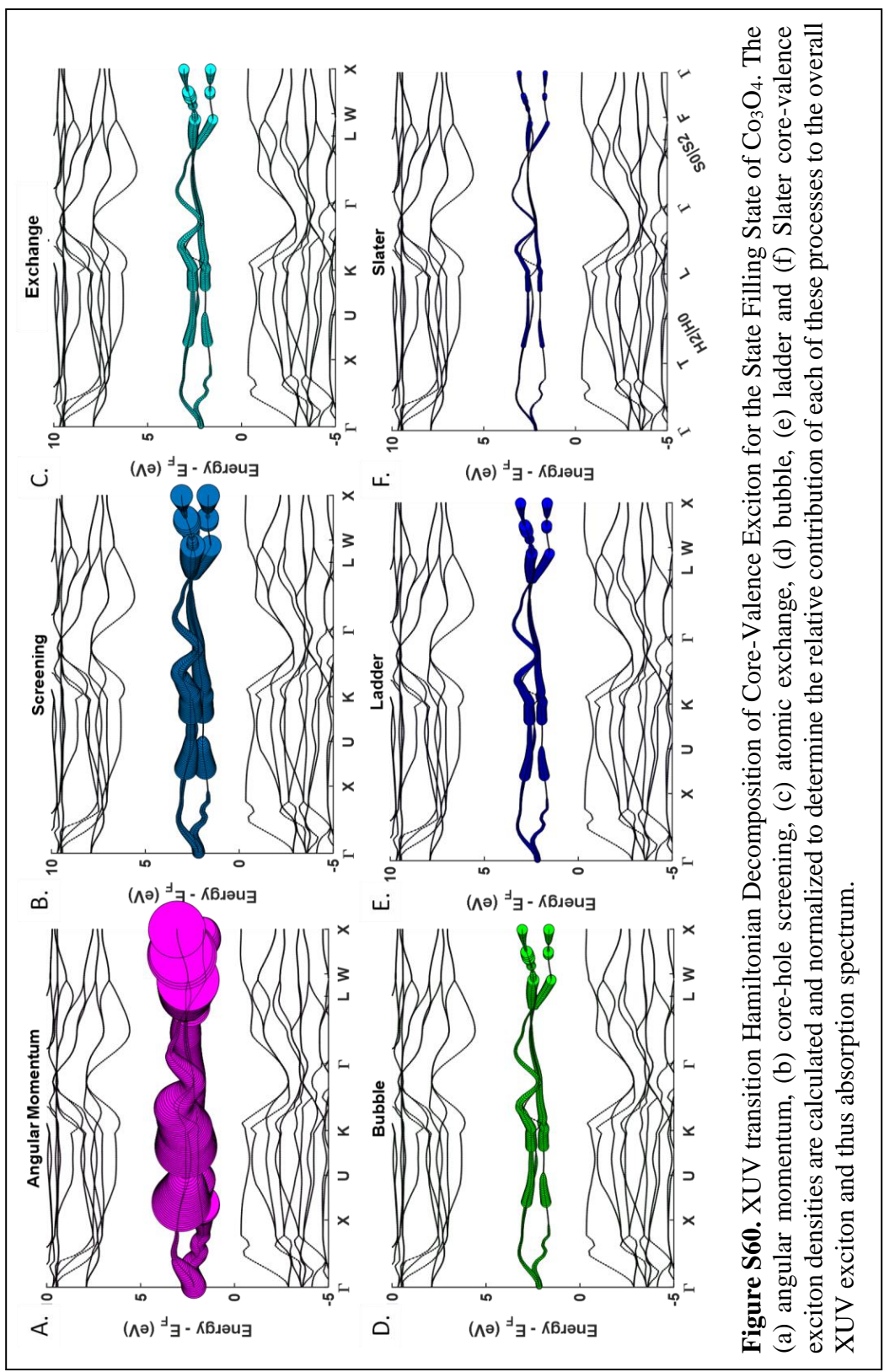

**Figure S60.** XUV transition Hamiltonian Decomposition of Core-Valence Exciton for the State Filling State of $Co_3O_4$. The (a) angular momentum, (b) core-hole screening, (c) atomic exchange, (d) bubble, (e) ladder and (f) Slater core-valence exciton densities are calculated and normalized to determine the relative contribution of each of these processes to the overall XUV exciton and thus absorption spectrum.

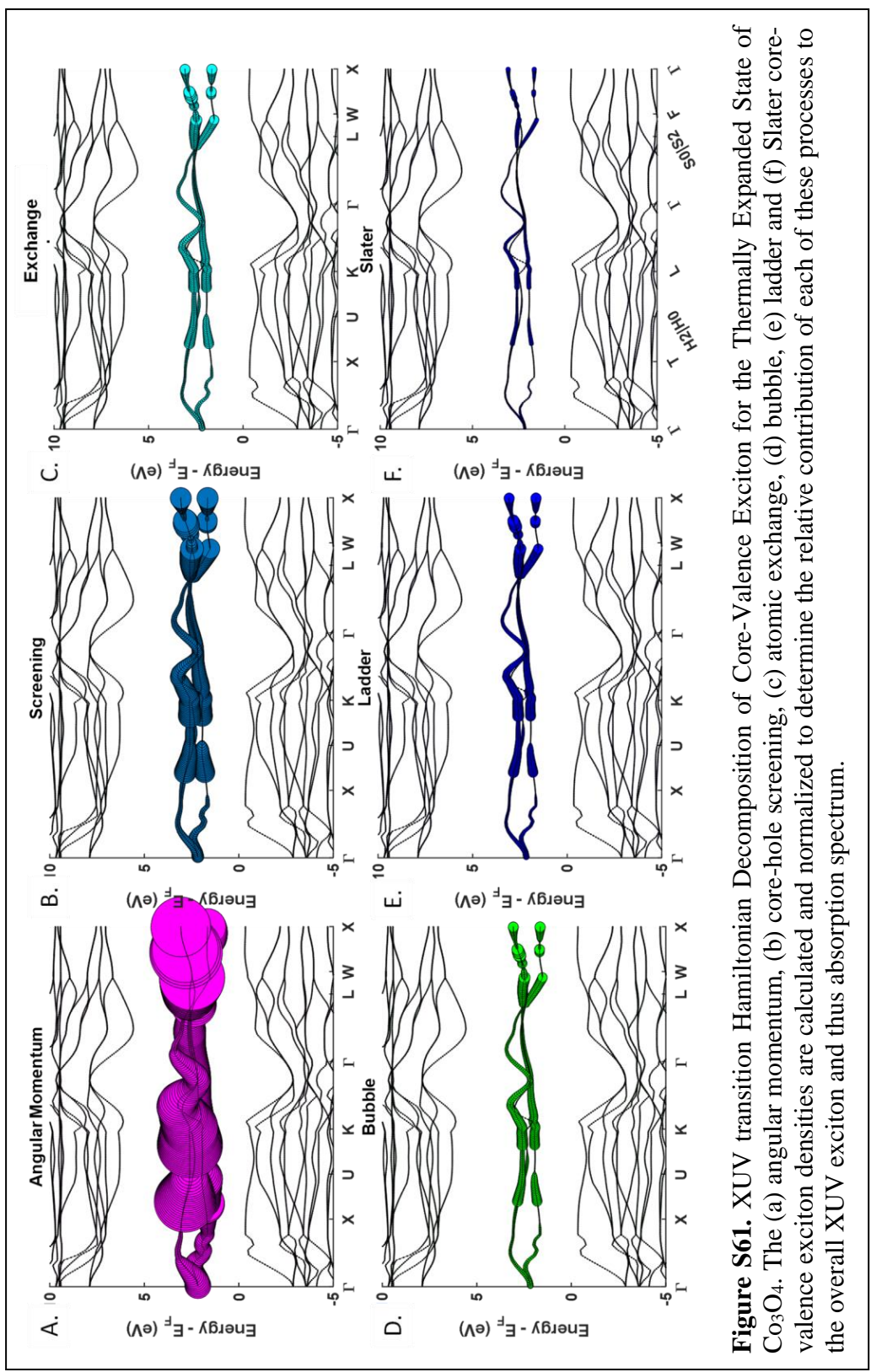

**Figure S61.** XUV transition Hamiltonian Decomposition of Core-Valence Exciton for the Thermally Expanded State of $Co_3O_4$. The (a) angular momentum, (b) core-hole screening, (c) atomic exchange, (d) bubble, (e) ladder and (f) Slater core-valence exciton densities are calculated and normalized to determine the relative contribution of each of these processes to the overall XUV exciton and thus absorption spectrum.

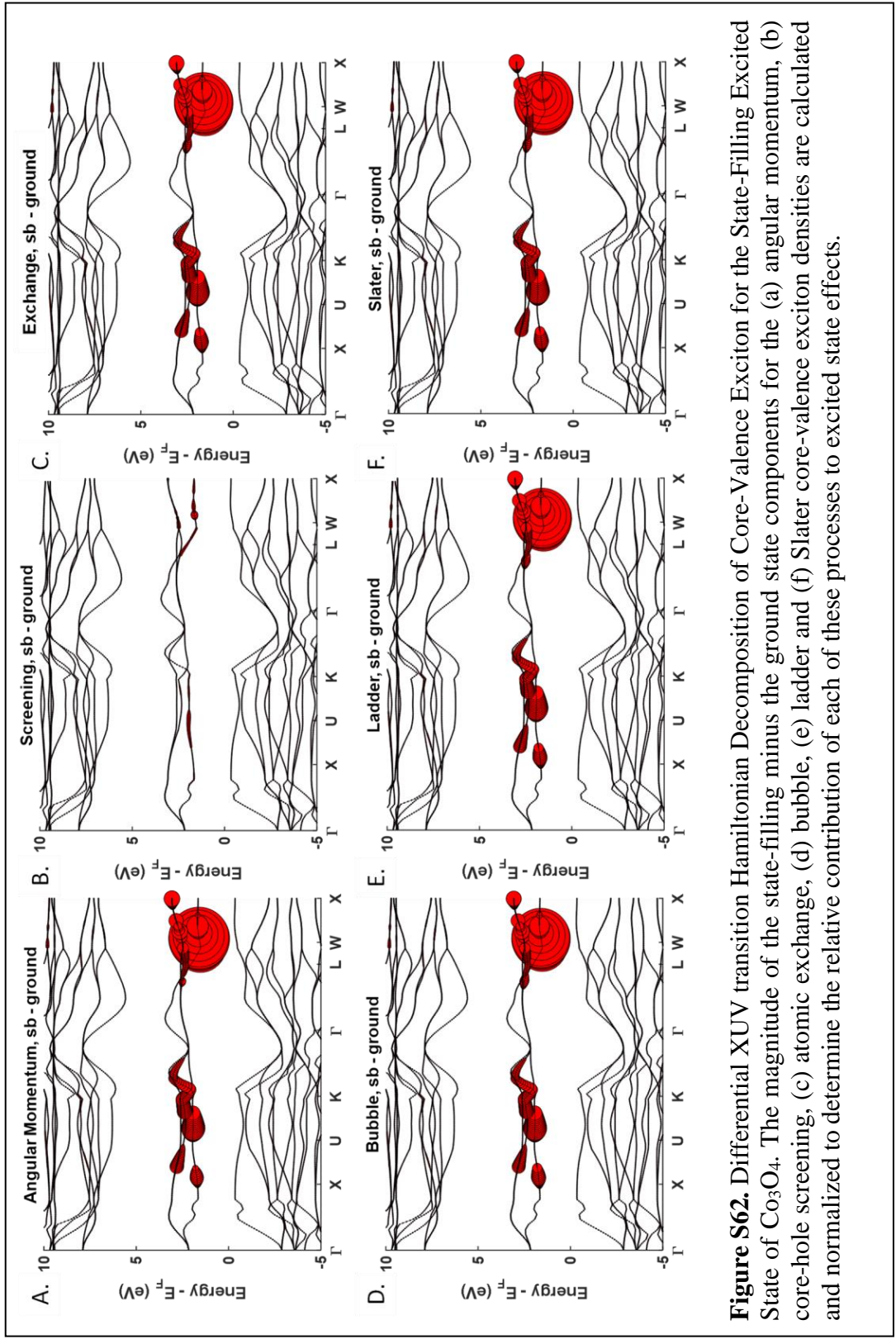

**Figure S62.** Differential XUV transition Hamiltonian Decomposition of Core-Valence Exciton for the State-Filling Excited State of $Co_3O_4$. The magnitude of the state-filling minus the ground state components for the (a) angular momentum, (b) core-hole screening, (c) atomic exchange, (d) bubble, (e) ladder and (f) Slater core-valence exciton densities are calculated and normalized to determine the relative contribution of each of these processes to excited state effects.

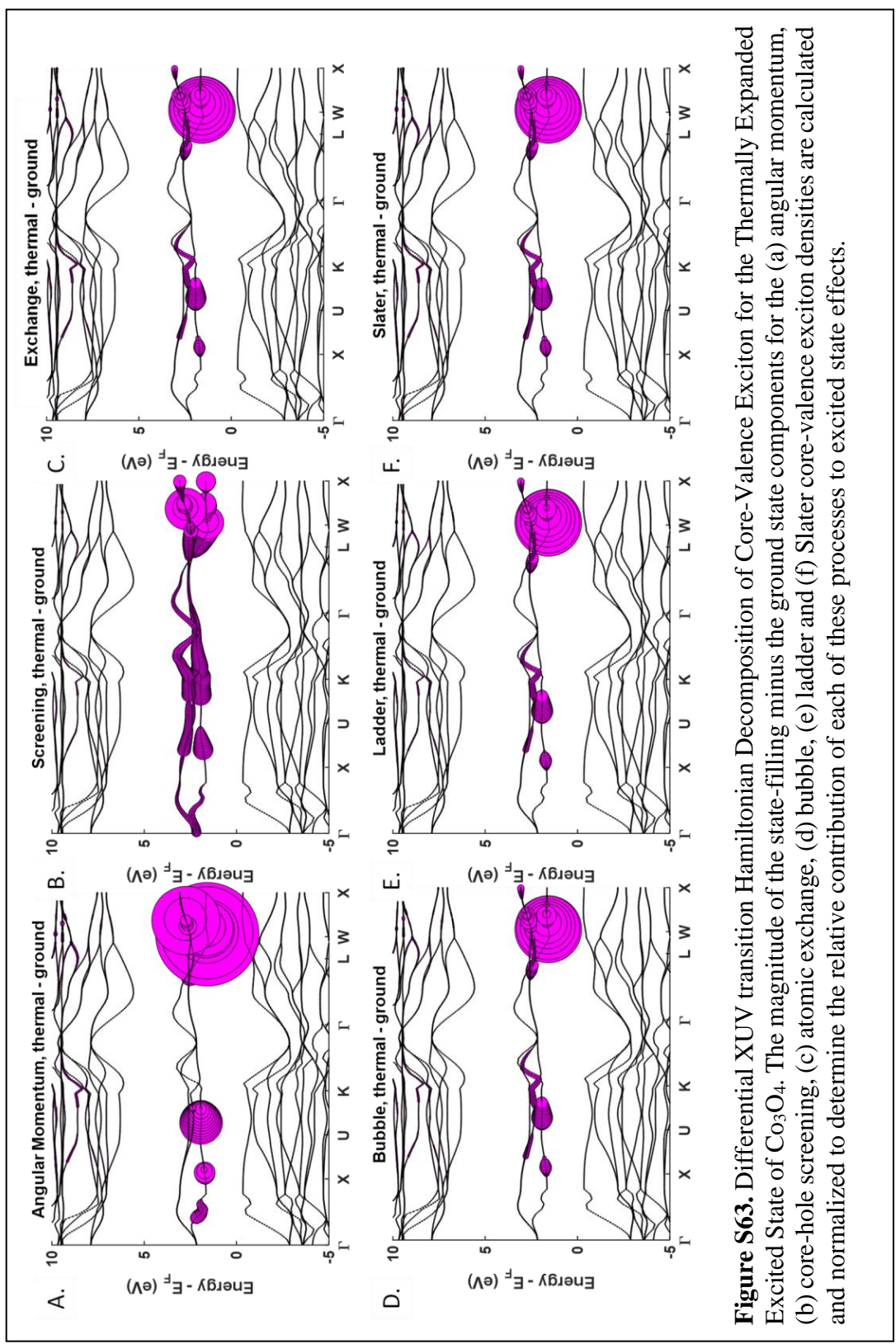

**Figure S63.** Differential XUV transition Hamiltonian Decomposition of Core-Valence Exciton for the Thermally Expanded Excited State of $Co_3O_4$. The magnitude of the state-filling minus the ground state components for the (a) angular momentum, (b) core-hole screening, (c) atomic exchange, (d) bubble, (e) ladder and (f) Slater core-valence exciton densities are calculated and normalized to determine the relative contribution of each of these processes to excited state effects.

## 7. NiO

a. Structural Data for Calculations

i. Ground State and State Blocking

**Unit Cell Parameters (bohr)**

{7.9689  8.1564  8.1564}

**Primitive Vectors**

{0.0  0.5  0.5

 0.5  0.0  0.5

 0.5  0.5  0.0}

**Reduced coordinates, ( x, y, z )**

Ni  0.5000  0.5000  0.5000

O   0.0000  0.0000  0.0000

ii. Thermally Expanded Lattice

**Unit Cell Parameters (bohr)**

{8.0646  8.254  8.254 }

**Primitive Vectors**

{0.0  0.5  0.5

 0.5  0.0  0.5

 0.5  0.5  0.0}

**Reduced coordinates, ( x, y, z )**

Ni  0.5000  0.5000  0.5000

O   0.0000  0.0000  0.0000

b. Ground State Calculations

i. Band Structure and DOS

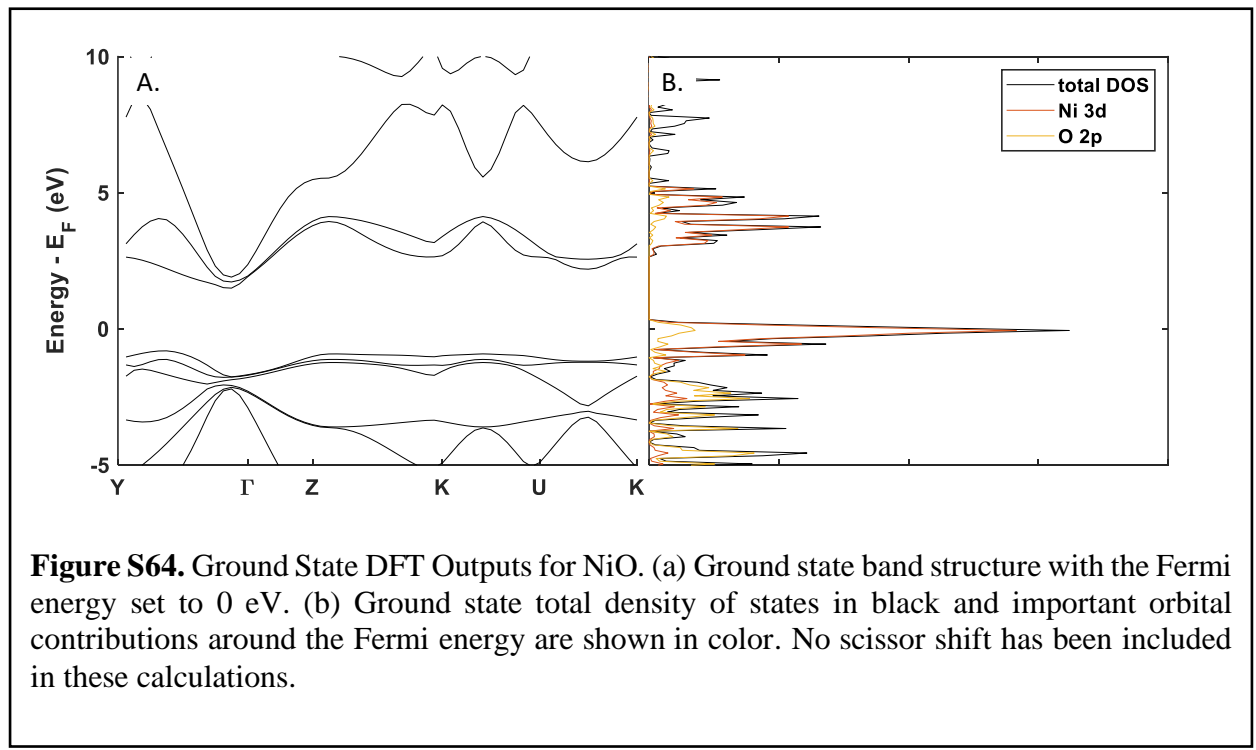

**Figure S64.** Ground State DFT Outputs for NiO. (a) Ground state band structure with the Fermi energy set to 0 eV. (b) Ground state total density of states in black and important orbital contributions around the Fermi energy are shown in color. No scissor shift has been included in these calculations.

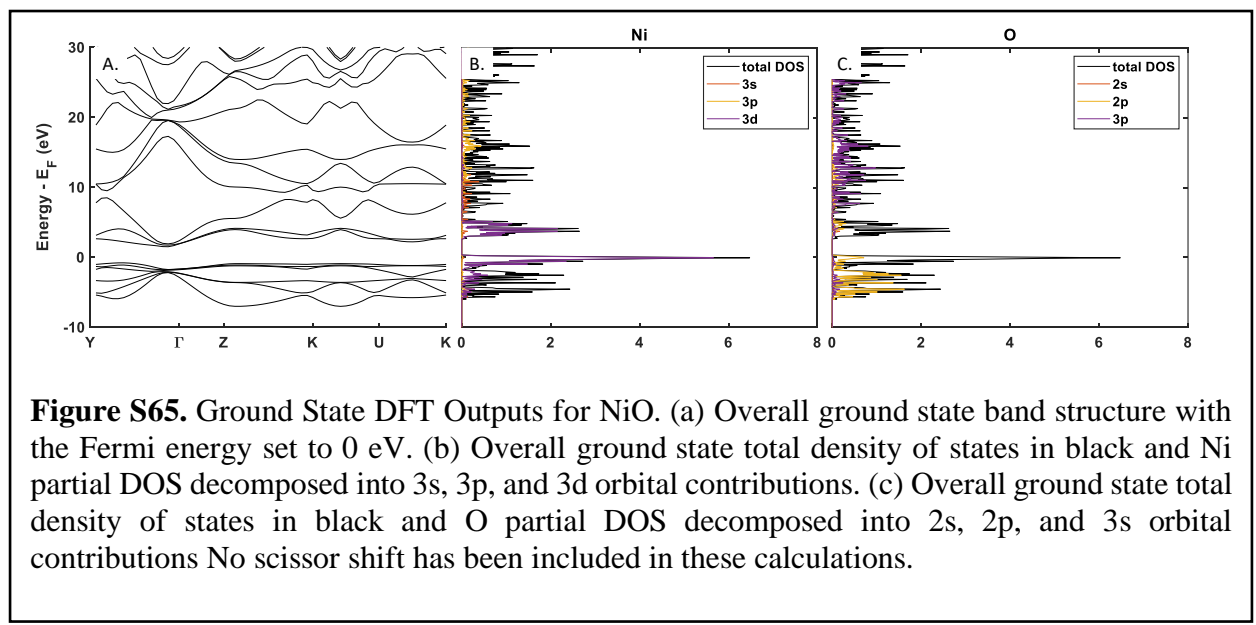

**Figure S65.** Ground State DFT Outputs for NiO. (a) Overall ground state band structure with the Fermi energy set to 0 eV. (b) Overall ground state total density of states in black and Ni partial DOS decomposed into 3s, 3p, and 3d orbital contributions. (c) Overall ground state total density of states in black and O partial DOS decomposed into 2s, 2p, and 3s orbital contributions No scissor shift has been included in these calculations.

## ii. Ground State Spectrum[10]

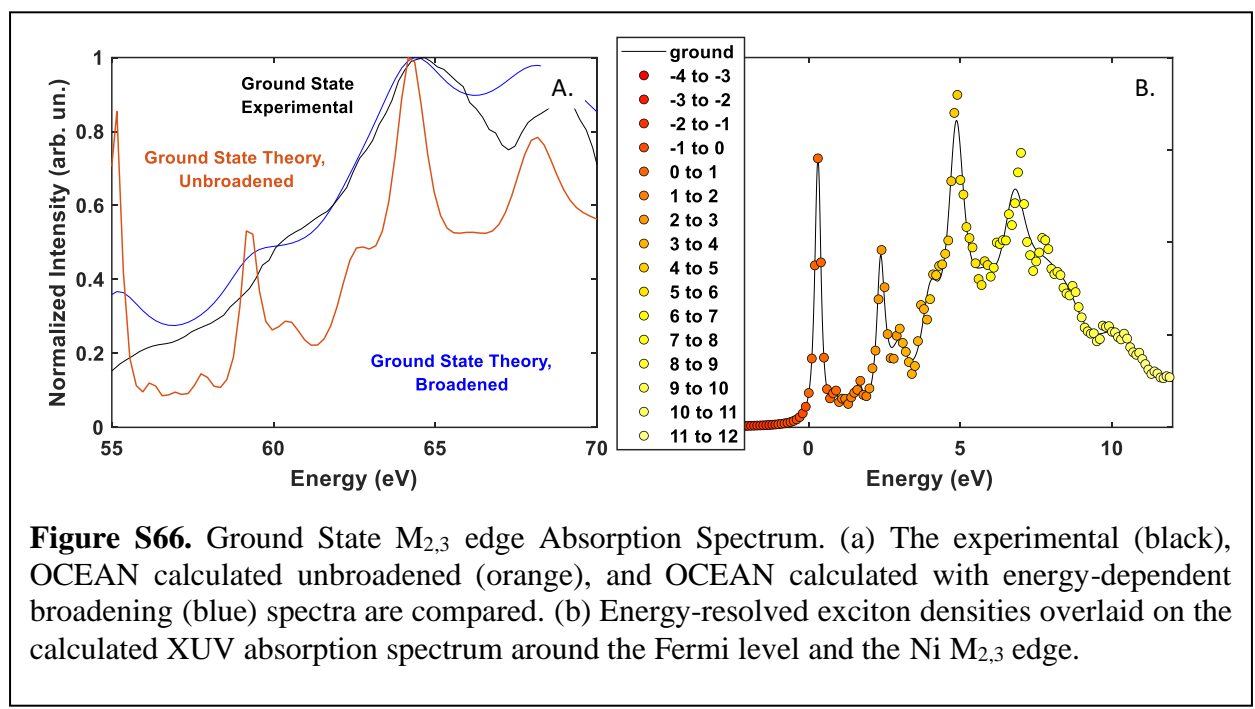

**Figure S66.** Ground State $M_{2,3}$ edge Absorption Spectrum. (a) The experimental (black), OCEAN calculated unbroadened (orange), and OCEAN calculated with energy-dependent broadening (blue) spectra are compared. (b) Energy-resolved exciton densities overlaid on the calculated XUV absorption spectrum around the Fermi level and the Ni $M_{2,3}$ edge.

iii. Ground State GMRES Energy Decomposition

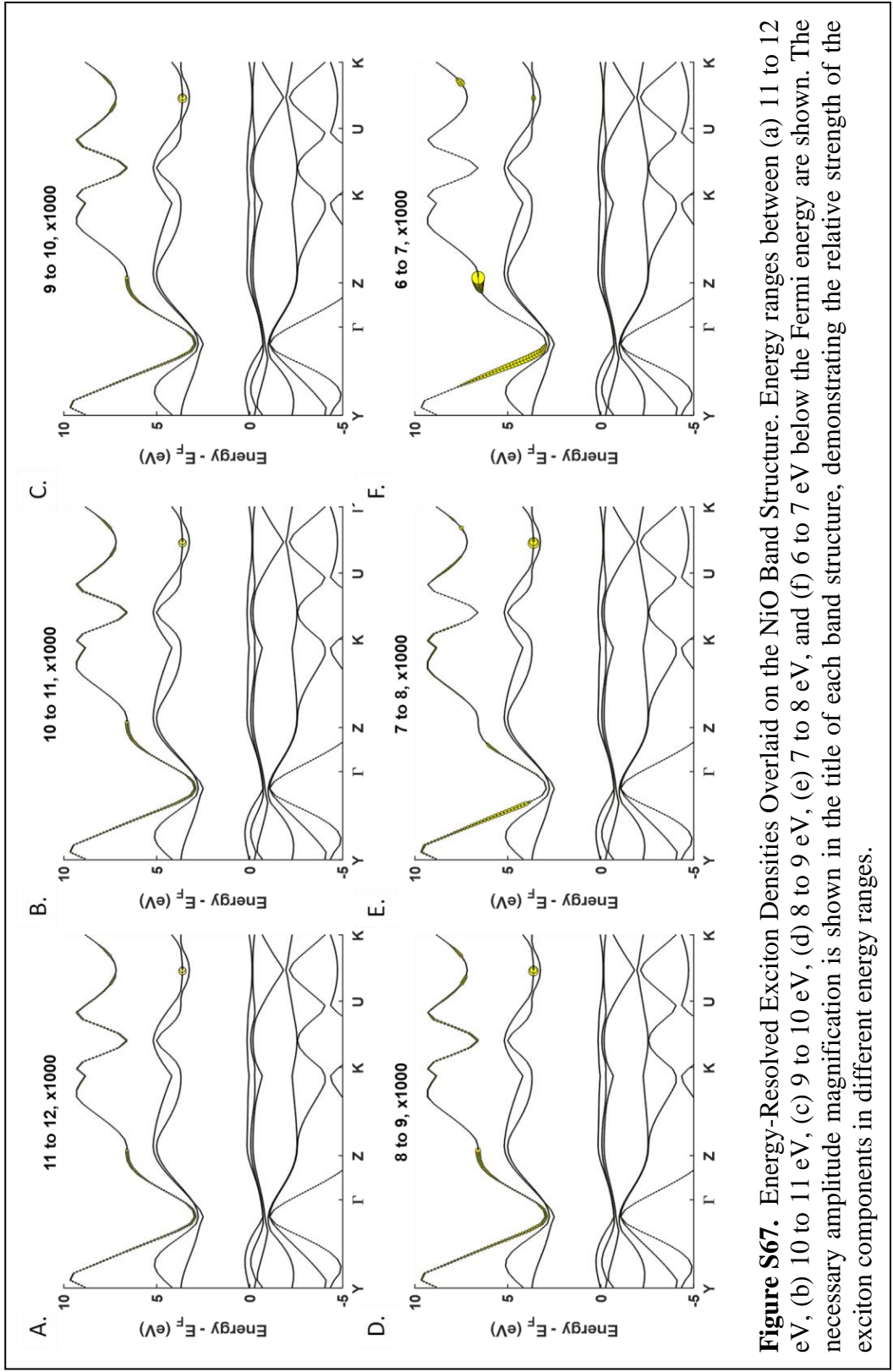

**Figure S67.** Energy-Resolved Exciton Densities Overlaid on the NiO Band Structure. Energy ranges between (a) 11 to 12 eV, (b) 10 to 11 eV, (c) 9 to 10 eV, (d) 8 to 9 eV, (e) 7 to 8 eV, and (f) 6 to 7 eV below the Fermi energy are shown. The necessary amplitude magnification is shown in the title of each band structure, demonstrating the relative strength of the exciton components in different energy ranges.

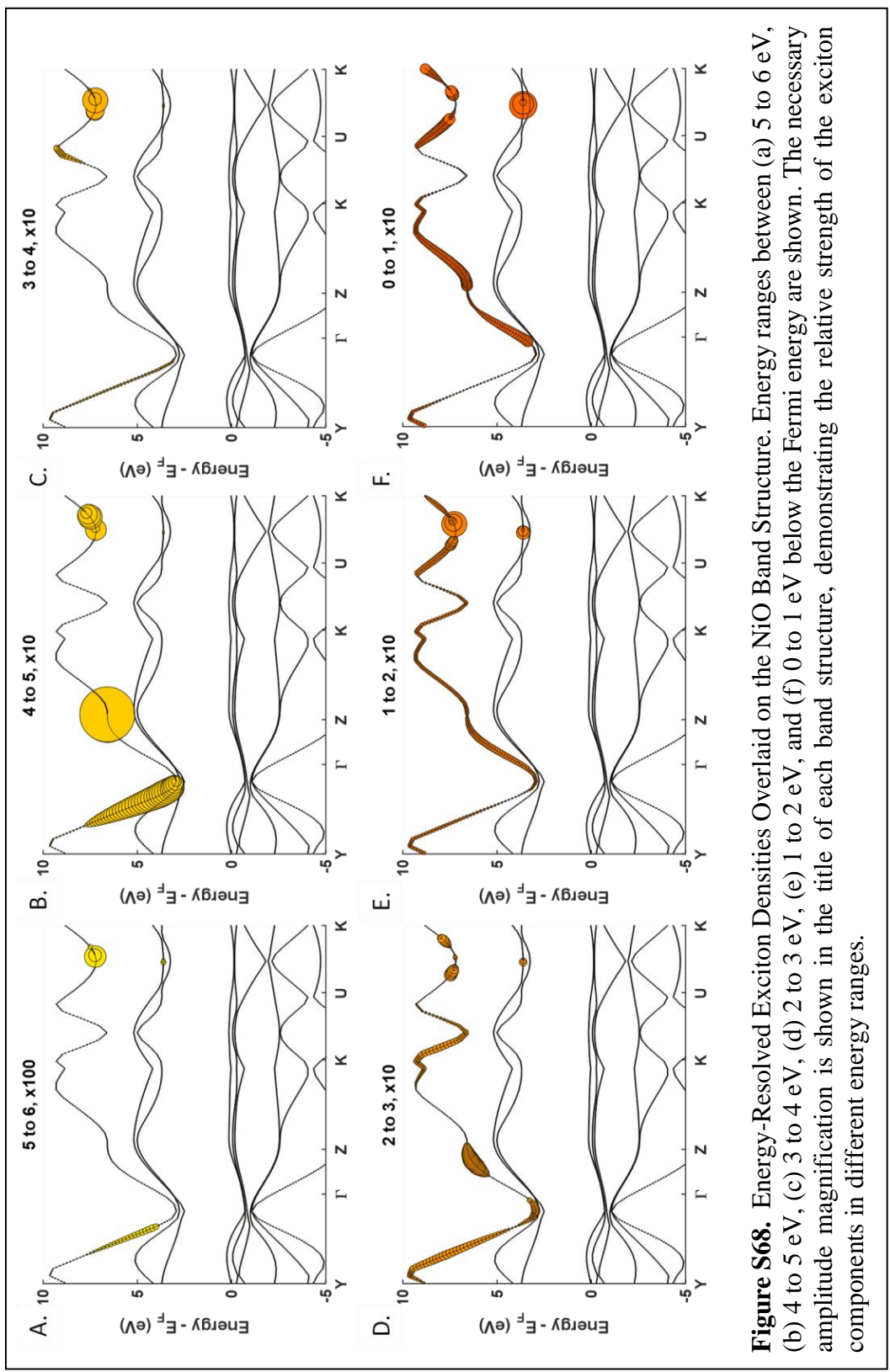

**Figure S68.** Energy-Resolved Exciton Densities Overlaid on the NiO Band Structure. Energy ranges between (a) 5 to 6 eV, (b) 4 to 5 eV, (c) 3 to 4 eV, (d) 2 to 3 eV, (e) 1 to 2 eV, and (f) 0 to 1 eV below the Fermi energy are shown. The necessary amplitude magnification is shown in the title of each band structure, demonstrating the relative strength of the exciton components in different energy ranges.

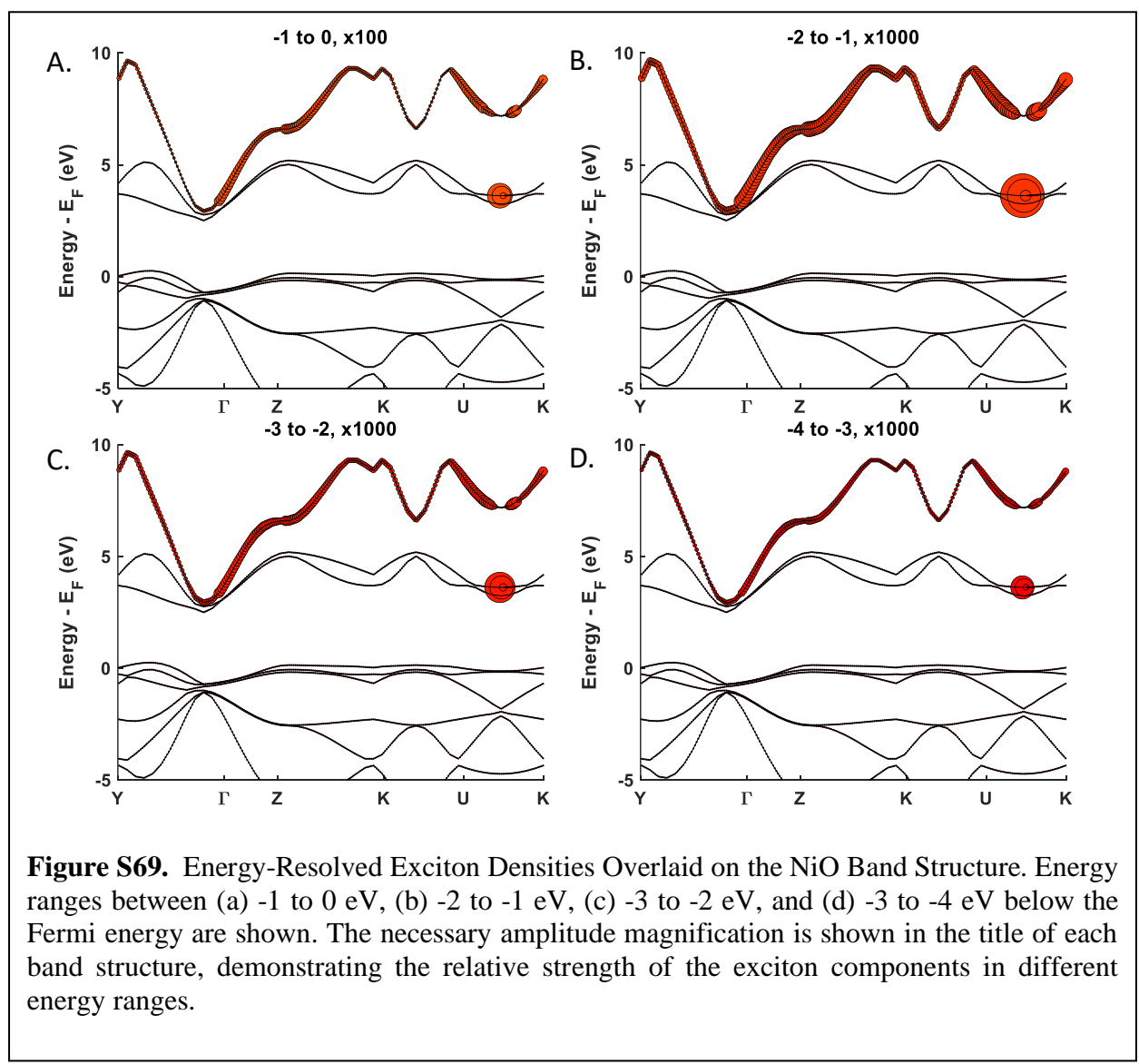

**Figure S69.** Energy-Resolved Exciton Densities Overlaid on the NiO Band Structure. Energy ranges between (a) -1 to 0 eV, (b) -2 to -1 eV, (c) -3 to -2 eV, and (d) -3 to -4 eV below the Fermi energy are shown. The necessary amplitude magnification is shown in the title of each band structure, demonstrating the relative strength of the exciton components in different energy ranges.

c. Excited State Calculations

    i. State filling band diagrams and Full Spectra

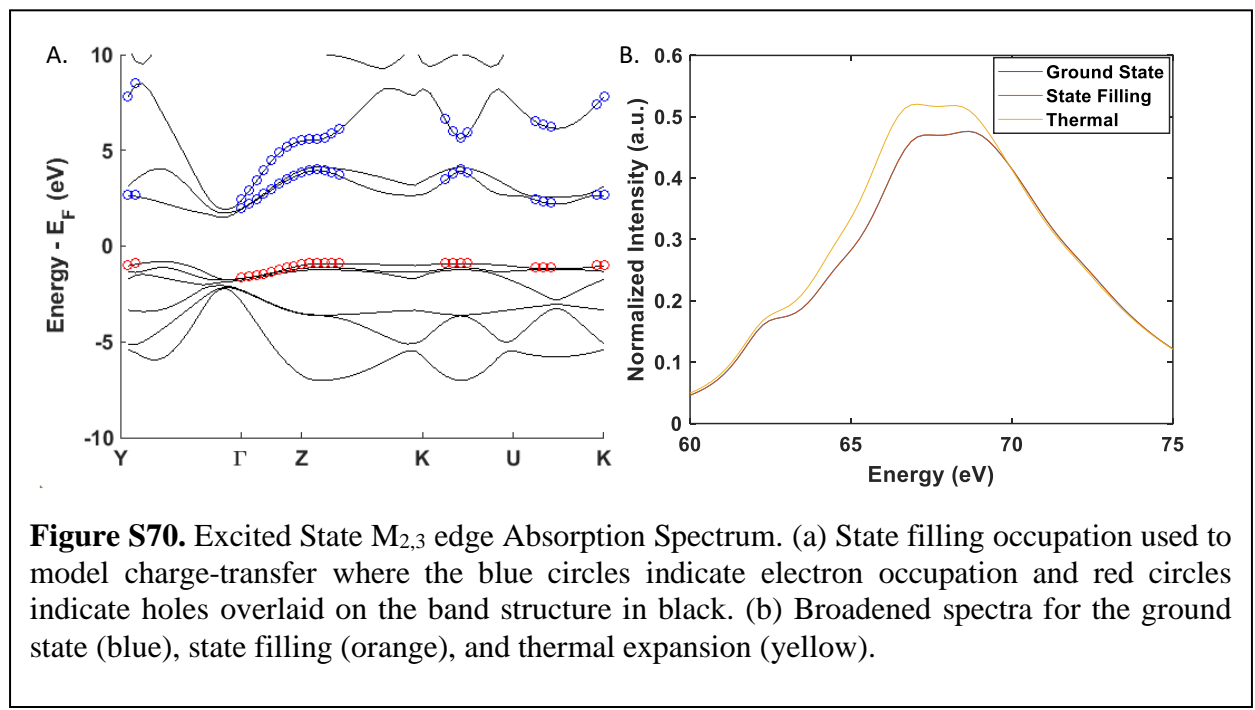

**Figure S70.** Excited State $M_{2,3}$ edge Absorption Spectrum. (a) State filling occupation used to model charge-transfer where the blue circles indicate electron occupation and red circles indicate holes overlaid on the band structure in black. (b) Broadened spectra for the ground state (blue), state filling (orange), and thermal expansion (yellow).

d. Hamiltonian Decompositions

    i. Total Exciton Comparisons

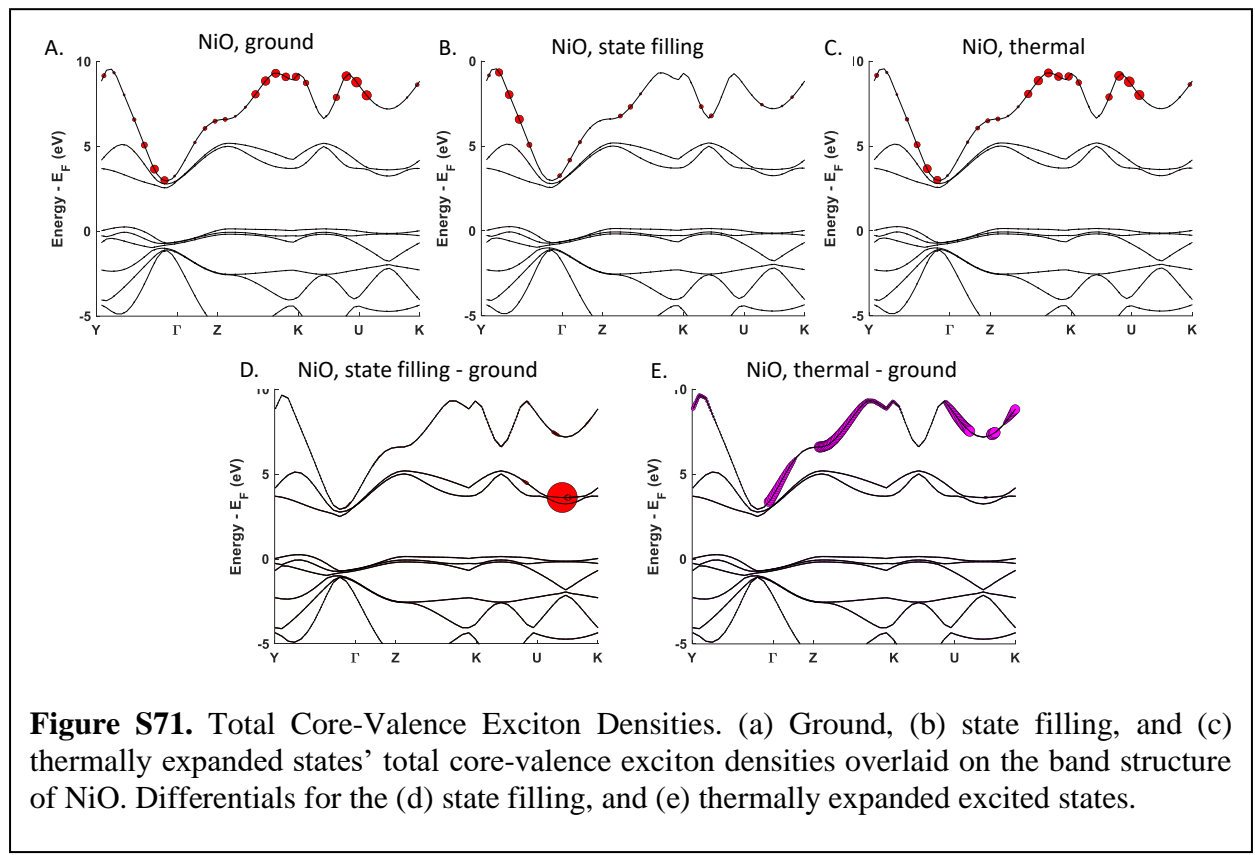

**Figure S71.** Total Core-Valence Exciton Densities. (a) Ground, (b) state filling, and (c) thermally expanded states' total core-valence exciton densities overlaid on the band structure of NiO. Differentials for the (d) state filling, and (e) thermally expanded excited states.

ii. Hamiltonian Decomposition of Exciton Components for ground, state filling, and

thermally expanded model.

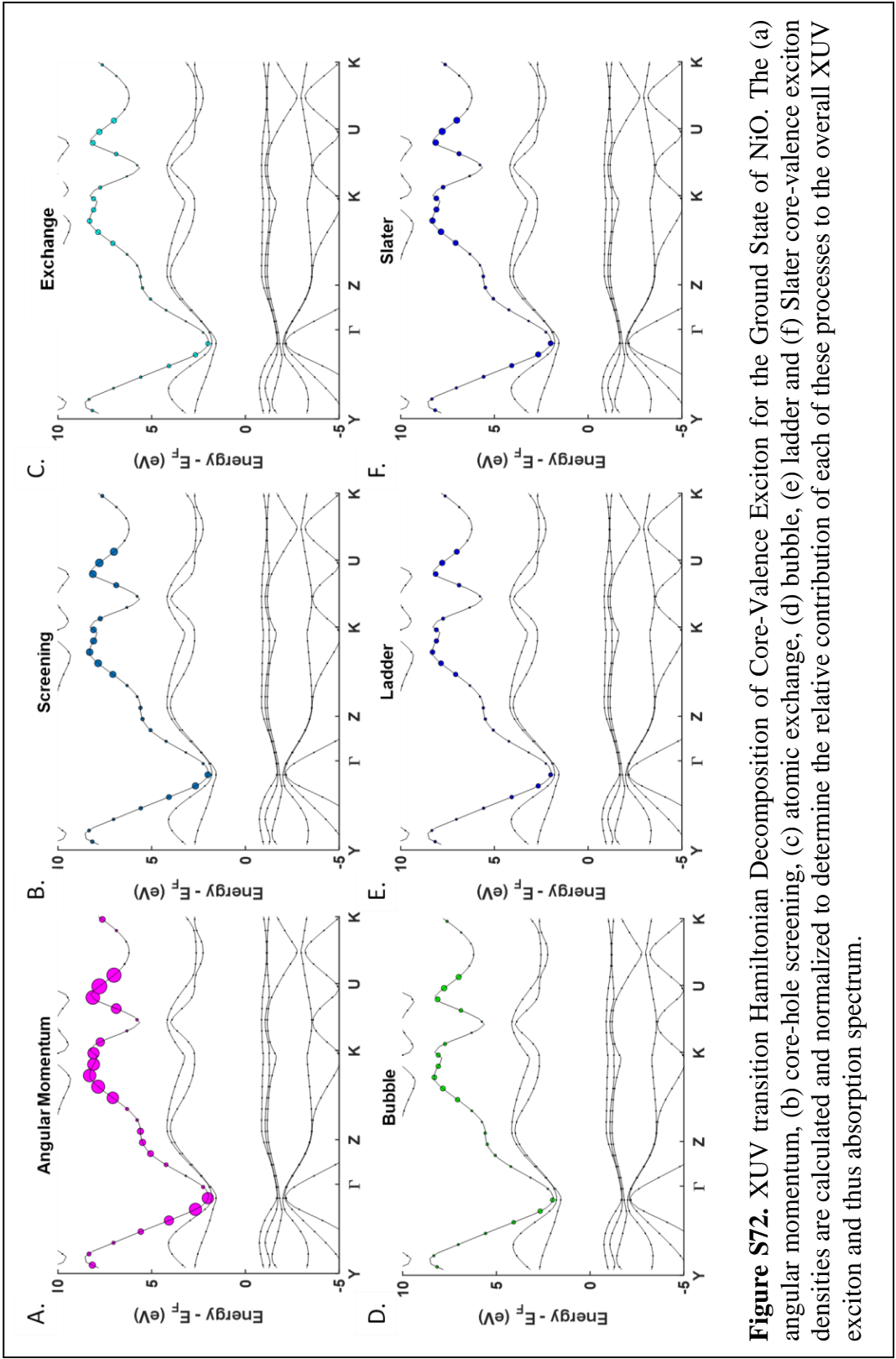

**Figure S72.** XUV transition Hamiltonian Decomposition of Core-Valence Exciton for the Ground State of NiO. The (a) angular momentum, (b) core-hole screening, (c) atomic exchange, (d) bubble, (e) ladder and (f) Slater core-valence exciton densities are calculated and normalized to determine the relative contribution of each of these processes to the overall XUV exciton and thus absorption spectrum.

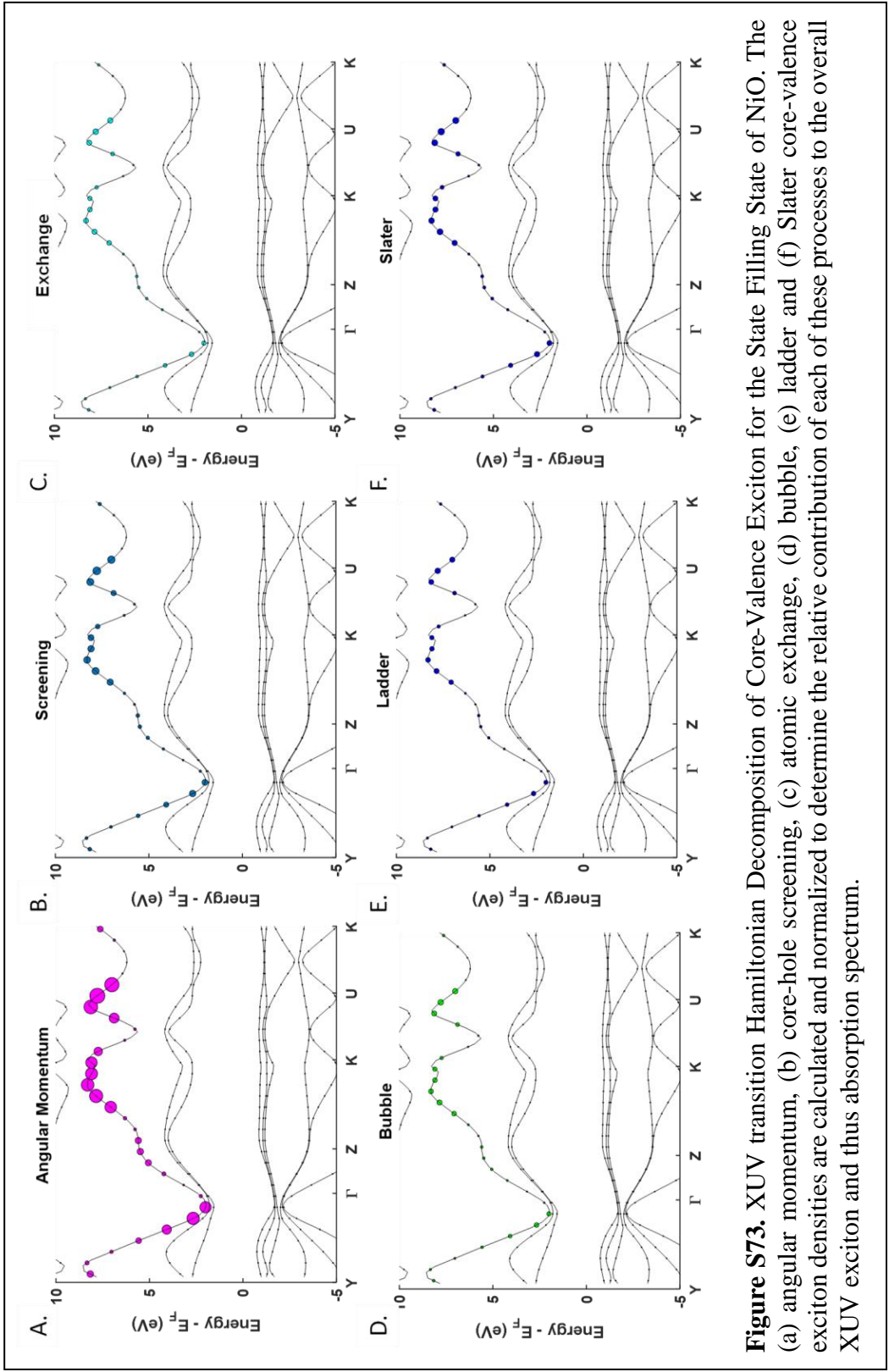

**Figure S73.** XUV transition Hamiltonian Decomposition of Core-Valence Exciton for the State Filling State of NiO. The (a) angular momentum, (b) core-hole screening, (c) atomic exchange, (d) bubble, (e) ladder and (f) Slater core-valence exciton densities are calculated and normalized to determine the relative contribution of each of these processes to the overall XUV exciton and thus absorption spectrum.

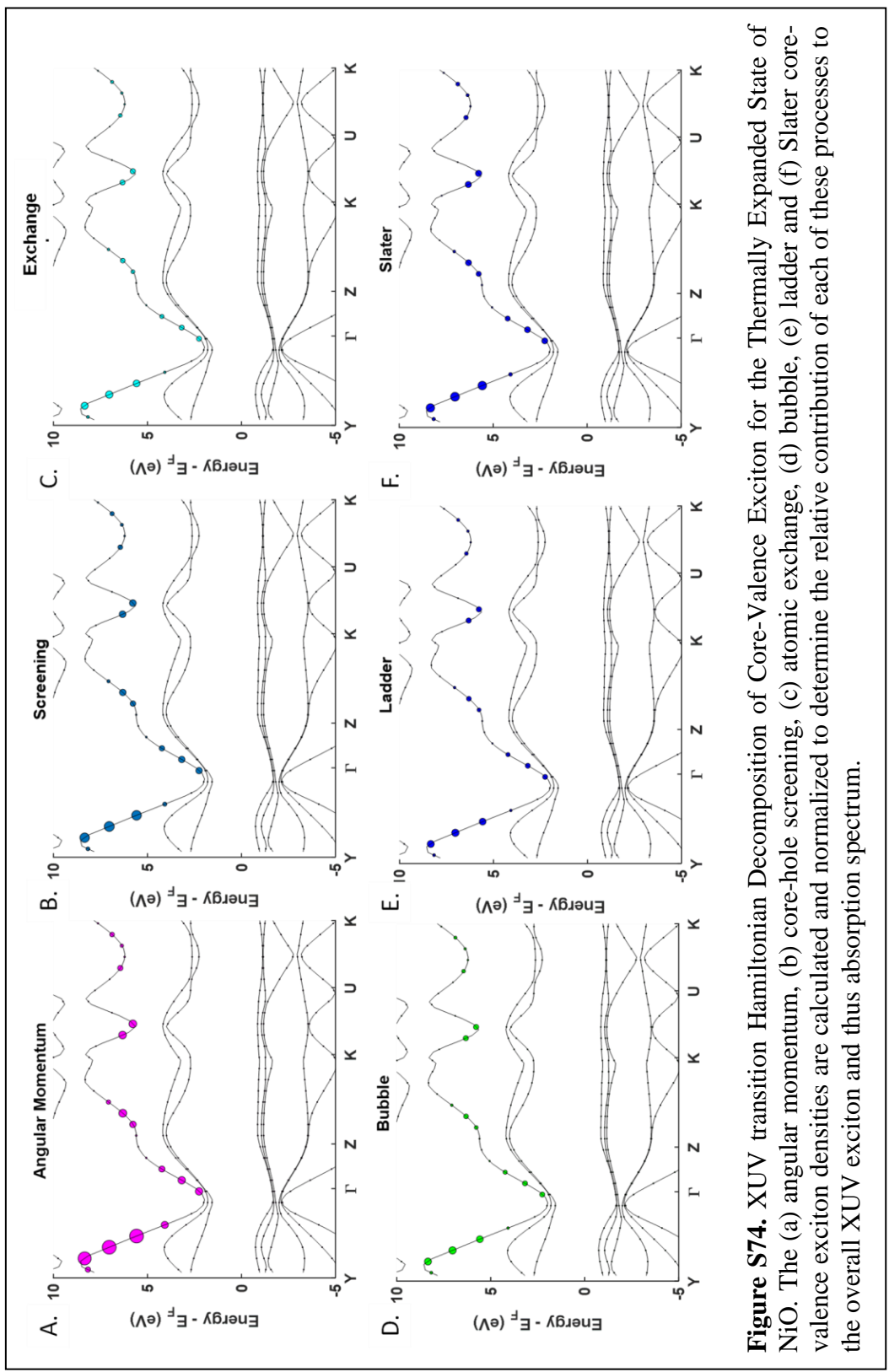

**Figure S74.** XUV transition Hamiltonian Decomposition of Core-Valence Exciton for the Thermally Expanded State of NiO. The (a) angular momentum, (b) core-hole screening, (c) atomic exchange, (d) bubble, (e) ladder and (f) Slater core-valence exciton densities are calculated and normalized to determine the relative contribution of each of these processes to the overall XUV exciton and thus absorption spectrum.

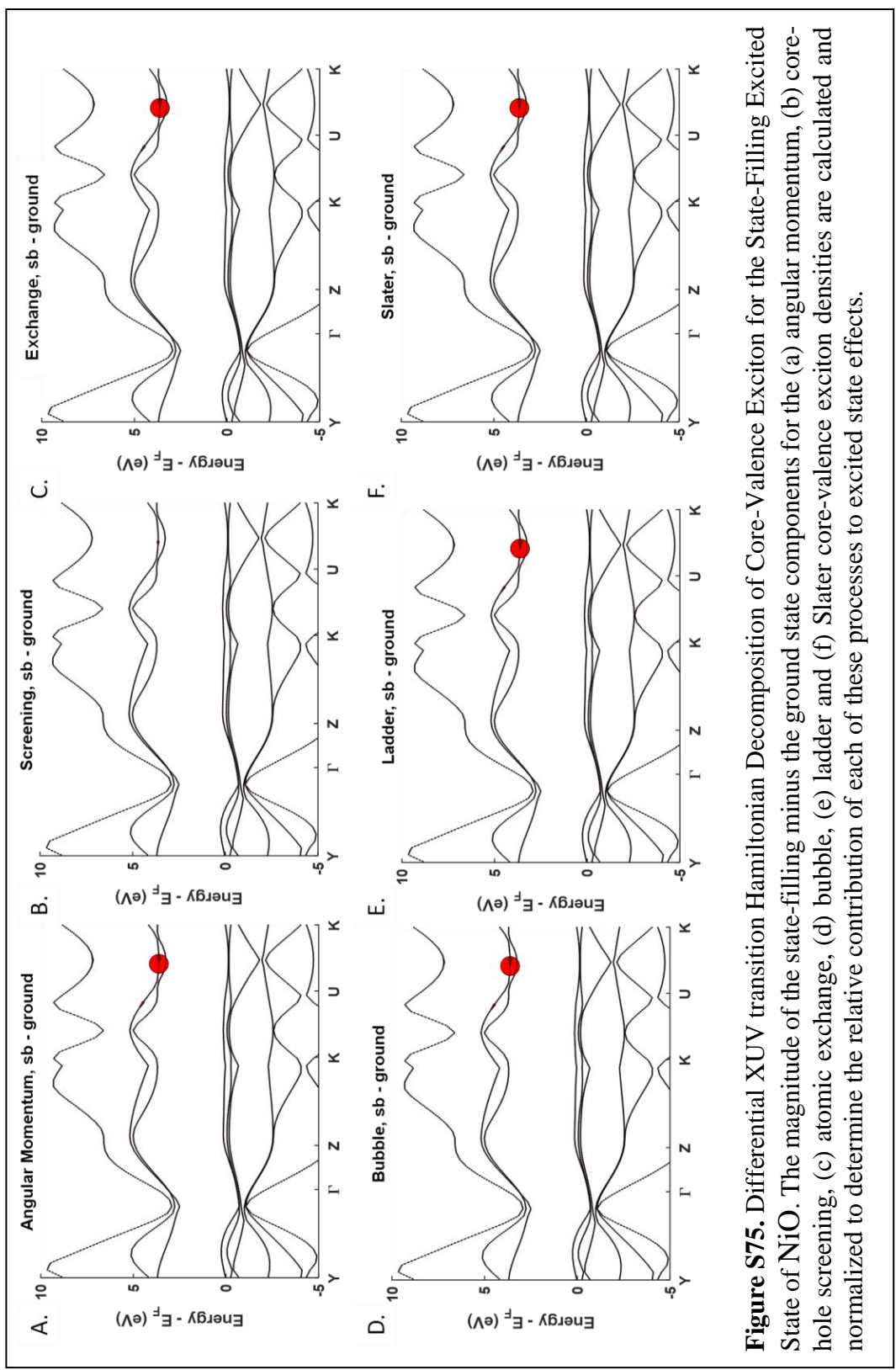

**Figure S75.** Differential XUV transition Hamiltonian Decomposition of Core-Valence Exciton for the State-Filling Excited State of NiO. The magnitude of the state-filling minus the ground state components for the (a) angular momentum, (b) core-hole screening, (c) atomic exchange, (d) bubble, (e) ladder and (f) Slater core-valence exciton densities are calculated and normalized to determine the relative contribution of each of these processes to excited state effects.

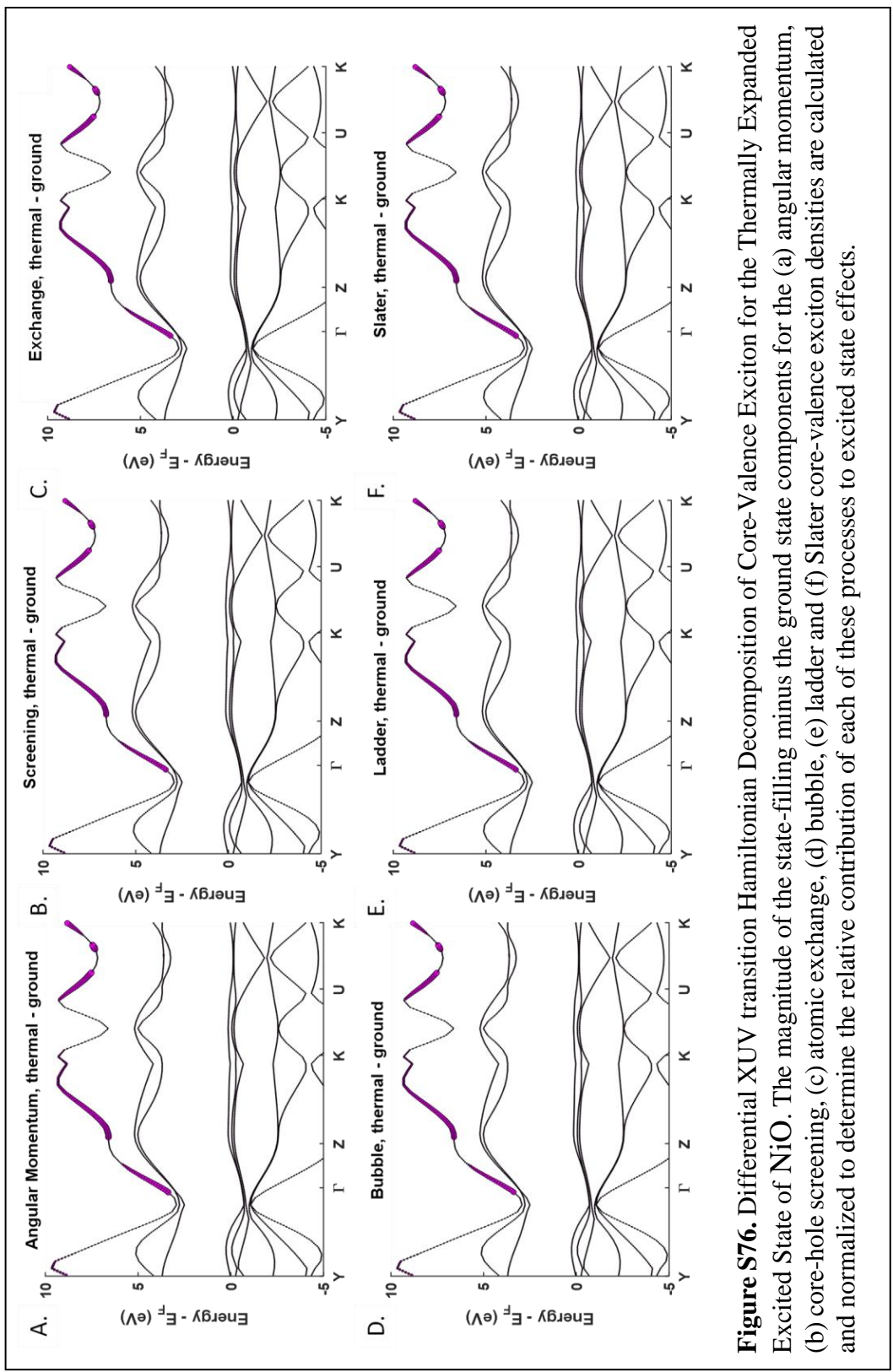

**Figure S76.** Differential XUV transition Hamiltonian Decomposition of Core-Valence Exciton for the Thermally Expanded Excited State of NiO. The magnitude of the state-filling minus the ground state components for the (a) angular momentum, (b) core-hole screening, (c) atomic exchange, (d) bubble, (e) ladder and (f) Slater core-valence exciton densities are calculated and normalized to determine the relative contribution of each of these processes to excited state effects.

## 8. CuO

a. Structural Data for Calculations

i. Ground State and State Blocking

**Unit Cell Parameters (bohr)**

{7.9984   7.6543 9.7940}

**Primitive Vector**

{0.5 -0.5   0.1

0.5   0.5   0.1

-0.305     0     0.9784 }

**Reduced coordinates, ( x, y, z )**

Cu  0.000000000   0.500000000   0.000000000

Cu  0.500000000   0.000000000   0.500000000

O   0.503680623   0.496319387   0.250000000

O   0.496319377   0.503680613   0.75000000

ii. Thermally Expanded Lattice

**Unit Cell Parameters (bohr)**

{8.0264 7.681 9.828}

**Primitive Vector**

{0.5    -0.5    0.1

0.5      0.5    0.1

-0.305   0     0.9784 }

**Reduced coordinates, ( x, y, z )**

Cu  0.000000000  0.500000000  0.000000000

Cu  0.500000000  0.000000000  0.500000000

O   0.503680623  0.496319387  0.250000000

O   0.496319377  0.503680613  0.75000000

b. Ground State Calculations

i. Band Structure and DOS

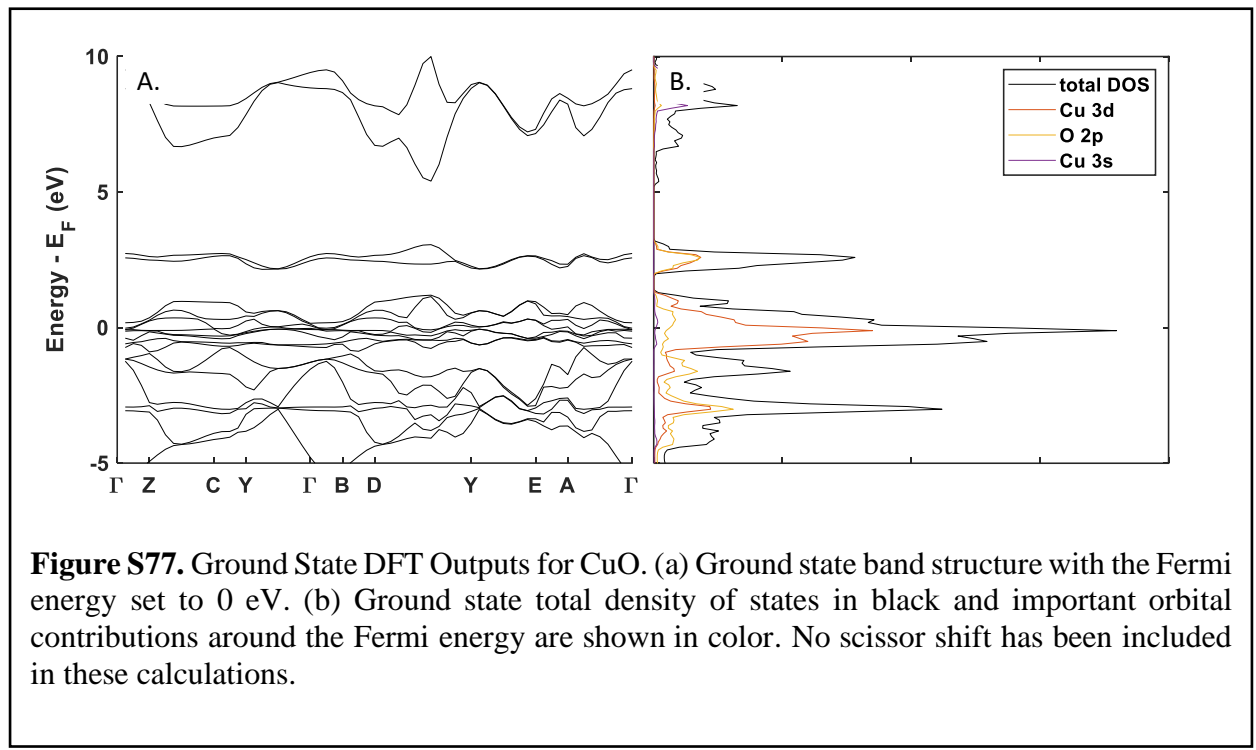

**Figure S77.** Ground State DFT Outputs for CuO. (a) Ground state band structure with the Fermi energy set to 0 eV. (b) Ground state total density of states in black and important orbital contributions around the Fermi energy are shown in color. No scissor shift has been included in these calculations.

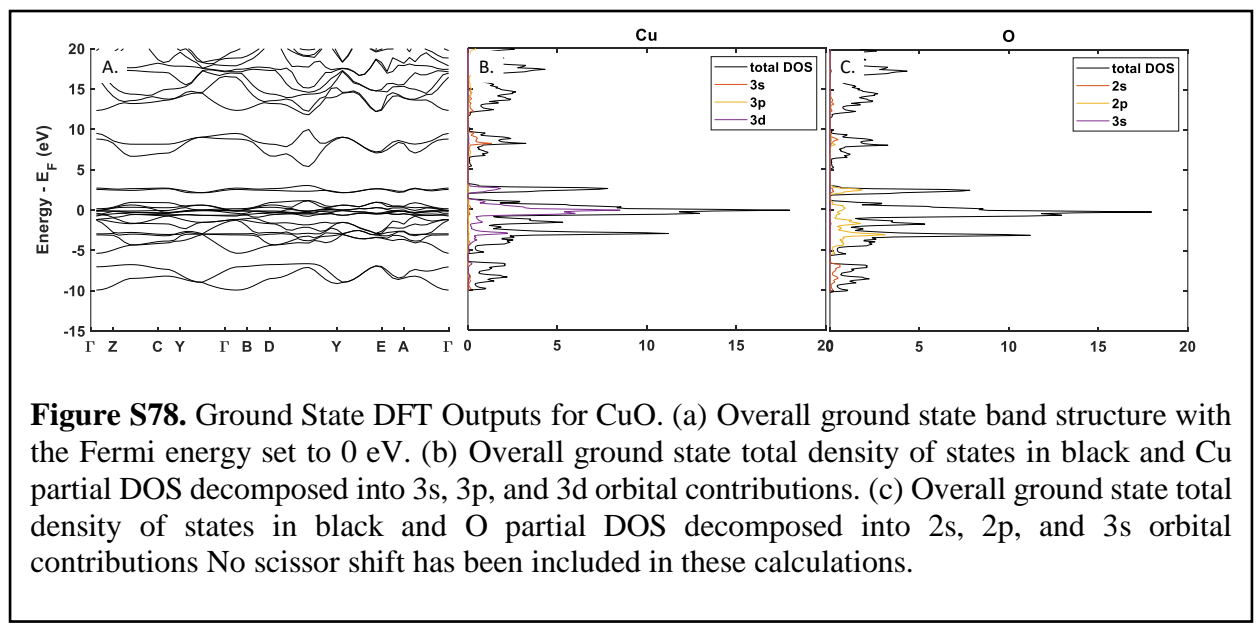

**Figure S78.** Ground State DFT Outputs for CuO. (a) Overall ground state band structure with the Fermi energy set to 0 eV. (b) Overall ground state total density of states in black and Cu partial DOS decomposed into 3s, 3p, and 3d orbital contributions. (c) Overall ground state total density of states in black and O partial DOS decomposed into 2s, 2p, and 3s orbital contributions No scissor shift has been included in these calculations.

## ii. Ground State Spectrum[10]

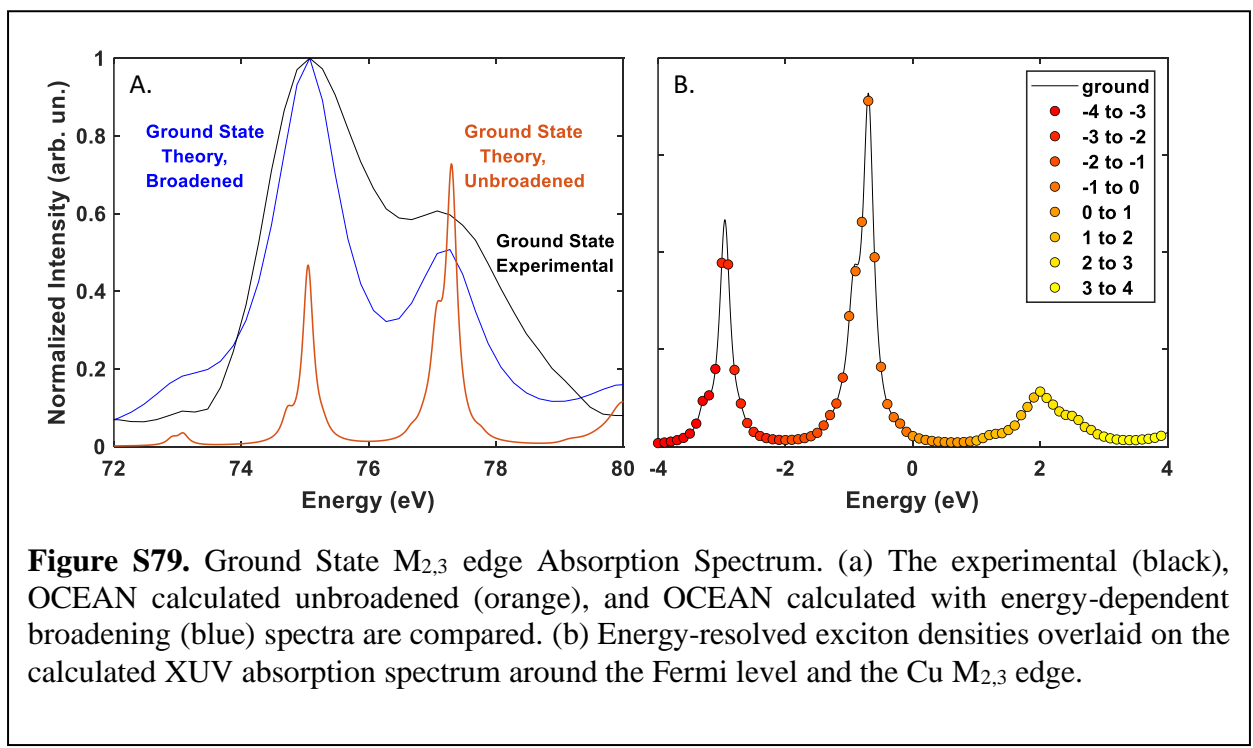

**Figure S79.** Ground State $M_{2,3}$ edge Absorption Spectrum. (a) The experimental (black), OCEAN calculated unbroadened (orange), and OCEAN calculated with energy-dependent broadening (blue) spectra are compared. (b) Energy-resolved exciton densities overlaid on the calculated XUV absorption spectrum around the Fermi level and the Cu $M_{2,3}$ edge.

iii. Ground State GMRES Energy Decomposition

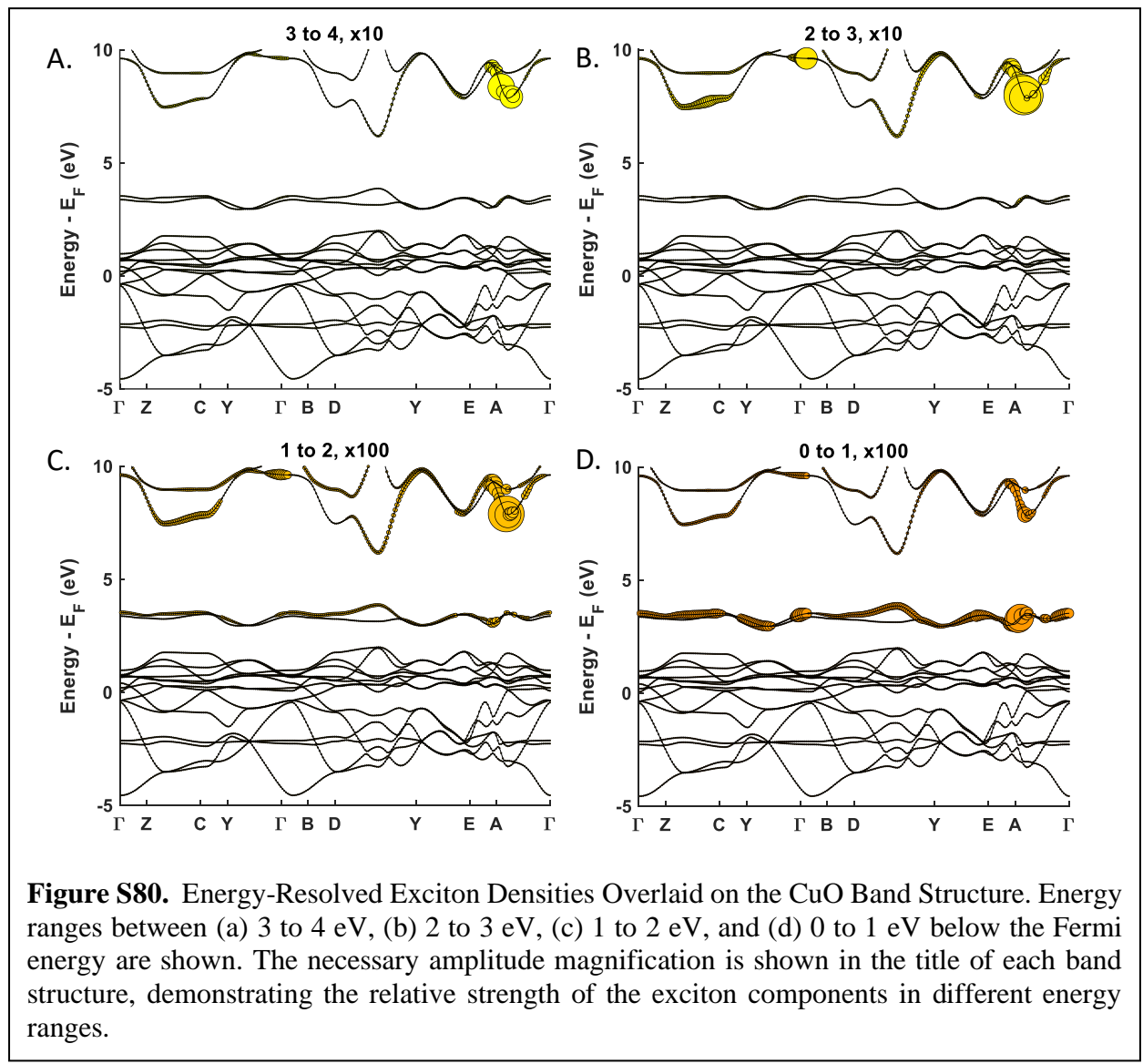

**Figure S80.** Energy-Resolved Exciton Densities Overlaid on the CuO Band Structure. Energy ranges between (a) 3 to 4 eV, (b) 2 to 3 eV, (c) 1 to 2 eV, and (d) 0 to 1 eV below the Fermi energy are shown. The necessary amplitude magnification is shown in the title of each band structure, demonstrating the relative strength of the exciton components in different energy ranges.

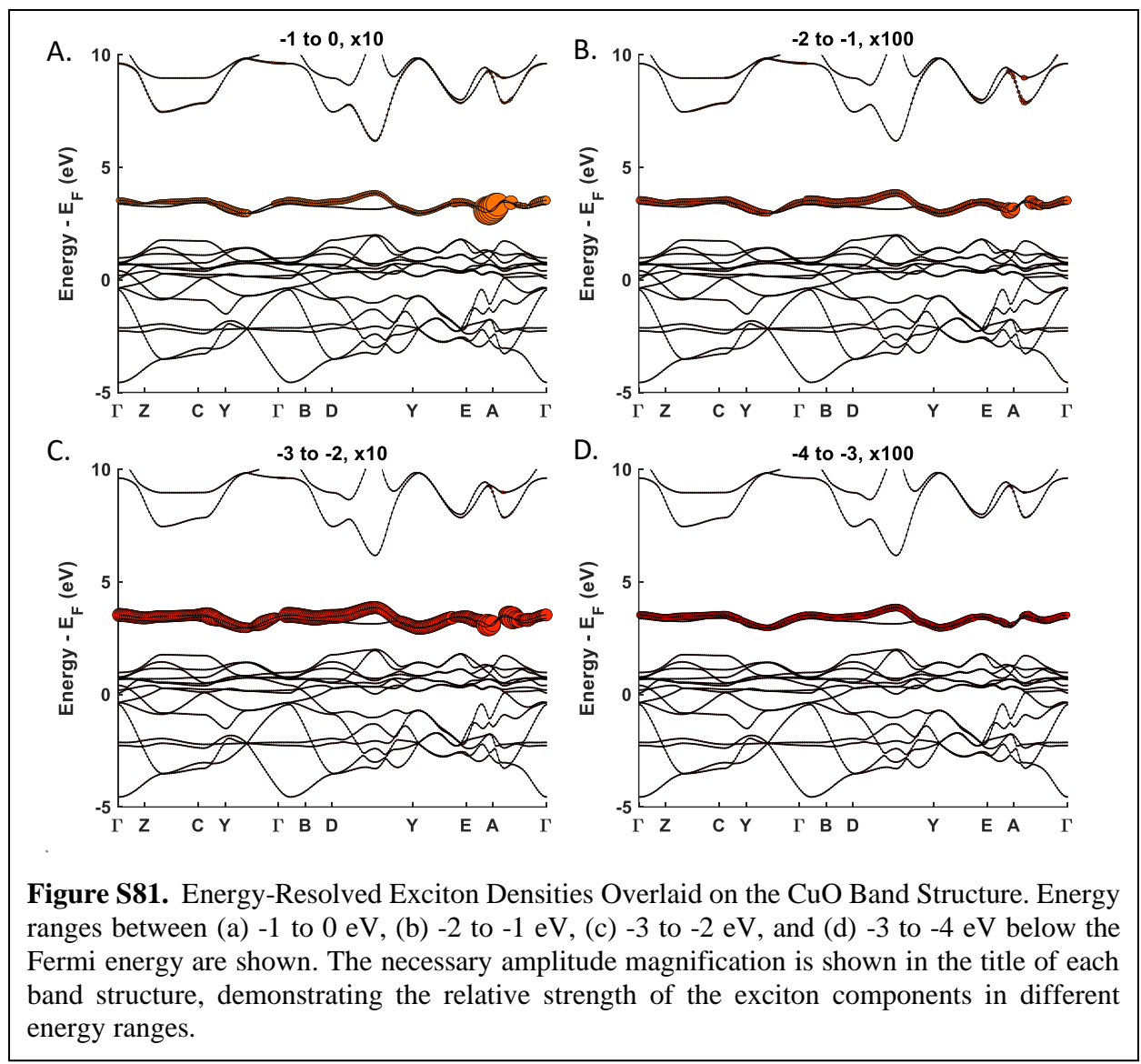

**Figure S81.** Energy-Resolved Exciton Densities Overlaid on the CuO Band Structure. Energy ranges between (a) -1 to 0 eV, (b) -2 to -1 eV, (c) -3 to -2 eV, and (d) -3 to -4 eV below the Fermi energy are shown. The necessary amplitude magnification is shown in the title of each band structure, demonstrating the relative strength of the exciton components in different energy ranges.

c. Excited State Calculations

    i. State filling band diagrams and Full Spectra

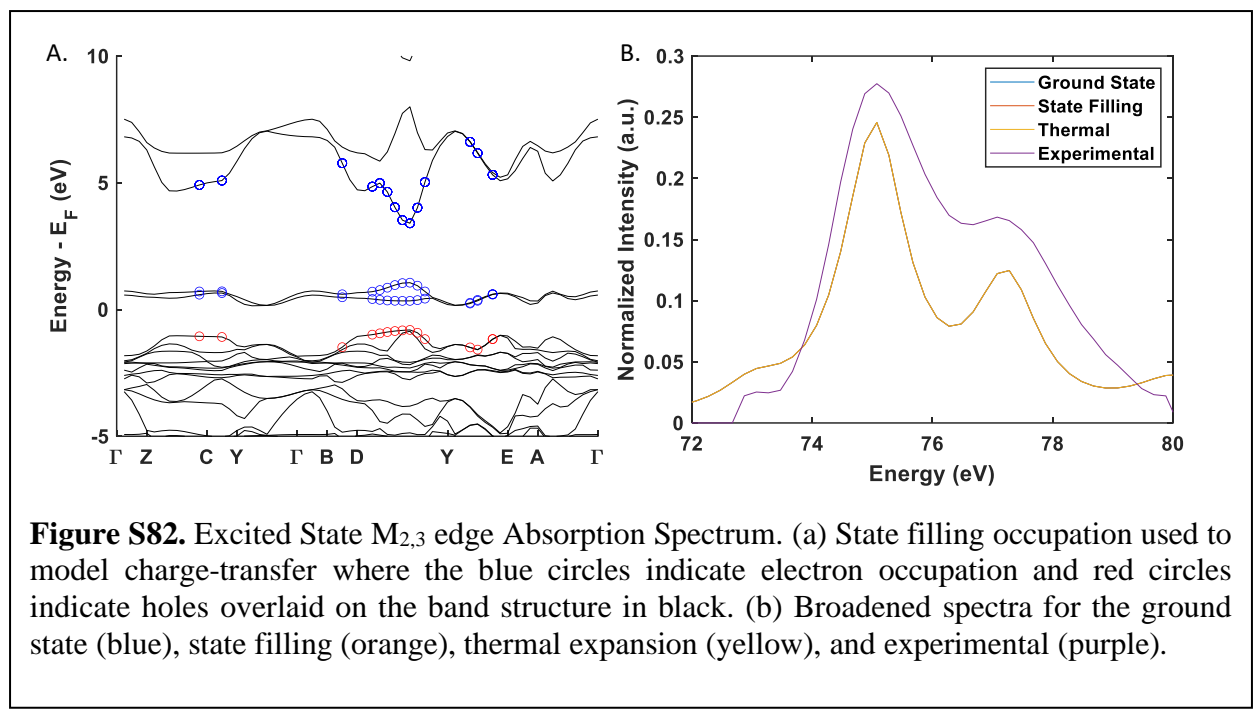

**Figure S82.** Excited State $M_{2,3}$ edge Absorption Spectrum. (a) State filling occupation used to model charge-transfer where the blue circles indicate electron occupation and red circles indicate holes overlaid on the band structure in black. (b) Broadened spectra for the ground state (blue), state filling (orange), thermal expansion (yellow), and experimental (purple).

### d. Hamiltonian Decompositions

#### i. Total Exciton Comparisons

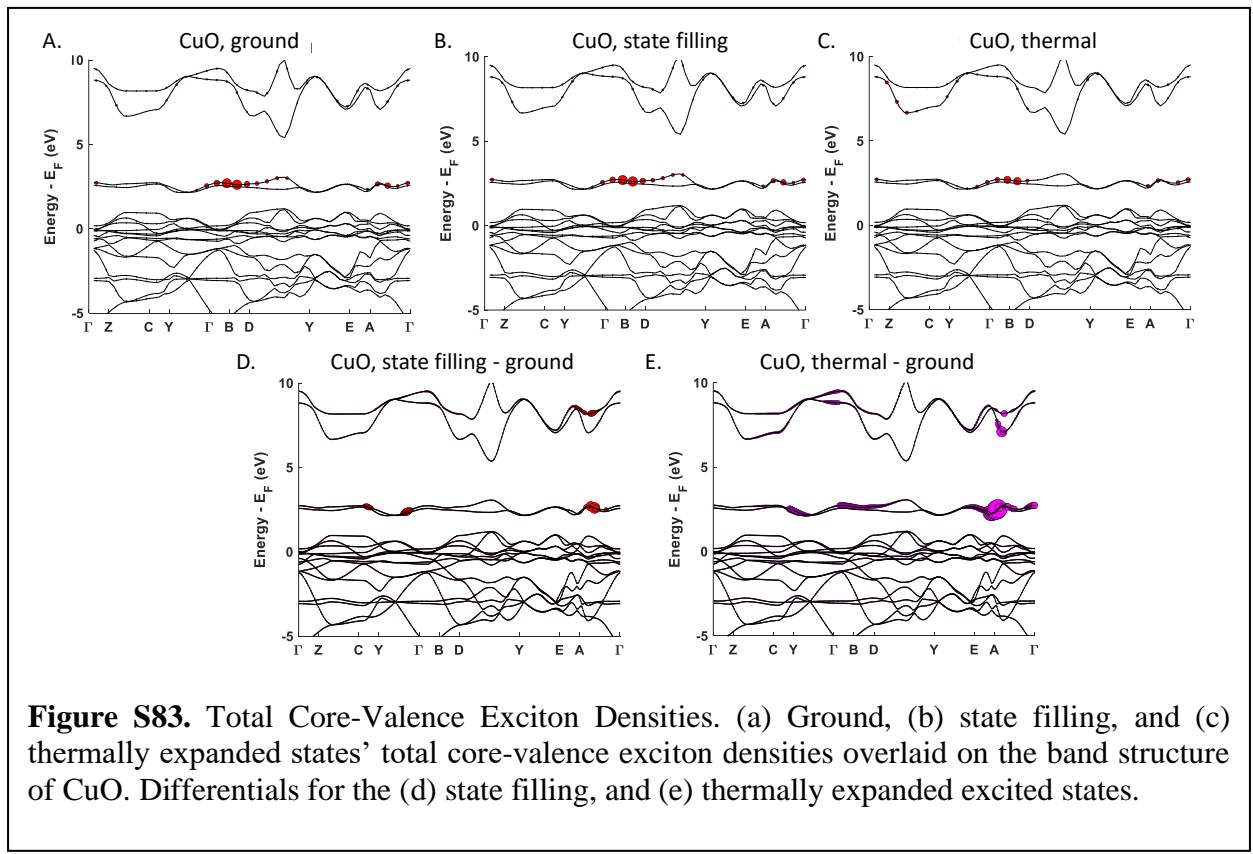

**Figure S83.** Total Core-Valence Exciton Densities. (a) Ground, (b) state filling, and (c) thermally expanded states' total core-valence exciton densities overlaid on the band structure of CuO. Differentials for the (d) state filling, and (e) thermally expanded excited states.

ii. Hamiltonian Decomposition of Exciton Components for ground, state filling, and thermally expanded models.

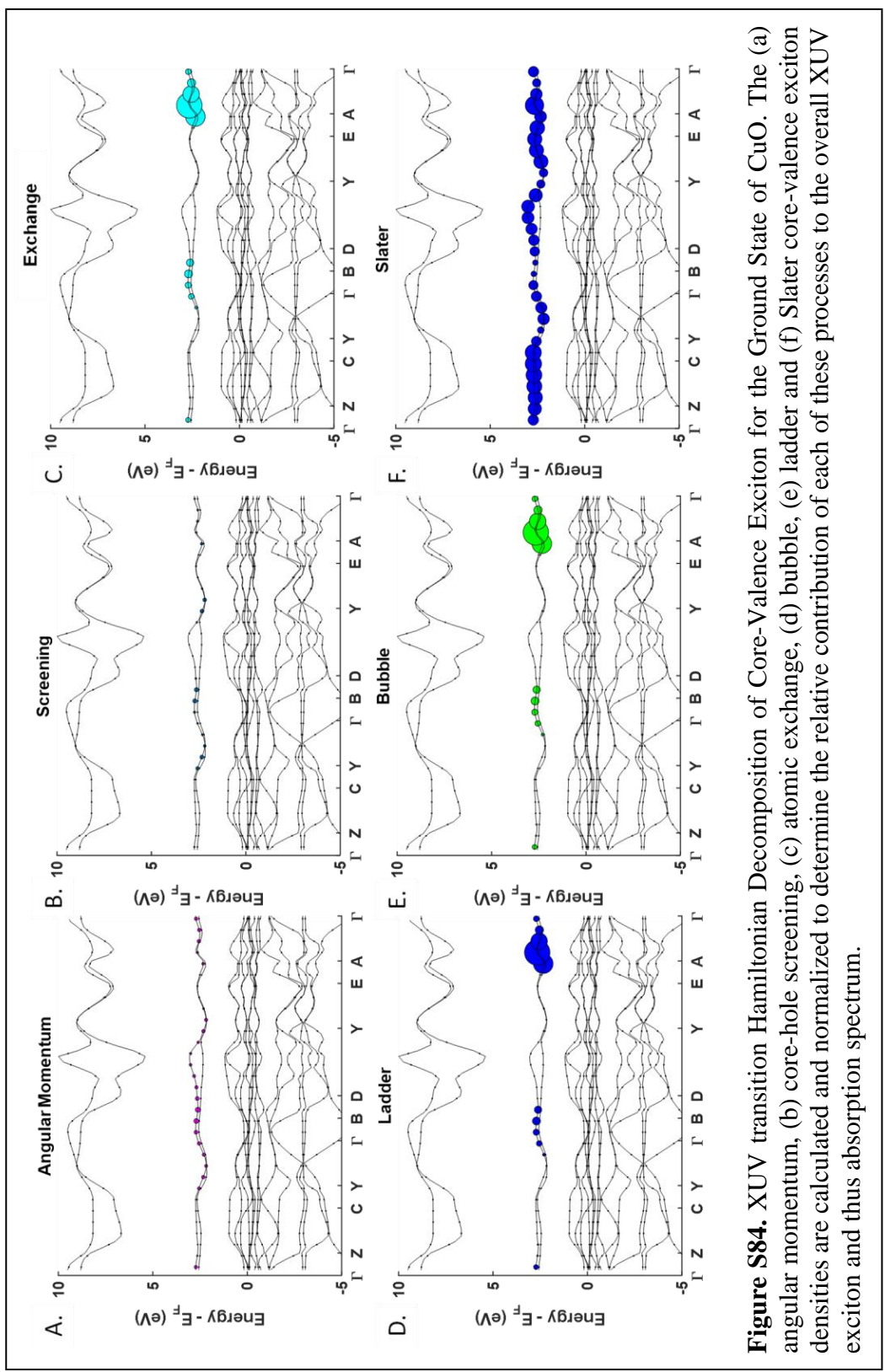

**Figure S84.** XUV transition Hamiltonian Decomposition of Core-Valence Exciton for the Ground State of CuO. The (a) angular momentum, (b) core-hole screening, (c) atomic exchange, (d) bubble, (e) ladder and (f) Slater core-valence exciton densities are calculated and normalized to determine the relative contribution of each of these processes to the overall XUV exciton and thus absorption spectrum.

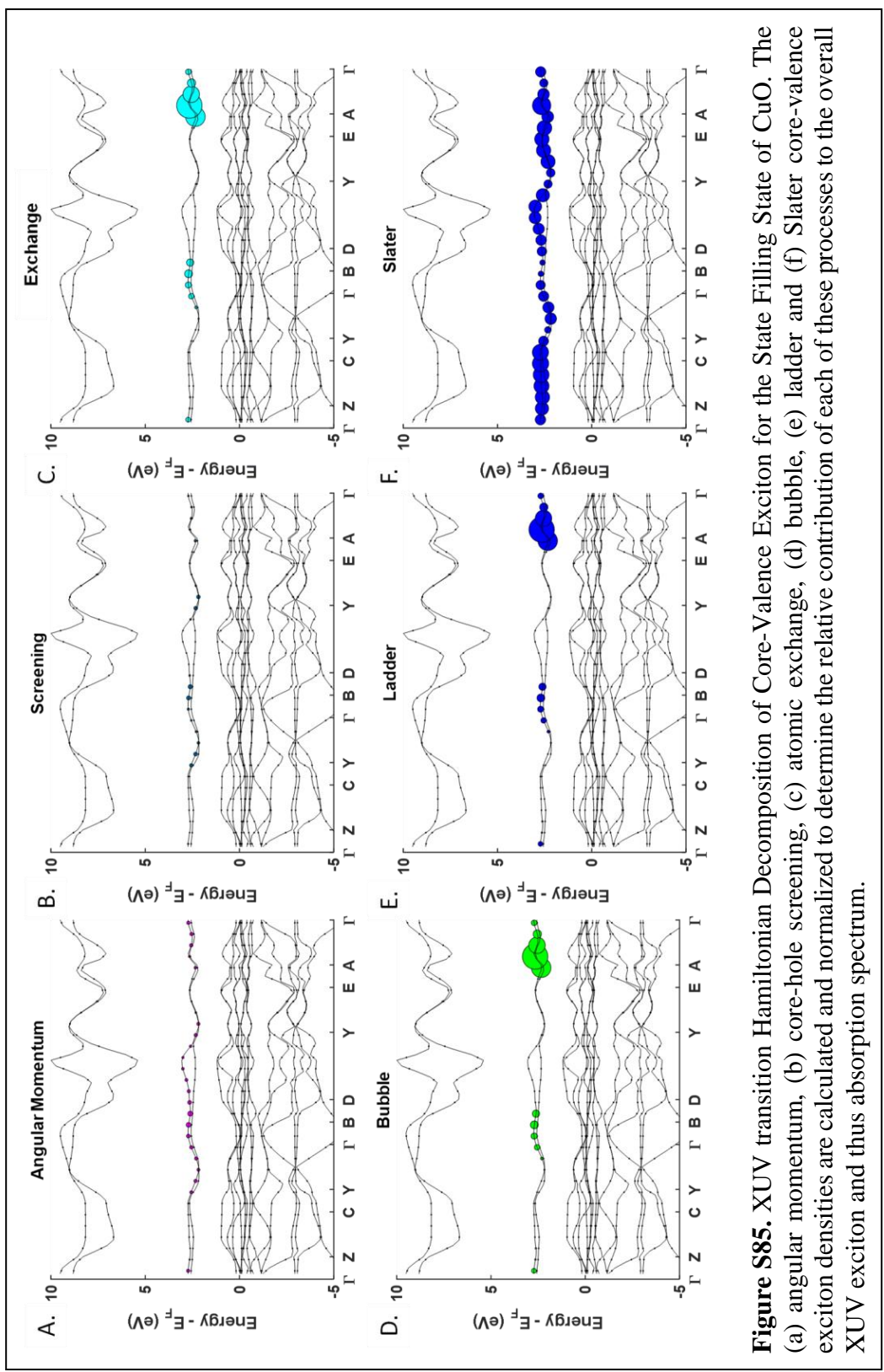

**Figure S85.** XUV transition Hamiltonian Decomposition of Core-Valence Exciton for the State Filling State of CuO. The (a) angular momentum, (b) core-hole screening, (c) atomic exchange, (d) bubble, (e) ladder and (f) Slater core-valence exciton densities are calculated and normalized to determine the relative contribution of each of these processes to the overall XUV exciton and thus absorption spectrum.

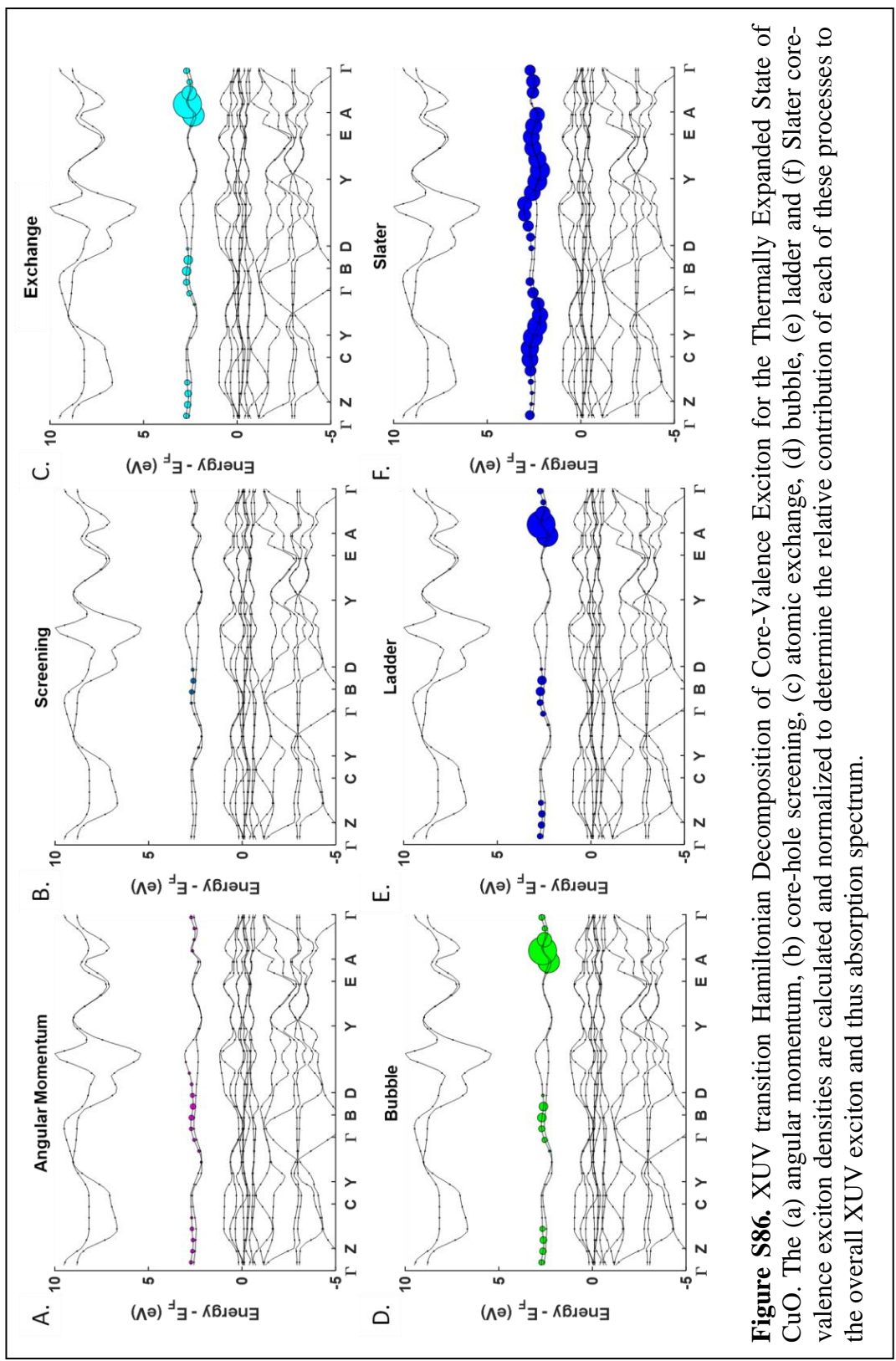

**Figure S86.** XUV transition Hamiltonian Decomposition of Core-Valence Exciton for the Thermally Expanded State of CuO. The (a) angular momentum, (b) core-hole screening, (c) atomic exchange, (d) bubble, (e) ladder and (f) Slater core-valence exciton densities are calculated and normalized to determine the relative contribution of each of these processes to the overall XUV exciton and thus absorption spectrum.

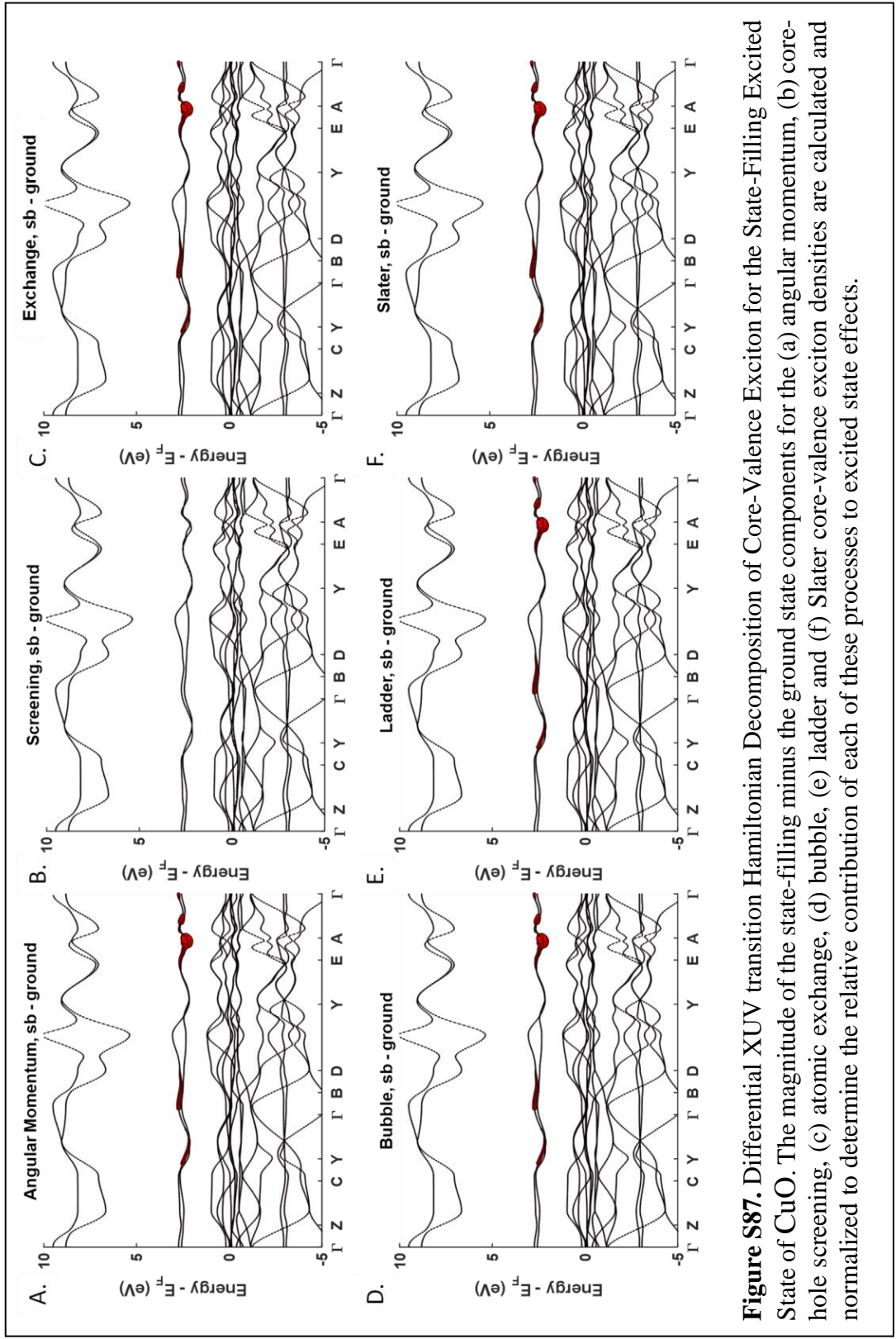

**Figure S87.** Differential XUV transition Hamiltonian Decomposition of Core-Valence Exciton for the State-Filling Excited State of CuO. The magnitude of the state-filling minus the ground state components for the (a) angular momentum, (b) core-hole screening, (c) atomic exchange, (d) bubble, (e) ladder and (f) Slater core-valence exciton densities are calculated and normalized to determine the relative contribution of each of these processes to excited state effects.

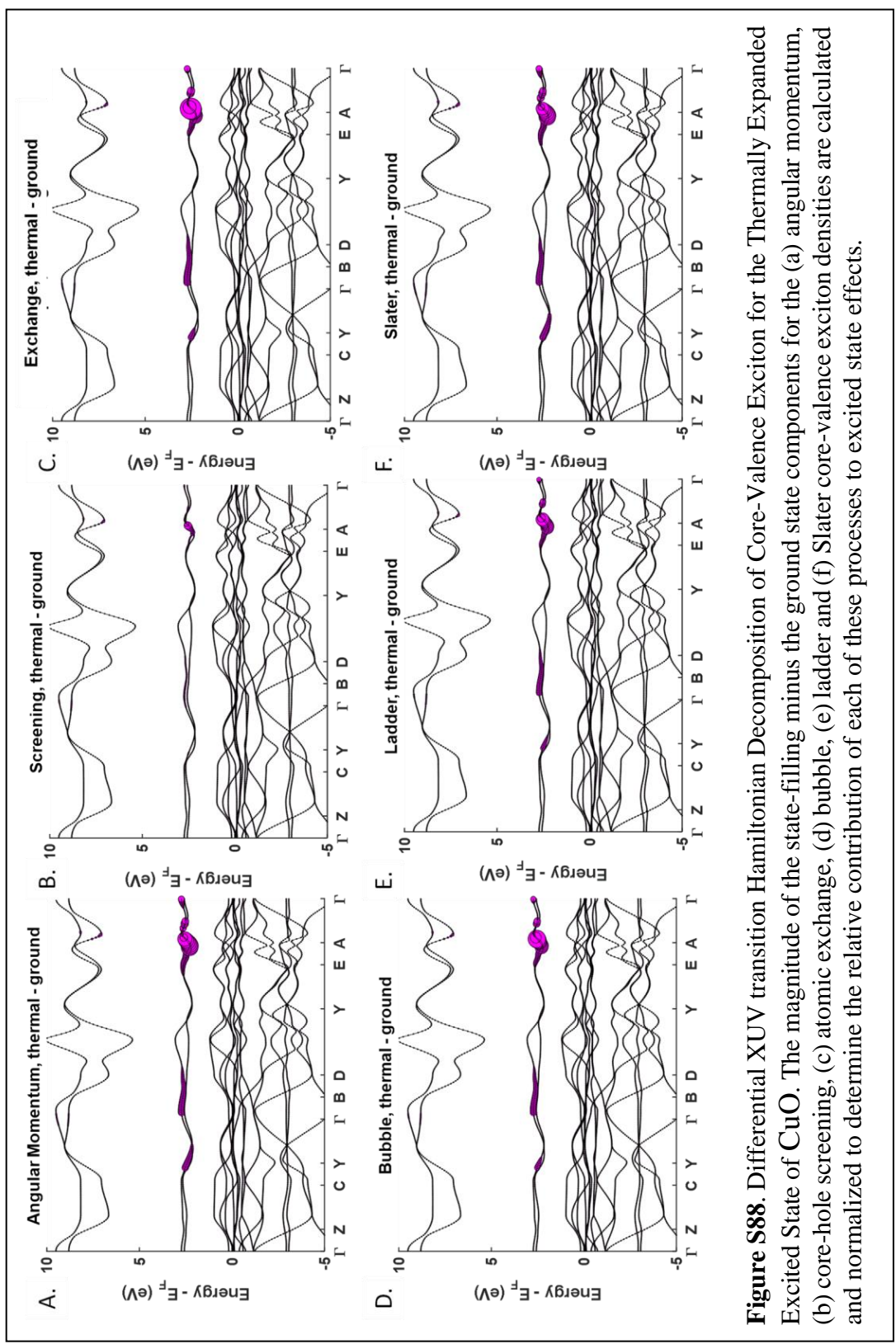

**Figure S88.** Differential XUV transition Hamiltonian Decomposition of Core-Valence Exciton for the Thermally Expanded Excited State of CuO. The magnitude of the state-filling minus the ground state components for the (a) angular momentum, (b) core-hole screening, (c) atomic exchange, (d) bubble, (e) ladder and (f) Slater core-valence exciton densities are calculated and normalized to determine the relative contribution of each of these processes to excited state effects.

## 9. ZnO

a. Structural Data for Calculations

i. Ground State and State Blocking

**Unit Cell Parameters (bohr)**

{ 6.146932044 6.146932044 9.978075761 }

**Primitive Vectors**

{0.5 -0.866 0

0.5 0.866 0

0 0 1}

**Reduced coordinates, ( x, y, z )**

Zn 0.333332565 0.666667435 0.001542181

Zn 0.666667435 0.333332565 0.501542181

O 0.333332311 0.666667689 0.378757819

O 0.666667689 0.333332311 0.878757819

ii. Thermally Expanded Lattice

**Unit Cell Parameters (bohr)**

{6.18074 6.18074 10.032955 }

**Primitive Vectors**

{0.5 -0.866 0

0.5 0.866 0

0 0 1}

**Reduced coordinates, ( x, y, z )**

Zn 0.333332565 0.666667435 0.001542181

Zn 0.666667435 0.333332565 0.501542181

O 0.333332311 0.666667689 0.378757819

O 0.666667689 0.333332311 0.878757819

b. Ground State Calculations

i. Band Structure and DOS

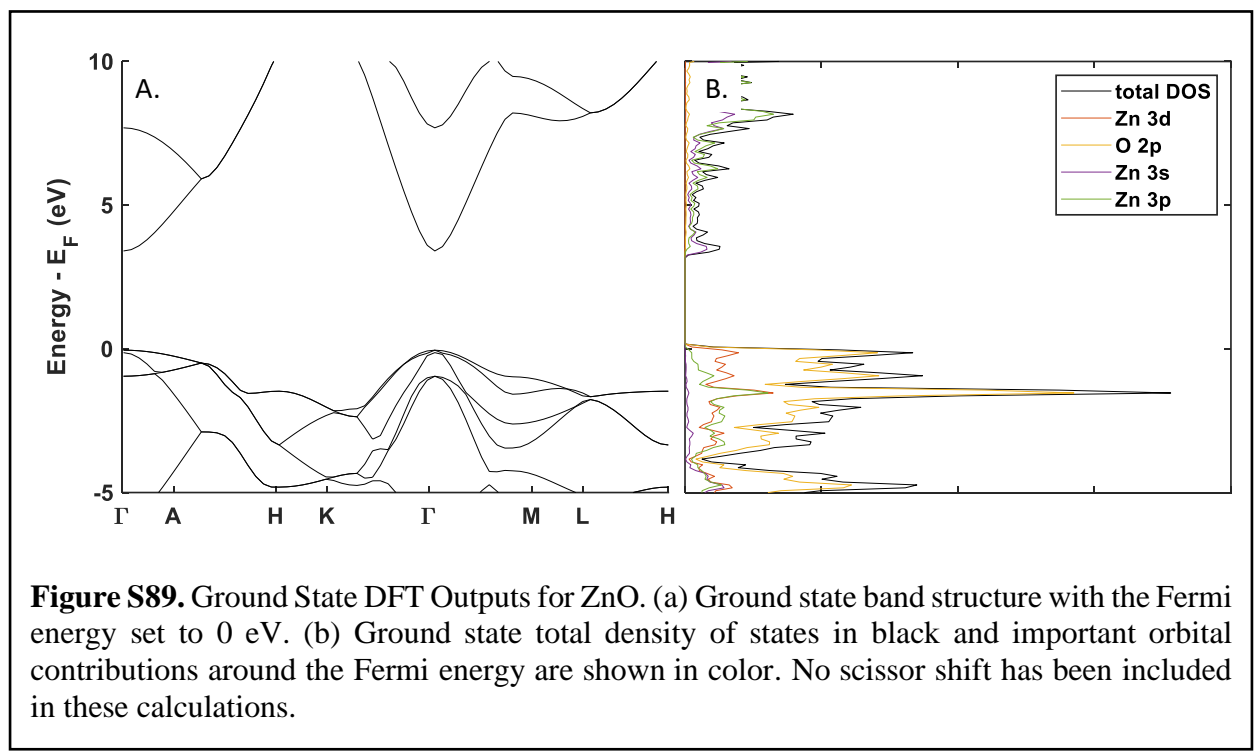

**Figure S89.** Ground State DFT Outputs for ZnO. (a) Ground state band structure with the Fermi energy set to 0 eV. (b) Ground state total density of states in black and important orbital contributions around the Fermi energy are shown in color. No scissor shift has been included in these calculations.

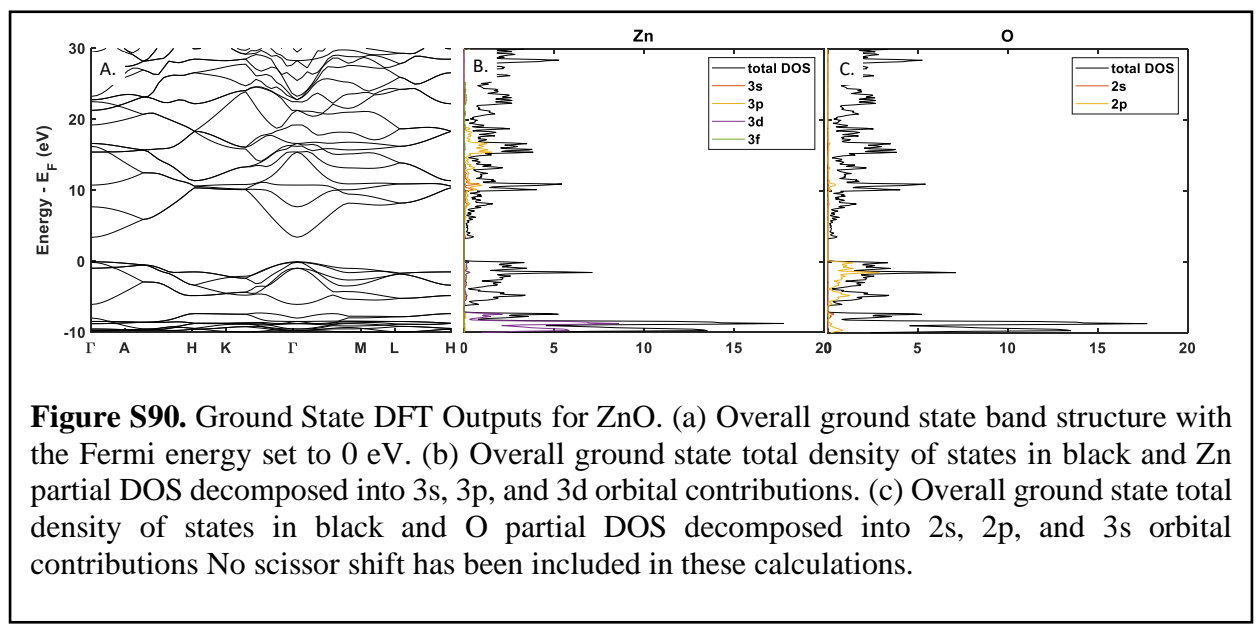

**Figure S90.** Ground State DFT Outputs for ZnO. (a) Overall ground state band structure with the Fermi energy set to 0 eV. (b) Overall ground state total density of states in black and Zn partial DOS decomposed into 3s, 3p, and 3d orbital contributions. (c) Overall ground state total density of states in black and O partial DOS decomposed into 2s, 2p, and 3s orbital contributions No scissor shift has been included in these calculations.

ii. Ground State Spectrum

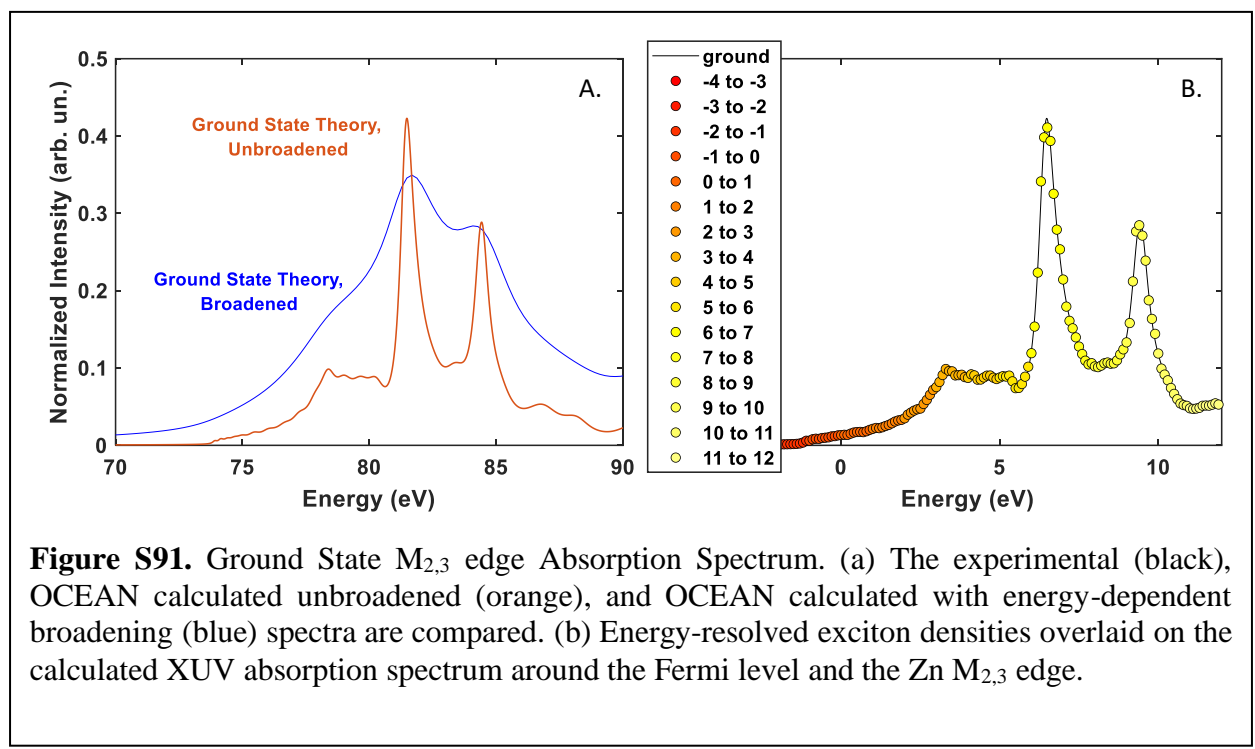

**Figure S91.** Ground State $M_{2,3}$ edge Absorption Spectrum. (a) The experimental (black), OCEAN calculated unbroadened (orange), and OCEAN calculated with energy-dependent broadening (blue) spectra are compared. (b) Energy-resolved exciton densities overlaid on the calculated XUV absorption spectrum around the Fermi level and the Zn $M_{2,3}$ edge.

iii. Ground State GMRES Energy Decomposition

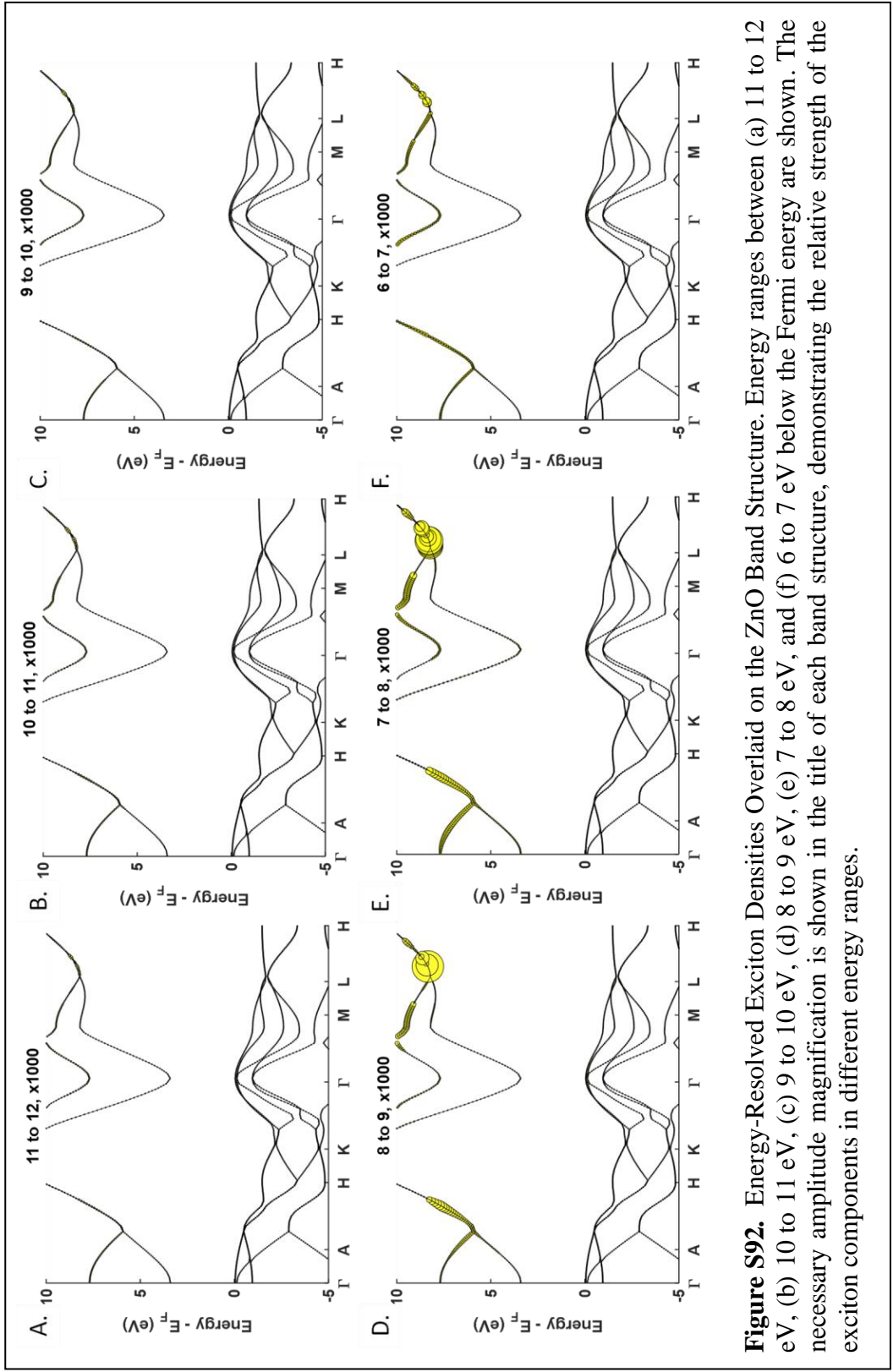

**Figure S92.** Energy-Resolved Exciton Densities Overlaid on the ZnO Band Structure. Energy ranges between (a) 11 to 12 eV, (b) 10 to 11 eV, (c) 9 to 10 eV, (d) 8 to 9 eV, (e) 7 to 8 eV, and (f) 6 to 7 eV below the Fermi energy are shown. The necessary amplitude magnification is shown in the title of each band structure, demonstrating the relative strength of the exciton components in different energy ranges.

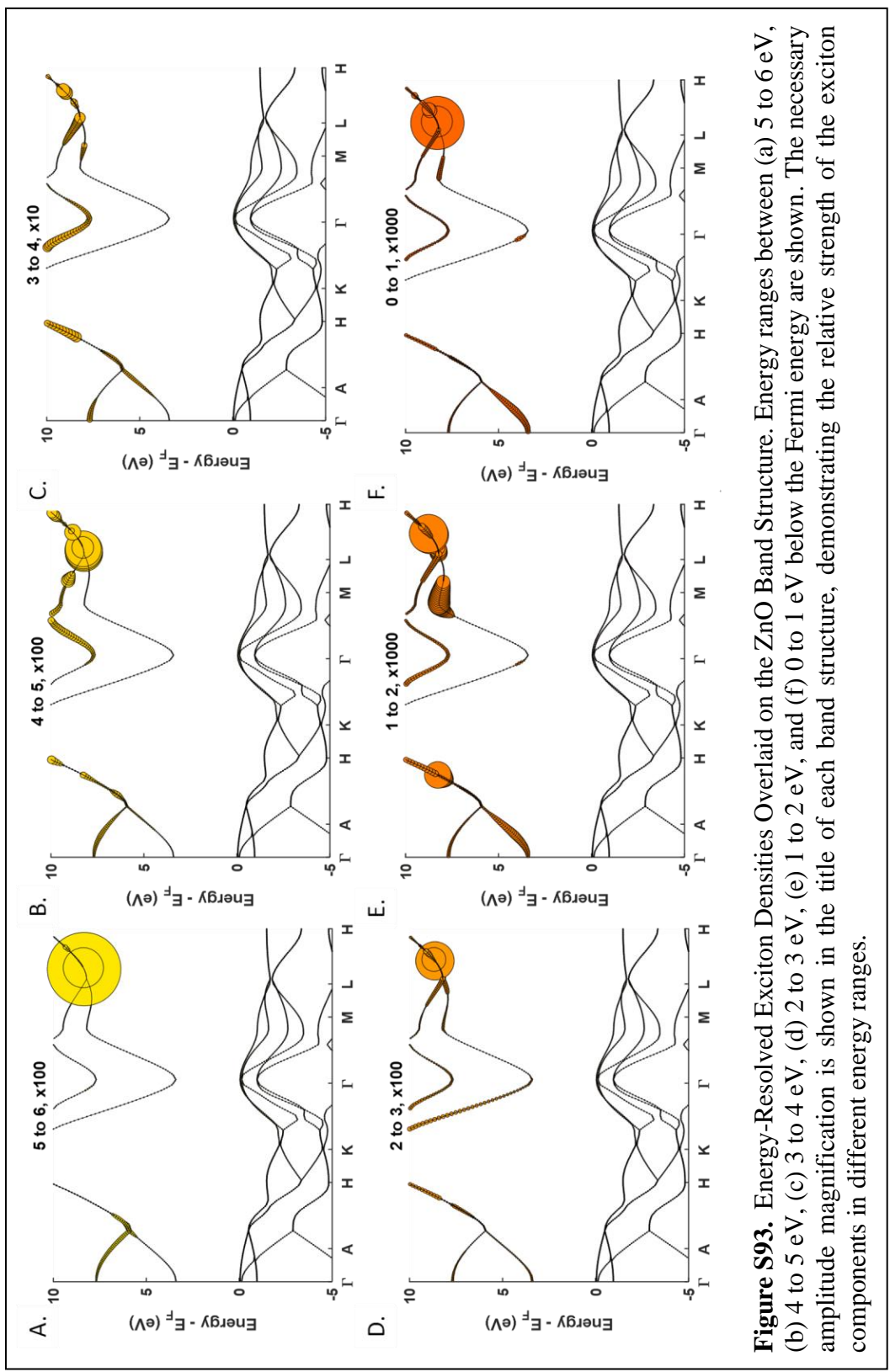

**Figure S93.** Energy-Resolved Exciton Densities Overlaid on the ZnO Band Structure. Energy ranges between (a) 5 to 6 eV, (b) 4 to 5 eV, (c) 3 to 4 eV, (d) 2 to 3 eV, (e) 1 to 2 eV, and (f) 0 to 1 eV below the Fermi energy are shown. The necessary amplitude magnification is shown in the title of each band structure, demonstrating the relative strength of the exciton components in different energy ranges.

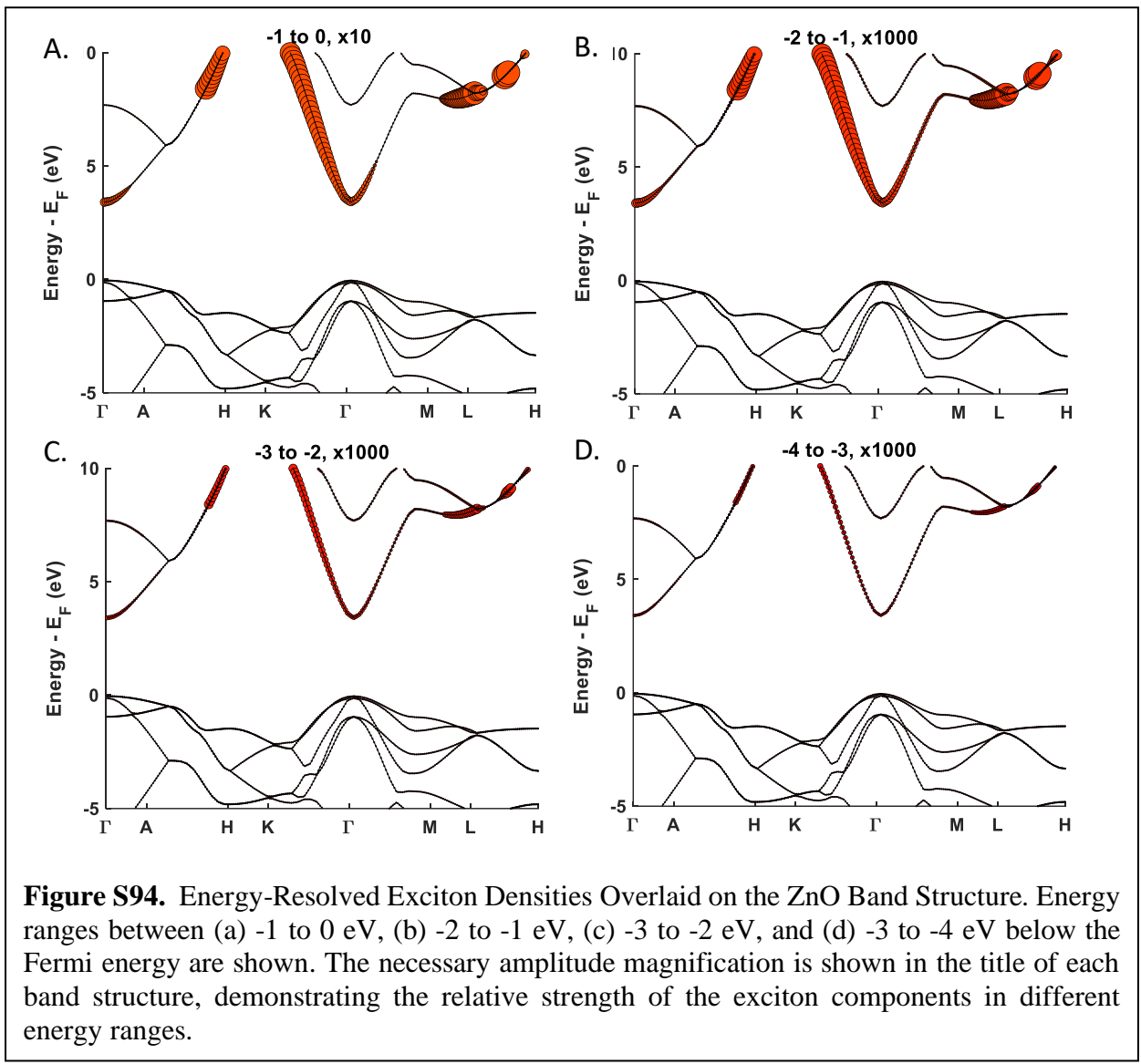

**Figure S94.** Energy-Resolved Exciton Densities Overlaid on the ZnO Band Structure. Energy ranges between (a) -1 to 0 eV, (b) -2 to -1 eV, (c) -3 to -2 eV, and (d) -3 to -4 eV below the Fermi energy are shown. The necessary amplitude magnification is shown in the title of each band structure, demonstrating the relative strength of the exciton components in different energy ranges.

c. Excited State Calculations

    i. State filling band diagrams and Full Spectra

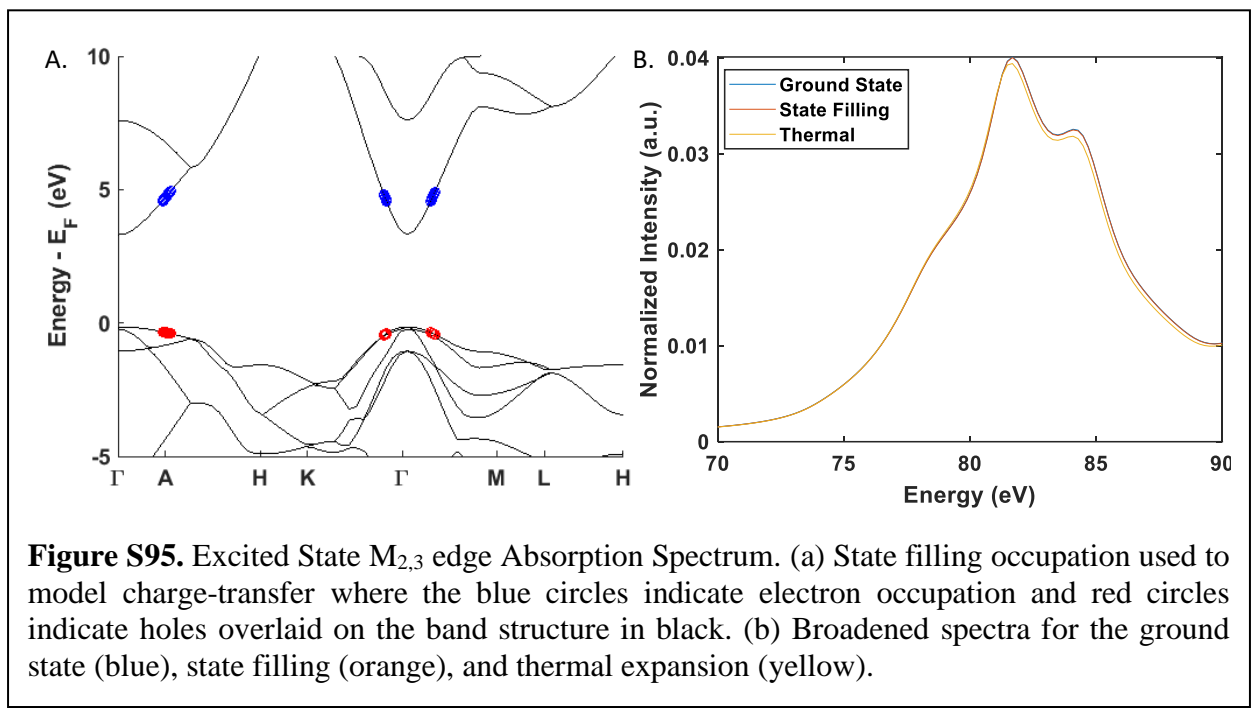

**Figure S95.** Excited State $M_{2,3}$ edge Absorption Spectrum. (a) State filling occupation used to model charge-transfer where the blue circles indicate electron occupation and red circles indicate holes overlaid on the band structure in black. (b) Broadened spectra for the ground state (blue), state filling (orange), and thermal expansion (yellow).

### d. Hamiltonian Decompositions

#### i. Total Exciton Comparisons

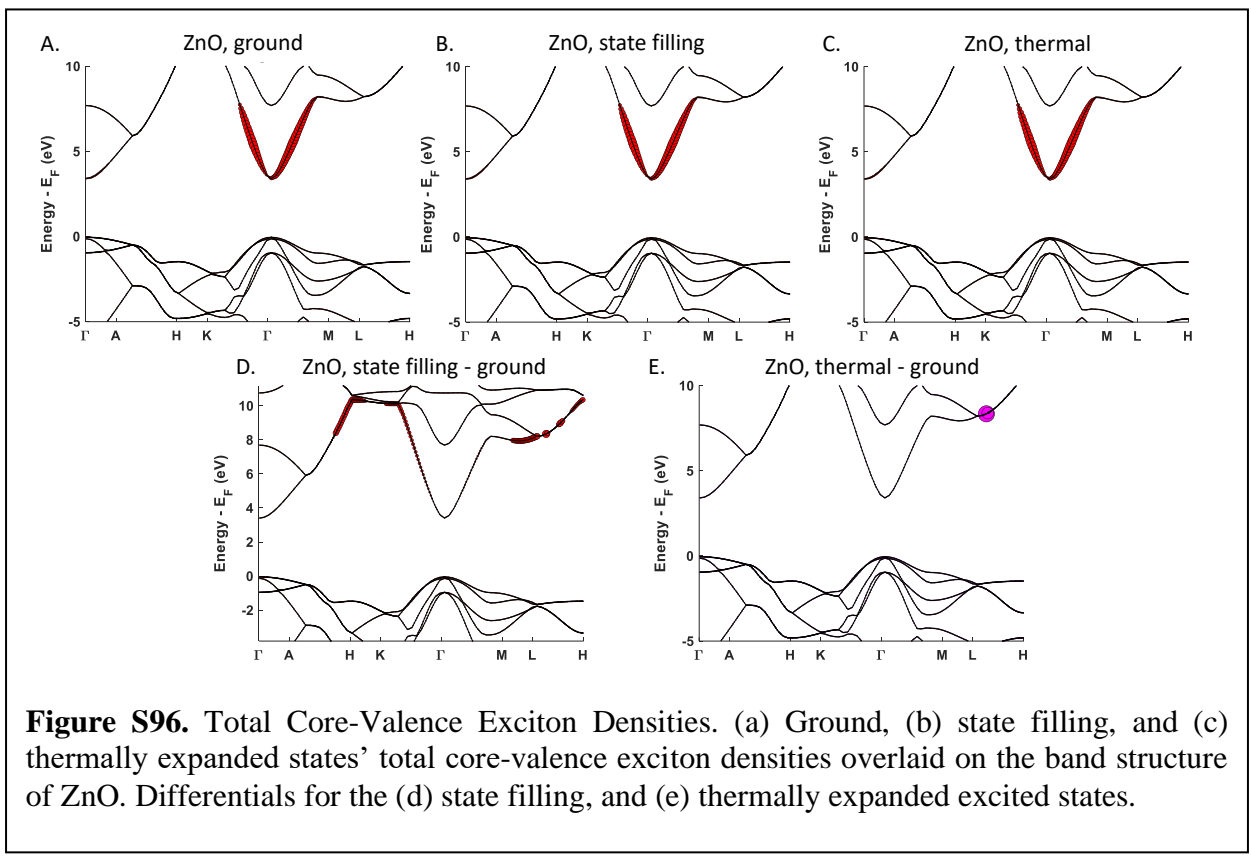

**Figure S96.** Total Core-Valence Exciton Densities. (a) Ground, (b) state filling, and (c) thermally expanded states' total core-valence exciton densities overlaid on the band structure of ZnO. Differentials for the (d) state filling, and (e) thermally expanded excited states.

ii. Hamiltonian Decomposition of Exciton Components for ground, state filling, and thermally expanded models.

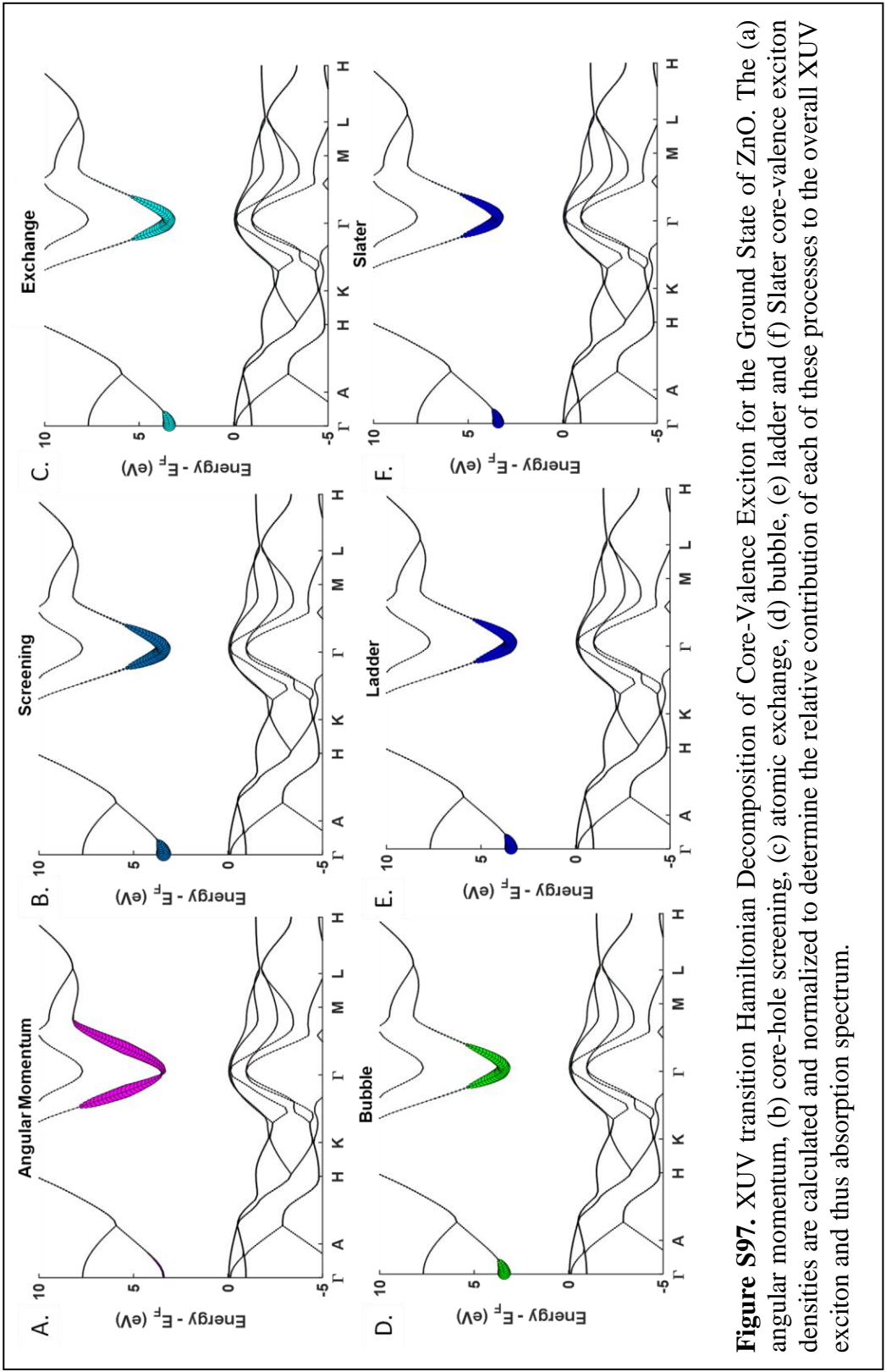

**Figure S97.** XUV transition Hamiltonian Decomposition of Core-Valence Exciton for the Ground State of ZnO. The (a) angular momentum, (b) core-hole screening, (c) atomic exchange, (d) bubble, (e) ladder and (f) Slater core-valence exciton densities are calculated and normalized to determine the relative contribution of each of these processes to the overall XUV exciton and thus absorption spectrum.

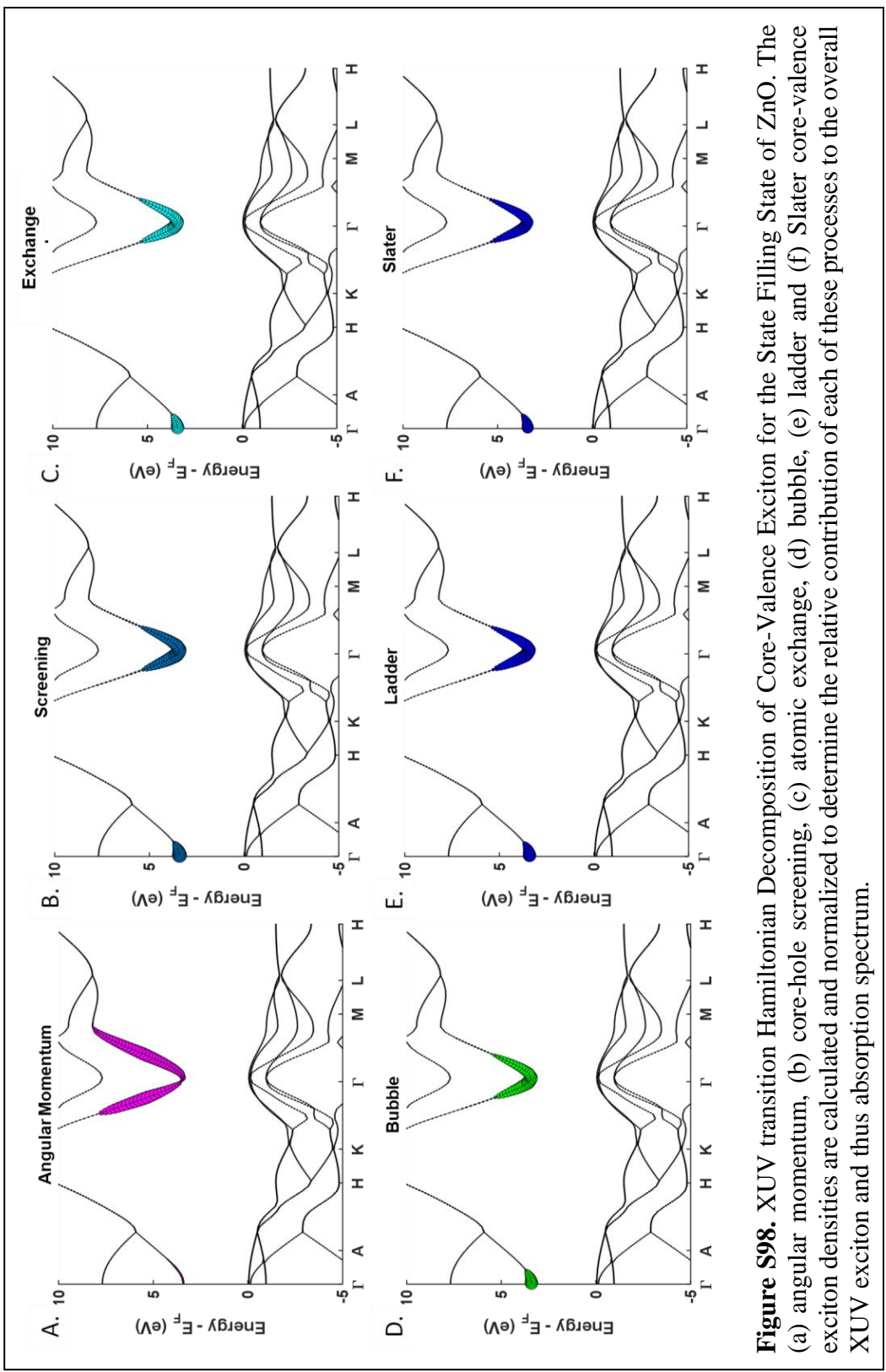

**Figure S98.** XUV transition Hamiltonian Decomposition of Core-Valence Exciton for the State Filling State of ZnO. The (a) angular momentum, (b) core-hole screening, (c) atomic exchange, (d) bubble, (e) ladder and (f) Slater core-valence exciton densities are calculated and normalized to determine the relative contribution of each of these processes to the overall XUV exciton and thus absorption spectrum.

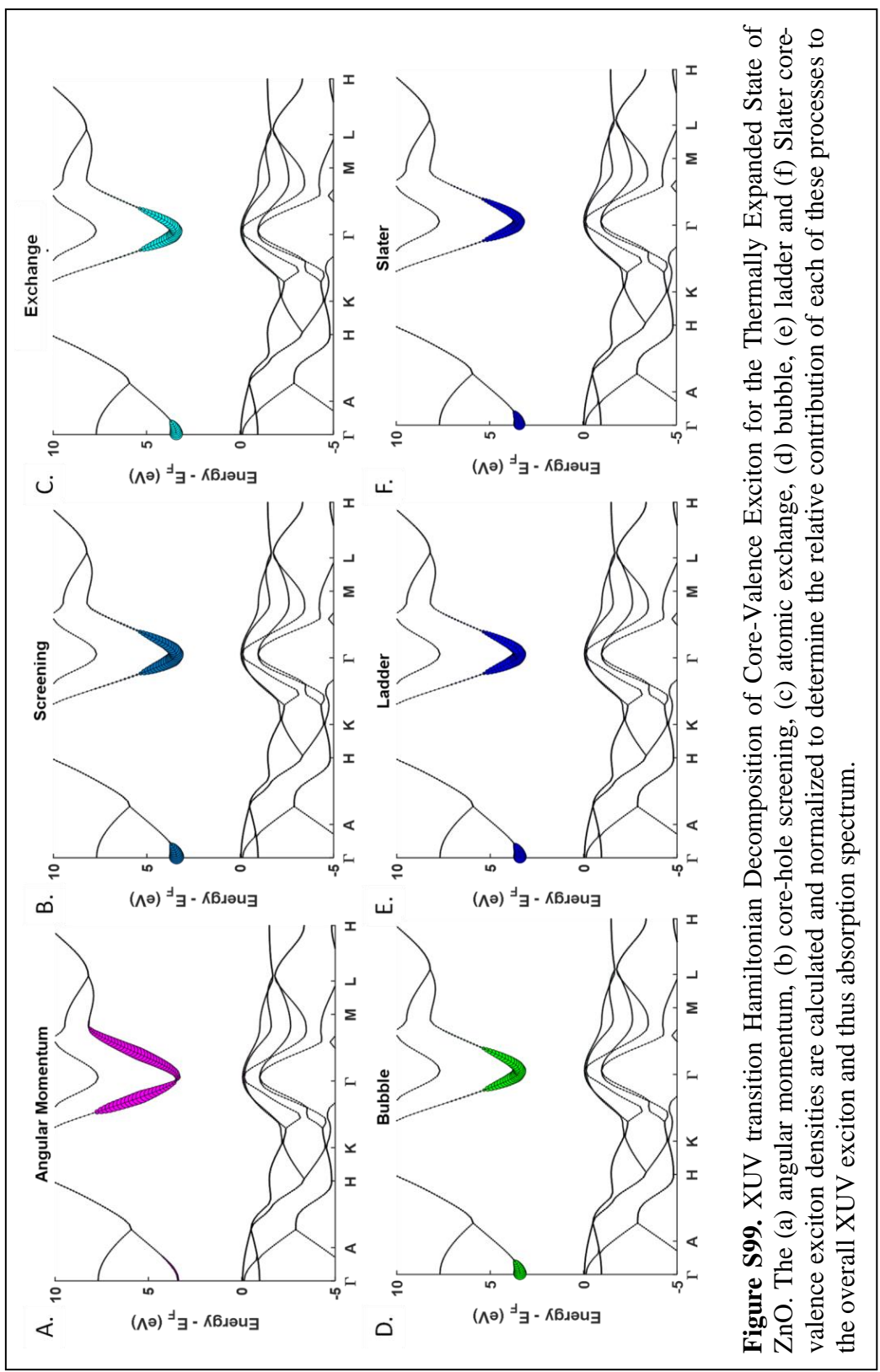

**Figure S99.** XUV transition Hamiltonian Decomposition of Core-Valence Exciton for the Thermally Expanded State of ZnO. The (a) angular momentum, (b) core-hole screening, (c) atomic exchange, (d) bubble, (e) ladder and (f) Slater core-valence exciton densities are calculated and normalized to determine the relative contribution of each of these processes to the overall XUV exciton and thus absorption spectrum.

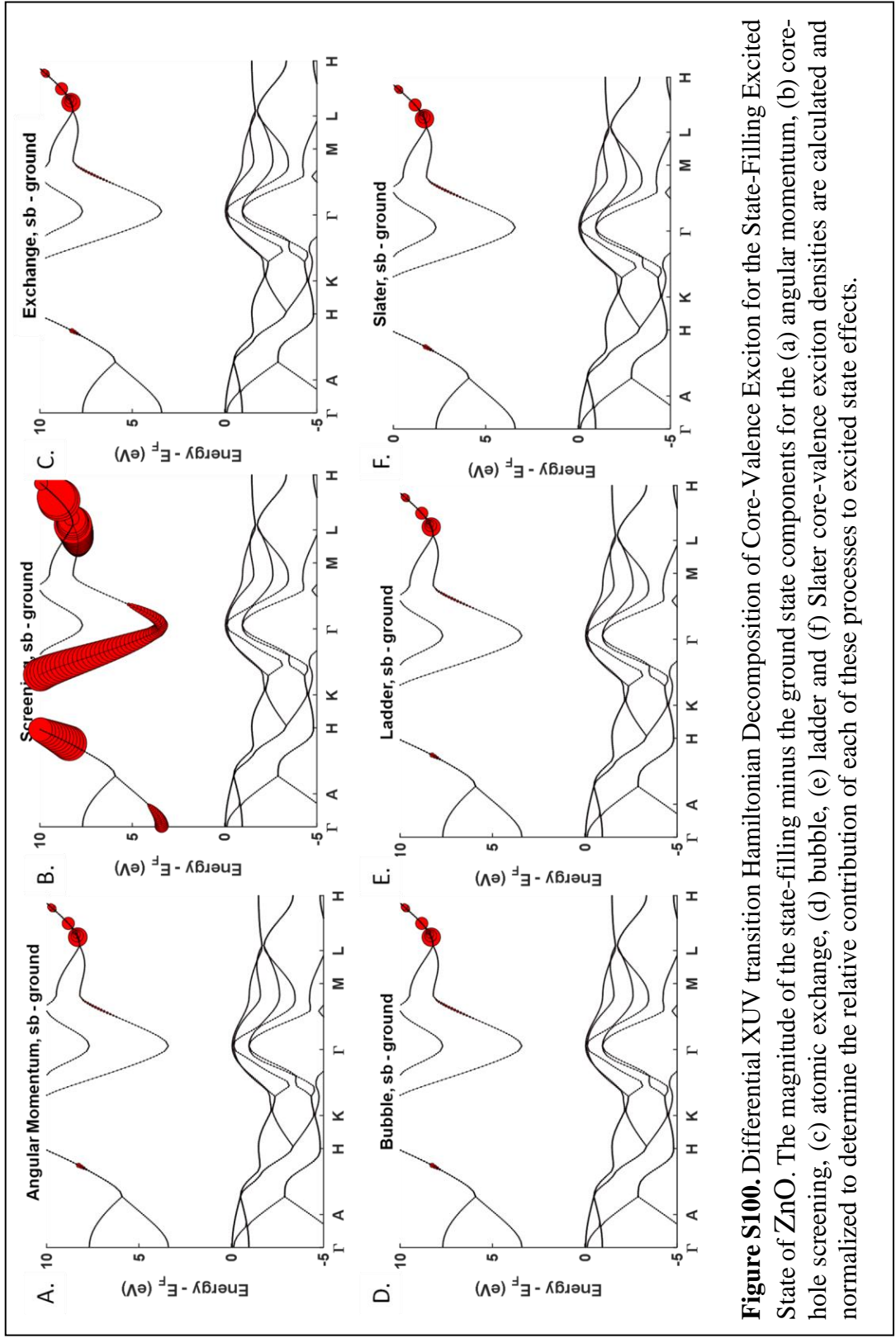

**Figure S100.** Differential XUV transition Hamiltonian Decomposition of Core-Valence Exciton for the State-Filling Excited State of ZnO. The magnitude of the state-filling minus the ground state components for the (a) angular momentum, (b) core-hole screening, (c) atomic exchange, (d) bubble, (e) ladder and (f) Slater core-valence exciton densities are calculated and normalized to determine the relative contribution of each of these processes to excited state effects.

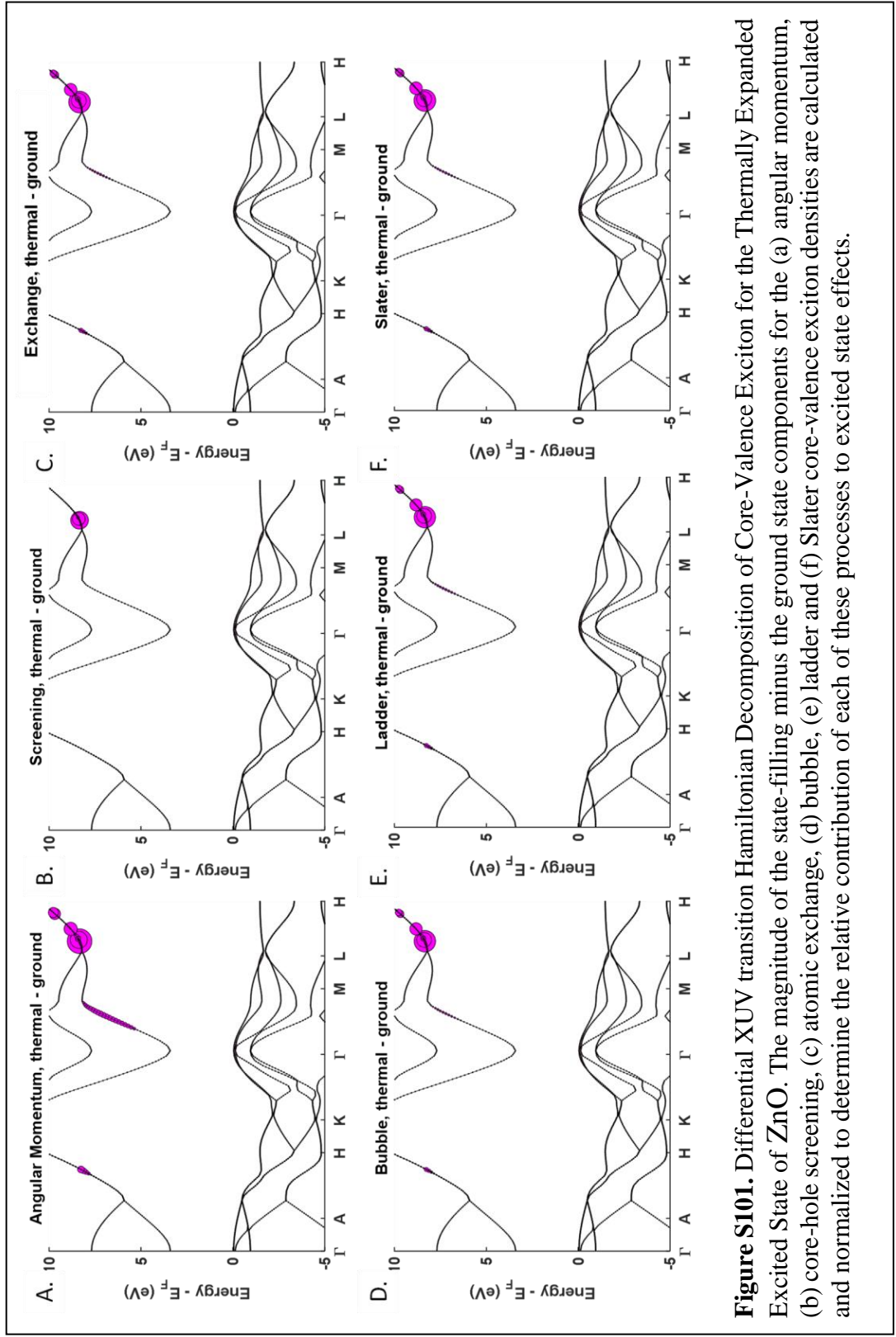

**Figure S101.** Differential XUV transition Hamiltonian Decomposition of Core-Valence Exciton for the Thermally Expanded Excited State of ZnO. The magnitude of the state-filling minus the ground state components for the (a) angular momentum, (b) core-hole screening, (c) atomic exchange, (d) bubble, (e) ladder and (f) Slater core-valence exciton densities are calculated and normalized to determine the relative contribution of each of these processes to excited state effects.

## 10. Comparisons

a. Thermal Expansion Calculations

**Table 3.** Percent Thermal Expansion for the Transition Metal Oxides, 300 K to 650 K

|              | TiO$_2$ | Cr$_2$O$_3$ | MnO$_2$ | Fe$_2$O$_3$ | Co$_3$O$_4$ | NiO | CuO | ZnO |
|--------------|---------|-------------|---------|-------------|-------------|-----|-----|-----|
| % Expansion  | 1       | 0.6         | 0.3     | 0.8         | 0.3         | 1.2 | 0.4 | 0.7 |

b. Physical Properties for Hamiltonian Contribution Explanations

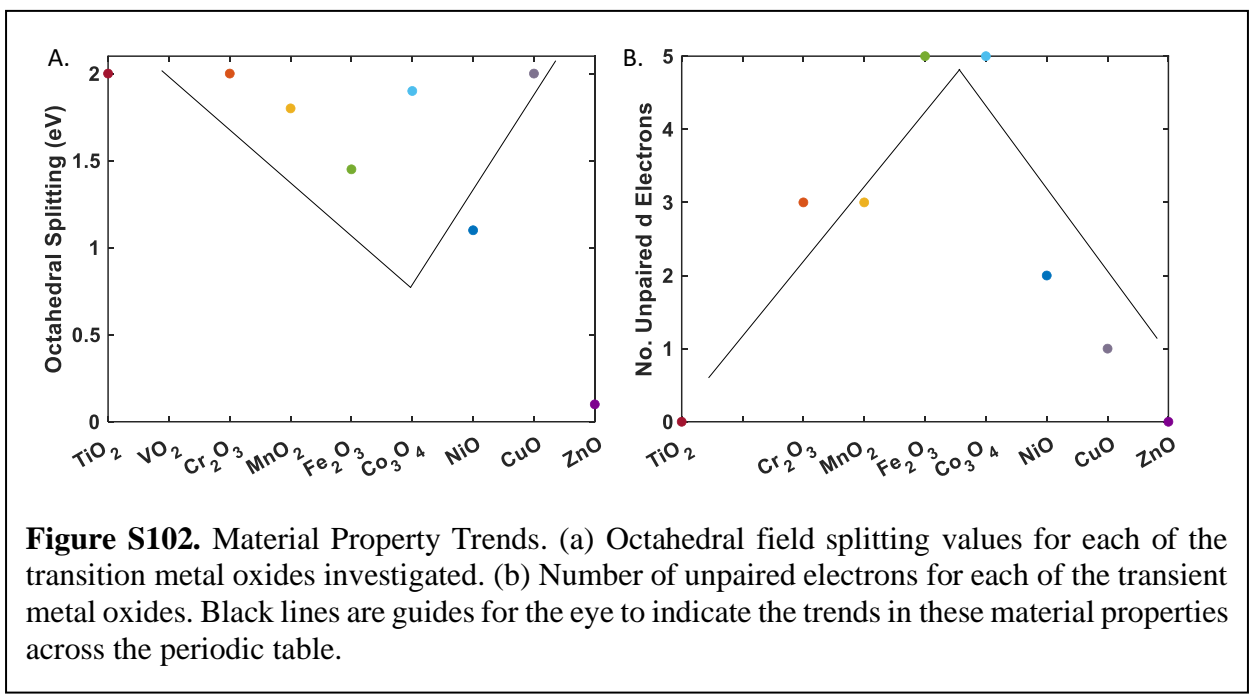

**Figure S102.** Material Property Trends. (a) Octahedral field splitting values for each of the transition metal oxides investigated. (b) Number of unpaired electrons for each of the transient metal oxides. Black lines are guides for the eye to indicate the trends in these material properties across the periodic table.

**References**

(1) Piccinin, S. The Band Structure and Optical Absorption of Hematite (α-$Fe_2O_3$): A First-Principles GW-BSE Study. *Phys. Chem. Chem. Phys.* **2019**, *21* (6), 2957–2967. https://doi.org/10.1039/C8CP07132B.

(2) Vura-Weis, J.; Jiang, C.-M.; Liu, C.; Gao, H.; Lucas, J. M.; de Groot, F. M. F.; Yang, P.; Alivisatos, A. P.; Leone, S. R. Femtosecond $M_{2,3}$-Edge Spectroscopy of Transition-Metal Oxides: Photoinduced Oxidation State Change in α-$Fe_2O_3$. *J. Phys. Chem. Lett.* **2013**, *4* (21), 3667–3671. https://doi.org/10.1021/jz401997d.

(3) Cushing, S. K.; Zürch, M.; Kraus, P. M.; Carneiro, L. M.; Lee, A.; Chang, H.-T.; Kaplan, C. J.; Leone, S. R. Hot Phonon and Carrier Relaxation in Si(100) Determined by Transient Extreme Ultraviolet Spectroscopy. *Structural Dynamics* **2018**, *5* (5), 054302. https://doi.org/10.1063/1.5038015.

(4) Attar, A. R.; Chang, H.-T.; Britz, A.; Zhang, X.; Lin, M.-F.; Krishnamoorthy, A.; Linker, T.; Fritz, D.; Neumark, D. M.; Kalia, R. K.; Nakano, A.; Ajayan, P.; Vashishta, P.; Bergmann, U.; Leone, S. R. Simultaneous Observation of Carrier-Specific Redistribution and Coherent Lattice Dynamics in 2H-MoTe$_{2}$ with Femtosecond Core-Level Spectroscopy. *arXiv:2009.00721 [cond-mat]* **2020**.

(5) Cushing, S. K.; Porter, I. J.; Roulet, B. R. de; Lee, A.; Marsh, B. M.; Szoke, S.; Vaida, M. E.; Leone, S. R. Layer-Resolved Ultrafast Extreme Ultraviolet Measurement of Hole Transport in a Ni-TiO2-Si Photoanode. *Science Advances* **2020**, *6* (14), eaay6650. https://doi.org/10.1126/sciadv.aay6650.

(6) Biswas, S.; Husek, J.; Baker, L. R. Elucidating Ultrafast Electron Dynamics at Surfaces Using Extreme Ultraviolet (XUV) Reflection–Absorption Spectroscopy. *Chem. Commun.* **2018**, *54* (34), 4216–4230. https://doi.org/10.1039/C8CC01745J.

(7) Agarwal, B. K.; Givens, M. P. Soft X-Ray Absorption by Manganese and Manganese Oxide. *Journal of Physics and Chemistry of Solids* **1958**, *6* (2–3), 178–179. https://doi.org/10.1016/0022-3697(58)90092-1.

(8) Carneiro, L. M.; Cushing, S. K.; Liu, C.; Su, Y.; Yang, P.; Alivisatos, A. P.; Leone, S. R. Excitation-Wavelength-Dependent Small Polaron Trapping of Photoexcited Carriers in α-Fe2O3. *Nature Mater* **2017**, *16* (8), 819–825. https://doi.org/10.1038/nmat4936.

(9) Jiang, C.-M.; Baker, L. R.; Lucas, J. M.; Vura-Weis, J.; Alivisatos, A. P.; Leone, S. R. Characterization of Photo-Induced Charge Transfer and Hot Carrier Relaxation Pathways in Spinel Cobalt Oxide ($Co_3O_4$). *J. Phys. Chem. C* **2014**, *118* (39), 22774–22784. https://doi.org/10.1021/jp5071133.

(10) Wang, H.; Young, A. T.; Guo, J.; Cramer, S. P.; Friedrich, S.; Braun, A.; Gu, W. Soft X-Ray Absorption Spectroscopy and Resonant Inelastic X-Ray Scattering Spectroscopy below 100 EV: Probing First-Row Transition-Metal M-Edges in Chemical Complexes. *J Synchrotron Radiat* **2013**, *20* (Pt 4), 614–619. https://doi.org/10.1107/S0909049513003142.